\documentclass[a4paper]{article}
\usepackage{graphicx}

\begin{document}

\begin{center}
{\Large \bf Description of the Totem experimental data on elastic $pp$-scattering
            at $\sqrt{s}=7$ TeV in the framework of unified systematic of elastic
            scattering data}
\end{center}

\begin{center}
{
V. Uzhinsky\footnote{CERN, Geneva, Switzerland}$^,$
           \footnote{LIT, JINR, Dubna, Russia}}
 A. Galoyan\footnote{VBLHEP, JINR, Dubna, Russia}\\
\end{center}

\begin{center}
Abstract
\end{center}

\begin{center}
\begin{minipage}{12cm}
An unified systematic of elastic (anti)proton-proton scattering data is proposed based on
a simple expression for the process amplitude --
$f(q)=A\ [(\pi dq)/sh(\pi dq)][i\ J_1(Rq)/(Rq)+\rho J_0(Rq)+ . . .]$.
The parameters $R$ and $d$ are obtained at a fitting of (anti)proton-proton experimental
data on differential cross sections from $P_{lab} \sim$ 1 GeV/c up to Tevatron energies. The
fitting gives extra-ordinary good results, $\chi^2/NoF \sim$ 1 of below at $|t|<$ 1.75 (GeV/c)$^2$.
An extrapolation of the parameter's energy dependencies to the LHC energies allows a good
description of the Totem data up to the second diffraction maximum. Predictions for other LHC
energies are presented also.\\
~~~\\
The amplitude provides one with parameterizations of total and elastic cross sections. Its
impact parameter representation corresponds to the 2-dimensional Fermi-function --
$1/[1+exp((b-R)/d)]$, which is very useful for Glauber calculations of nucleus-nucleus
cross sections at super high energies.\\
~~~\\
It is shown for the first time that experimental high $|t|$ elastic scattering data have a weak
energy dependency. This allows to describe high $|t|$ tail of the Totem data.
\end{minipage}
\end{center}

\section*{Introduction}
According to the experimental data on elastic (anti)proton-proton scattering data in the
Coulomb-nuclear interference region, the nuclear elastic scattering amplitude in the momentum
representation, $f(\vec q)$, has a small real part and a large imaginary one. Correspondingly,
the amplitude in the impact parameter representation, $\gamma (\vec b)$, has a large real part
and a small imaginary one. The amplitudes are connected by the Fourier-Bessel transform:
\begin{equation}
f(\vec q)=\frac{i}{2\pi}\int \ e^{i\ \vec q \vec b}\gamma (\vec b)d^2b=
i\int^{\infty}_0 b\ J_0(\vec q \vec b)\ \gamma (\vec b)db,
\label{Eq1}
\end{equation}
\begin{equation}
\gamma (\vec b)=\frac{1}{2\pi i}\int \ e^{i\vec q \vec b} f(\vec q)d^2q=
\frac{1}{i}\int^{\infty}_0 \ q\ J_0(\vec q \vec b)\ f(\vec q)dq,
\label{Eq2}
\end{equation}
\begin{equation}
\sigma^{tot}=4\pi \ Im f(0) =4\pi\ \int^{\infty}_0 b\ Re\ \gamma (b)db,
\label{Eq3}
\end{equation}
\begin{equation}
\sigma^{el}=2\pi \int^{\infty}_0 b\  |\gamma (b)|^2 db,
\label{Eq4}
\end{equation}
\begin{equation}
d^2\sigma/dq^2=|f(\vec q)|^2,\ \ \ Re f(0)\sim 0 \ \ \ Im \gamma(b)\sim 0,
\label{Eq5}
\end{equation}
where $\vec q$ is the momentum transfer, the corresponding 4-momentum transfer -- $t=-\vec q^2$.

Any function existing on the semi-infinite interval, $[0 - \infty]$, can be represented in
a first approximation as
$$
\gamma (b)\simeq a_1\ [\Theta(R_1-b)-\Theta(0-b)],
$$
where
$$
R_1^2=2\ \int^{\infty}_0 b^3\ \gamma (b)db/\int^{\infty}_0 b\ \gamma (b)db,
$$
$$
a_1=(2/R^2)\ \int^{\infty}_0 b\ \gamma (b)db.
$$

In a next approximation it is needed to consider 2 functions:
$$
\gamma_{(L1)}(b)=\gamma(b)-\Theta(R_1-b), \ \ \ b \leq R_1,
$$
$$
\gamma_{(R1)}(b)=\gamma(b), \ \ \ \ \ \ \ \ \ \ \ \ \ \ \ \ \ \ \ \ b > R_1.
$$

The function $\gamma_{(R1)}(b)$ can be approximated as,
$$
\gamma_{(R1)}(b)=a_2\ [\Theta(R_2-b)-\Theta(R_1-b)],
$$
$$
R_2^2=2I_3/I_1 - R_1^2, \ \ \ a_2=\frac{I_1}{I_3/I_1 - R_1^2},
$$
$$
I_1=\int_{R_1}^\infty b \gamma(b) db, \ \ \ I_3=\int_{R_1}^\infty b^3 \gamma(b) db.
$$

The procedure can be repeated one more giving new functions -- $\gamma_{(L2)}(b)$ and
$\gamma_{(R2)}(b)$. One can obtain the results continuing the procedure,
\begin{equation}
\gamma(b)=\sum_{i=1}^\infty a_i\ [\Theta(R_i-b)-\Theta(R_{i-1}-b)]\ + \
          \sum_{i=1}^\infty \gamma_{(Li)}(b), \ \ \ R_0=0.
\label{Eq6}
\end{equation}

The corresponding $f(q)$ is (see \cite{GR}),
\begin{equation}
Im f(q)= \sum_{i=1}^\infty a_i [R_i^2\frac{J_1(R_i q)}{R_i q}-R_{i-1}^2\frac{J_1(R_{i-1} q)}{R_{i-1} q}]\ + \ ...
\label{Eq7}
\end{equation}

Each term in the second series in Eq. \ref{Eq6} can be decomposed on the wavelets \cite{Wavelets}, especially,
on the Haar wavelets. This will lead to an appearance of new $\Theta$-functions. Each term of the series in
Eq. \ref{Eq7} can be decomposed in the Talor series on $(R_i-R_1)^n$ in the vicinity of $R_1$.
Because derivatives  of the Bessel functions, $J_0$ and $J_1$, are expressed through each other, the final
expression will be:
\begin{equation}
f(q)= f_1(q)\frac{J_1(R_1 q)}{R_1 q)}\ + \ f_2(q) J_0(R_1 q) \ + \ ...,
\label{Eq8}
\end{equation}
where $f_1$, $f_2$, ..., are non-oscillating functions.

The first term of the Eq. \ref{Eq8} was considered \cite{SAM,Frahn} many years ago in the Strong Absorption Model
(SAM) in application to hadron-nucleus and nucleus-nucleus scattering at low energies. Various form of
the smearing function, $f_1$, were proposed in that time without solid physical foundation.
$f_1(q)=\pi d q/sh(\pi d q)$ was used in papers \cite{Frahn} by W.E. Frahn et al. Recently,
this form of the smearing function was obtained \cite{PJA} at the Fourier-Bessel transform of
the symmetrized 2-dimensional Fermi-function,
$$
\gamma(b)=\frac{1}{1+e^{(b-R)/d}}+\frac{1}{1+e^{-(b+R)/d}}-1
$$
\begin{equation}
Im f_{SFF}(q)=R^2\frac{\pi d q}{sinh(\pi d q)}\frac{J_1(R q)}{R q}\ +\
\frac{1}{2q^2}\frac{\pi d q}{sinh(\pi d q)}\left( \frac{\pi d q}{tanh(\pi d q)}-1\right)\ J_0(R q)+...
\label{Eq9}
\end{equation}

The symmetrized function is very close to the ordinary Fermi-function,
\begin{equation}
\gamma(b)=\frac{A}{1+e^{(b-R)/d}},
\label{Eq10}
\end{equation}
where we introduce the coefficient, $A$, to have more general expression.
Usually it is assumed that $A=1$.
The function (\ref{Eq10}) was used in the paper \cite{DiasDeDeus} by P. Brogueira and
J. Dias de Deus for a description of elastic $pp$-data at $\sqrt{s}=$ 14, 20, 53 GeV,
and $\bar pp$-data at $\sqrt{s}=$ 546, 630, 1800 GeV. "Unexpected qualitative agreement with the data
was found" by the authors.

Because the expression \ref{Eq8} is general, we accept as a working hypothesis that the elastic
scattering amplitude can be described by the expression,
\begin{equation}
f(q)=A \frac{\pi d q}{sinh(\pi d q)}\left\{ i\left[ R^2\frac{J_1(R q)}{R q}\ + \
\frac{1}{2}R^2\left( \frac{\pi d q}{tanh(\pi d q)}-1\right)\ \frac{J_0(R q)}{(Rq)^2}\right]\ + \
\right.
\label{Eq11}
\end{equation}
$$
\left. \left( R^2+\pi^2 d^2/3\right) \left(\rho - a_1(\pi d q)^2\right) J_0(R q)/2\right\}.
$$

The parameter $A$ is a normalization constant.
We consider it as a free parameter in order to take into account possible normalization error
of experimental data.

$\rho$ is a ratio of the real to imaginary parts of the elastic amplitude at zero momentum transfer.
The functional form of the real part, $J_0(Rq)$, is obtained with a help of the derivative
dispersion relations (see \cite{Block}) applying them to the Eq. \ref{Eq9} at $q \rightarrow 0$.
The relations give also other terms which go to $0$ as $(\pi d q)^2$ at $q \rightarrow 0$. Thus
we include them efficiently  having the last term in Eq. \ref{Eq11}.

The parameter $R$ is responsible for a position of the first diffractional minimum. The parameter
$d$ determines the slope of the differential cross section. A filling of the diffractional dip
is connected  with the parameters $\rho$ and $a_1$. $\rho$ parameter is not enough to do this. Thus,
we introduce a dependence of the ratio of the real to imaginary parts of the amplitude on the
momentum transfer - $q$.

Below we present in Sec. 1 results of a fitting which we make to find the parameters $R$, $d$ and $a_1$.
We have used here a lot of experimental data on the differential cross sections of elastic $\bar pp$-
and $pp$-scatterings. It is found that the parameters have rather simple energy dependencies.

The dependencies are applied in Sec. 2 at an extension to the LHC energies. The extrapolated values
of $R$, $d$ and $a_1$ together with the expression \ref{Eq11} allow to describe the Totem data. According
to our estimation (see Sec. 1) the parameterization is valid at $|t| <$ 1.75 (GeV/c)$^2$, thus the data
are described up to the second diffraction maximum.

In Sec. 3 we consider a behaviour of the differential cross sections of elastic scattering at large
$|t|$. Here we present a collection of various experimental data which shows that the cross sections
have a weak energy dependencies. We propose a simple parameterization of the high $|t|$ tail of the
cross sections. This allows to describe the Totem data at large $|t|$.

At last, in Sec. 4 we consider possible applications of the proposed approach. As seems to us, the
most important one is its application in calculation of the Glauber cross sections of
hadron-nucleus and nucleus-nucleus interactions needed for cosmic ray studies. Usually,
it is assumed in the calculations, that $\gamma(b)$ has the gaussian form. It
would be well to use the Eq. \ref{Eq10}.

An another application is a usage of the Eq. \ref{Eq10} in a calculation of quark-gluon string
multiplicity distributions in hadron-nucleon, hadron-nucleus and nucleus-nucleus interactions within
the framework of the Quark-Gluon-String Model (QGSM) \cite{QGSM}. The analogous application can
be foreseen in the high energy model like HIJING \cite{HIJING} combined soft and hard interactions.

The list of the possible applications is not completed. We hope it will be extended in the future.

\section{Fitting of experimental data}
It is obvious that the proposed expression \ref{Eq11} cannot be applied at all $q$ because cross
sections predicted by it decrease exponentially with a $q$ growth. At the same time it is known that
experimental data show a fall down as $1/|t|^{6-8}$ where, as it is expected, hard QCD-processes
dominate ($|t|=q^2$). Thus we have to restrict the application region of our approach. To find
a "border" between hard and soft interactions we have undertook a research, results of which are
presented in Fig. 1.

The solid (black) line there shows a fitting results without restriction on $|t|$ with 5 free
parameters ($A$, $R$, $d$, $\rho$ and $a_1$). As seen, the fitting reproduces the data at large $|t|$,
but underestimates the data rather strongly at small $|t|$. At the maximum allowed $|t|=2$ (GeV/c)$^2$
we have the result presented by the dashed (green) line. In the case, the fitting curve starts to
deviate from the data only at $|t|=1.75$ (GeV/c)$^2$. At $|t|_{max}=1.5$ (GeV/c)$^2$ there is no problems
(see dotted (red) line). Thus, as a compromise, we have estimated $|t|_{max}$ as 1.75 (GeV/c)$^2$.

In the following we use a lot of experimental data from 1 GeV up to the Tevatron energies. Some part of
them were taken from data-base \cite{CudellDB} created by J.R. Cudell, A. Lengyel and E. Martynov.
It was described in the paper \cite{Complete}.
A complete list of references is given in the Appendix. Because a correlation between the parameters
was very strong at a fitting of the small scattering angle region, we selected data where the dip
region was presented. This restricted the set of the experimental data.
\begin{figure}[cbth]
\includegraphics[width=160mm,height=70mm,clip]{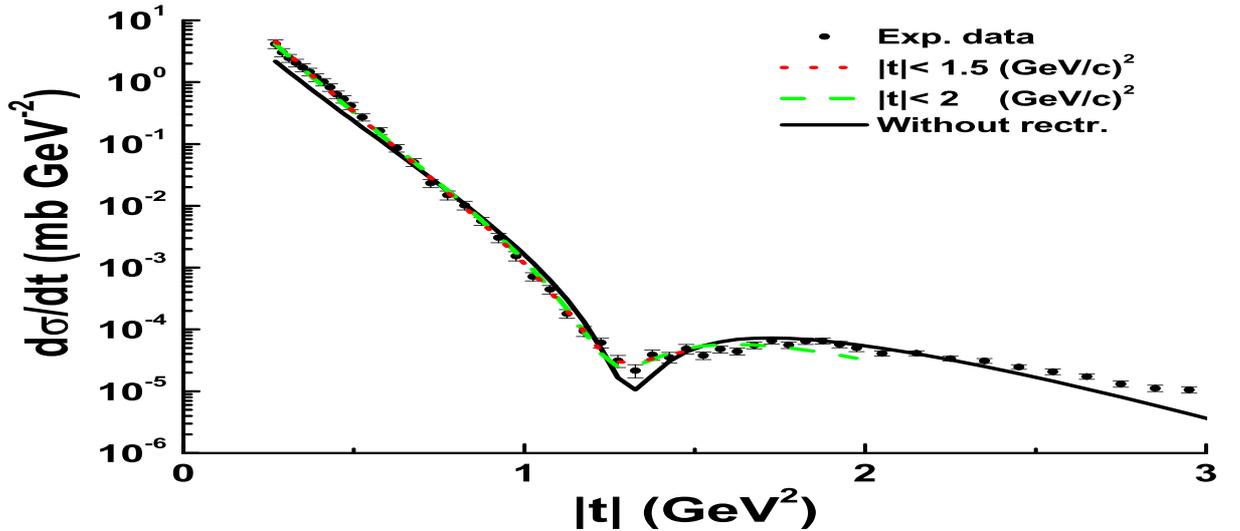}
\caption{Differential cross section of the elastic $pp$- scattering at $\sqrt{s}=62.3$ GeV.
         The points are the experimental data \protect{\cite{SS62}}. Lines are the fitting results.}
\label{Fig1}
\end{figure}

A good results of the fitting were obtained for $\bar pp$-interactions \cite{JETPLett}. More
complicated situation took place with the fitting of the $pp$-data. In order to reduce the number
of the free parameters, we have fixed $\rho$ using the following parameterizations of the corresponding
experimental data from the PDG data-base~\cite{PDGrho}:
\begin{equation}
\rho_{pp}=0.135-\frac{3}{\sqrt{s}}+\frac{4}{s}+\frac{80}{s^3}.
\label{Eq12}
\end{equation}

\begin{equation}
\rho_{\bar pp}=0.135-\frac{2.26}{\sqrt{s}}.
\label{Eq13}
\end{equation}

We assume that this allows us to attract indirectly an additional experimental information
because $\rho$ values were measured in the independent experiments -- in the Coulomb-nuclear
interference region at very small $q$. In principle, $\rho$ can be obtained at the fitting using
very exact measurements in the dip region. Such data were presented by the EDDA collaboration
\cite{EDDA_RPL} for projectile proton momenta, $P_{lab}$, from 1.1 GeV/c up to 3.3 GeV/c. A clear change
of the slope of the experimental cross sections was observed at $P_{lab} >$ 2 GeV/c and
$\theta_{cms} > 60^o$. Our obtained $\rho$ values for the data at $P_{lab} >$ 2 GeV/c are in a reasonable
agreement with other experimental data \cite{PDGrho}. Below 2 GeV/c the 4-parameter fit was unstable
due to the parameter's correlations. Thus we estimate the lower energy boundary of the application region of
the present approach as 2 GeV/c for $pp$-interactions. For $\bar pp$-interactions the boundary can be
 smaller.

\begin{figure}[cbth]
\includegraphics[width=80mm,height=60mm,clip]{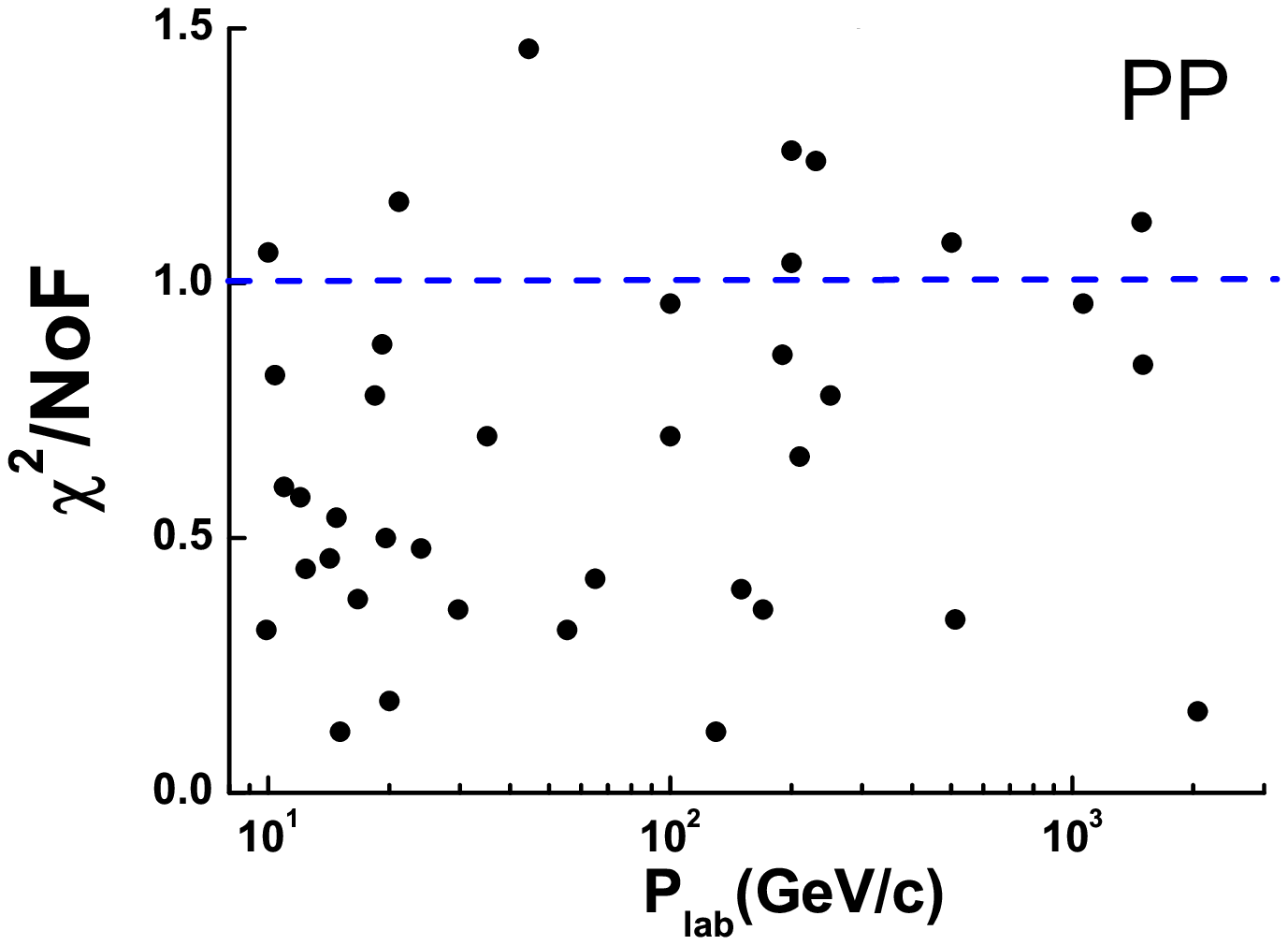}\includegraphics[width=80mm,height=60mm,clip]{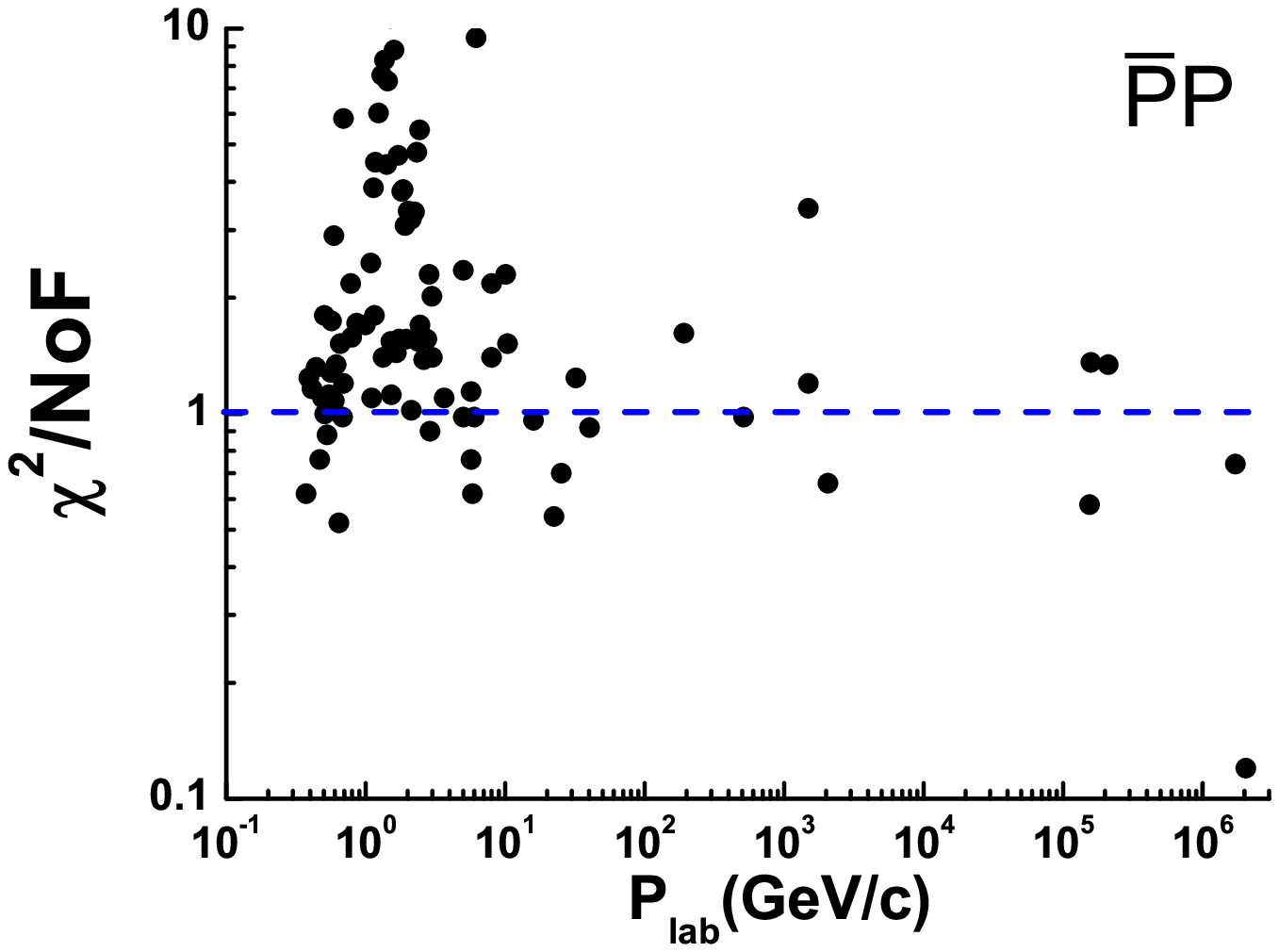}
\caption{Quality of the fitting of $\bar pp$ and $pp$ experimental data.}
\label{Fig2}
\end{figure}

With all the above given restrictions we have extra-ordinary good fit.
$\chi^2/NoF=1156/1489 \simeq  0.78$ for $pp$-interactions at $P_{lab}\geq 9.9$ GeV/c,
and $\chi^2/NoF=856/675 \simeq 1.27$ for $\bar pp$-ones at $P_{lab}\geq 8$ GeV/c.
Thus, it seems to us, that we can say about the unified systematic of all high energy
(anti)baryon-baryon elastic scattering data.

A quality of the fit is shown in
Fig.~2. As seen, most of the $pp$-interactions data are described quite well. The situation is more complicated
for the $\bar pp$-data especially at low energies. At high energies, the quality of the fitting of
$\bar pp$-interaction data becomes better.

Some fitting results in a comparison with experimental data are presented in Fig.~3, 4. We show there
the experimental data at all measured values of $|t|$ and our results extended outside the fitting
region ($|t|<$ 1.75 (GeV/c)$^2$) in order to demonstrate a necessary  to include a description of
the large angle scattering. As seen, we reproduce the cross sections for $\bar pp$-interactions in
the fitted region of $|t|$. The dip position is reproduced also.

\begin{figure}[cbth]
\includegraphics[width=80mm,height=35mm,clip]{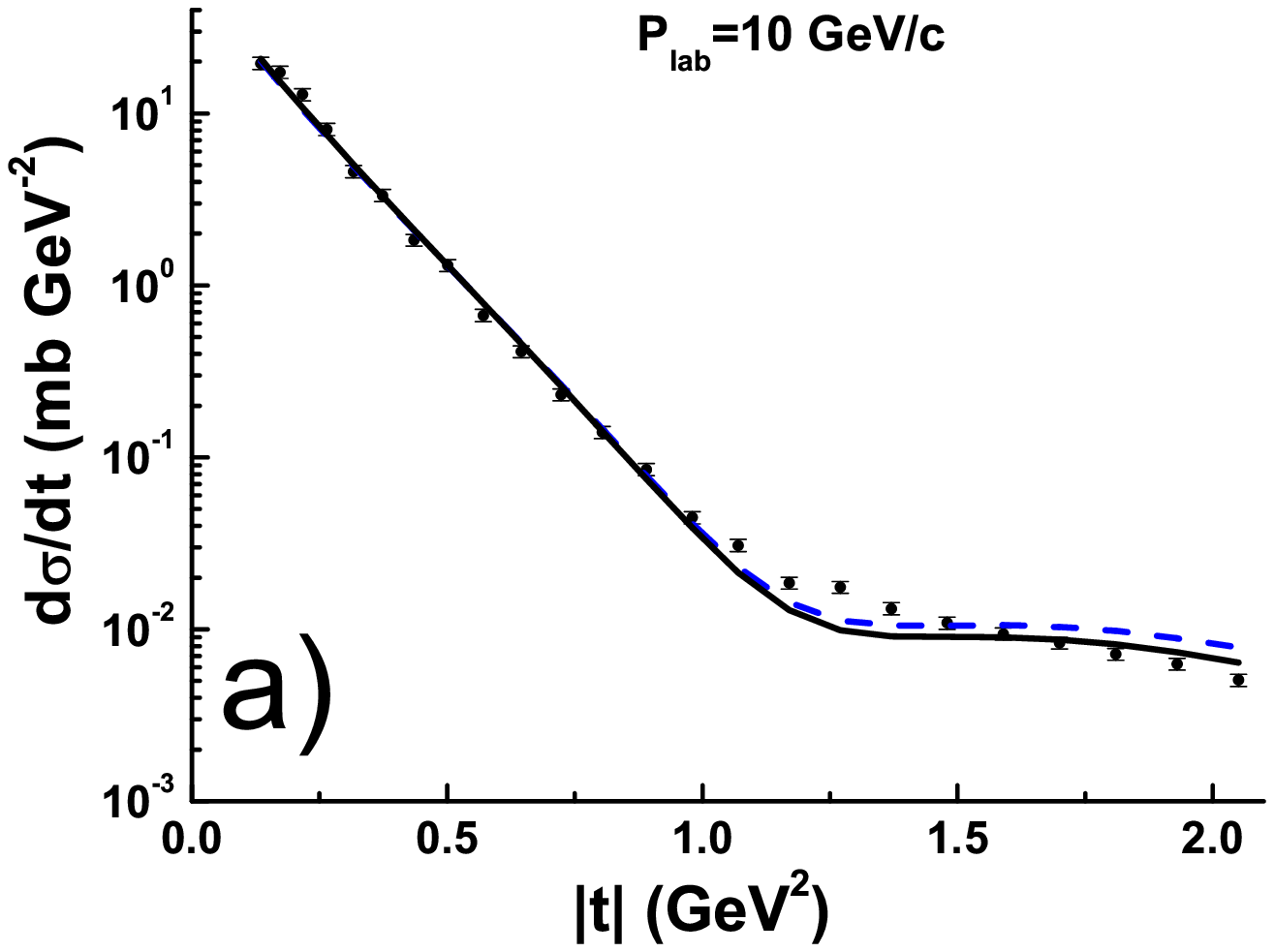}\includegraphics[width=80mm,height=35mm,clip]{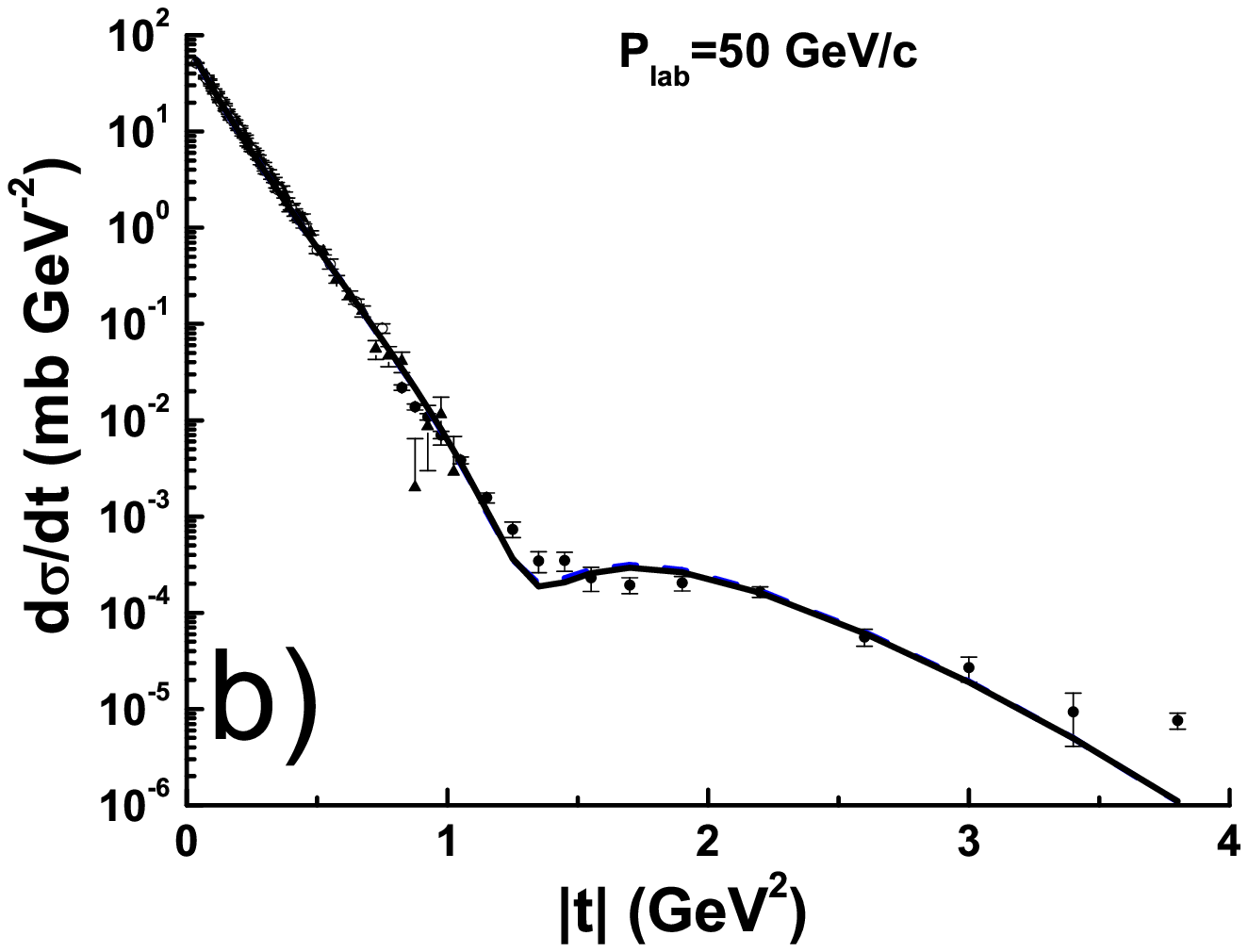}
\includegraphics[width=80mm,height=35mm,clip]{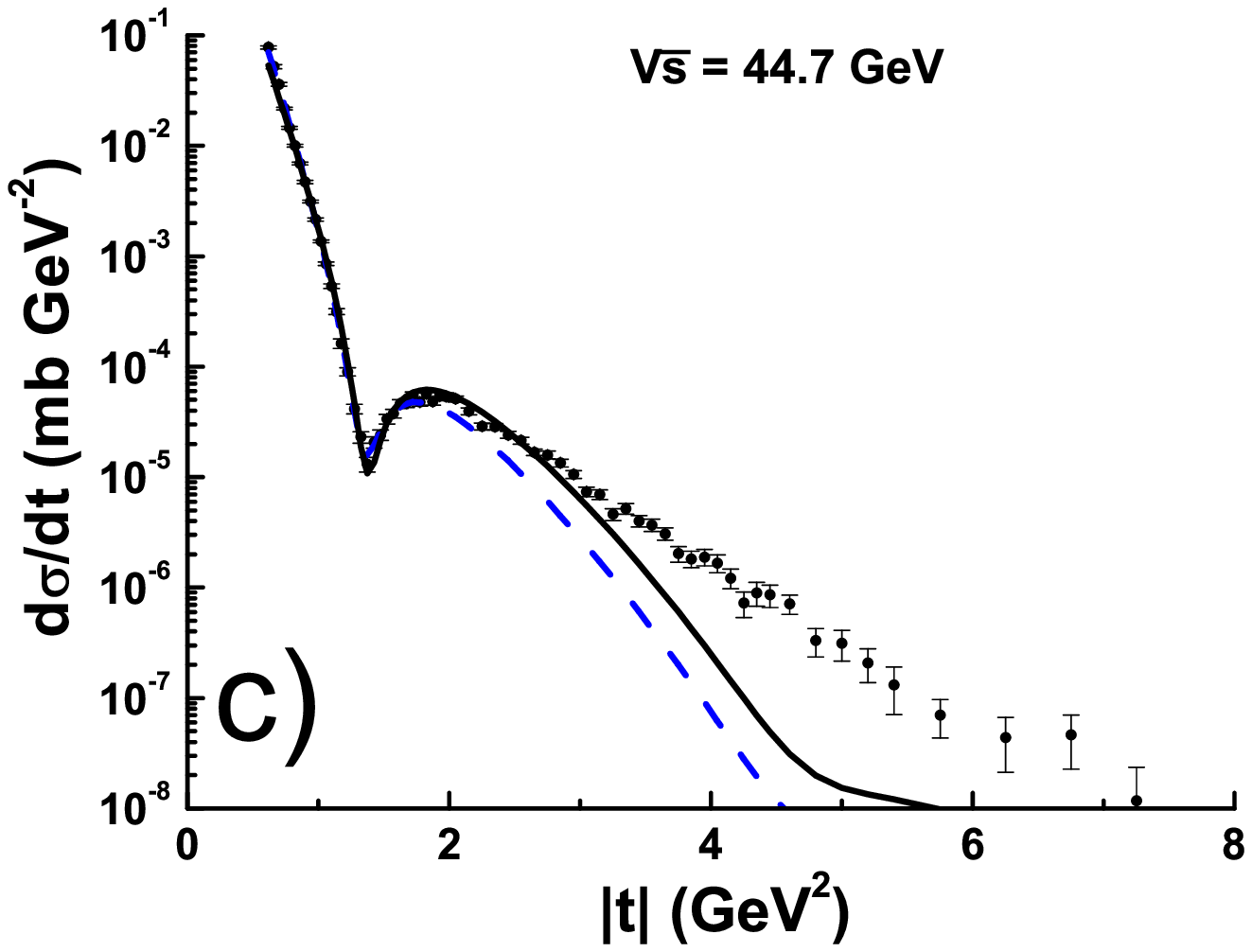}\includegraphics[width=80mm,height=35mm,clip]{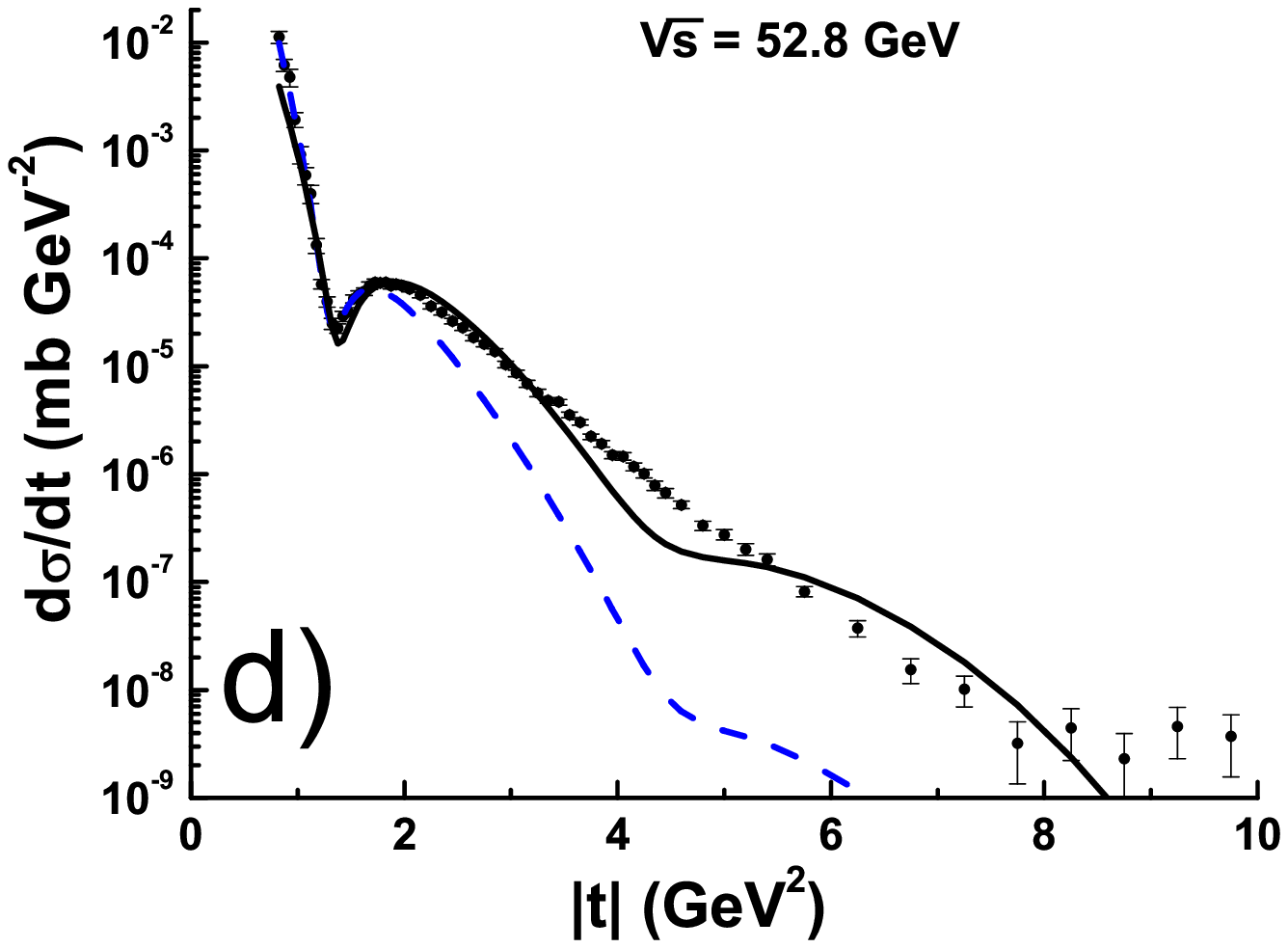}
\caption{Differential cross sections of elastic proton-proton scatterings. The points are
the experimental data (a - \protect{\cite{P10}}, b - \protect{\cite{P50}},
c - \protect{\cite{SS62}}, d - \protect{\cite{SS52_8}}). The solid lines are the fitting results without the
restriction on $|t|$.}
\label{Fig3}

\includegraphics[width=80mm,height=35mm,clip]{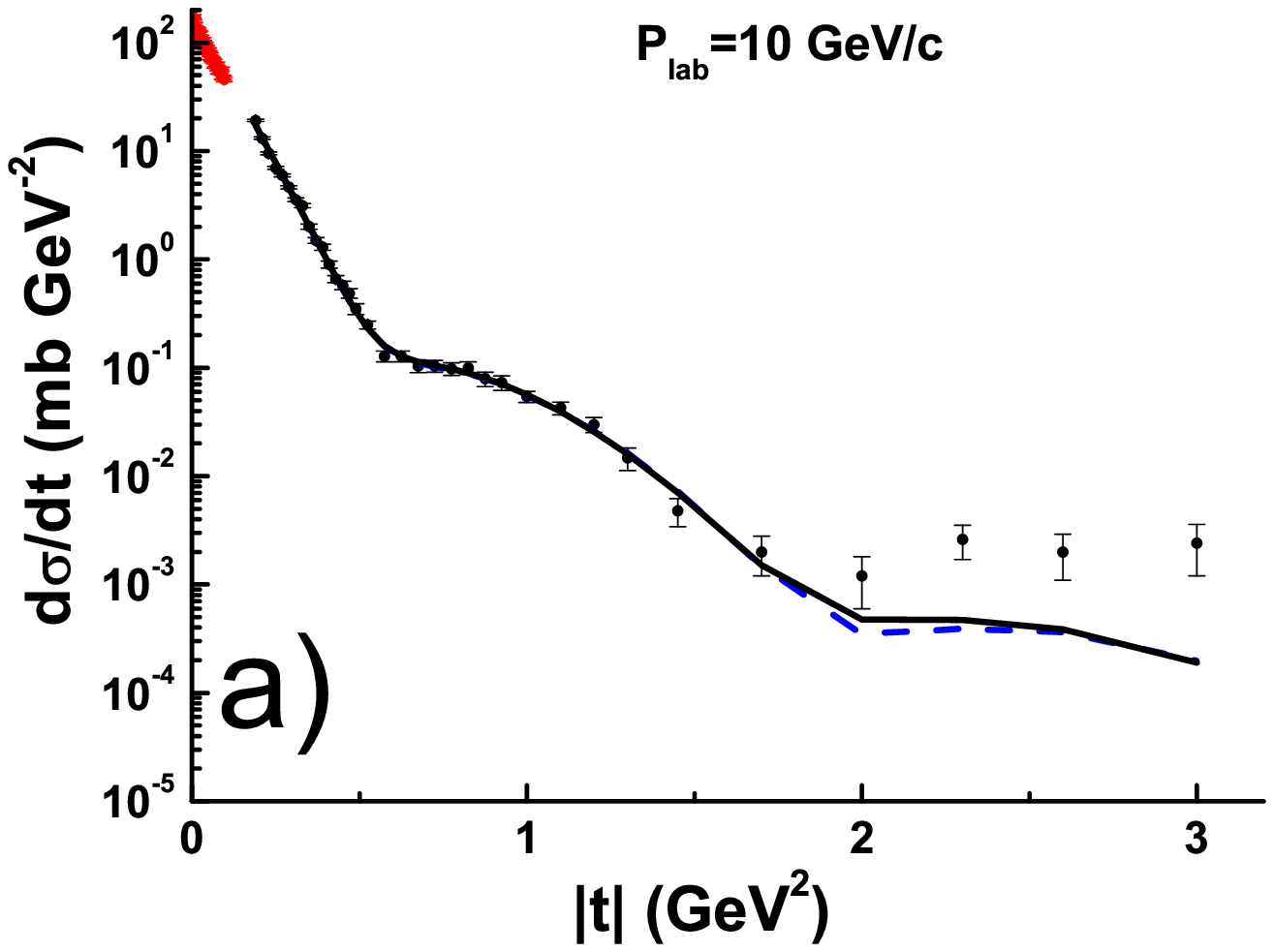}\includegraphics[width=80mm,height=35mm,clip]{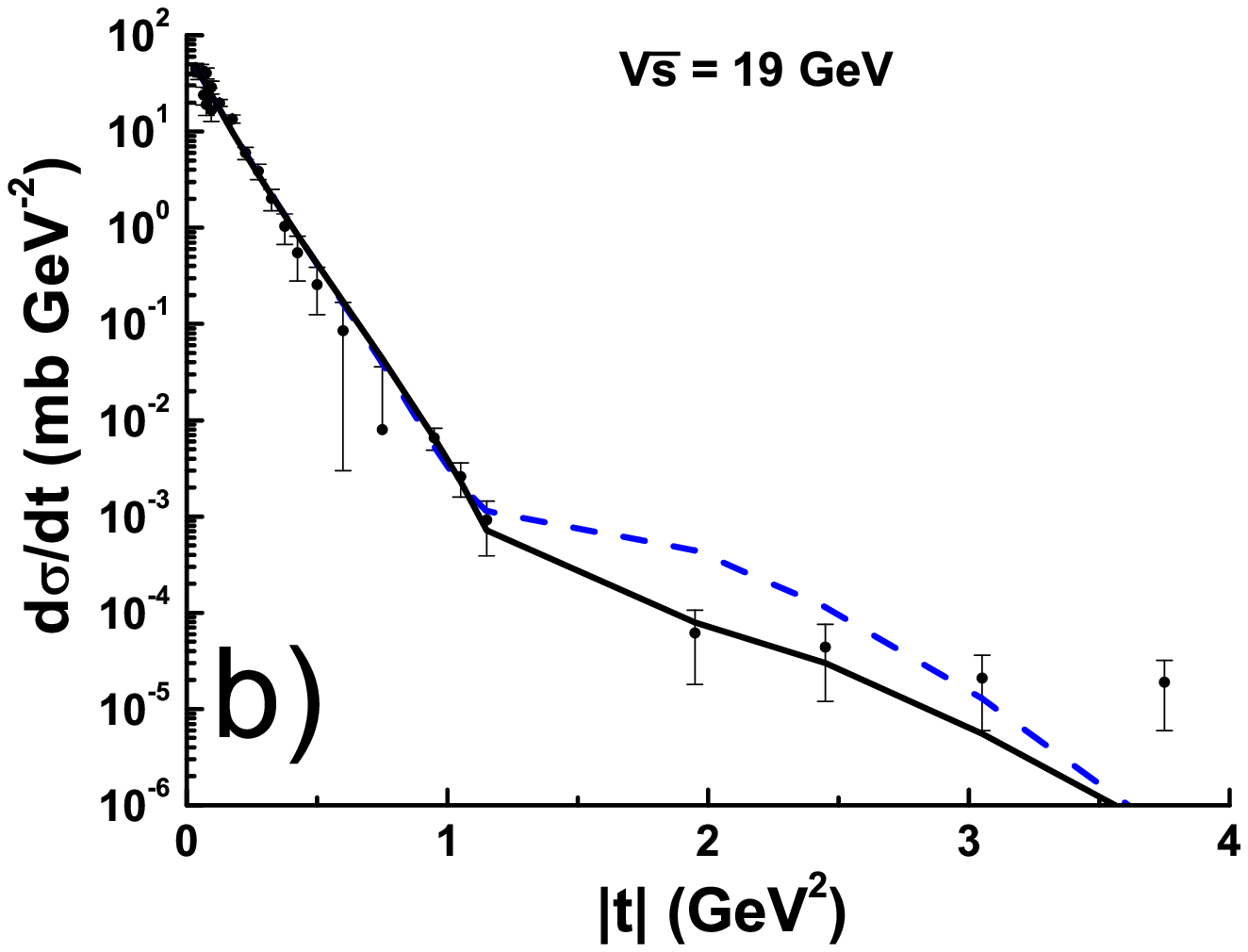}
\includegraphics[width=80mm,height=35mm,clip]{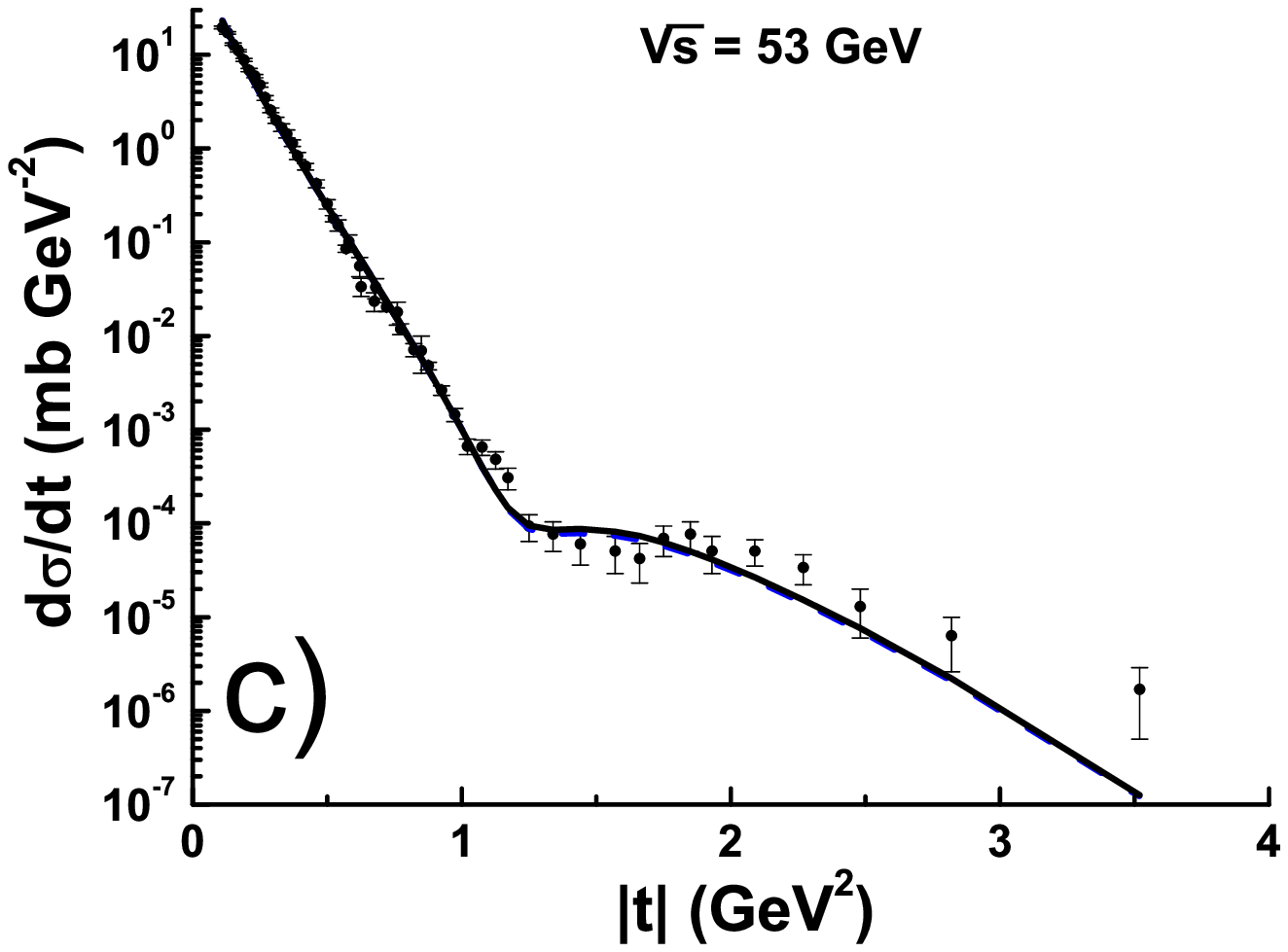}\includegraphics[width=80mm,height=35mm,clip]{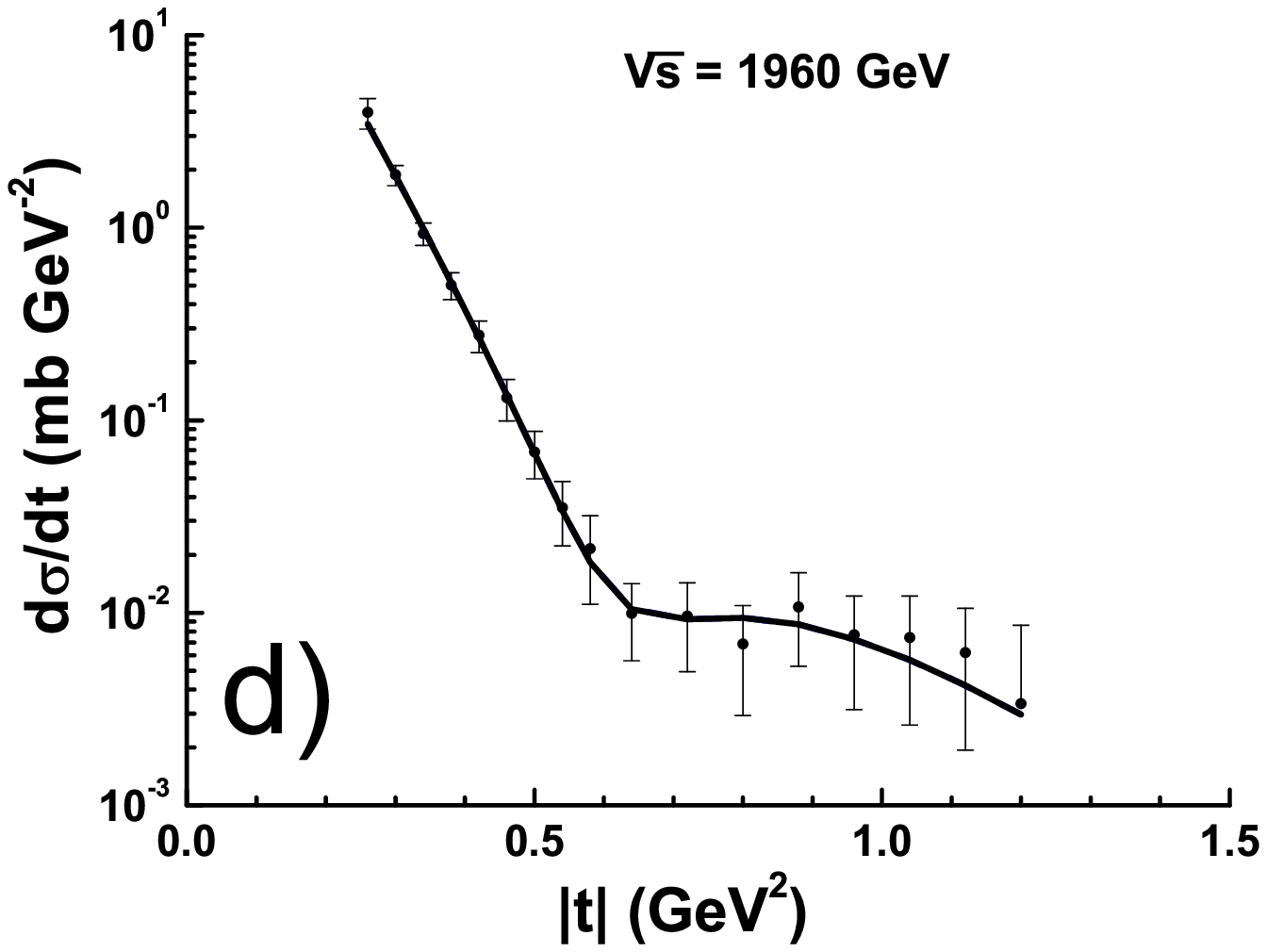}
\caption{Differential cross sections of elastic antiproton-proton scatterings. The points are
the experimental data??? (a - \protect{\cite{P10_1}}, b - \protect{\cite{SS19}},
c - \protect{\cite{SS53}}, d - \protect{\cite{SS1960}}). The solid lines are the fitting results without the
restriction on $|t|$.}
\label{Fig4}
\end{figure}

One can see that the diffraction minimum in $pp$-interactions connected with the first zero of $J_1(x)$
shifts to low values of $|t|$ with energy growth. This signals that $R$ is an increasing function of
$s$. The filling of the dip is caused by a variation of $\rho$. At $P_{lab}>$ 1.5 GeV/c and
$P_{lab} <$ 200 GeV/c $\rho$ is negative. At $P_{lab} \sim$ 200 GeV/c $\rho$ is closed to zero.
At larger energy it is positive. So, the filling of the dip depend on energy.

As known, the slope parameter, $B=d\ ln(d\sigma/dt)/dt|_{t\rightarrow 0}$, is increasing function.
It is mainly connected with $\pi dq/sinh(\pi dq)$, and with the parameter $d$.
So, the parameter $d$ must be increasing function also. The same regularities can be seen for
$\bar pp$-interactions.

The fitting results for $R$ and $d$ are presented in Fig. 5. They show that $R$ for $\bar pp$-interactions
decreases with the energy growth starting from low energy, reaches a minimum at $\sqrt{s} \sim 30$ GeV,
and continues the growth at higher energies. $R$ for $pp$-interactions in the studied energy range is
practically constant.

The energy dependence of $d$ is more complicated. For $pp$-interactions in the considered energy range
it is the increasing functions. $d$ for $\bar pp$-scattering has an interesting irregularity at
very low energies. At high energies it reaches a constant value.
\begin{figure}[cbth]
\includegraphics[width=80mm,height=80mm,clip]{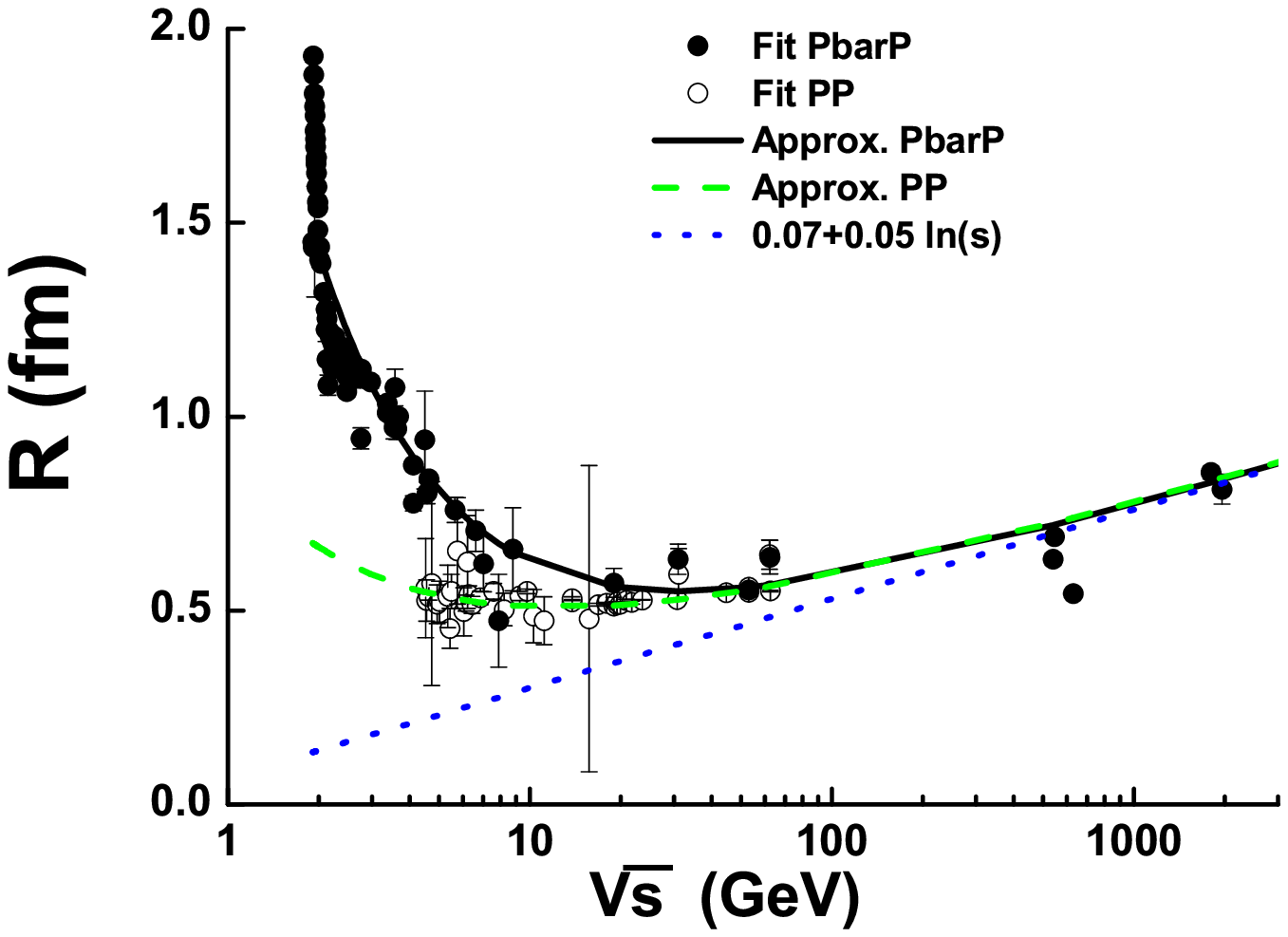}\includegraphics[width=80mm,height=80mm,clip]{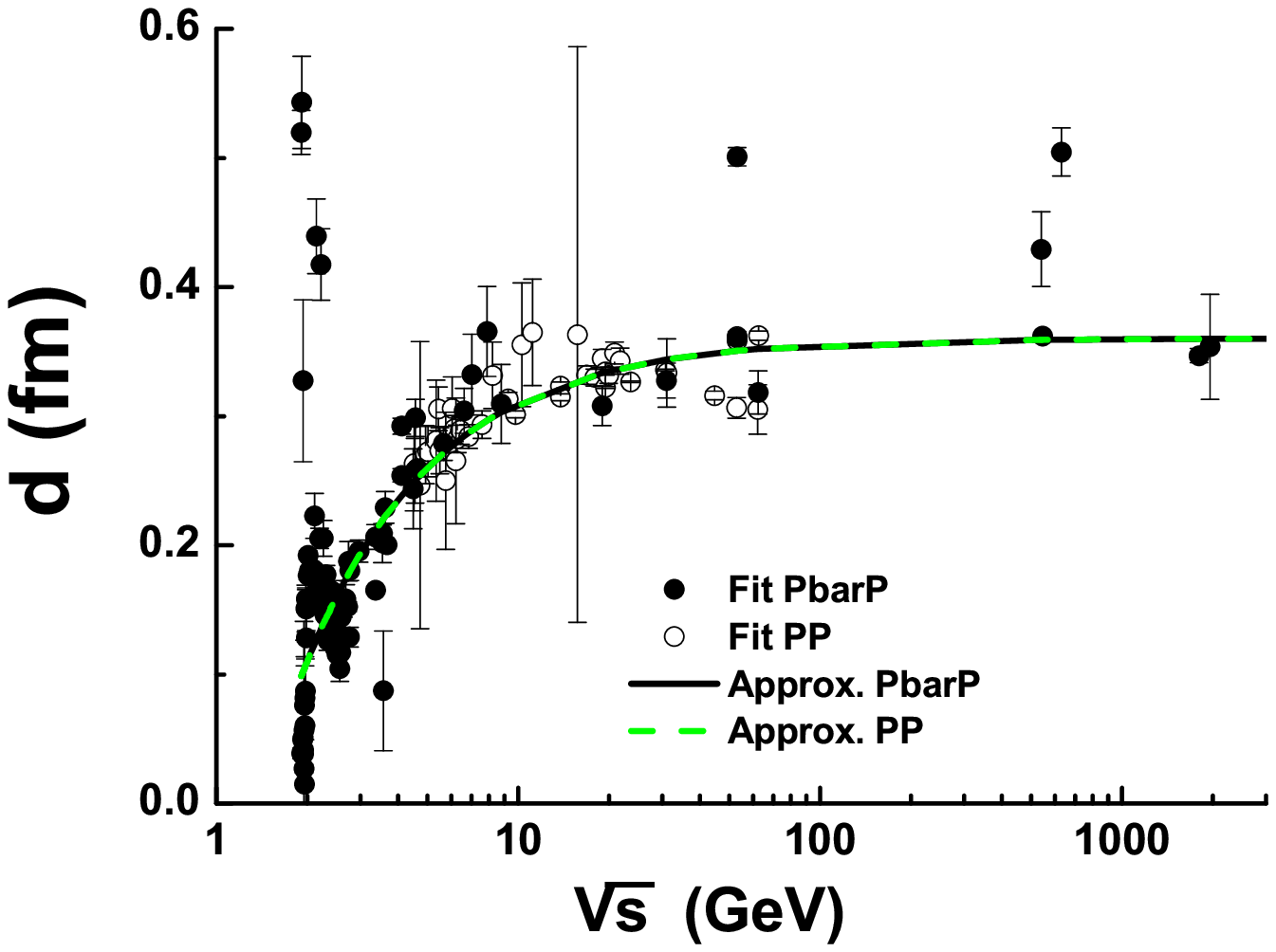}
\caption{The fitting results for the parameters $R$ and $d$. The black points present the results
for $\bar pp$-interactions, the open points -- for $pp$-interactions.}
\label{Fig5}
\end{figure}

For future applications we approximate the dependencies as:
\begin{equation}
R_{pp}=0.07 + 0.05\ln{s}+\frac{0.6}{s^{1/4}}+\frac{0.2}{s^{1/2}}, \ \ \ (fm),
\label{Eq14}
\end{equation}
\begin{equation}
R_{\bar pp}=0.07 + 0.05\ln{s}+\frac{0.4}{s^{1/4}}+\frac{2}{s^{1/2}}, \ \ \ (fm),
\label{Eq15}
\end{equation}
\begin{equation}
d_{pp}=d_{\bar pp}=0.36 - \frac{0.5}{\sqrt{s}}. \ \ \ (fm).
\label{Eq16}
\end{equation}

The dependencies are shown in Fig. 5 by the solid and dashed lines. The asymptotical part,
$0.07 + 0.05\ln{s}$, is presented by the dotted line.

The fitting results for the parameters $A$ and $a_1$ are shown in Fig. 6. As seen, the parameter $A$
fluctuates within $\pm 30$ \% at low energies. It is close to unity at high energies, and there is
a defined energy dependence of the parameter. Thus, we believe that the parameter does not reflect only
uncertainty of the experimental data normalization, but it contains some information on physics of
the processes.

If the parameter $A$ is below unity, it points out on a possible influence of the inelastic shadowing on
the elastic scattering due to the processes of excitations and deexcitations of low mass difractive states
during the scattering. A value of the parameter above the unity can be interpreted as a presence of
additional processes like $\pi$-meson exchange, annihilation and so on which are not taken into
account directly. The value above unity can violate the unitarity requirement according to which
$|\gamma(b)|$ must be below unity. If $A \leq 1+e^{-R/d}$ there is no problem with the unitariry,
but it will mean that the amplitude reaches the black disk limit in the central interactions.
If $A > 1+e^{-R/d}$ the simplest solution can be an application of any unitarisation scheme. We are
going to study the subject in the future.

We show in Fig. 6a the function $1+e^{-R/d}$ for $pp$-interactions by the dashed line. It seems to us
that taking into account the fluctuation of the fitting results for the $pp$-interactions at high energies
we can assume that the black disk limit is reached in the $pp$-collisions in the central region.

\begin{figure}[cbth]
\includegraphics[width=80mm,height=80mm,clip]{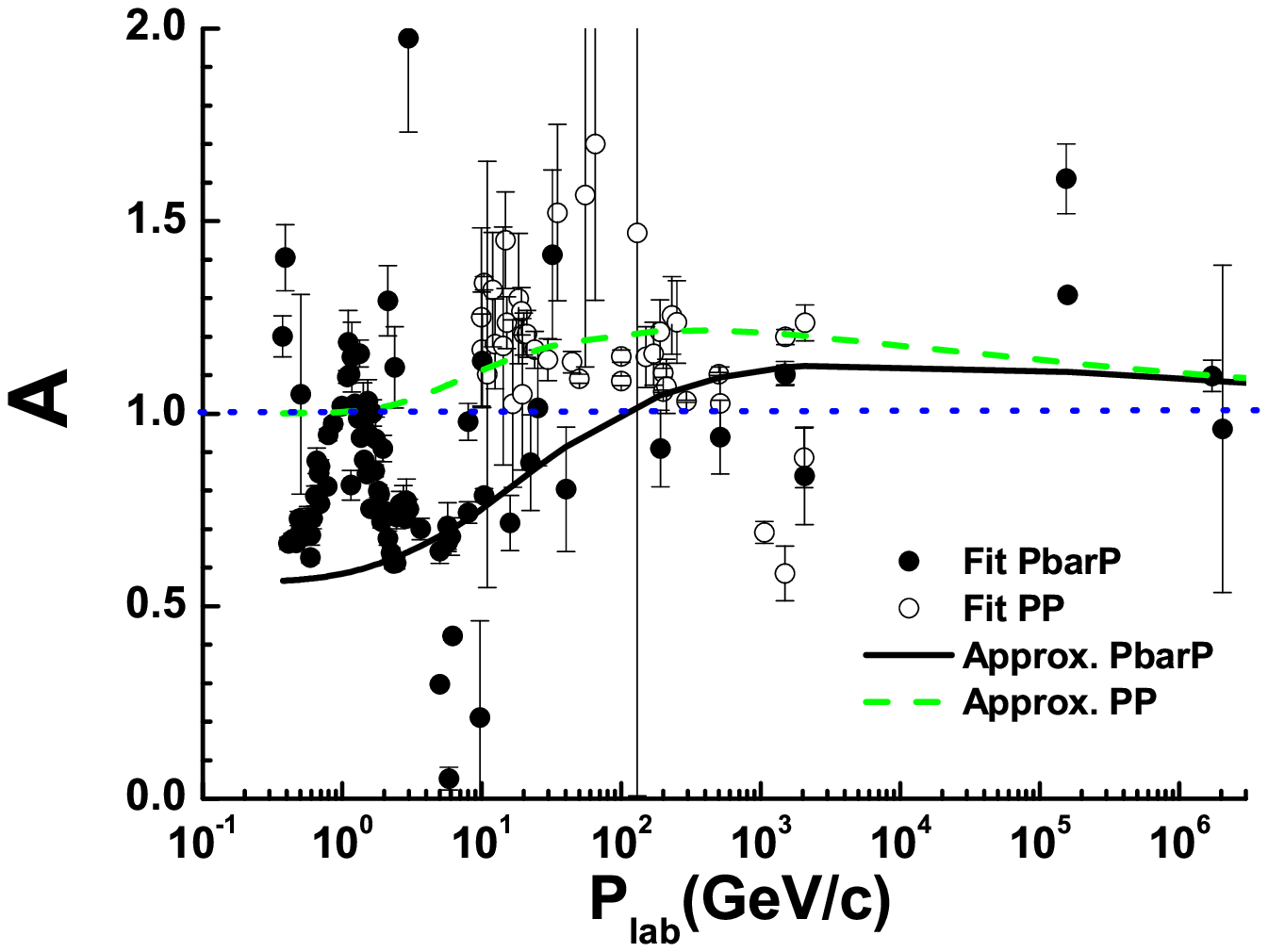}\includegraphics[width=80mm,height=80mm,clip]{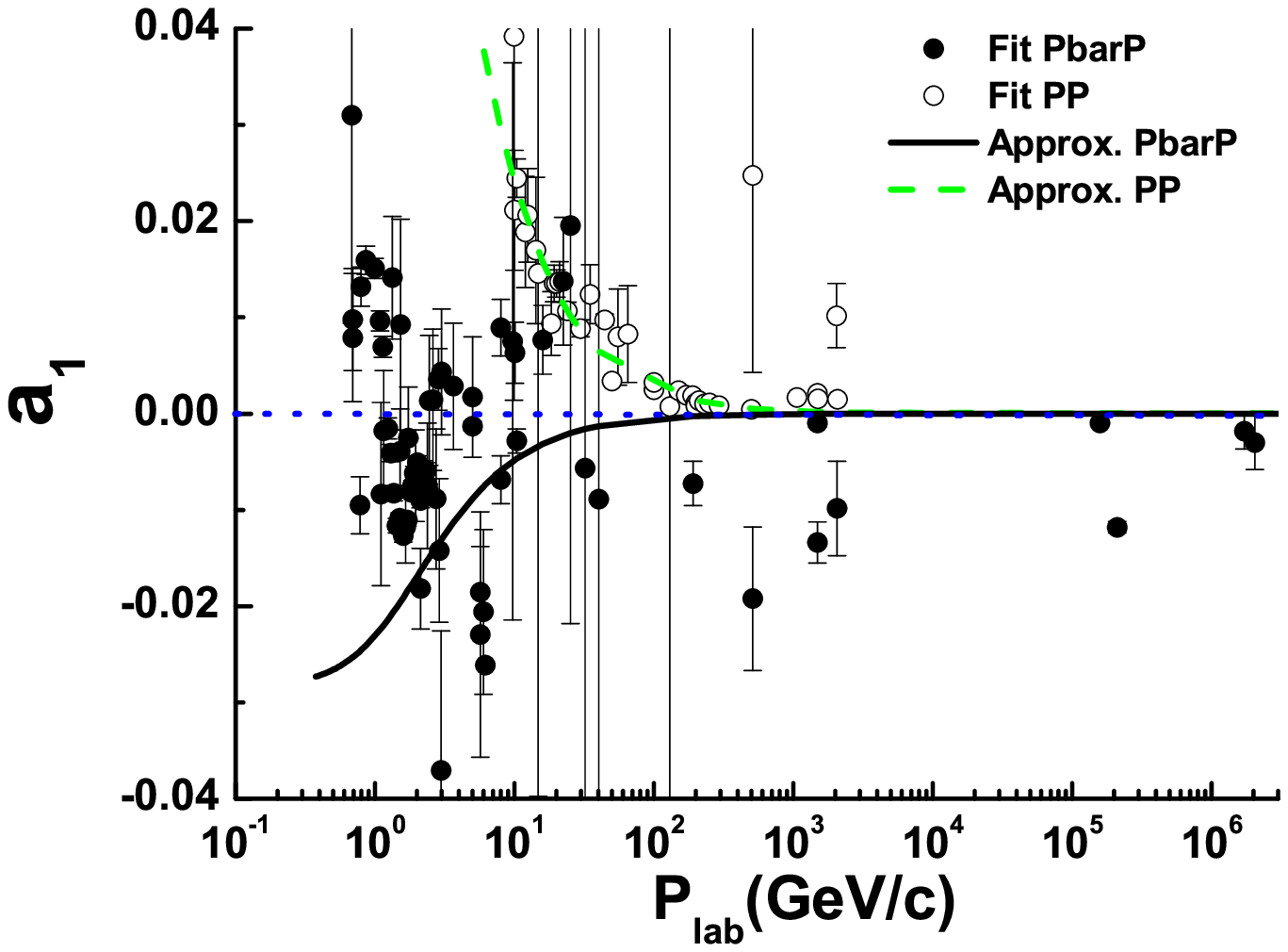}
\caption{The fitting results for the parameters $A$ and $a_1$. The black points present the results
for $\bar pp$-interactions, the open points -- for $pp$-interactions.}
\label{Fig6}
\end{figure}

As seen in Fig. 6, energy dependence of the parameter $a_1$ for $pp$-interaction is rather regular. We
cannot say this for $\bar pp$-interactions.

For future applications we approximate the dependencies of the parameters $A$ and $a_1$ as:
\begin{equation}
A_{pp}=1+e^{-R/d}, \ \ \ A_{\bar pp}=1+e^{-R/d} - 0.6/s^{0.25},
\label{Eq17}
\end{equation}
\begin{equation}
a_{1,\ pp}=0.5/s, \ \ \ a_{1,\ \bar pp}=-0.1/s.
\label{Eq18}
\end{equation}

Using the expressions \ref{Eq12} -- \ref{Eq18} we obtain a good description of the $\bar pp$-
and $pp$-interaction data (see Appendixes). Especially, we have
for $pp$-data $\chi^2/NoF=4866/1489 \simeq 3.26$ at $P_{lab}\geq 9.9$ GeV/c, and for $\bar pp$-data --
$\chi^2/NoF=3620/675 \simeq 5.36$ at $P_{lab}\geq 8$ GeV/c having only one fitting parameter -- $A$.

\section{Description of the Totem data}
The expressions \ref{Eq12}, \ref{Eq14}, \ref{Eq16} predict for the LHC energies the following values
of the parameters:

\begin{table}[cbth]
\begin{center}
\begin{tabular}{|c|c|c|c|c|c|}
\hline
$\sqrt{s}$(GeV)&     $R$ (fm) &    $d$ (fm) & $\rho$ &$\sigma_{tot}$ (mb)& $\sigma_{el}$ (mb) \\ \hline
       900     &    0.770     &   0.359     &  0.1316&    71.5           &       16.9         \\ \hline
      7000     &    0.963     &   0.360     &  0.1346&    90.9           &       22.7         \\ \hline
     10000     &    0.997     &   0.360     &  0.1347&    94.8           &       24.0         \\ \hline
     14000     &    1.030     &   0.360     &  0.1348&    98.8           &       25.3         \\ \hline
\end{tabular}
\caption{Parameters and estimated value for LHC}
\end{center}
\end{table}
Using them we calculate total and elastic cross sections, $\sigma_{tot}$ and $\sigma_{el}$.
\begin{equation}
\sigma_{tot}=2\pi\ A\ (R^2+\pi^2d^2/3),
\label{Eq19}
\end{equation}
\begin{equation}
\sigma_{el}\simeq \pi\ A^2\ (R^2-19Rd/10+2\pi^2d^2/7).
\label{Eq20}
\end{equation}

The Totem collaboration \cite{Totem2} published the total and elastic cross sections which are
$\sigma_{tot}=98.3 \pm 0.2^{stat} \pm 2.8^{sys}$ (mb) and
$\sigma_{el}=24.8 \pm 0.2^{stat} \pm 1.2^{sys}$ (mb). They are above our predictions. To understand
the difference, we calculate differential elastic scattering cross section according to our approach,
and the cross section using the simple gaussian parameterization and the value of the slope parameter
given by the Collaboration, $B=20.1 \pm 0.2^{stat} \pm 0.3^{sys}$ (GeV$^{-2}$). They are presented
in Fig. 7 by the solid and dashed lines, respectively.
\begin{figure}[cbth]
\begin{center}
\includegraphics[width=90mm,height=60mm,clip]{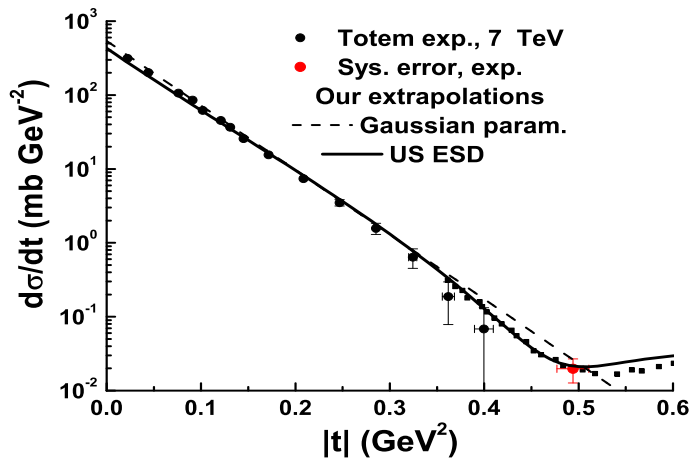}
\begin{minipage}{80mm}{
\caption{Differential cross section of $pp$ elastic scattering at 7 TeV. The points are
the experimental data \protect{\cite{Totem1,Totem2}}. The lines are our calculation.}
} \end{minipage}
\end{center}
\label{Fig7}
\end{figure}

As seen, first of all, the Collaboration fitted the differential cross section at $|t|<$ 0.35 (GeV$^{-2})$ to
obtain the slope and the cross sections (dashed line). Our prediction (solid line) catches
the points at $|t|>$ 0.15 (GeV$^{-2}$, especially, in the region of the minimum. At $|t|$ below 0.15 (GeV$^{-2})$
the prediction deviates regularly from the corresponding data\footnote{We could not be able to digitize quite
well the experimental points presented in \cite{Totem2}. Thus experimental errors are not shown. We cannot
guarantee that exactness of the shown points is sufficiently high.}. The data are above our curve.
Maybe additional expansion terms are needed to be included in Eq. \ref{Eq11}. The can give corrections at small
$|t|$.

Our description of the data \cite{Totem1,Totem2} in the whole measured values of $t$ is presented in Fig. 8.
\begin{figure}[cbth]
\includegraphics[width=160mm,height=65mm,clip]{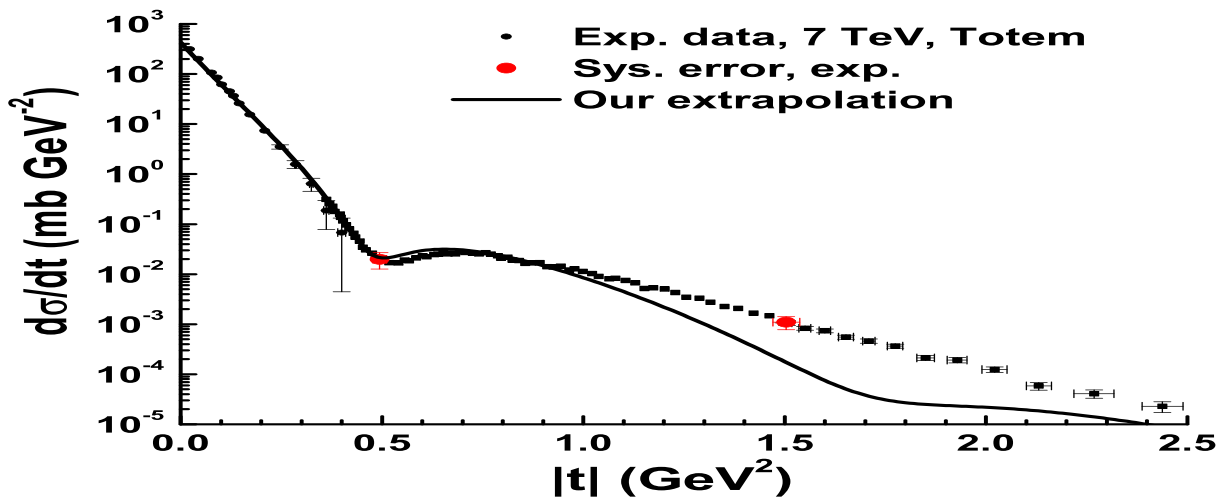}
\caption{Differential cross section of $pp$ elastic scattering at 7 TeV. The points are
the experimental data \protect{\cite{Totem1,Totem2}}. The line is our calculation.}
\label{Fig8}
\end{figure}

One can see that the forward scattering data are reproduced quite well. The dip is filled rather well.
Its position is right. At the same time, the calculations deviate from the data starting from lower
values of $|t|$ than it was at other energies. The high of the second diffractional maximum is
mainly determined by the parameter $d$. The slope of the forward part of the spectra is connected
with the parameter also. At chosen value of the parameter we overestimate a little bit the high
of the maximum. We can describe better the forward part of the date varying $d$ in its accuracy
limits making worse the description of the dip region, and vice-versa. We expect that an exactness
of the parameter determination will be improved when the Totem collaboration will publish final
data.

We have to note that an accuracy of the parameters entering in Eqs. \ref{Eq14} -- \ref{Eq16}, \ref{Eq18}
is equal to $\pm 5$ \%. We expect the same accuracy for the calculated differential cross section.
The accuracy can be improved when new Totem data at other energies will be appeared.

In order to understand a quality of the calculations, let us compare our calculations with predictions
of other models \cite{Models} presented by the Totem collaboration in the paper \cite{Totem1}. For this, we
show the model predictions in the dip region and in the region of large $|t|$ in Fig. 9. Because we could
not be able to take experimental errors in \cite{Totem1}, we plotted the points without errors. Though,
the systematic errors are rather large for a correct discrimination of the model, one can see that only our
approach gives results that are quite closed to the data at small angles and in the dip region.
The high of the second maximum is reproduced also in the approach. But instead of other models we
predict too fast decreasing of the cross section above the diffraction maximum. The other models predict
much slower decreasing of the cross sections. Here we have to note, that the models were tuned using much less
set of experimental data. Additional to this, they included, directly or indirectly, the high momentum
transfer scattering.
\begin{figure}[cbth]
\includegraphics[width=80mm,height=60mm,clip]{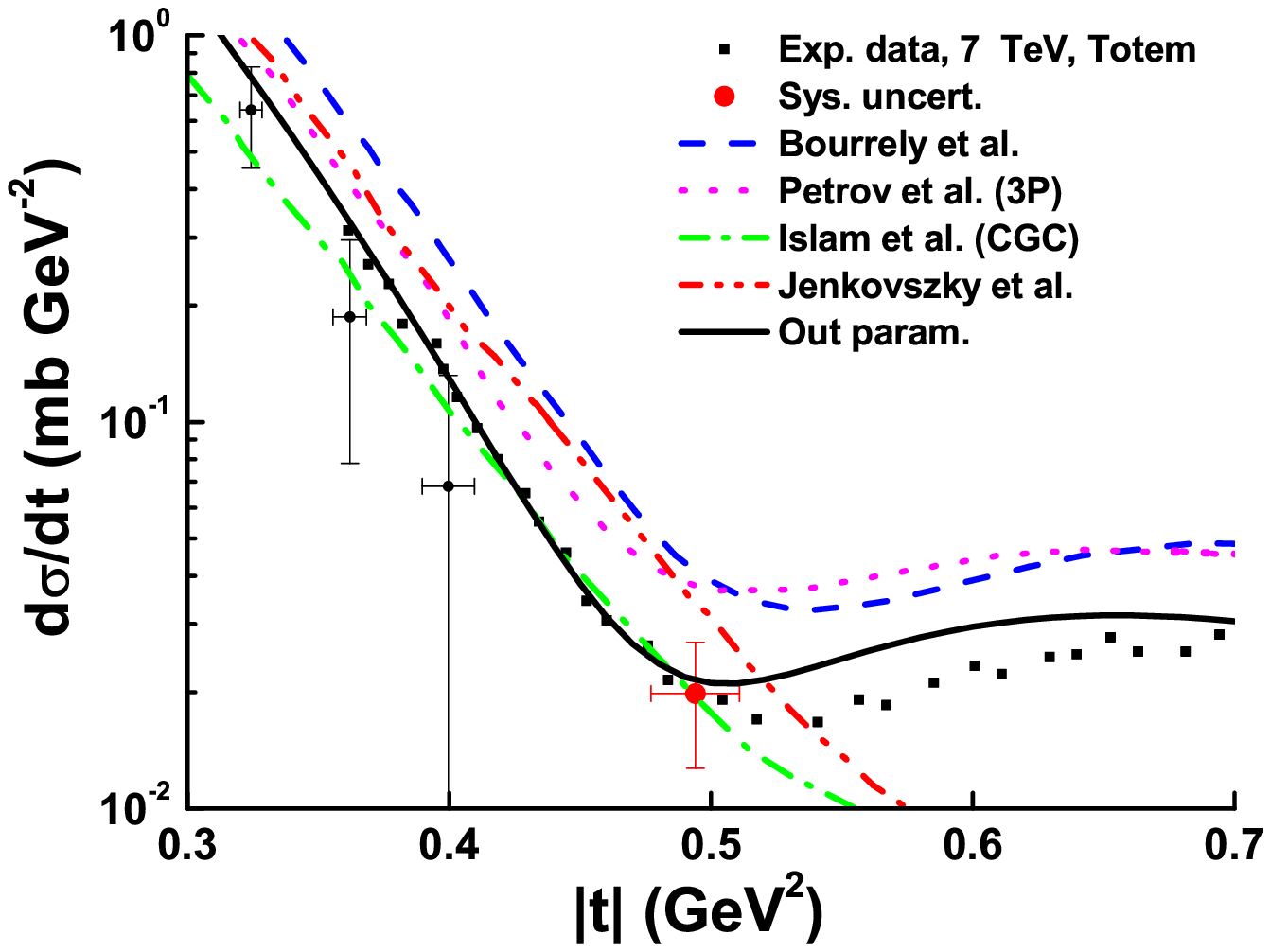}\includegraphics[width=80mm,height=60mm,clip]{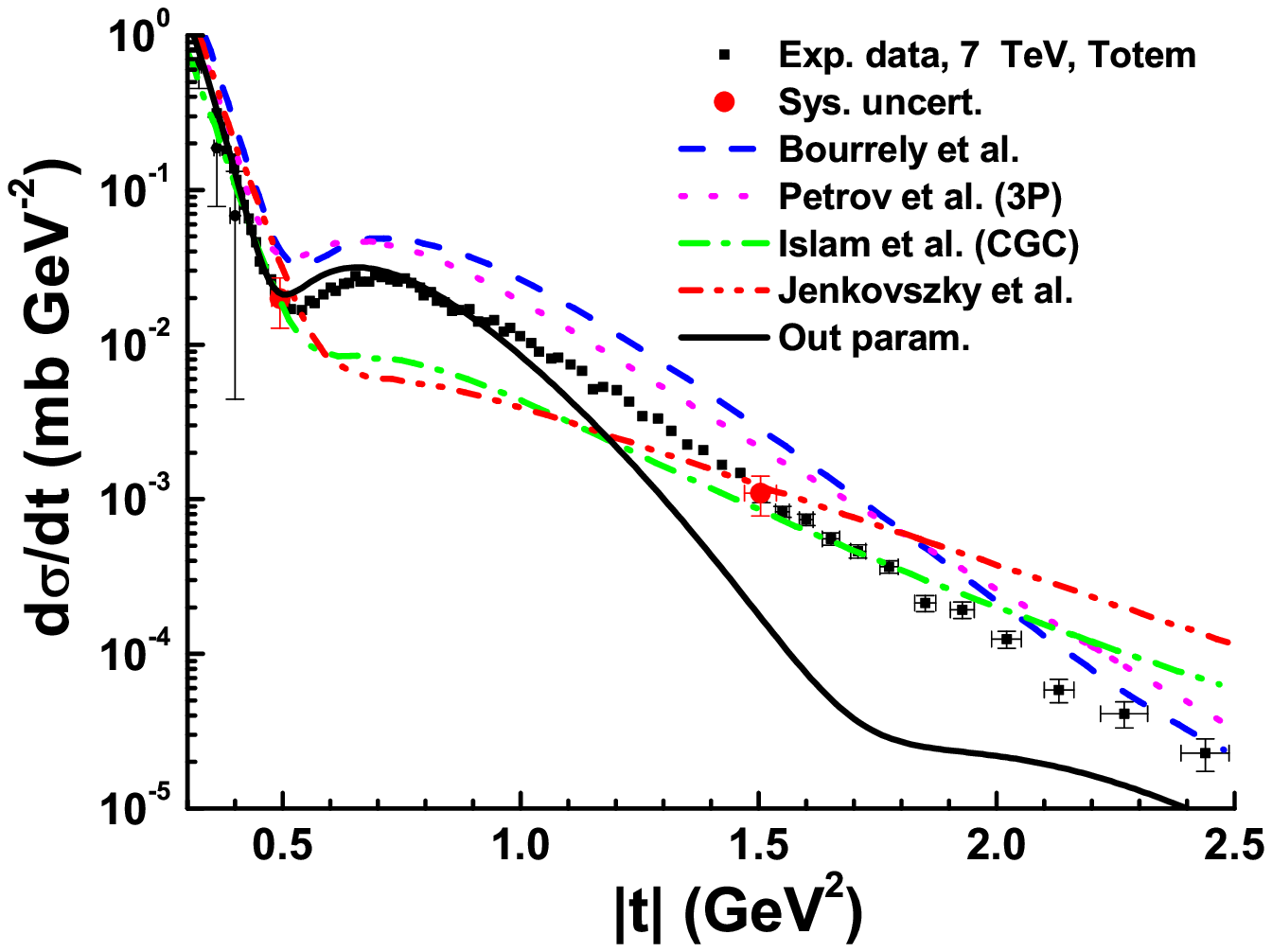}
\caption{Comparison of model predictions presented in \protect{\cite{Totem1}} with the Totem
experimental data. The solid (black) lines show our calculations.}
\label{Fig9}
\end{figure}

\section{Description of the high momentum part of the data}
The nature of processes with high momentum transfer are debated until now. It is commonly accepted that
they can be described in QCD. There are a lot of publications on the subject. Instead of analyzing all of
them in order to select a reliable one we turn to experimental data. In Fig. 10 we present some experimental
data on elastic scattering at large $|t|$.

As seen, at $|t|>$ 2 (GeV/c)$^2$ all the cross sections have the same shape at $\sqrt{s}>$ 10 GeV. At
the projectile momentum below 200 GeV/c they have strong energy dependence. To reproduce the high energy
behavior of the cross sections we add to the imaginary part of the amplitude (\ref{Eq11}) a "hard"
scattering amplitude:
\begin{equation}
Im f_{hard}(q)=-0.05\ [1+tanh(R q -5.5)]\frac{1}{(1+|t|/0.71)^{4}}.
\label{Eq21}
\end{equation}

The sign "-" is needed to increase the cross sections in the second maximum where $J_1(R q)$ is negative.
The hyperbolic tangent imitates a smooth transition from soft to hard scatterings. According to Fig. 10
the border of the hard processes slowly moves to small momentum transfer with an energy growth. It can be
if the border is connected with the radius of the soft interactions. Thus, we assume that the tangent
argument is $R q$. The value "5.5" gives the exact position of the border. The last factor in Eq. \ref{Eq21}
is the proton form-factor in a tuned power. All values in (\ref{Eq21}) are sampled only in order to
reproduce the cross section behavior qualitative at $\sqrt{s}>$ 10 GeV. With all of these we have a description
of the Totem data presented in Fig. 11.
\begin{figure}[cbth]
\includegraphics[width=50mm,height=40mm,clip]{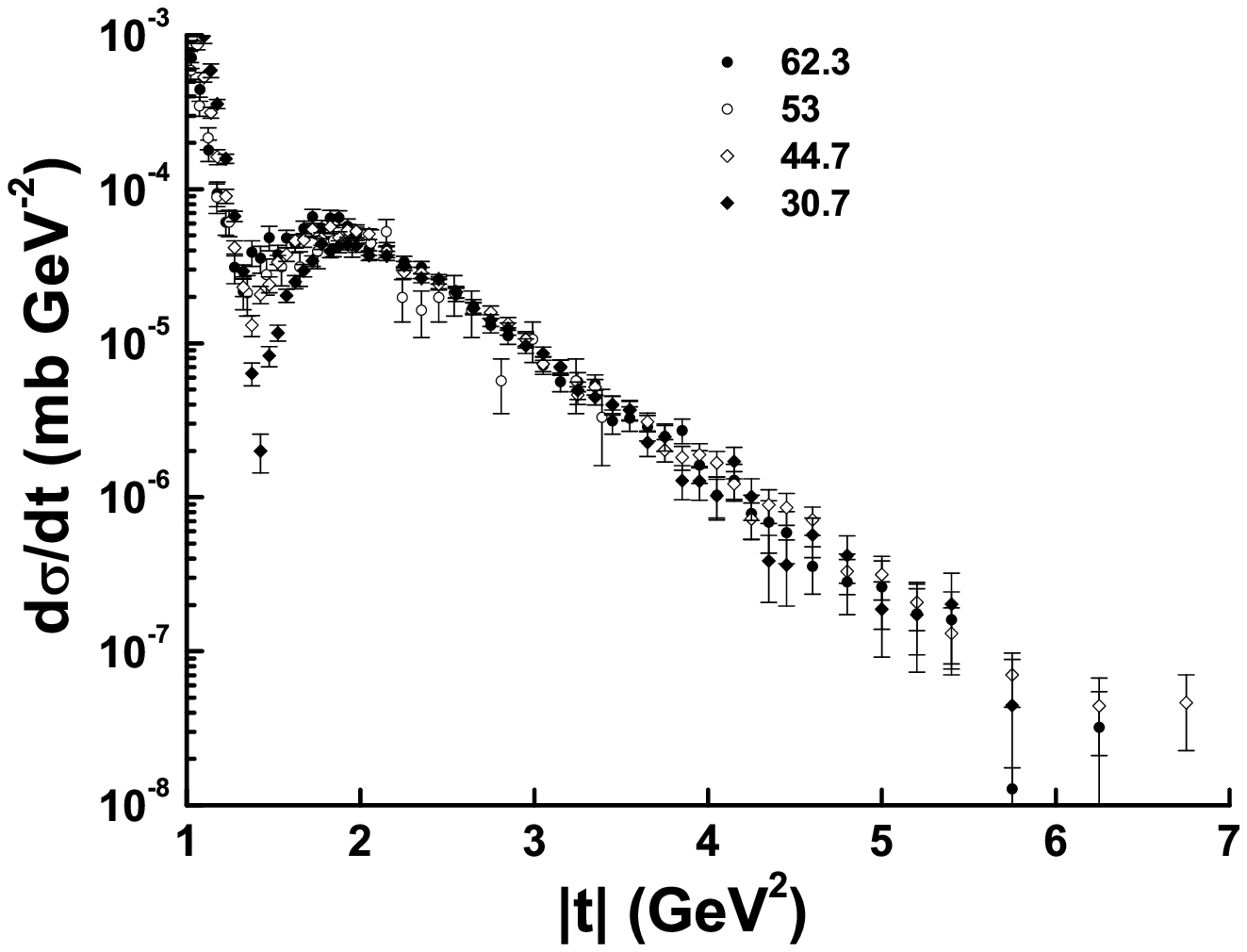}\includegraphics[width=50mm,height=40mm,clip]{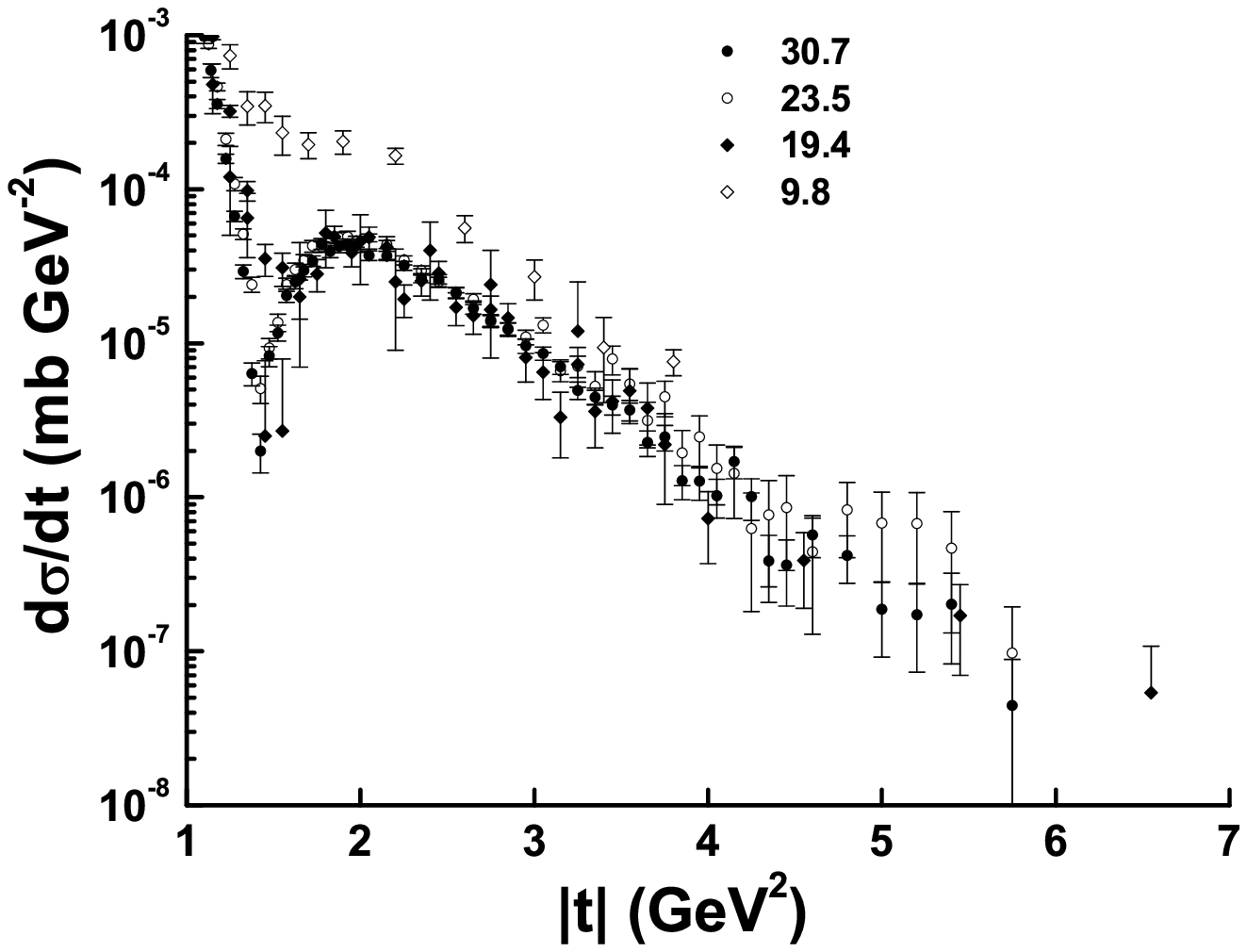}\includegraphics[width=50mm,height=40mm,clip]{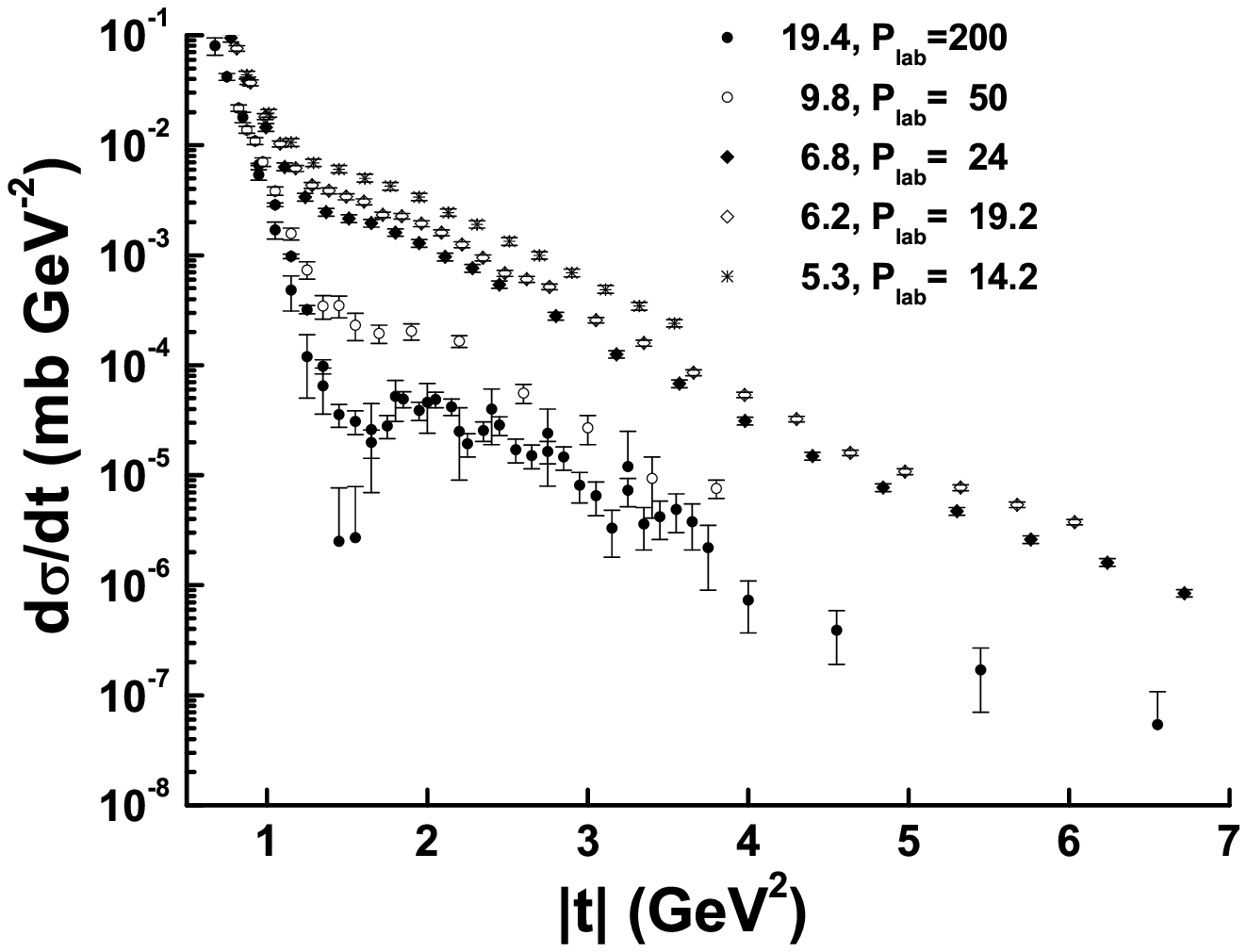}
\caption{$pp$ elastic scattering data at large momentum transfer. Points are experimental data
\protect{\cite{SS62},\cite{SS53},\cite{SS52_8},\cite{P200b},\cite{P50},\cite{P10},\cite{P19_20}}.}
\label{Fig10}
\end{figure}

Of course, our parameterization of the high momentum part is not perfect one. But at least, it
describes the previous experimental data, and we cannot simple disregard it. The behavior of the
predictions in Fig. 11 is explained by the variation of $R$. As energy increases, $R$ is increased
also, and the yield of the soft part in the high $|t|$ region decreases, the dip is shifted to the lower
$|t|$, and the second maximum increases. If we are right, the future measurement of the Totem collaboration
can show this. The measurement will give us more information about interplay of the soft and hard interactions.
\begin{figure}[cbth]
\includegraphics[width=160mm,height=100mm,clip]{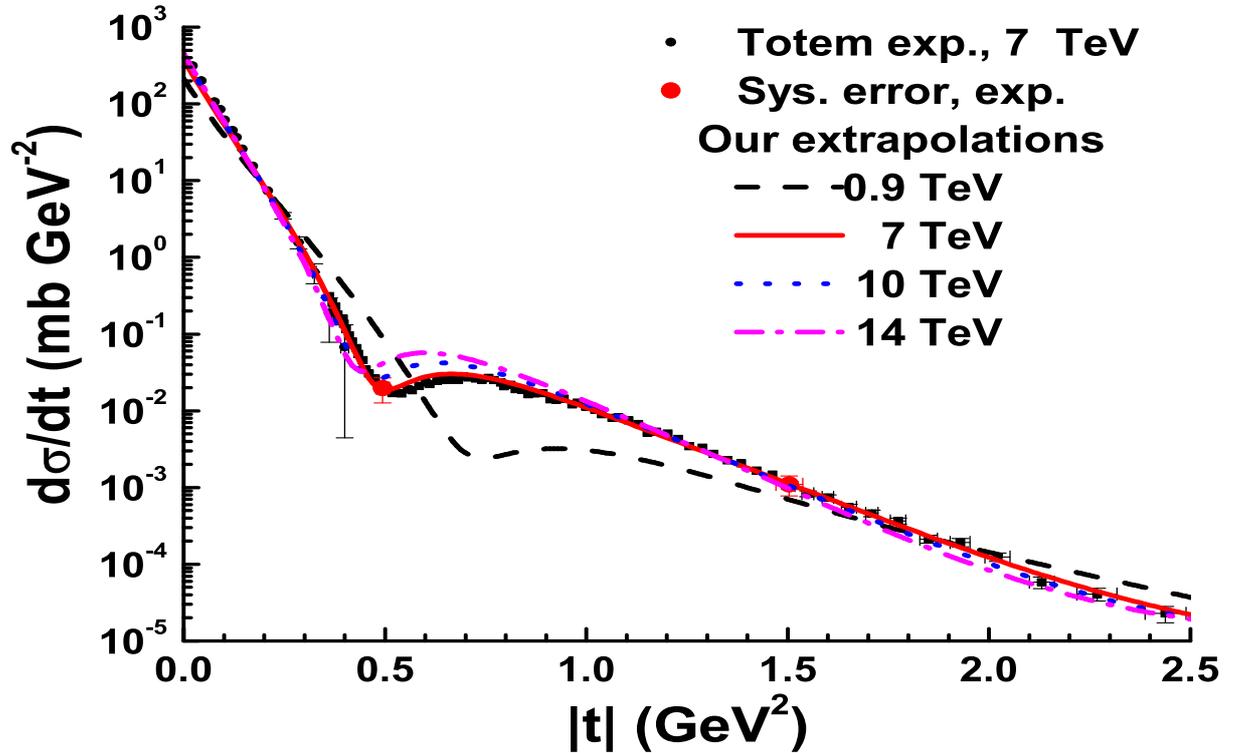}
\caption{Description of the Totem data and our predictions for other LHC energies.}
\label{Fig11}
\end{figure}

\section{Concluding remarks}
\subsection{Total and elastic cross sections}
According to Exps. (\ref{Eq3}) and (\ref{Eq9}) the total cross section is given by the expression:
\begin{equation}
\sigma_{tot}\simeq 2\pi\ A (R^2+\pi^2d^2/3).
\label{Eq22}
\end{equation}

\noindent For the elastic cross section we have the following expression an exactness of which is
about few percent.
\begin{equation}
\sigma_{el}\simeq \pi\ A^2 (R^2-19 R d/20 +\pi^2d^2/7).
\label{Eq23}
\end{equation}

\noindent They together with Exps. ({\ref{Eq14}) -- ({\ref{Eq17}) provide one with a good parameterization of
the cross sections.

\subsection{Eikonal representation}
The profile-function ({\ref{Eq10}) can be represented as an eikonal one:
\begin{equation}
\gamma (b)=1-\left[1- \frac{A}{1+e^{(b-R)/d}}\right]=
1- e^{-\ln[(e^{R/d}+e^{b/d})/(e^{b/d}-1)]}.
\label{Eq24}
\end{equation}
Here we use the assumption that $A=1+e^{-R/d}$.

It can be used in the Quark-Gluon String Model \cite{QGSM} for a calculations of string multiplicity
distributions.
\begin{equation}
P_n=\frac{1}{C}\int b \frac{\{2\ln[(e^{R/d}+e^{b/d})/(e^{b/d}-1)]\}^n}{n!}\ \
 e^{-2\ln[(e^{R/d}+e^{b/d})/(e^{b/d}-1)]}\ db.
\label{Eq25}
\end{equation}

\subsection{Application of USESD in calculations of hadron-nucleus and nucleus-nucleus properties.}
An amplitude for an elastic scattering of an nucleus containing $B$ baryons on a target
nucleus with mass number $A$ is given as \cite{Franco68}:
\begin{equation}
F_{BA}(\vec q) =\frac{i}{2\pi}\ \int d^2b\ e^{i\vec q \vec b}\
\left\{ 1-\prod^B_{i=1}\prod^A_{j=1}\left[ 1- \gamma(\vec b + \vec \tau_i - \vec s_j)\right]
\right\}|\Psi_B|^2 |\Psi_A|^2\ \left( \prod^B_{i=1}d^3\ t_i\right)\ \left( \prod^A_{j=1}d^3\ r_j\right)
\label{Eq26}
\end{equation}
$$
=i\ \int^{\infty}_0 bP_{BA}(b)\ J_0(qb)db,
$$
where $\gamma$ is an amplitude of an elastic nucleon-nucleon scattering in the impact
parameter representation, averaged over the spin and isospin degrees of freedom,
$$
\gamma (\vec b)=\frac{1}{2\pi i}\int d^2q\ e^{i\vec q \vec b} f(\vec q).
$$
$\Psi_A\ (\Psi_B)$ is the wave function of the target (projectile) nucleus in the ground state.
Taking the origin of the
coordinate system to coincide with the center of the nucleus, the nucleon coordinates
($\{ \vec r_A\}$, $\{ \vec t_B\}$) are decomposed into longitudinal (\{$z_i$\}) and transverse
($\{\vec s_j\}$, $\{\vec \tau_i\}$) components. The $z$-axis is directed along the projectile
momentum. $\vec b$ is the impact parameter vector orthogonal to the momentum. $P_{BA}(b)$
is the profile function and $J_0$ is the Bessel function of zero order.

Quite often $\gamma$ is parameterized as\footnote{The amplitude must be corrected at low
energies in order to take into account the unitarity requirement ($Re \gamma (0)\ \leq 1$)
and a restriction of the phase space.}:
\begin{equation}
\gamma(\vec b)=\frac{\sigma^{tot}_{NN}\ (1-i\rho)} {4\pi\ \beta}\ e^{-\vec b^2/2\beta},
\label{Eq27}
\end{equation}
where $\sigma^{tot}_{NN}$ is the total cross section of the nucleon-nucleon
interactions, $\rho$ is the ratio of the real to imaginary parts of the $NN$ elastic
scattering amplitude at zero momentum transfer, and $\beta$ is a slope parameter of the
$NN$ differential elastic scattering cross section. Then
\begin{equation}
f(\vec q)=\frac{i}{4\pi}\sigma^{tot}_{NN}\ (1-i\rho)\ e^{-\beta \vec q^2/2}, \ \ \
\label{Eq28}
\end{equation}

\noindent $\beta$ can be found as:
$$
\beta =(\sigma^{tot}_{NN})^2(1+|\rho|^2)/(16\ \pi\ \sigma^{el}_{NN}\ 0.3897).
$$
Here, $\sigma^{el}_{NN}$ is the $NN$ elastic cross section and $0.3897$ is a
coefficient required in order to express $\beta$ in units of $(GeV/c)^{-2}$, if the cross
sections are given in millibarns.

The squared modulus of the wave function is usually written as:
\begin{equation}
|\Psi_A|^2=\delta(\sum_{i=1}^A\vec r_i/A)\ \prod_{i=1}^A\rho_A(\vec r_i).
\label{Eq29}
\end{equation}
$\rho_A$ coincides with the one-particle density of the nucleus if one neglects the
center-of-mass correlation connected with the $\delta$ function.

We have used Exp. (\ref{Eq27}) at the calculation of the differential elastic scattering cross section
at $\sqrt{s}=7$ TeV, results of which are shown in Fig. 7 as the dashed line (Gaussian param.). It is
obvious, that it can describe the cross sections only at small value of $|t|$. It would be better
to use Exp. (\ref{Eq10}) for more exact calculations as nucleon-nucleon interaction properties, as well
as baryon-nucleus and nucleus-nucleus ones.

The task is very actual for calculations of wounded nucleon multiplicities in nucleus-nucleus interactions
at high energies.

\subsection{Comparison of USESD with other approaches}
Many years ago T.-Y. Cheng, S.-Y. Chu and A.W. Hendry \cite{CCH} successfully used 2-dimensional Fermi-function
to describe $pp$ elastic scattering at all angles from 3 to 24 GeV/c. They analyzed polarized proton
scattering by proton. Most probably is that they used numerical integration. We have used analytical
expressions, and we have considered only unpolarized proton scattering.

In 2000 M. Kawasaki, T. Maehara and M. Yonezawa analyzing the general structure of the elastic scattering
amplitude in the impact parameter representation proposed the following expression for the amplitude in
the momentum representation \cite{KMY1}:
\begin{equation}
Im\ f(q) \propto R^2\left[ \frac{J_1(Rq)}{Rq}\Phi_0(R,q)+ \frac{J_0(Rq)}{\mu q}\Phi_1(R,q)\right],
\label{Eq30}
\end{equation}
where $\mu$ is a parameter. They proposed also a concrete form of the dumping functions:
$$
\Phi_0(R,q)=Re\ \Gamma\left(1+i\frac{q}{\mu}\right), \ \ \
\Phi_1(R,q)=-Im\ \Gamma\left(1+i\frac{q}{\mu}\right).
$$

They were trying to fit the high energy $\bar pp$- and $pp$-data \cite{KMY2,KMY3} but results were not
impressive. Especially, in the paper \cite{KMY3} the authors fitted  the differential cross sections
of $pp$ elastic scattering at $\sqrt{s} \geq 23.5$ GeV and the $\bar pp$-data at $\sqrt{s} \geq 546$ GeV.
Only small momentum transfer region, $0.02 \leq |t| \leq 0.2$ (GeV/c)$^2$, was included in the fit.

In our paper we have considered much larger energy region and a larger region of the momentum transfer.
We assume our results as promising ones. Thus a choice of the dumping functions is very important for
a correct reproduction of experimental data.

P. Gauron, B. Nicolescu and E. Leader proposed in the paper \cite{CauronNicolescu} the following expressions for
asymptotic parts of (anti) proton-proton elastic scattering amplitude:
\begin{equation}
\frac{1}{is}F_+(s,t)=F_1\ ln^2(\bar s) \frac{2J_1(R_+\bar \tau)}{R_+\bar \tau}\ e^{b^+_1 t} +
                     F_2\ ln  (\bar s) J_0(R_+\bar \tau)\ e^{b^+_2 t}                         +
\label{Eq31}
\end{equation}
$$
                  F_3\left[ J_0(R_+\bar \tau) - R_+\bar \tau\ J_1(R_+\bar \tau)\right]\ \ e^{b^+_3 t},
$$

\begin{equation}
\frac{1}{s}F_-(s,t)=O_1\ ln^2(\bar s) \frac{sin(R_-\bar \tau)}{R_+\bar \tau}\ e^{b^-_1 t} +
                    O_2\ ln  (\bar s) cos(R_-\bar \tau)\ e^{b^-_2 t} + O_3 \ e^{b^-_3 t},
\label{Eq32}
\end{equation}

\noindent where $\bar s=\frac{s}{s_0}e^{-i\pi /2}$; $\bar \tau = \left(-\frac{t}{t_0}\right)^{1/2} ln(\bar s)$;
$s_0=t_0=1$ GeV$^2$; $F_i$, $O_i$, $b_i$, $R_+$ and $R_-$ are constants.
$$
F_{pp}=F_+\ + \ F_-, \ \ \ F_{\bar pp}=F_+\ - \ F_-.
$$

The appearance of the exponents in the expressions is connected with the simplest assumption about
the functional form of the residue functions. If one replaces the exponents by
$\pi d \sqrt{-t}/sinh(\pi d \sqrt{-t})$ then the leading term of $F_+$ will be coincided with
the first term of our expression (\ref{Eq11}). Thus our approach corresponds to the approach
of the authors at defined assumptions on the residue functions. But instead of the authors we did not
assume any energy dependence of our parameter $R$. We obtained it at the fitting of the experimental
data.

If we suppose that $R=R_0+\Delta R$ where $\Delta R \propto s^{-\alpha_R}$, and expand
the first term of Eq. \ref{Eq11} we will have yields of "non-dominant Regge pole contributions"
into the scattering amplitude.

Very often in a Regge-like analysis\footnote{A typical successfully Regge analysis of $pp$ elastic
scattering and polarization at $P_{lab}=$ 3 -- 50 GeV/c see in \protect{\cite{Sibirtsev}}.}
the elastic scattering amplitude of a process $1+2\rightarrow 1+2$
is represented as:
\begin{equation}
\gamma(b)=\frac{1}{C_1C_2}(1-e^{-\omega(b)}),
\label{Eq33}
\end{equation}
where
\begin{equation}
\omega(b)=\frac{\eta C_1 C_2 g_1(0) g_2(0)}{R_1^2+R_2^2+\alpha '\xi '}\
e^{\Delta \xi}\ e^{-b^2/4(R_1^2+R_2^2+\alpha '\xi ')},
\label{Eq34}
\end{equation}
$\eta=1+icotan(\pi \alpha_P(0)/2)$ is a signature factor; $C_1$ ($C_2$) -- shower enhancement
coefficient in the interaction vertex of the first (second) particle with a reggeon/pomeron;
$\alpha_P(0)=1+\Delta$ is an intercept of the reggeon/pomeron; $\xi=ln\ s_{12}$ is a logarithm
of CMS energy squared; $\xi '=\xi -i\pi/2$. It is assumed that Regge  trajectories are linear,
$\alpha(t)=1+\Delta+\xi 't$. Non-linear trajectories were considered in \cite{Godizov}.
It is assumed also that the residue functions have the gaussian shape -- $g_i(t)=g_i(0)exp(R^2_i\ t)$,
for simplicity.

It is complicated to find a correspondence between the eikonal \ref{Eq34} and our eikonal \ref{Eq24}.
Though, the structure of the eikonal \ref{Eq34} is rather simple -- it is a product of a function of $b$,
and a factor strongly dependented on the energy -- $e^{\Delta \xi}$. For our eikonal at large $b$ and
$C_1=C_2=1$, we have:
$$
\omega(b)=\ln[(e^{R/d}+e^{b/d})/(e^{b/d}-1)]|_{b\rightarrow \infty}\simeq e^{-b/d}[e^{R/d}+1].
$$
Thus the energy dependence of the eikonal is determined by $e^{R/d}\simeq s^{0.1/d}\simeq s^{0.28}$.
Here we take into account that $R\propto 0.1\ln s$, and $d\propto 0.36$. So, an effective intercept
in our model is 1.28. It is in a correspondence with results of papers \cite{Godizov,Petrov}.

Summing up, we can say that our model is in the main stream of phenomenological analysis of the
elastic scattering data. The model assumes the defined choice of the dumping functions, or the
residue functions. A correct form of the functions is very important for high energy phenomenology.

\vspace{5mm}
The authors are thankful to the Geant4 hadronic working group for interest in the work.

\noindent{\bf Appendix A: Comparison of experimental data on $pp$-interactions with USESD parameterization}

\begin{figure}[cbth]
\includegraphics[width=75mm,height=70mm,clip]{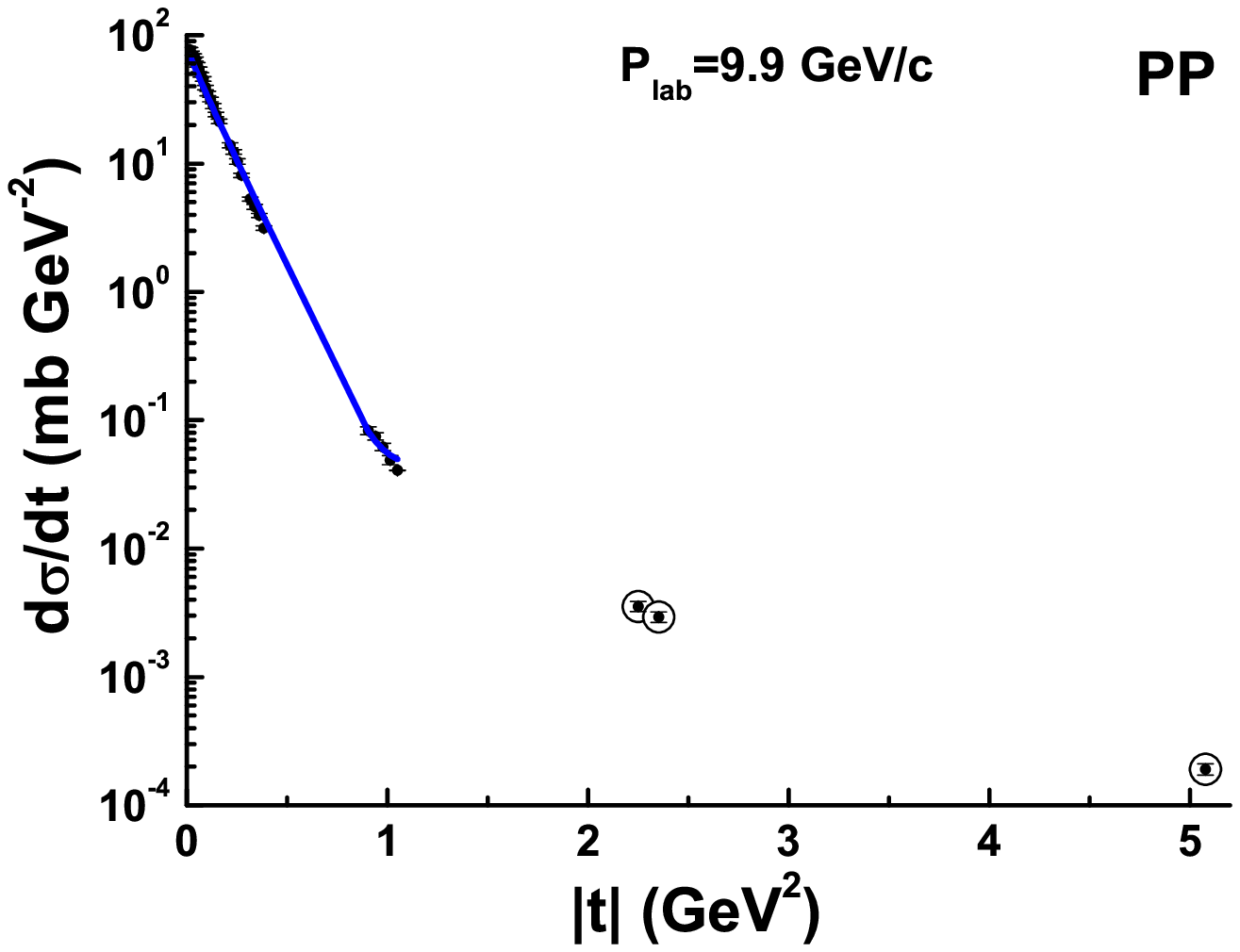}\hspace{5mm}\includegraphics[width=75mm,height=70mm,clip]{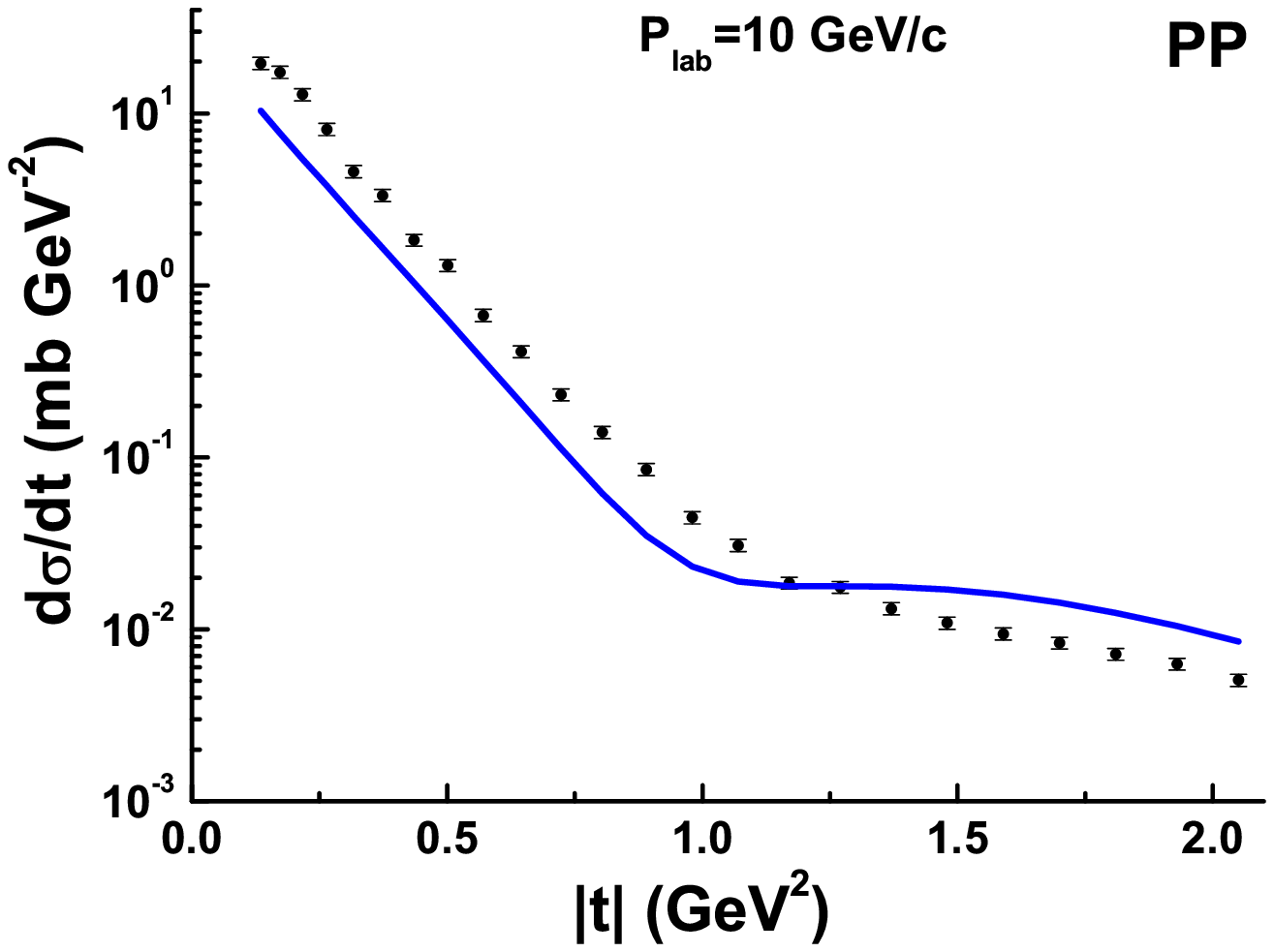}
\begin{minipage}{75mm}
{
\caption{The points are the experimental data by R.M. Edelstein et al., Phys. Rev. {\bf D5} (1972) 1073.}
}
\end{minipage}
\hspace{5mm}
\begin{minipage}{75mm}
{
\caption{The points are the experimental data by J.V. Allaby et al., Nucl. Phys. {\bf B52} (1973) 316.}
}
\end{minipage}
\includegraphics[width=75mm,height=70mm,clip]{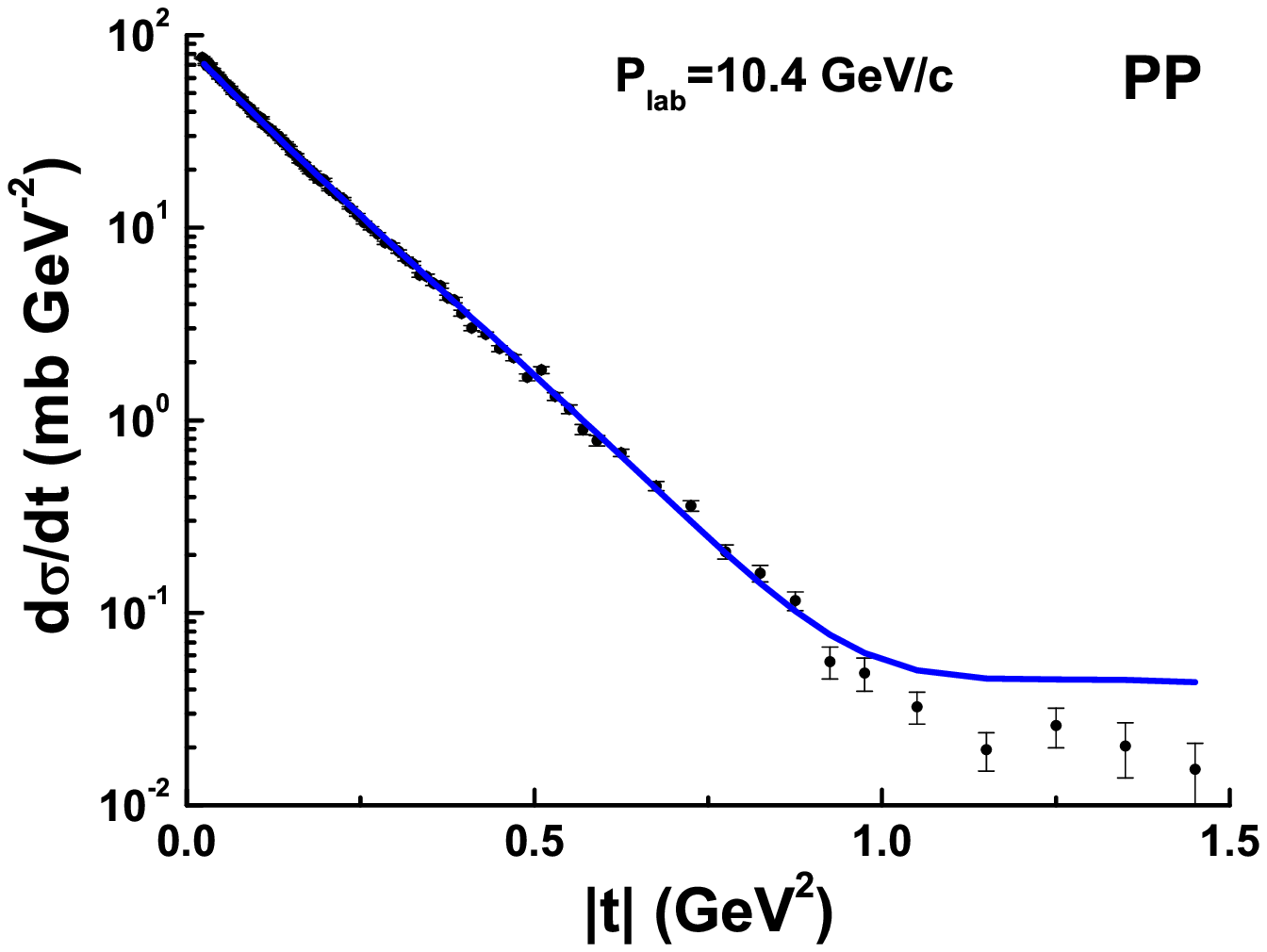}\hspace{5mm}\includegraphics[width=75mm,height=70mm,clip]{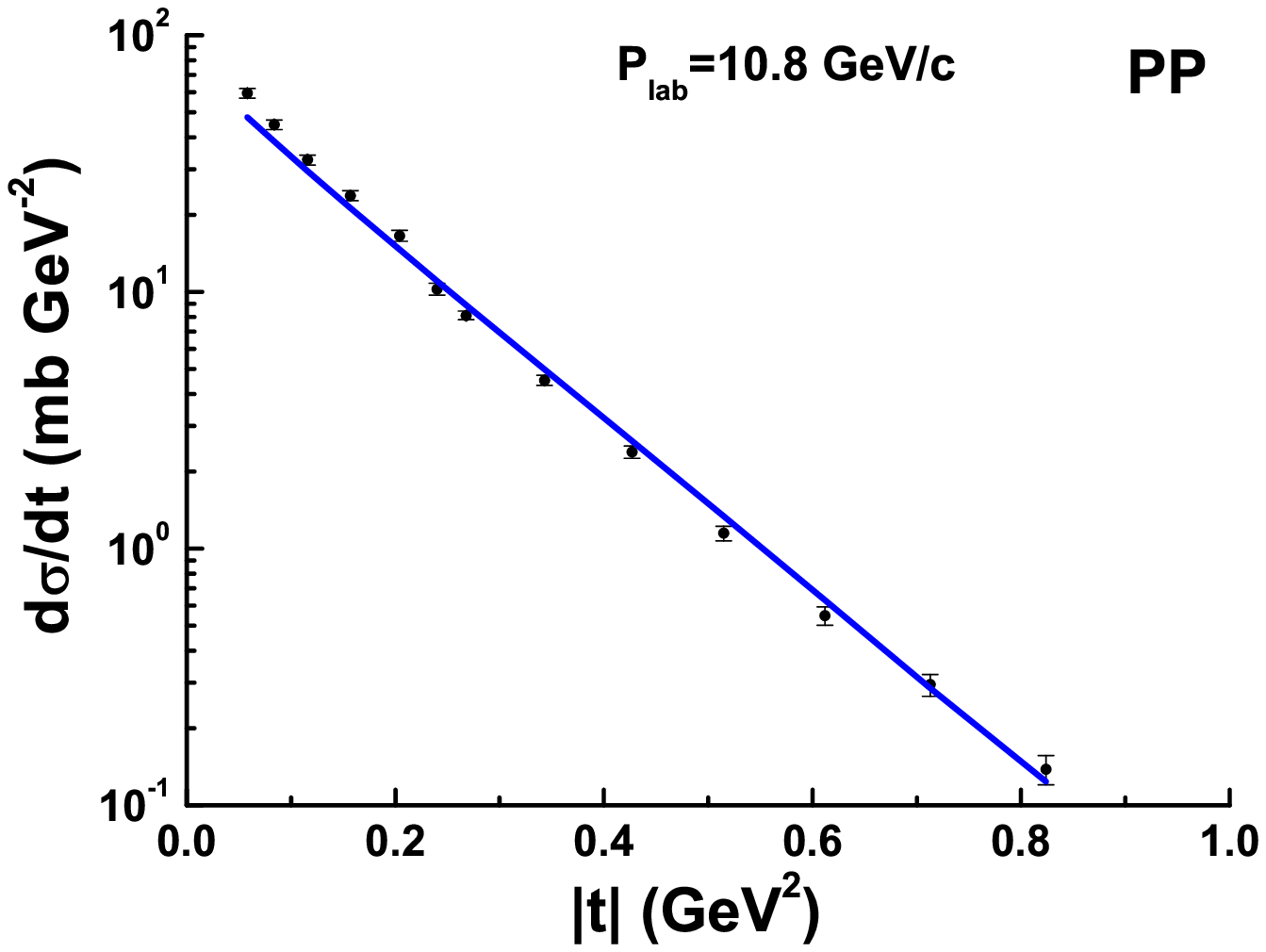}
\begin{minipage}{75mm}
{
\caption{The points are the experimental data by G.W. Brandenburg et al., Phys. Lett. {\bf 58B} (1975) 367.}
}
\end{minipage}
\hspace{5mm}
\begin{minipage}{75mm}
{
\caption{The points are the experimental data by K.J. Foley et al., Phys. Rev. Lett. {\bf 11} (1963) 425.}
}
\end{minipage}

\end{figure}

\begin{figure}[cbth]

\includegraphics[width=75mm,height=66mm,clip]{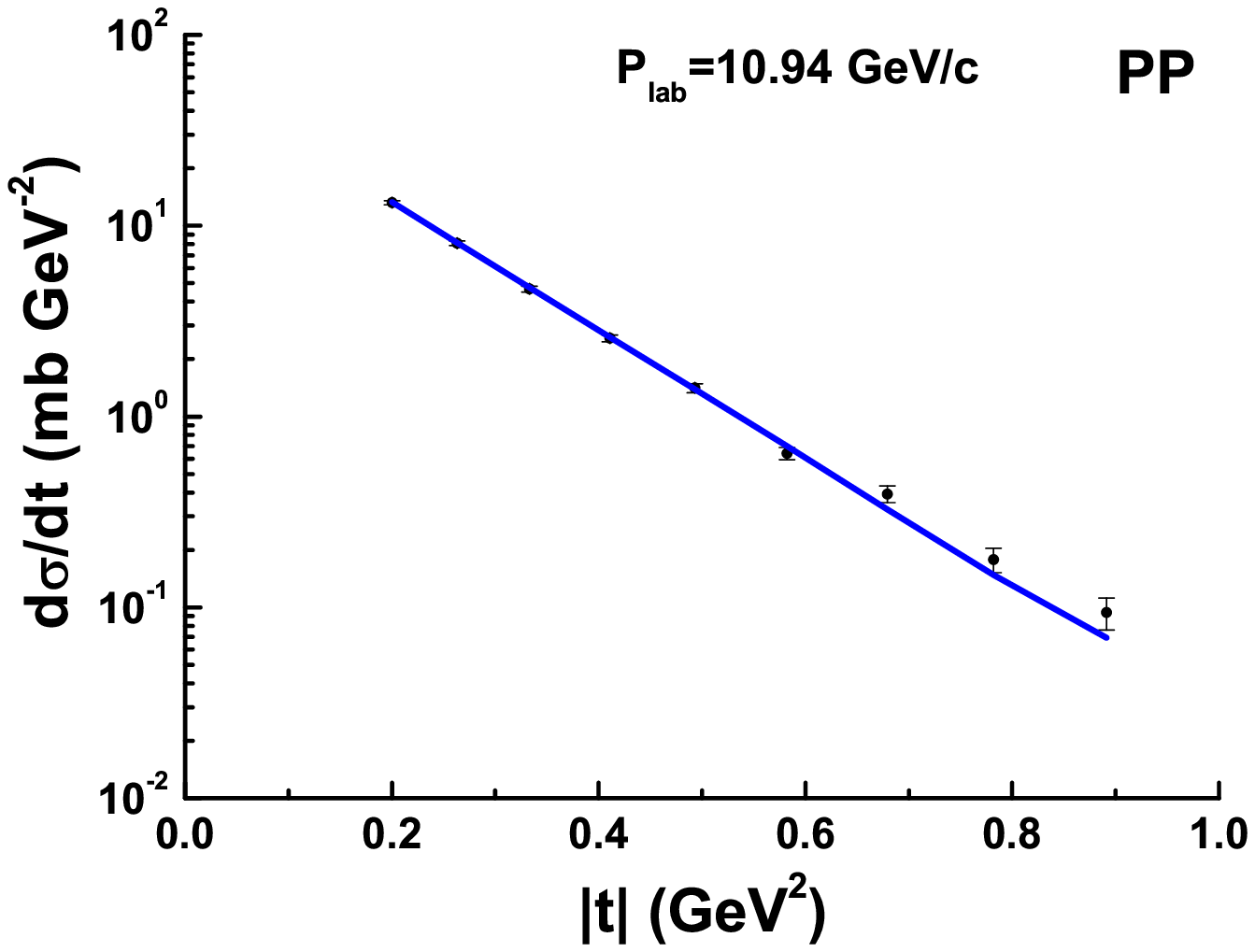}\hspace{5mm}\includegraphics[width=75mm,height=66mm,clip]{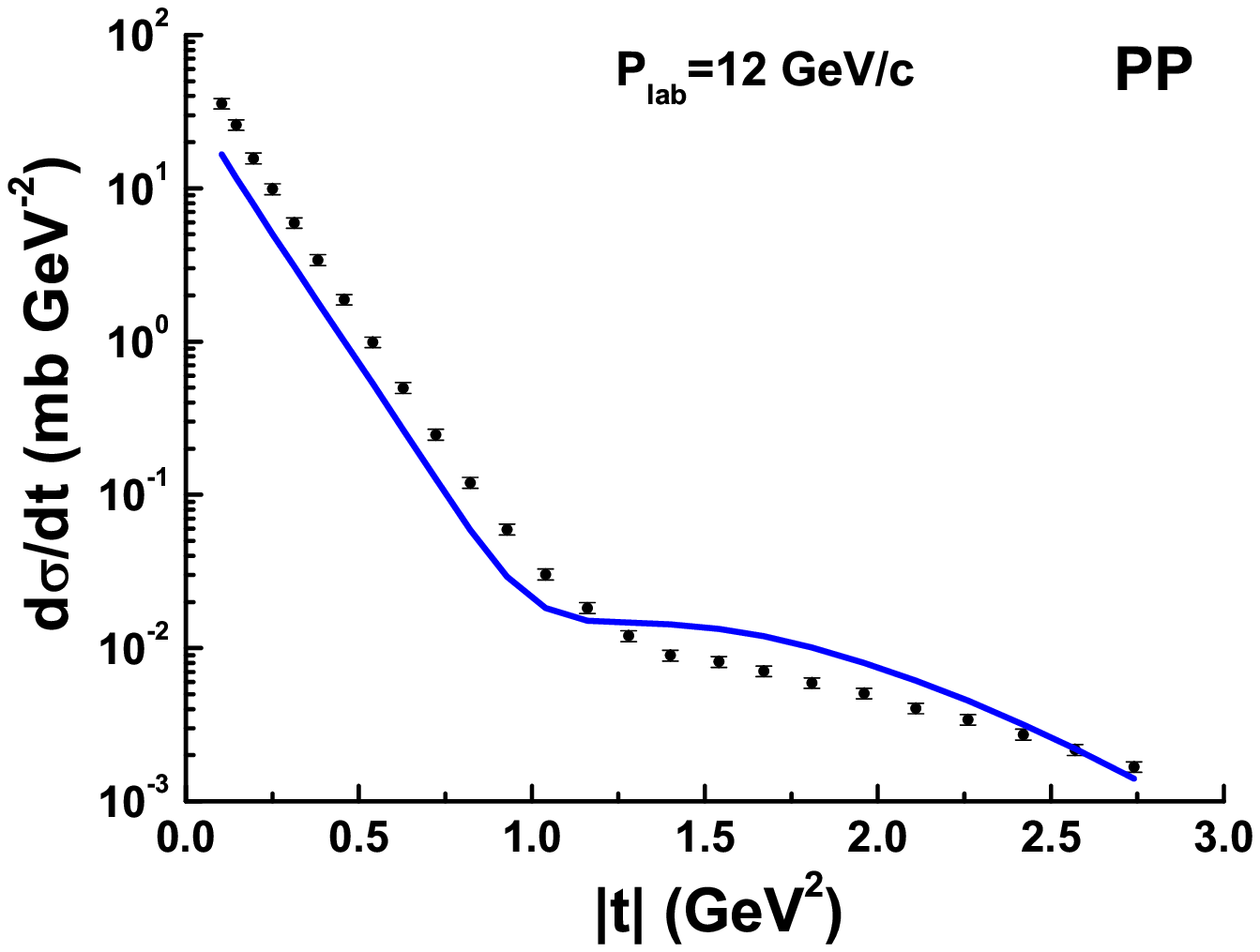}
\begin{minipage}{75mm}
{
\caption{The points are the experimental data by K.J. Foley et al., Phys. Rev. Lett. {\bf 15} (1965) 45.}
}
\end{minipage}
\hspace{5mm}
\begin{minipage}{75mm}
{
\caption{The points are the experimental data by J.V. Allaby et al., Nucl. Phys. {\bf B52} (1973) 316.}
}
\end{minipage}
\includegraphics[width=75mm,height=66mm,clip]{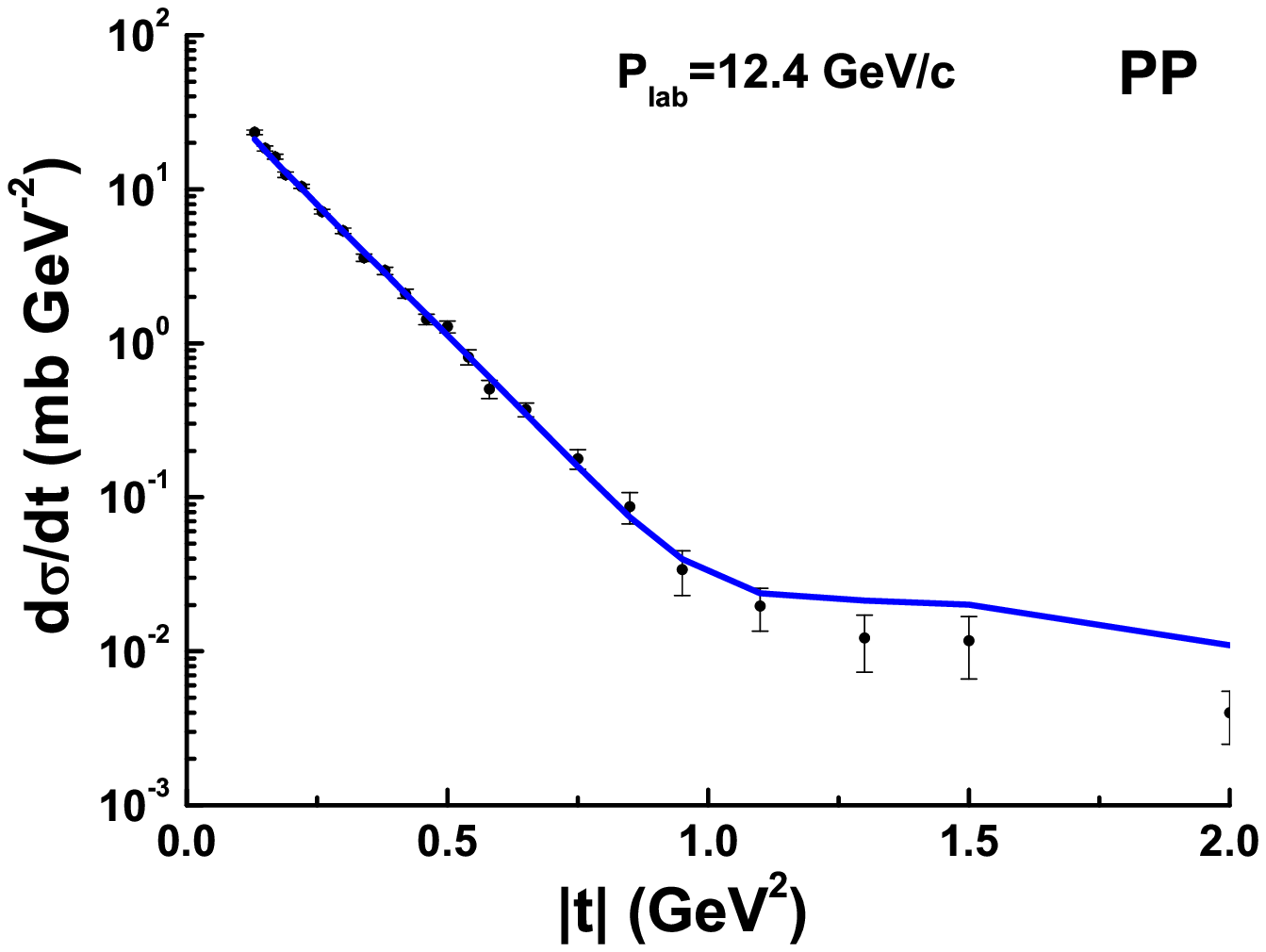}\hspace{5mm}\includegraphics[width=75mm,height=66mm,clip]{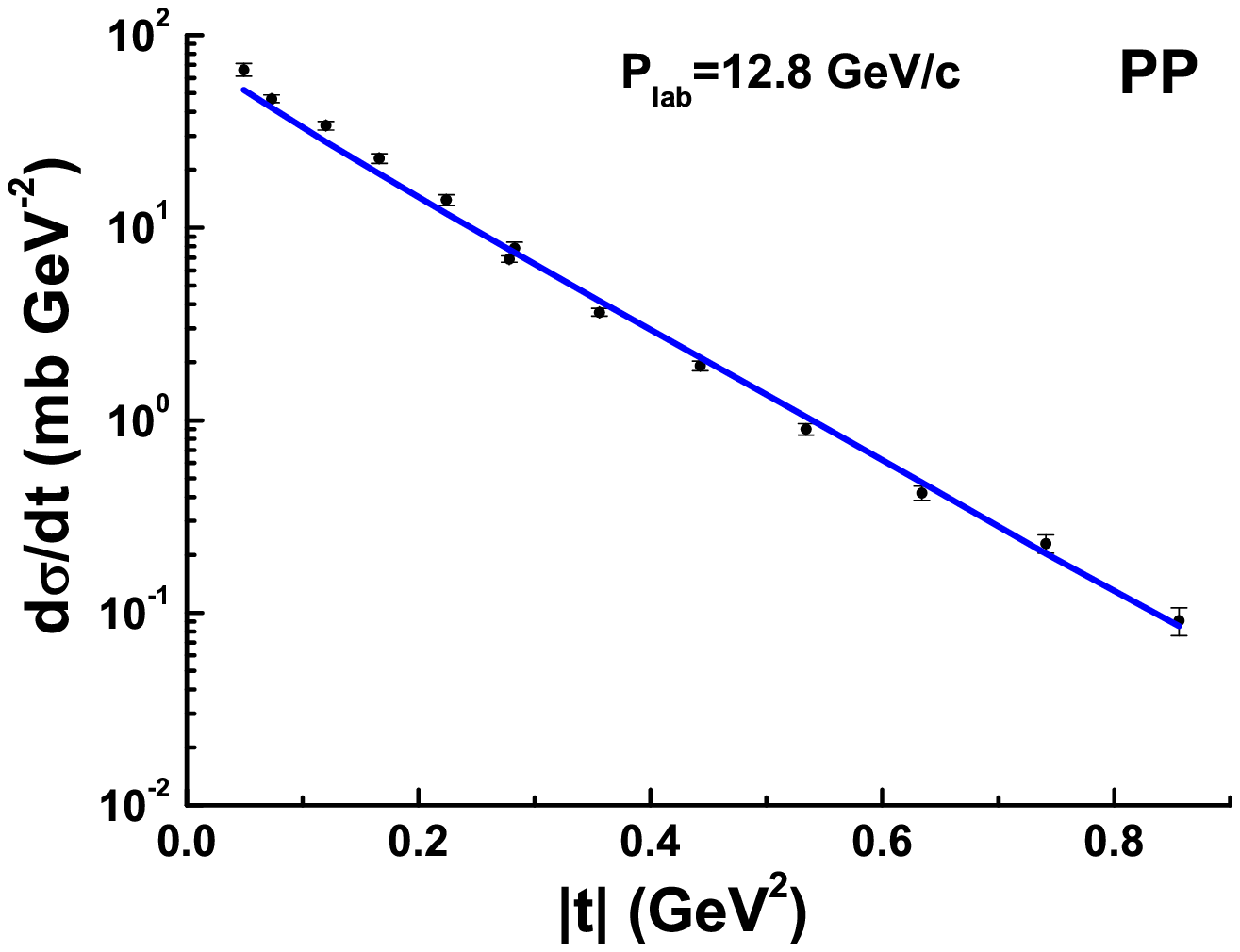}
\begin{minipage}{75mm}
{
\caption{The points are the experimental data by D. Harting, Nuovo Cimento {\bf 38} (1965) 60.}
}
\end{minipage}
\hspace{5mm}
\begin{minipage}{75mm}
{
\caption{The points are the experimental data by K.J. Foley et al., Phys. Rev. Lett. {\bf 11} (1963) 425.}
}
\end{minipage}
\includegraphics[width=75mm,height=66mm,clip]{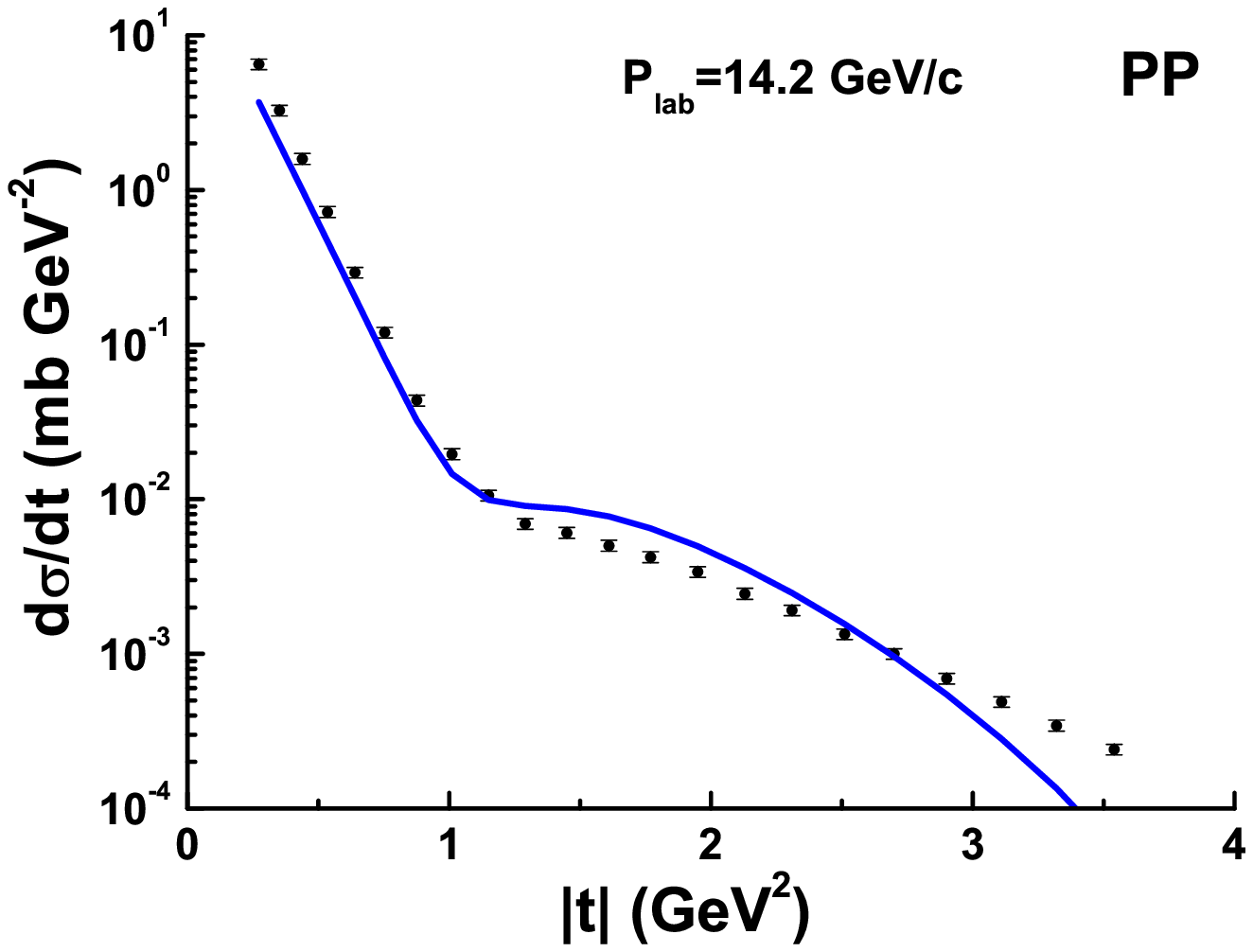}\hspace{5mm}\includegraphics[width=75mm,height=66mm,clip]{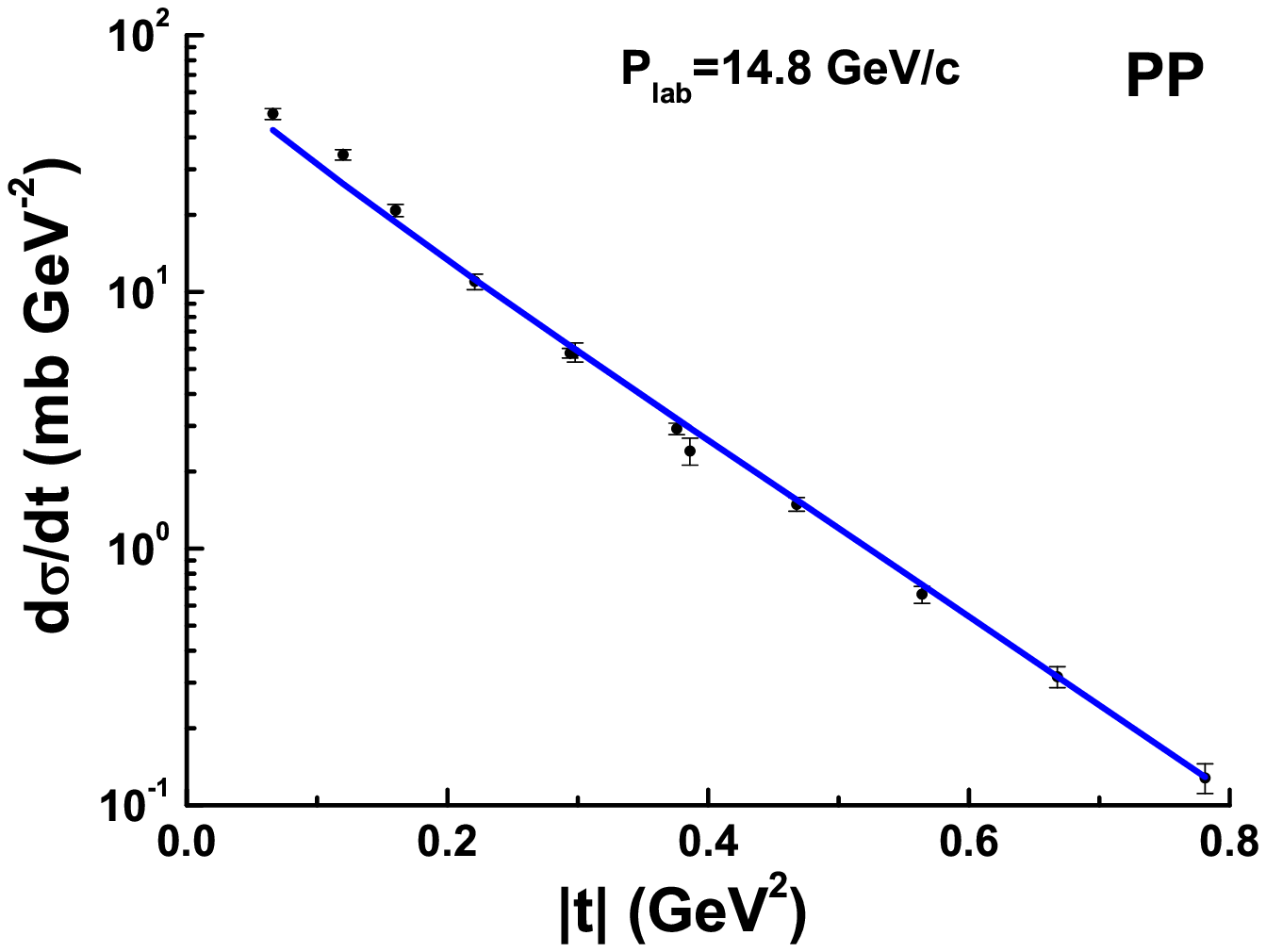}
\begin{minipage}{75mm}
{
\caption{The points are the experimental data by J.V. Allaby et al., Nucl. Phys. {\bf B52} (1973) 316.}
}
\end{minipage}
\hspace{5mm}
\begin{minipage}{75mm}
{
\caption{The points are the experimental data by K.J. Foley et al., Phys. Rev. Lett. {\bf 11} (1963) 425.}
}
\end{minipage}
\end{figure}

\begin{figure}[cbth]
\includegraphics[width=75mm,height=66mm,clip]{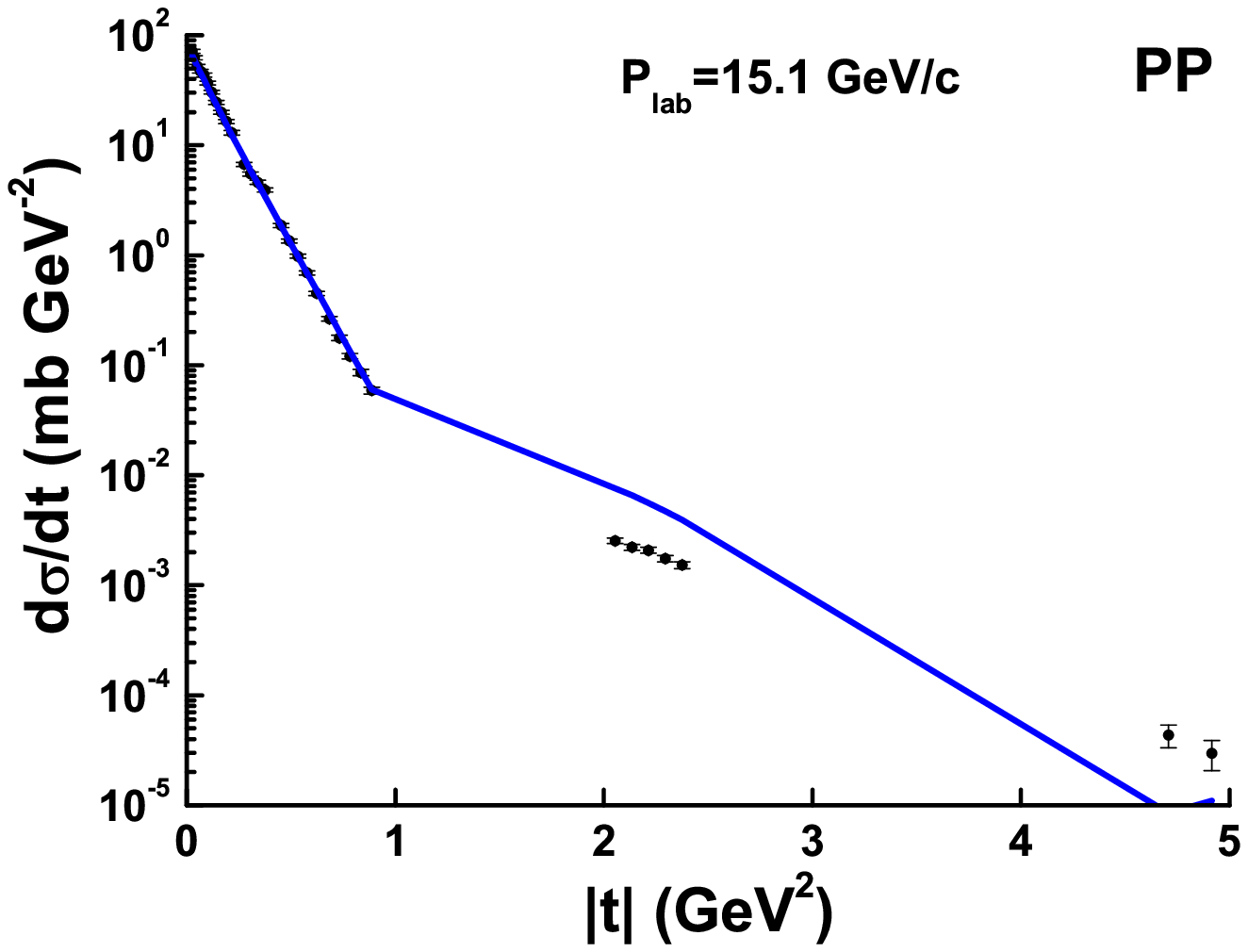}\hspace{5mm}\includegraphics[width=75mm,height=66mm,clip]{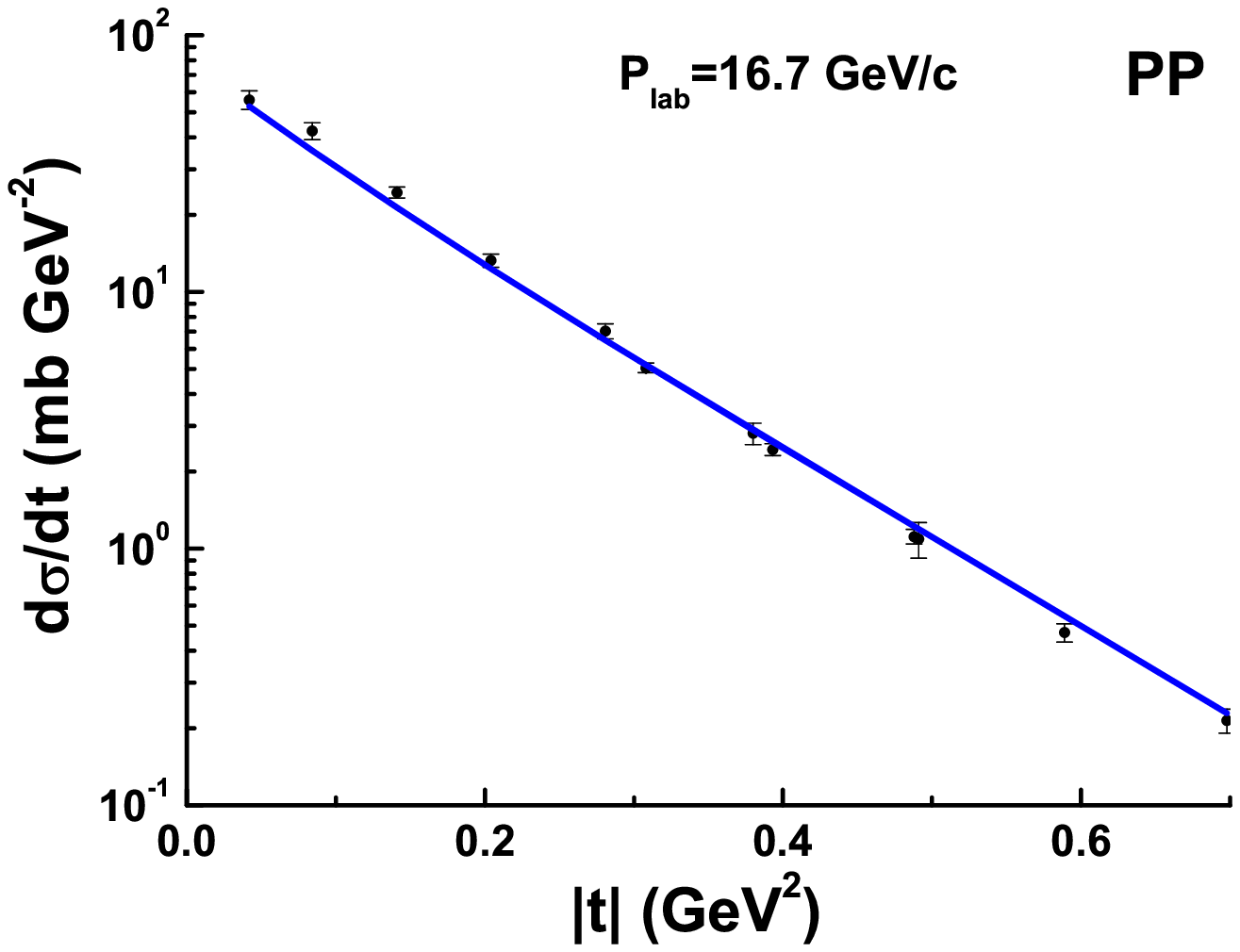}
\begin{minipage}{75mm}
{
\caption{The points are the experimental data by R.M. Edelstein et al., Phys. Rev. {\bf D5} (1972) 1073.}
}
\end{minipage}
\hspace{5mm}
\begin{minipage}{75mm}
{
\caption{The points are the experimental data by .K.J. Foley et al., Phys. Rev. Lett. {\bf 11} (1963) 425}
}
\end{minipage}
\includegraphics[width=75mm,height=66mm,clip]{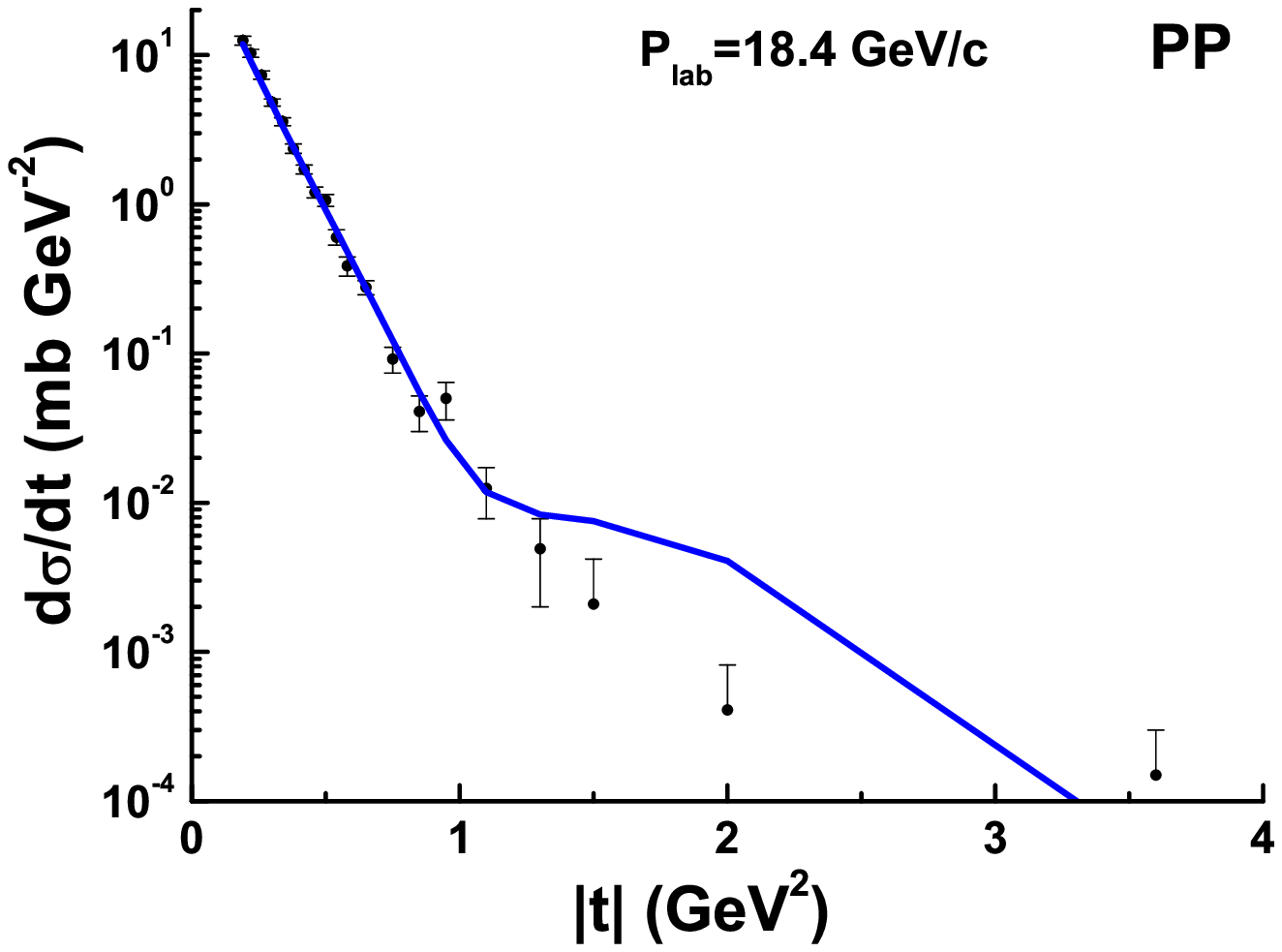}\hspace{5mm}\includegraphics[width=75mm,height=66mm,clip]{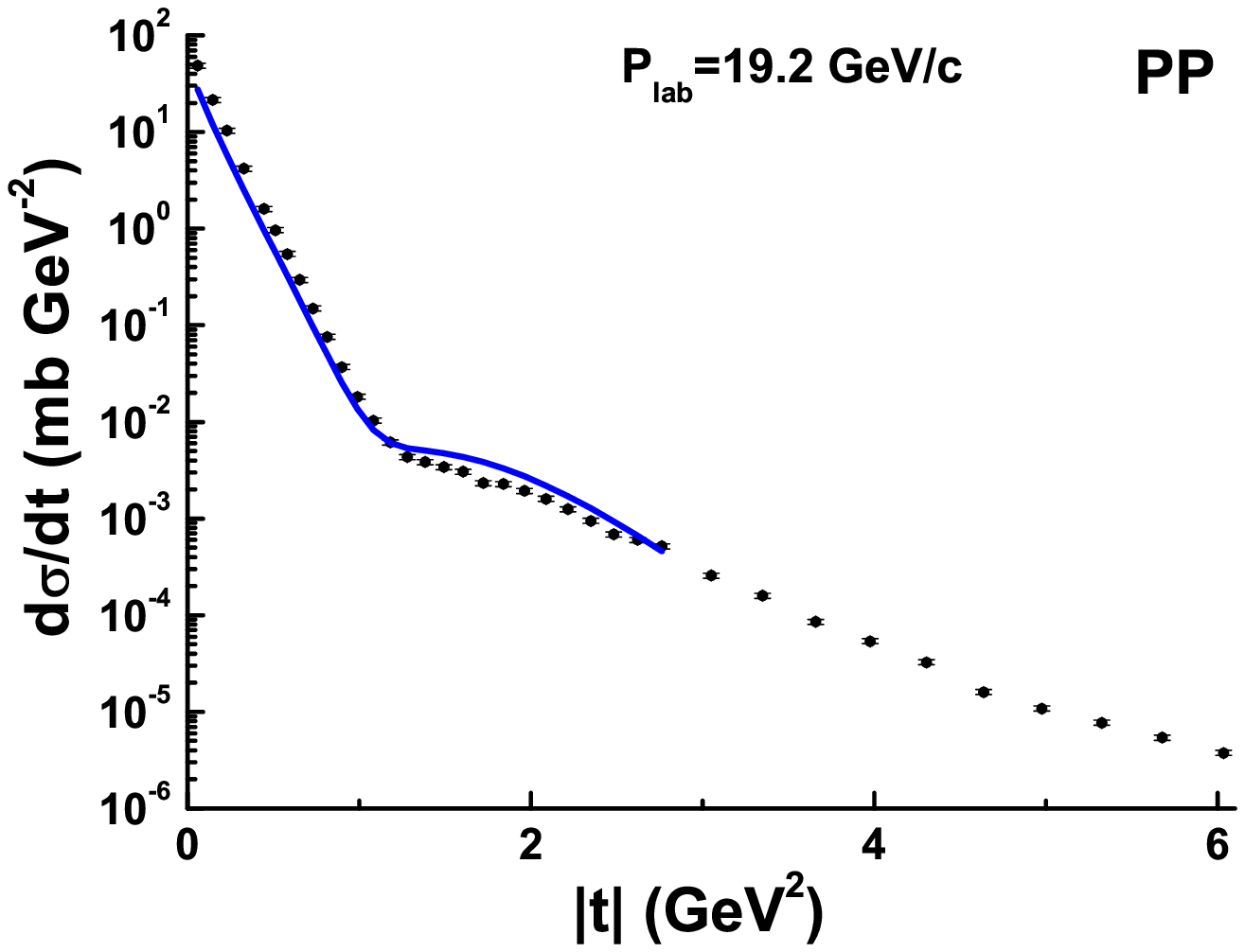}
\begin{minipage}{75mm}
{
\caption{The points are the experimental data by D. Harting, Nuovo Cimento {\bf 38} (1965) 60.}
}
\end{minipage}
\hspace{5mm}
\begin{minipage}{75mm}
{
\caption{The points are the experimental data by J.V. Allaby et al., Phys. Lett. {\bf B28} (1968) 67.}
}
\end{minipage}
\includegraphics[width=75mm,height=66mm,clip]{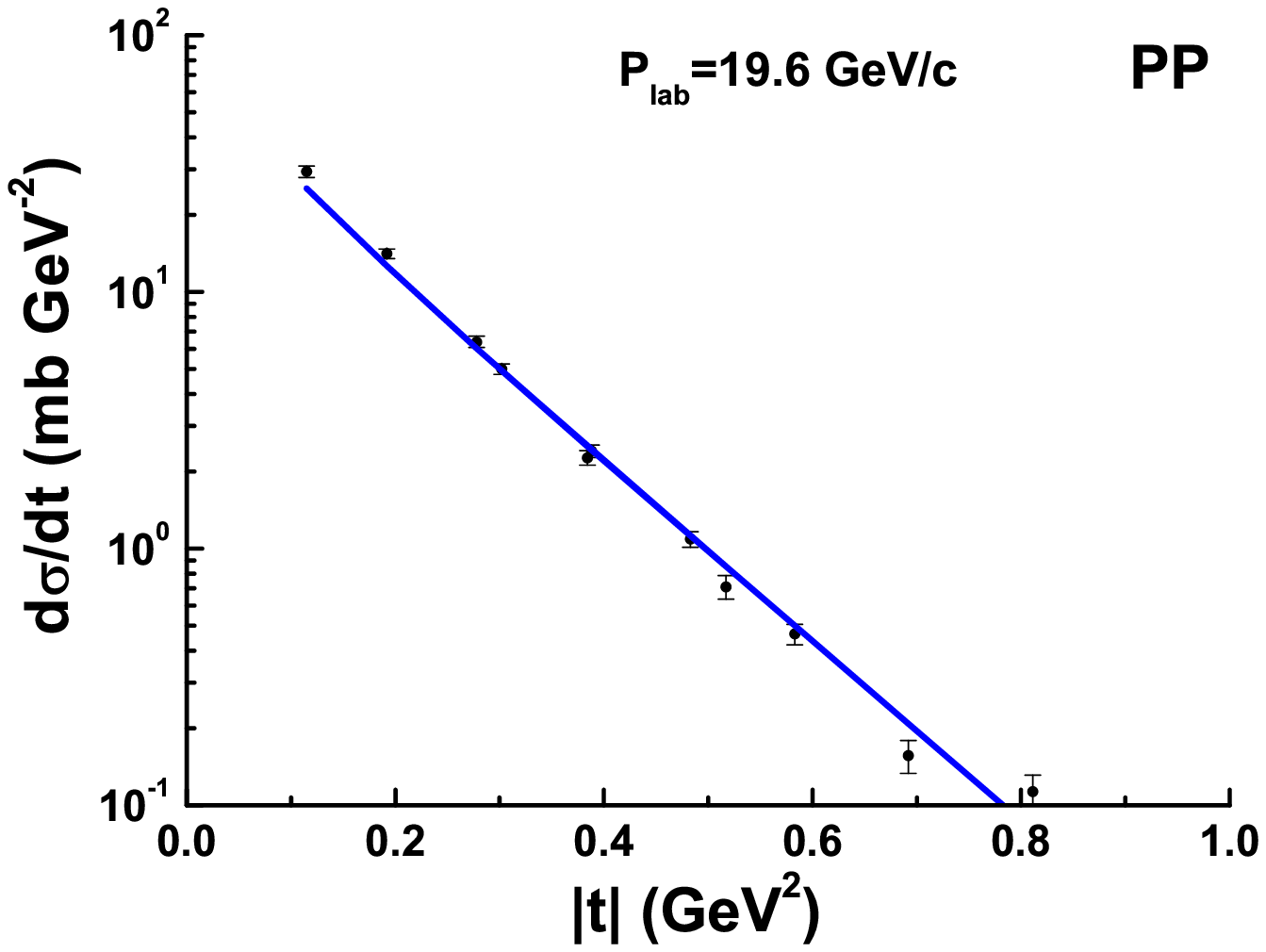}\hspace{5mm}\includegraphics[width=75mm,height=66mm,clip]{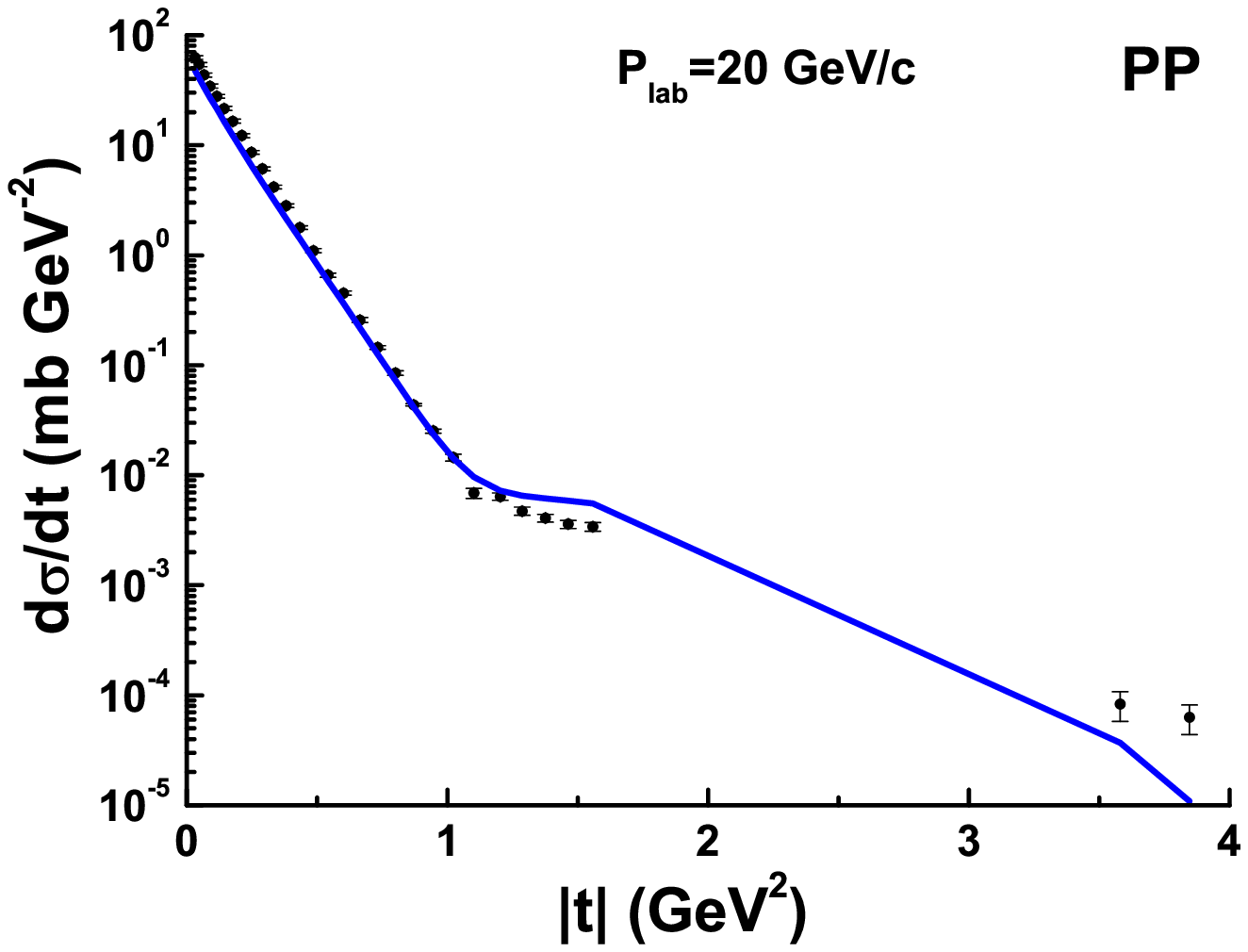}
\begin{minipage}{75mm}
{
\caption{The points are the experimental data by .K.J. Foley et al., Phys. Rev. Lett. {\bf 11} (1963) 425}
}
\end{minipage}
\hspace{5mm}
\begin{minipage}{75mm}
{
\caption{The points are the experimental data by R.M. Edelstein et al., Phys. Rev. {\bf D5} (1972) 1073.}
}
\end{minipage}
\end{figure}

\begin{figure}[cbth]
\includegraphics[width=75mm,height=66mm,clip]{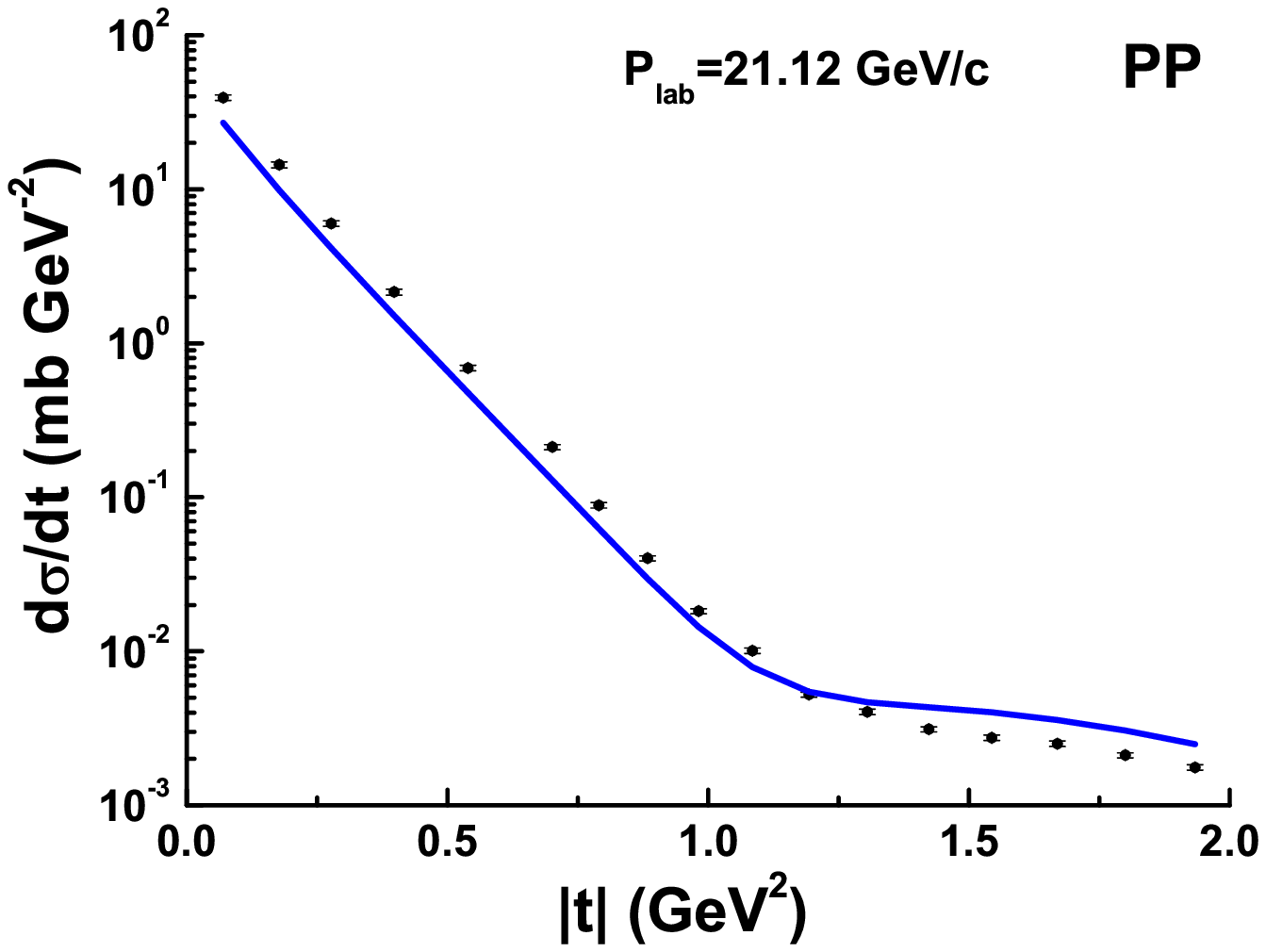}\hspace{5mm}\includegraphics[width=75mm,height=66mm,clip]{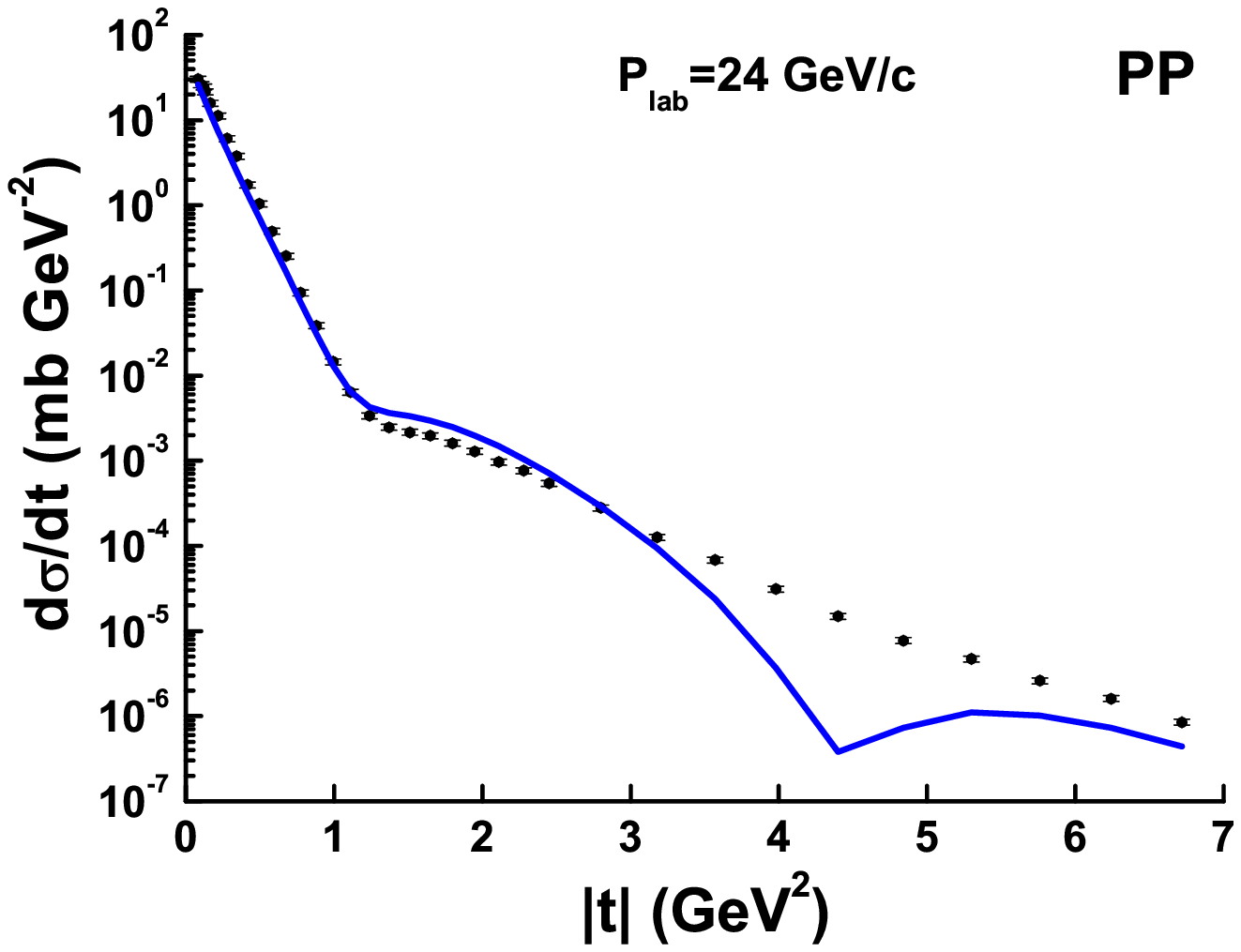}
\begin{minipage}{75mm}
{
\caption{The points are the experimental data by J.V. Allaby et al., Phys. Lett. {\bf B28} (1968) 67.}
}
\end{minipage}
\hspace{5mm}
\begin{minipage}{75mm}
{
\caption{The points are the experimental data by J.V. Allaby et al., Nucl. Phys. {\bf B52} (1973) 316.}
}
\end{minipage}
\includegraphics[width=75mm,height=66mm,clip]{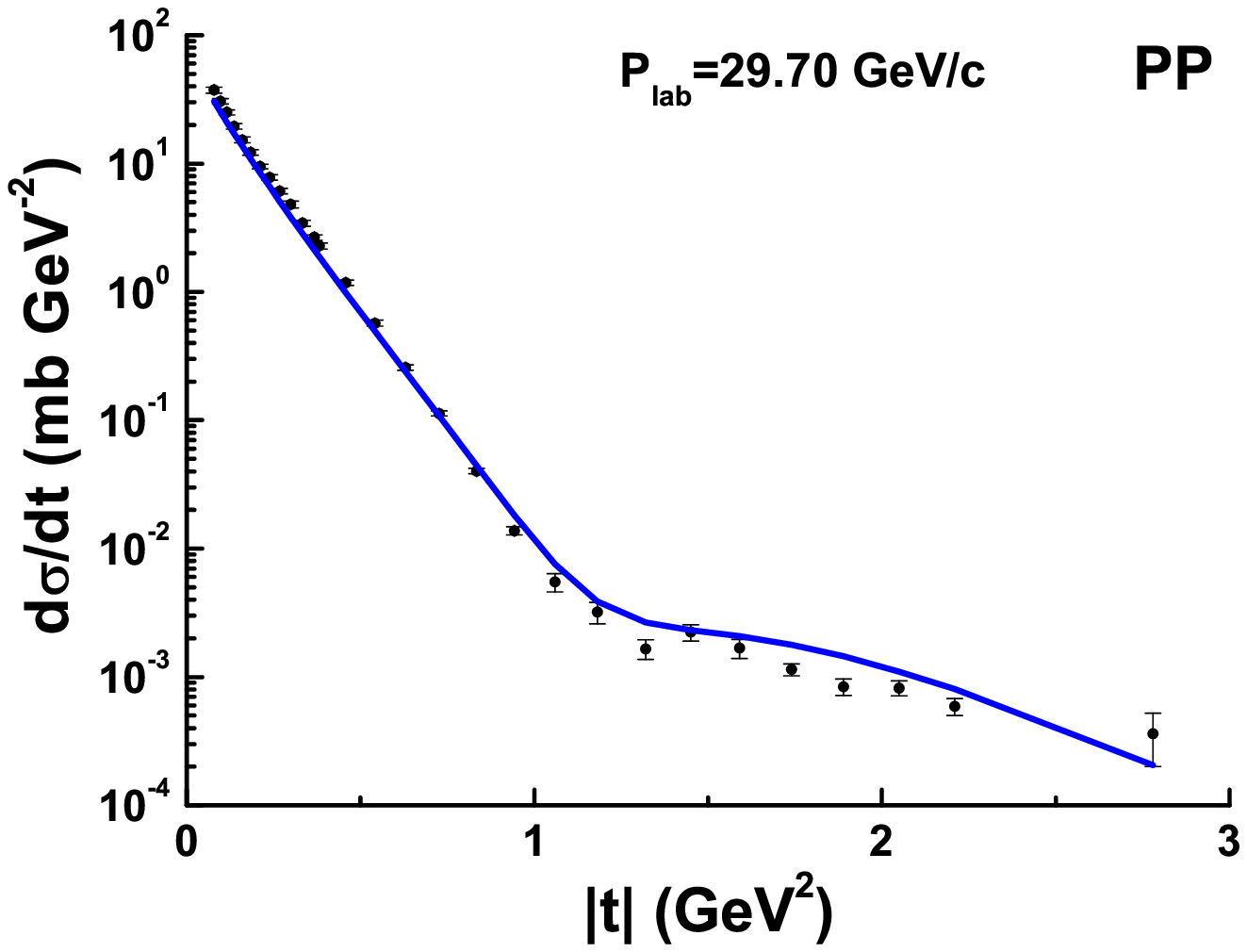}\hspace{5mm}\includegraphics[width=75mm,height=66mm,clip]{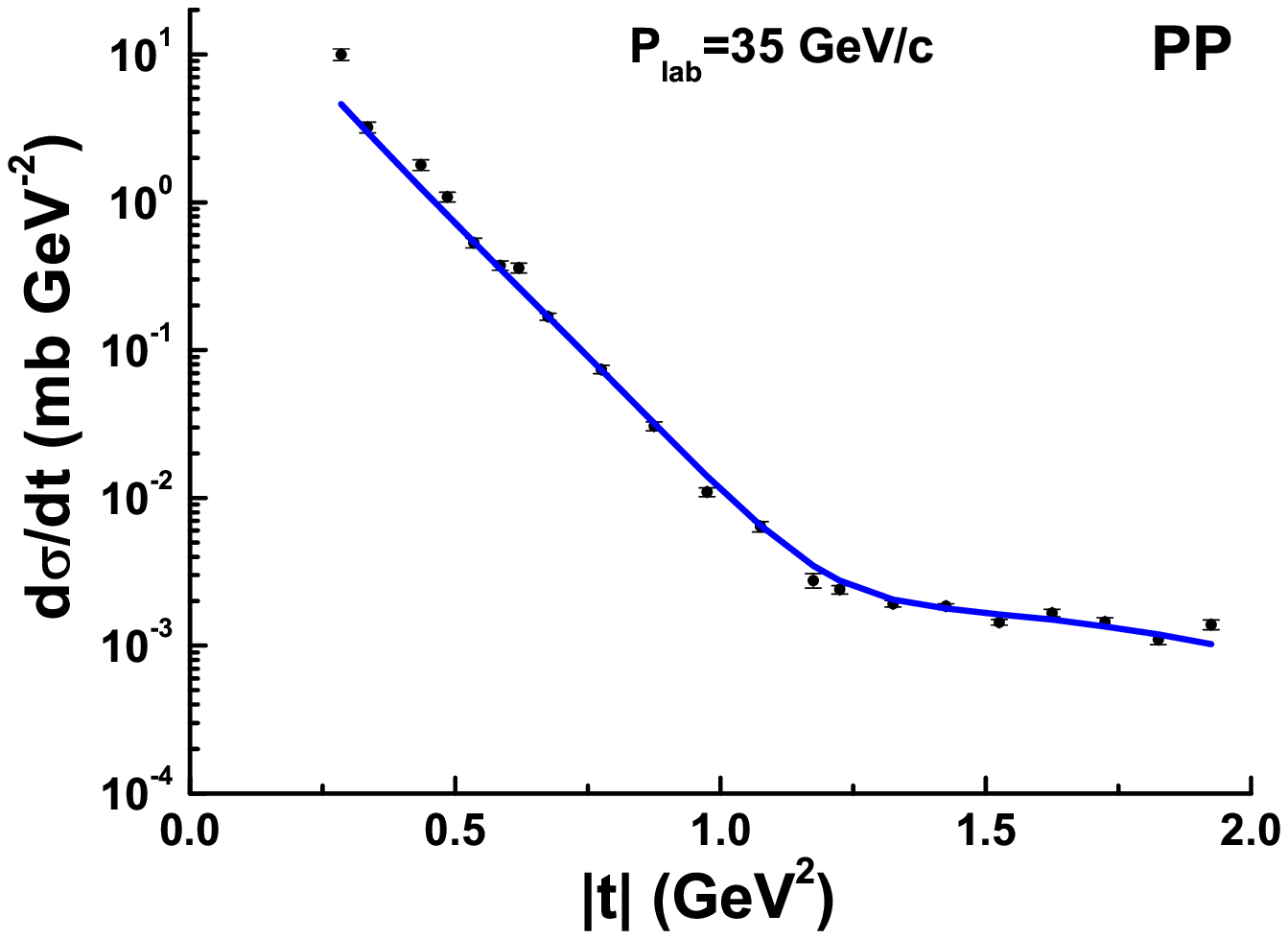}
\begin{minipage}{75mm}
{
\caption{The points are the experimental data by R.M. Edelstein et al., Phys. Rev. {\bf D5} (1972) 1073.}
}
\end{minipage}
\hspace{5mm}
\begin{minipage}{75mm}
{
\caption{The points are the experimental data by R. Rusack et al., Phys. Rev. Lett. {\bf 41} (1978) 1632.}
}
\end{minipage}
\includegraphics[width=75mm,height=66mm,clip]{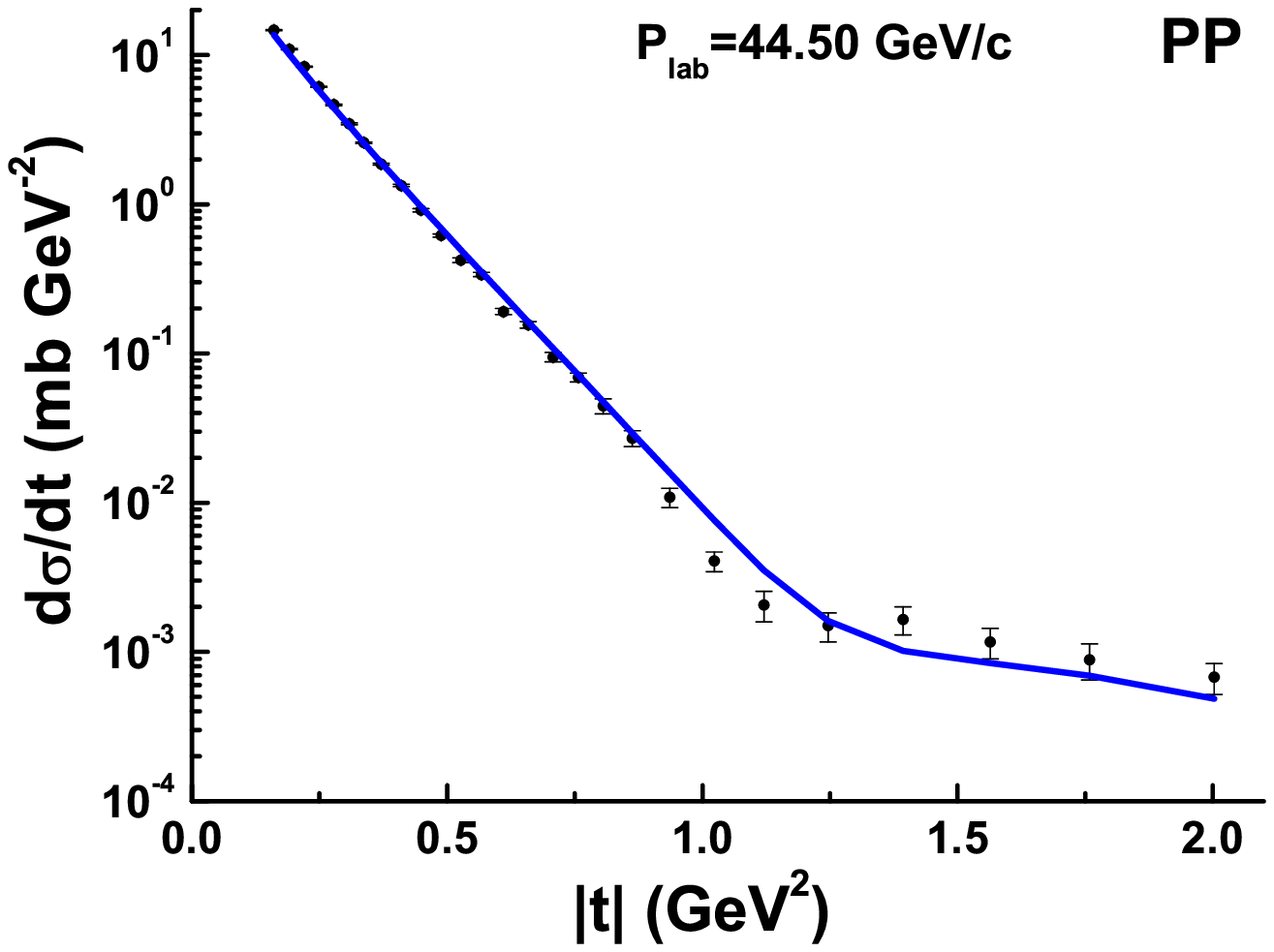}\hspace{5mm}\includegraphics[width=75mm,height=66mm,clip]{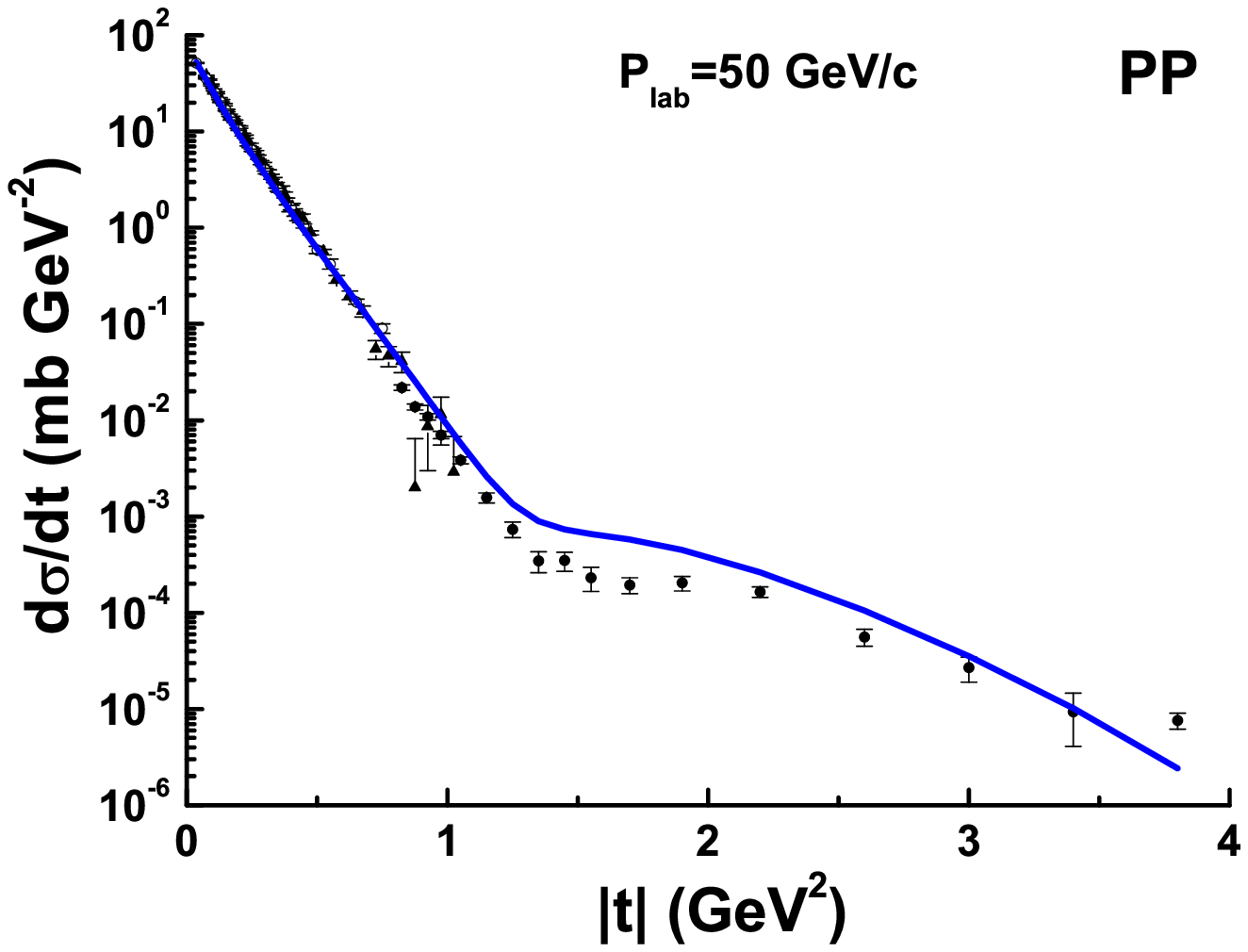}
\begin{minipage}{75mm}
{
\caption{The points are the experimental data by C. Bruneton et al., Nucl. Phys. {\bf B124} (1977) 391.}
}
\end{minipage}
\hspace{5mm}
\begin{minipage}{75mm}
{
\caption{The points are the experimental data by
             Z. Asad et al., Nucl. Phys. {\bf B255} (1984) 273;
             C.W. Akerlof et al., Phys. Rev. {\bf D14} (1976) 2864;
             D.S. Ayres et al., Phys. Rev. {\bf D15} (1977) 3105.}
}
\end{minipage}
\end{figure}

\begin{figure}[cbth]
\includegraphics[width=75mm,height=66mm,clip]{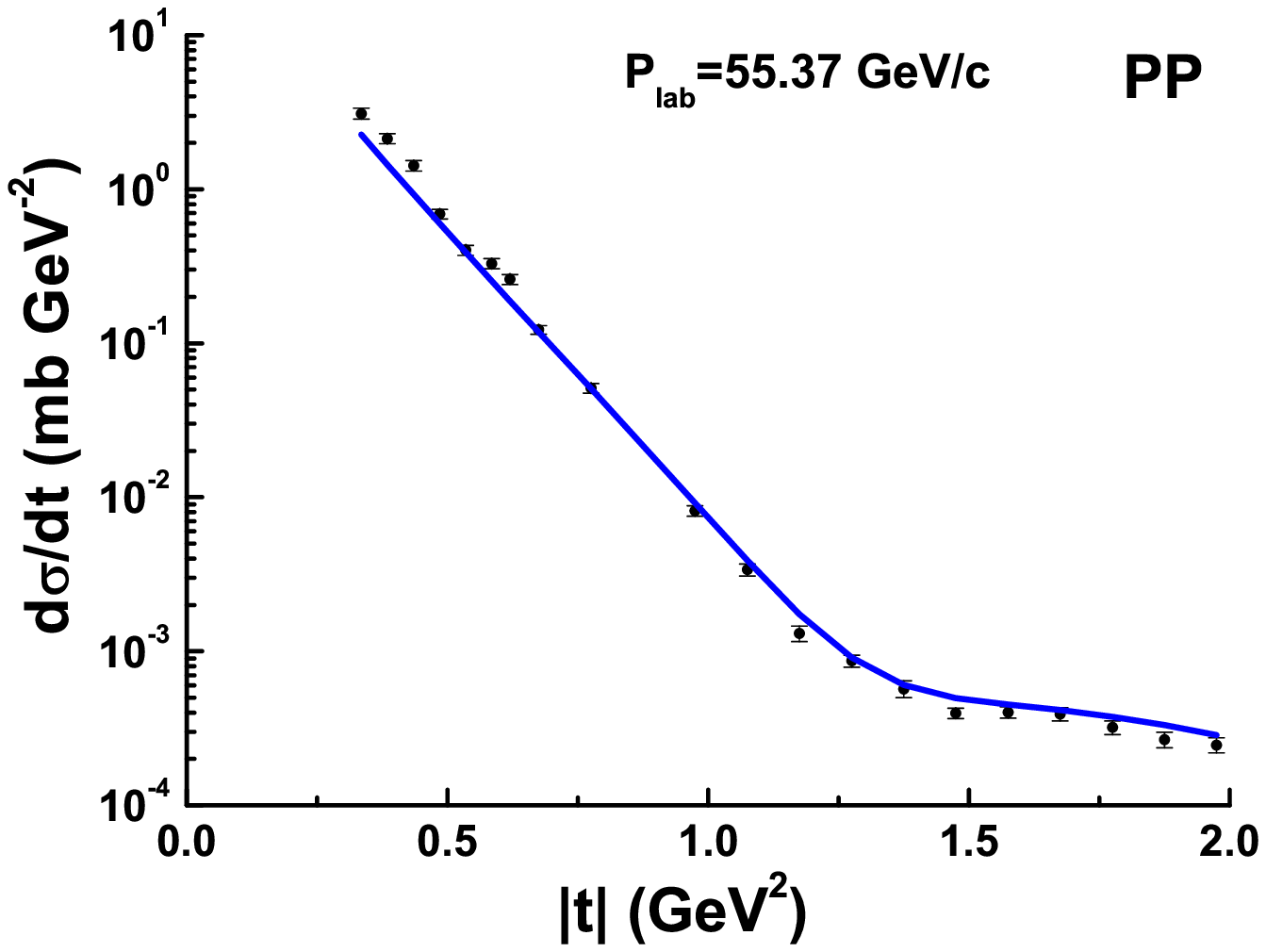}\hspace{5mm}\includegraphics[width=75mm,height=66mm,clip]{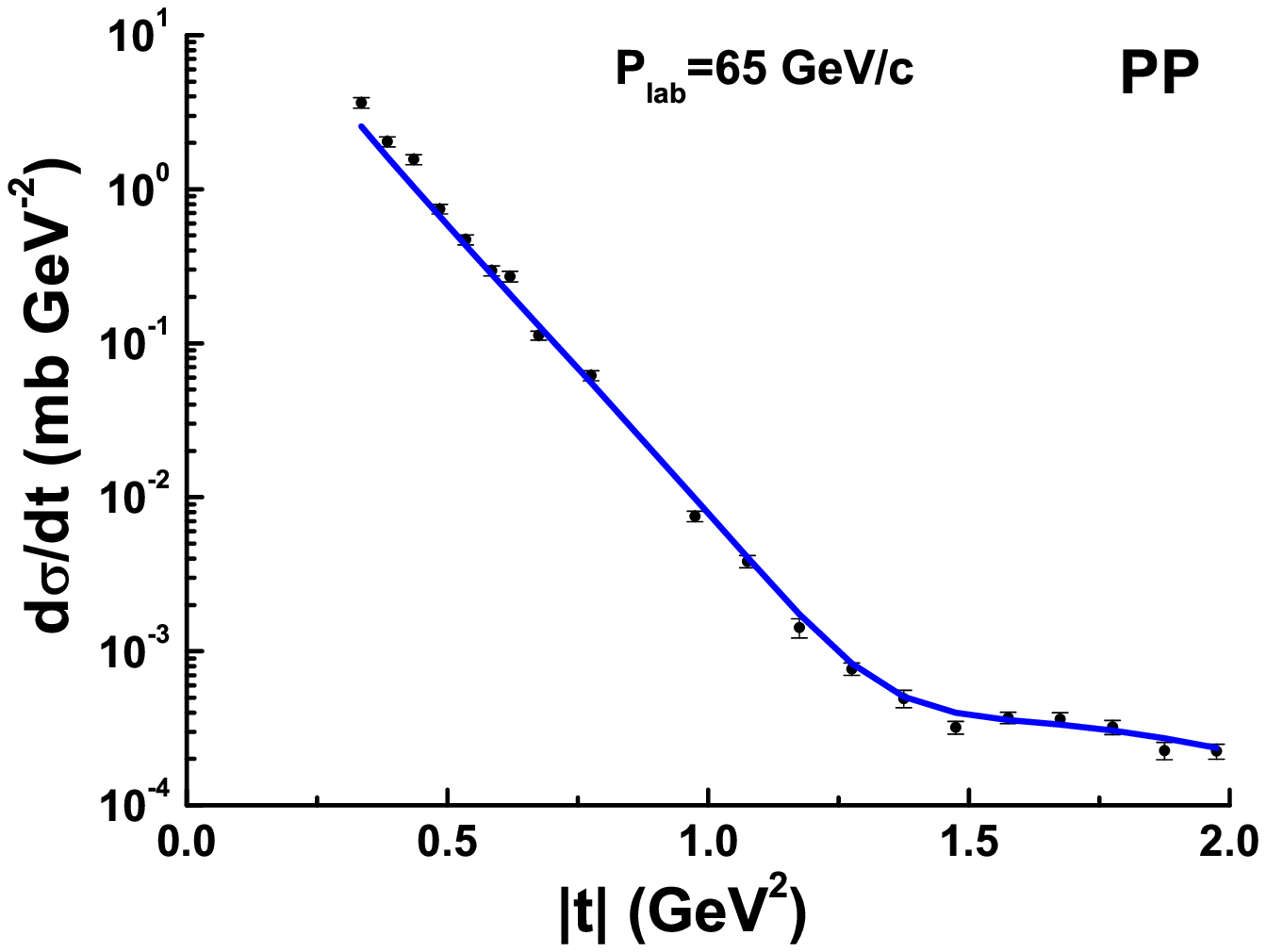}
\begin{minipage}{75mm}
{
\caption{The points are the experimental data by R. Rusack et al., Phys. Rev. Lett. {\bf 41} (1978) 1632.}
}
\end{minipage}
\hspace{5mm}
\begin{minipage}{75mm}
{
\caption{The points are the experimental data by R. Rusack et al., Phys. Rev. Lett. {\bf 41} (1978) 1632.}
}
\end{minipage}

\includegraphics[width=75mm,height=66mm,clip]{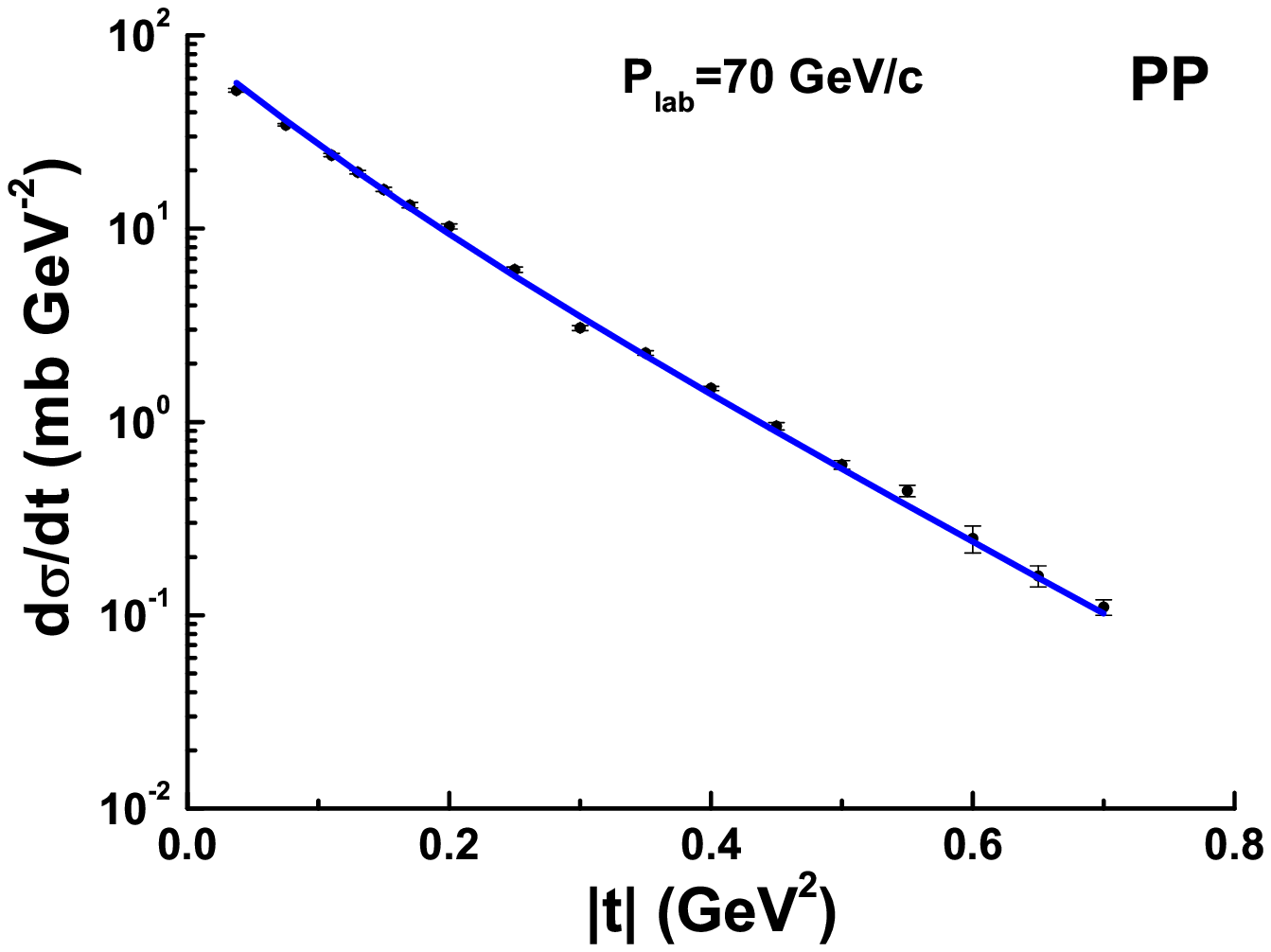}\hspace{5mm}\includegraphics[width=75mm,height=66mm,clip]{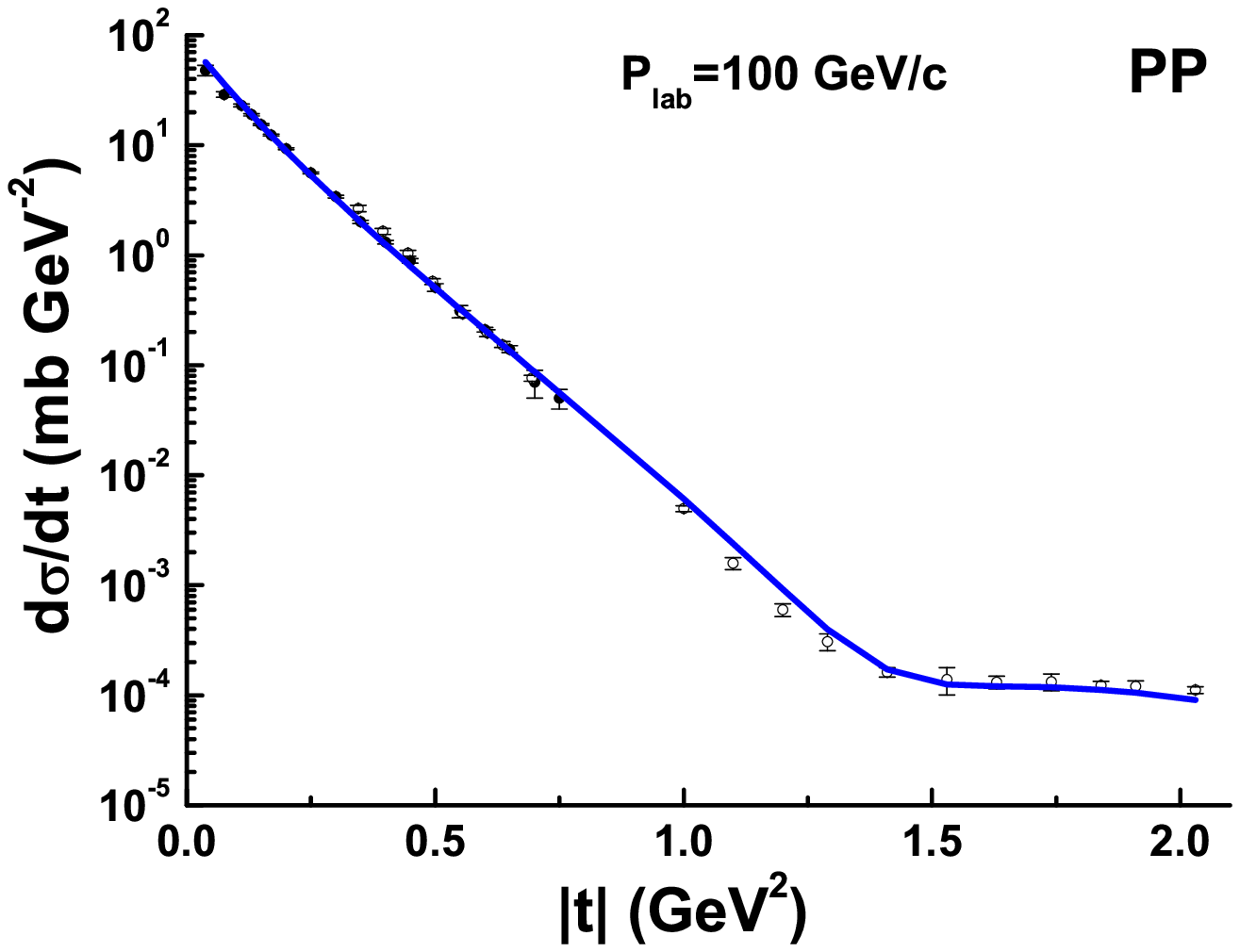}
\begin{minipage}{75mm}
{
\caption{The points are the experimental data by D.S. Ayres et al., Phys. Rev. {\bf D15} (1977) 3105.}
}
\end{minipage}
\hspace{5mm}
\begin{minipage}{75mm}
{
\caption{The points are the experimental data by C.W. Akerlof et al., Phys. Rev. {\bf D14} (1976) 2864;
                                           R. Rubinstein et al., Phys. Rev.{\bf D30} (1984) 1413.}
}
\end{minipage}
\includegraphics[width=75mm,height=66mm,clip]{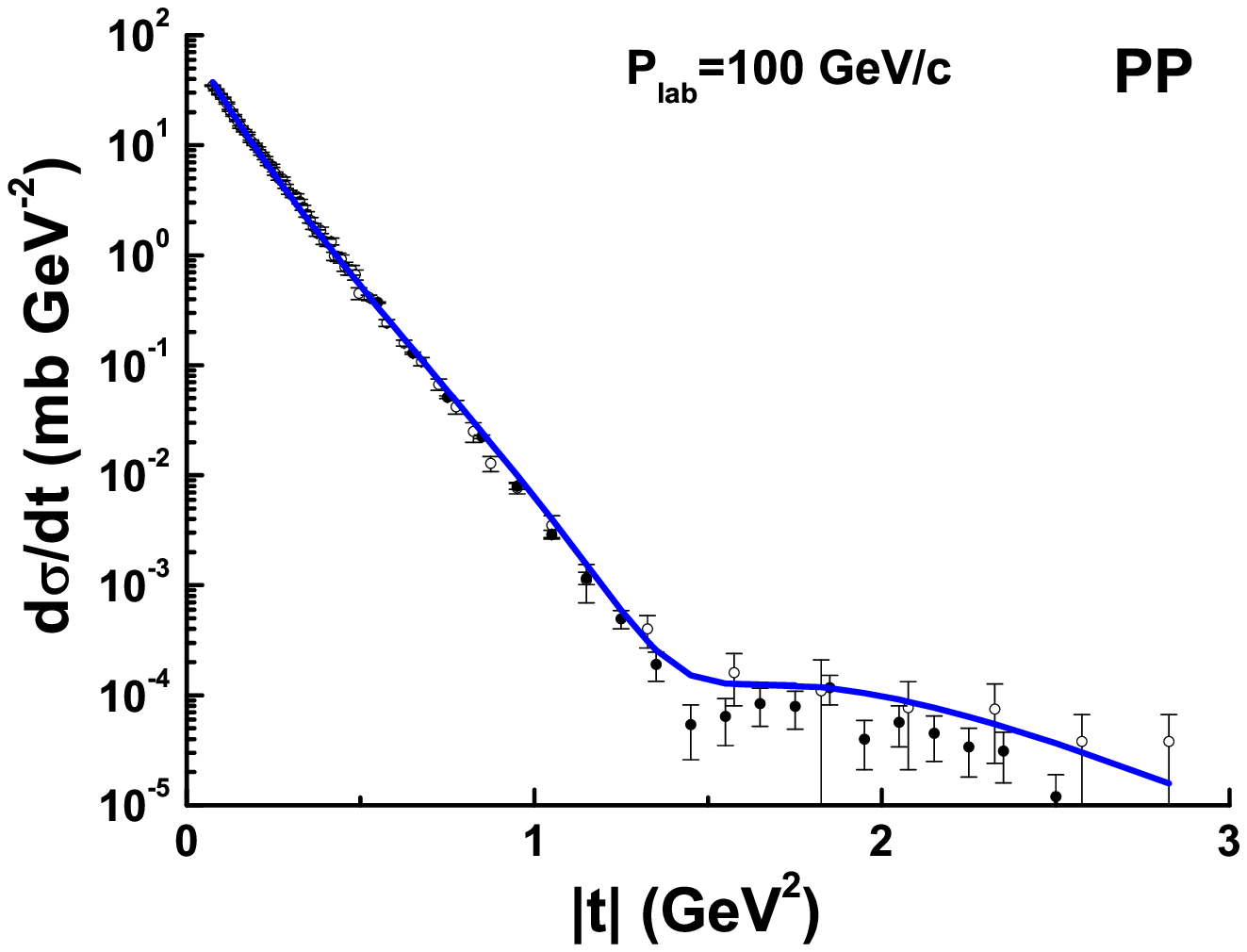}\hspace{5mm}\includegraphics[width=75mm,height=66mm,clip]{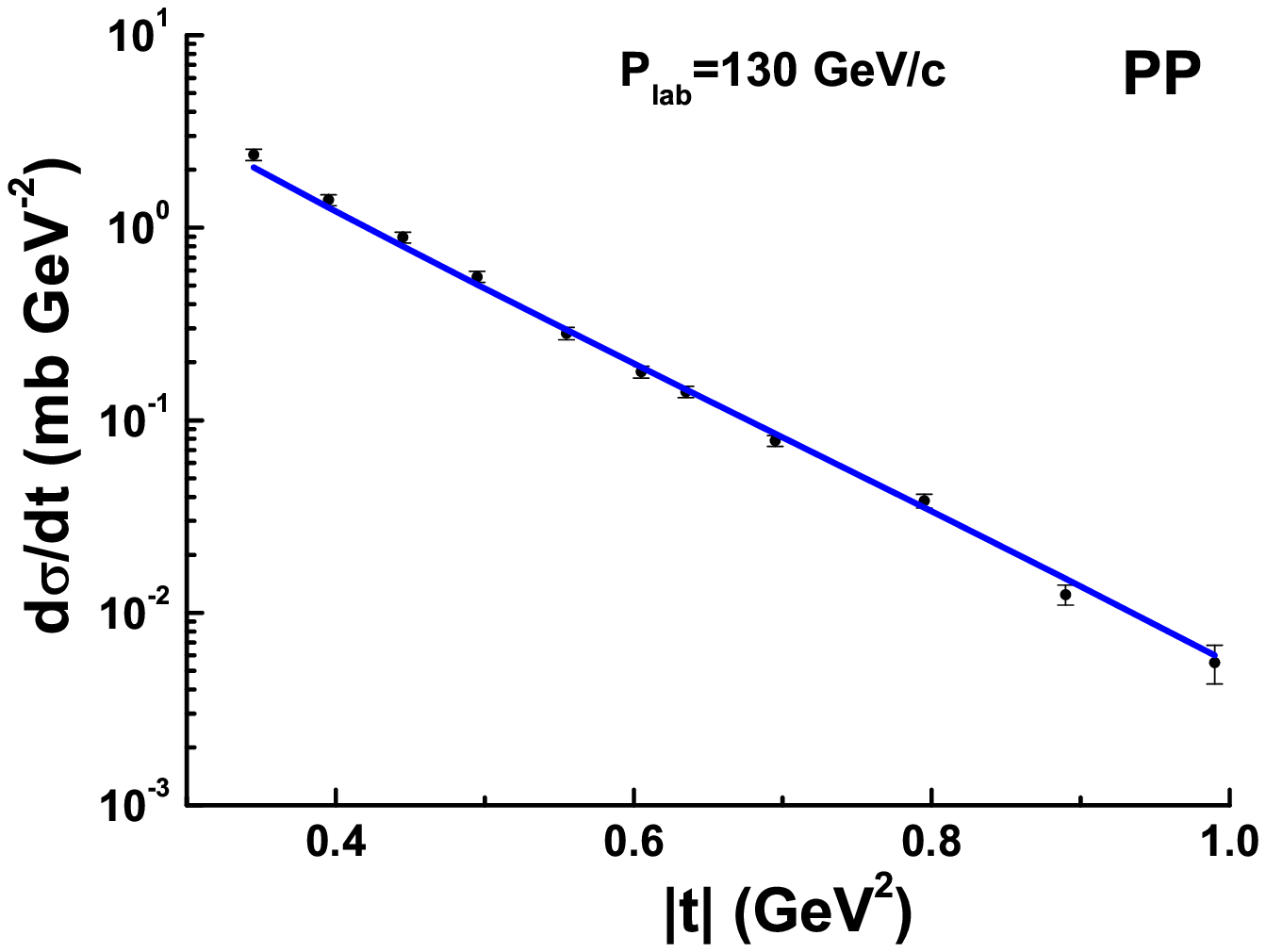}
\begin{minipage}{75mm}
{
\caption{The points are the experimental data by D.S. Ayres et al., Phys. Rev. {\bf D15} (1977) 3105;
R. Rusack et al., Phys. Rev. Lett. {\bf 41} (1978) 1632.}
}
\end{minipage}
\hspace{5mm}
\begin{minipage}{75mm}
{
\caption{The points are the experimental data by R. Rusack et al., Phys. Rev. Lett. {\bf 41} (1978) 1632.}
}
\end{minipage}
\end{figure}

\begin{figure}[cbth]
\includegraphics[width=75mm,height=66mm,clip]{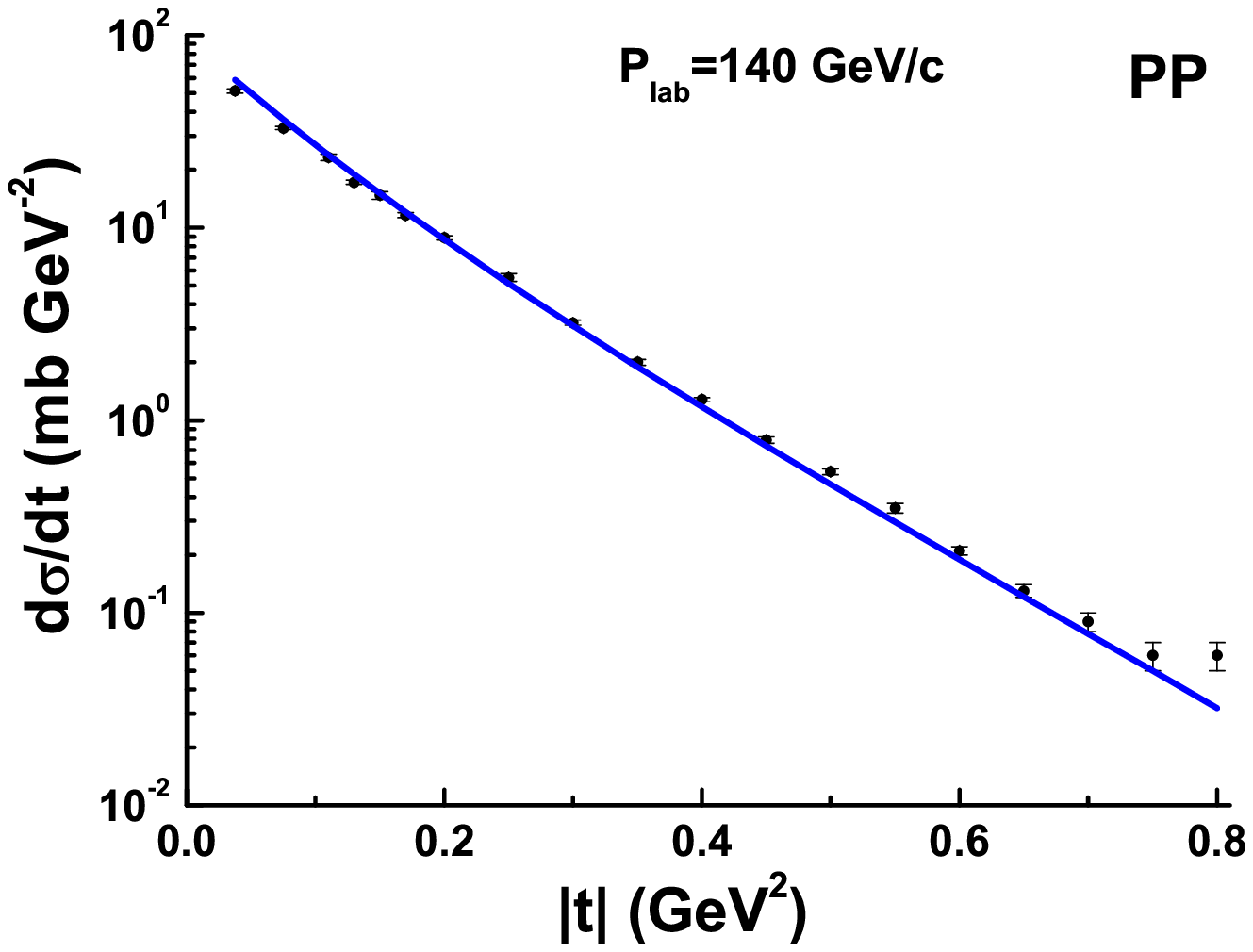}\hspace{5mm}\includegraphics[width=75mm,height=66mm,clip]{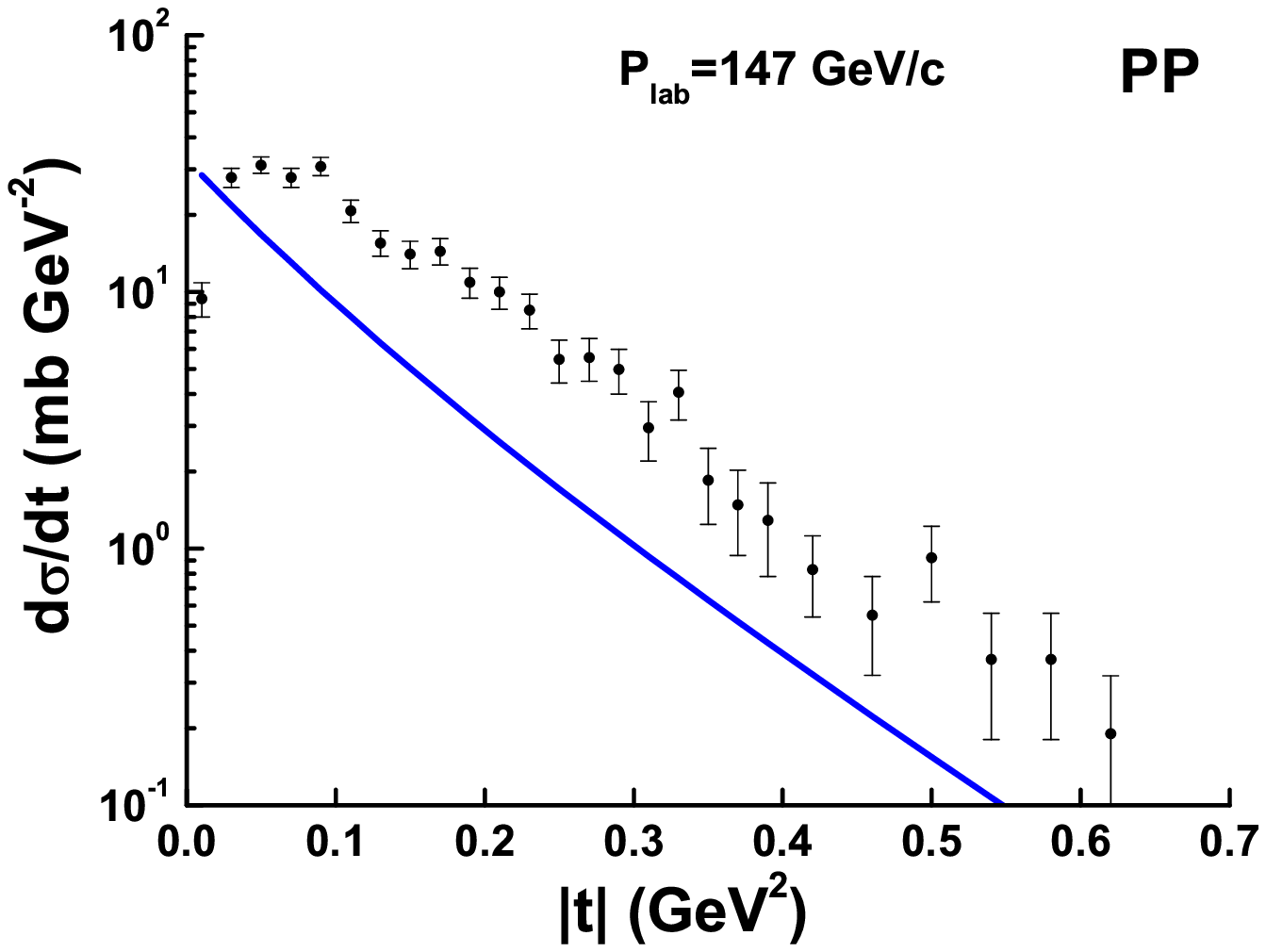}
\begin{minipage}{75mm}
{
\caption{The points are the experimental data by D.S. Ayres et al., Phys. Rev. {\bf D15} (1977) 3105.}
}
\end{minipage}
\hspace{5mm}
\begin{minipage}{75mm}
{
\caption{The points are the experimental data by D. Brick et al., Phys. Rev. {\bf D25} (1982) 2794.}
}
\end{minipage}
\includegraphics[width=75mm,height=66mm,clip]{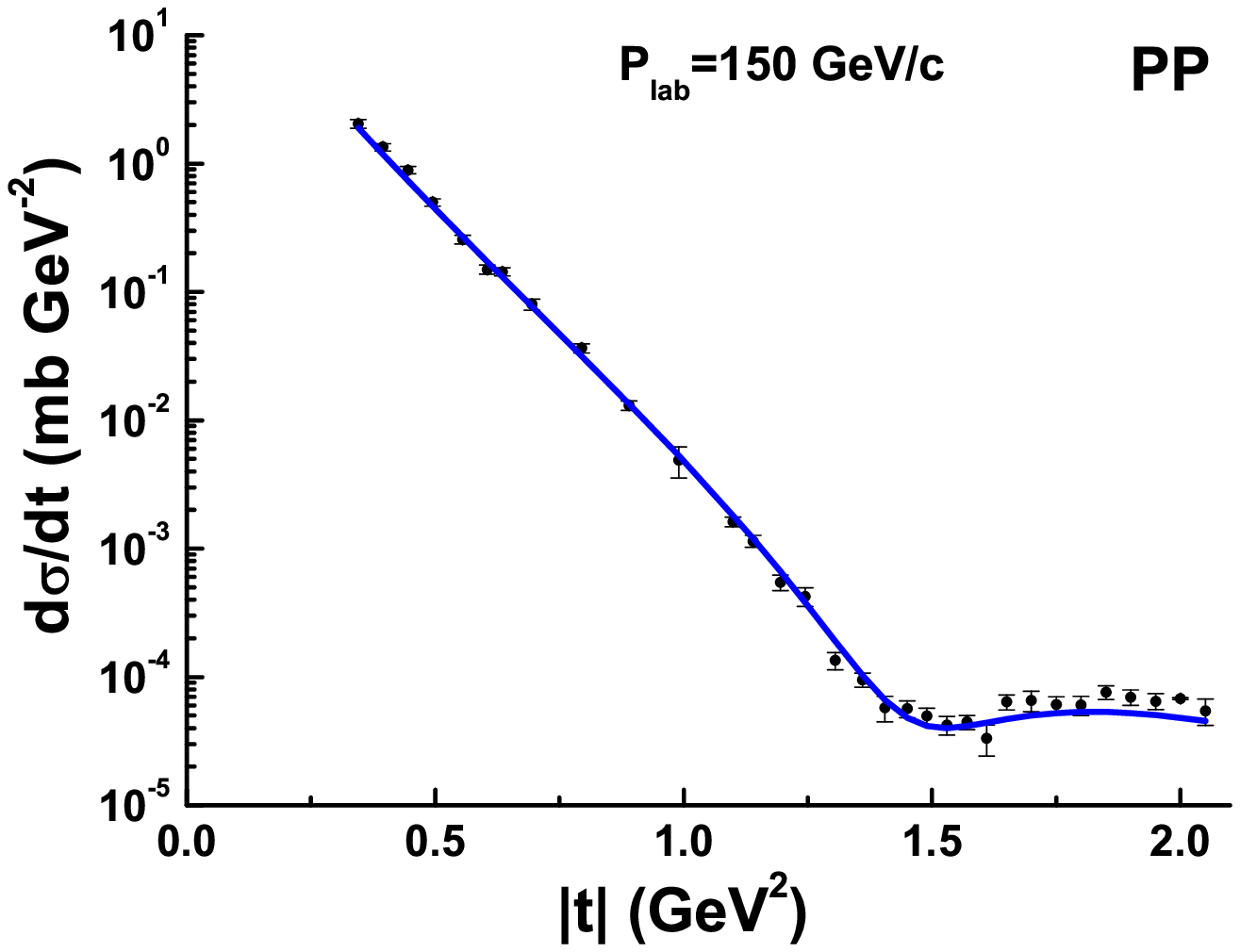}\hspace{5mm}\includegraphics[width=75mm,height=66mm,clip]{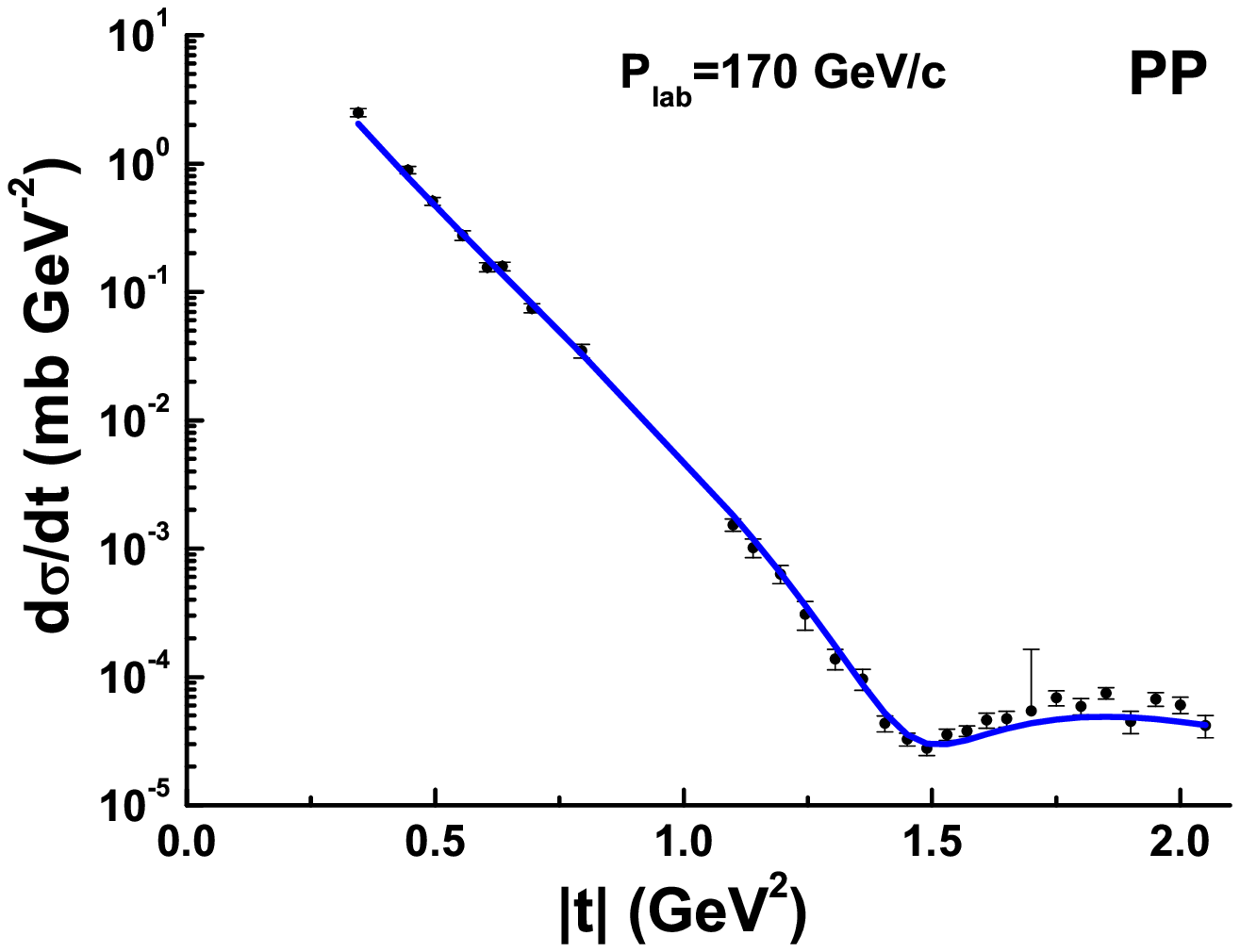}
\begin{minipage}{75mm}
{
\caption{The points are the experimental data by R. Rusack et al., Phys. Rev. Lett. {\bf 41} (1978) 1632.}
}
\end{minipage}
\hspace{5mm}
\begin{minipage}{75mm}
{
\caption{The points are the experimental data by R. Rusack et al., Phys. Rev. Lett. {\bf 41} (1978) 1632.}
}
\end{minipage}
\includegraphics[width=75mm,height=66mm,clip]{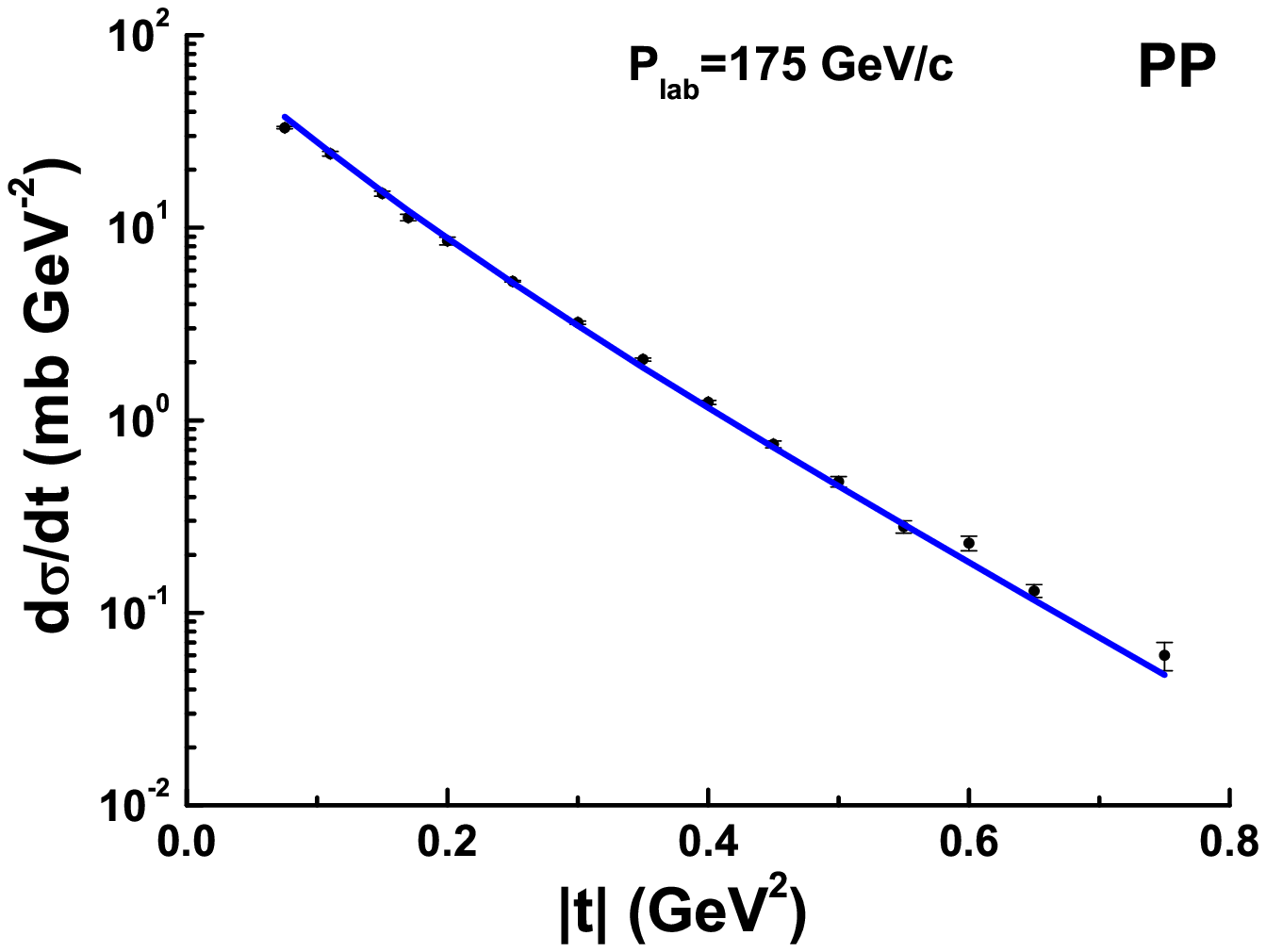}\hspace{5mm}\includegraphics[width=75mm,height=66mm,clip]{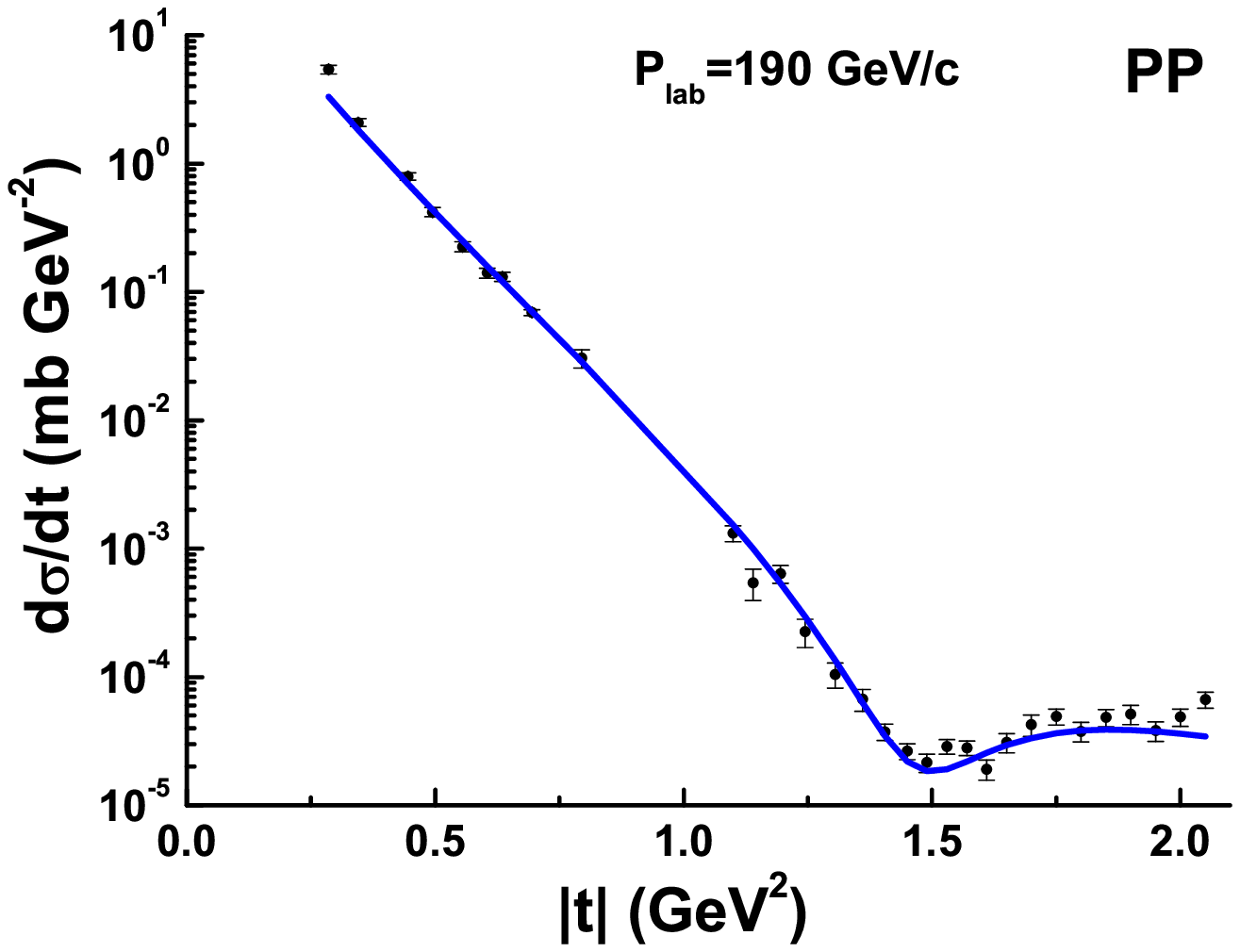}
\begin{minipage}{75mm}
{
\caption{The points are the experimental data by D.S. Ayres et al., Phys. Rev. {\bf D15} (1977) 3105.}
}
\end{minipage}
\hspace{5mm}
\begin{minipage}{75mm}
{
\caption{The points are the experimental data by R. Rusack et al., Phys. Rev. Lett. {\bf 41} (1978) 1632.}
}
\end{minipage}
\end{figure}

\begin{figure}[cbth]

\includegraphics[width=75mm,height=66mm,clip]{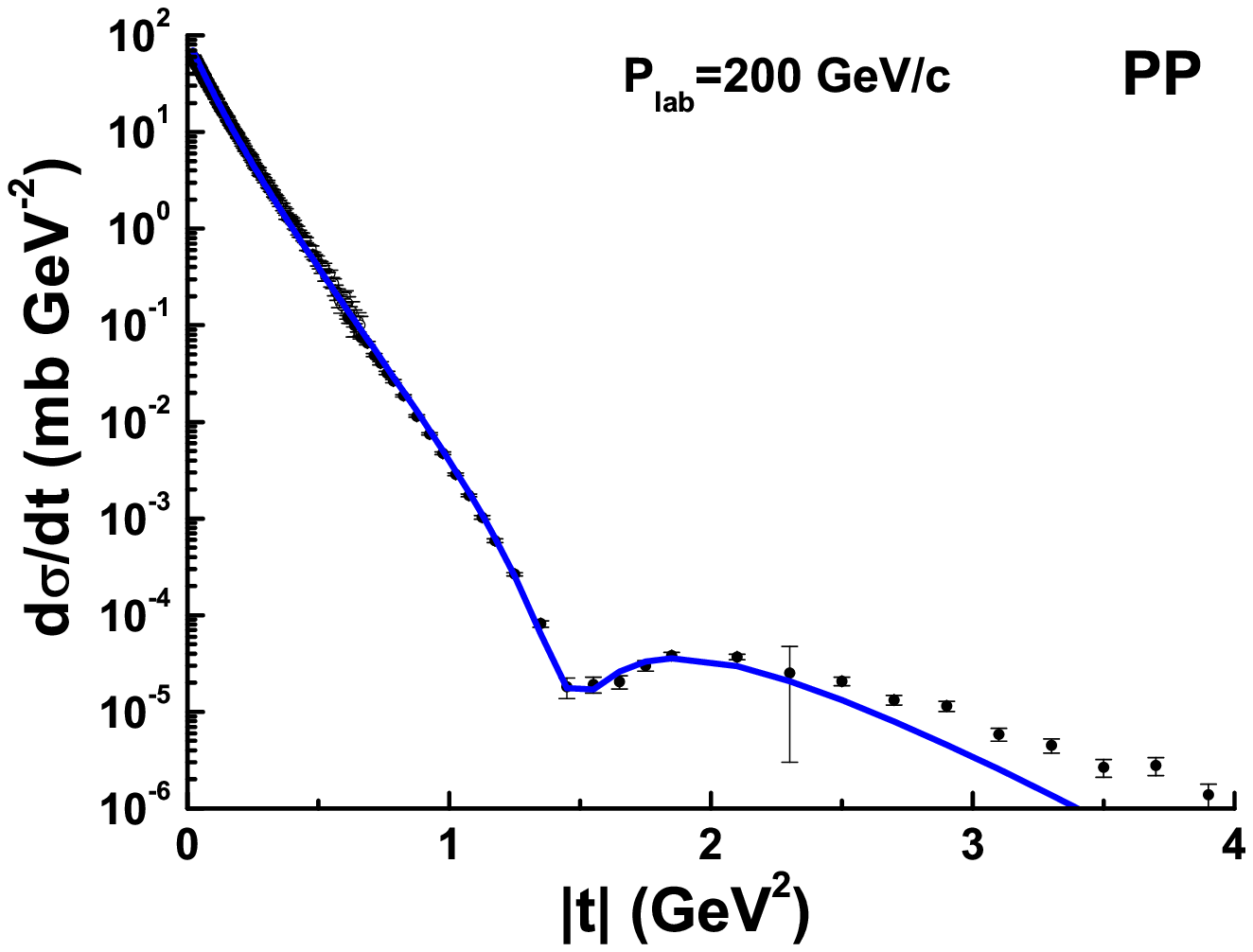}\hspace{5mm}\includegraphics[width=75mm,height=66mm,clip]{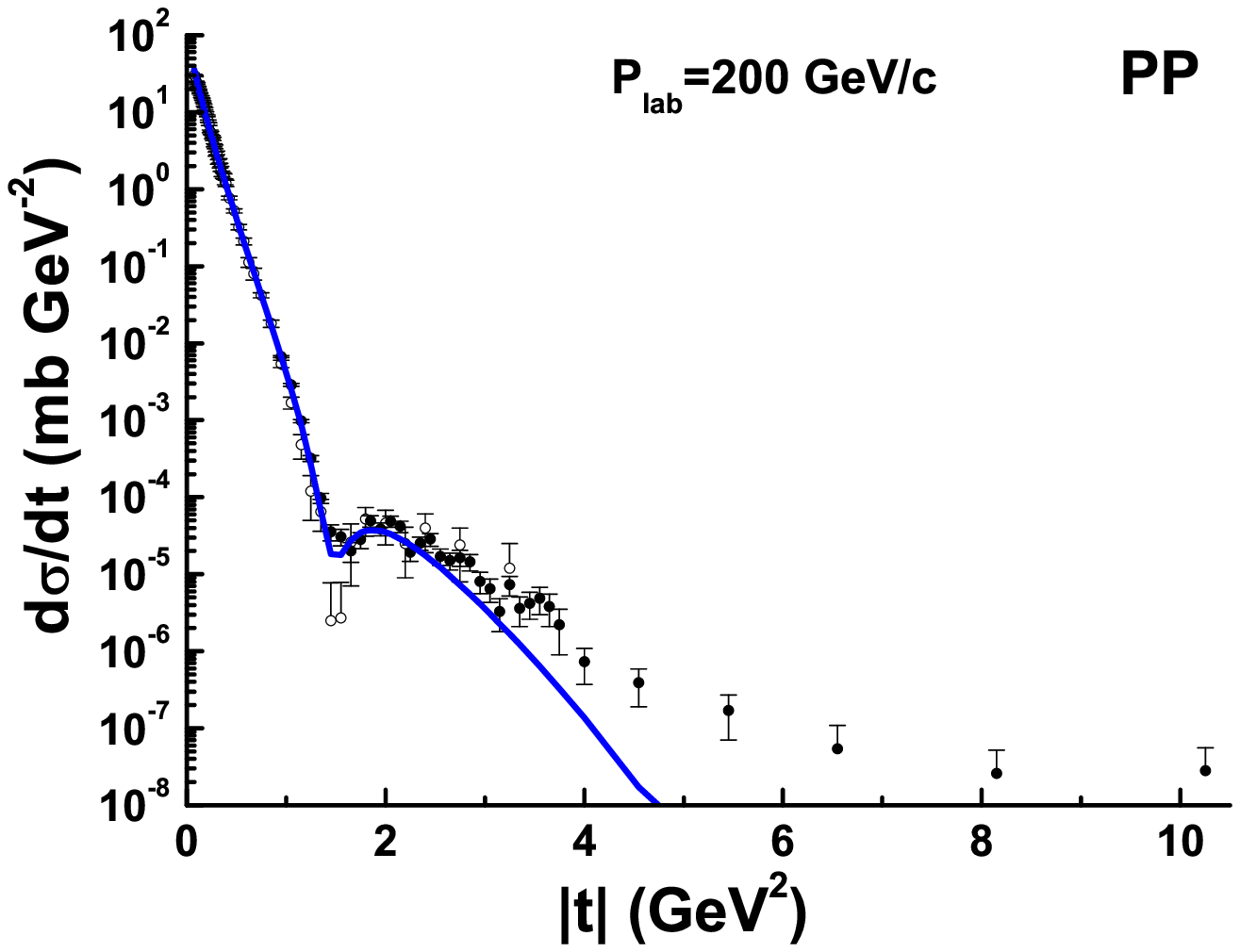}
\begin{minipage}{75mm}
{
\caption{The points are the experimental data by C.W. Akerlof et al., Phys. Rev. {\bf D14} (1976) 2864;
R. Rubinstein et al., Phys. Rev.{\bf D30} (1984) 1413.}
}
\end{minipage}
\hspace{5mm}
\begin{minipage}{75mm}
{
\caption{The points are the experimental data by A. Schiz et al., Phys. Rev. {\bf D24} (1981) 26;
G. Fidecaro et al., Nucl. Phys. {\bf B173} (1980) 513.}
}
\end{minipage}
\includegraphics[width=75mm,height=66mm,clip]{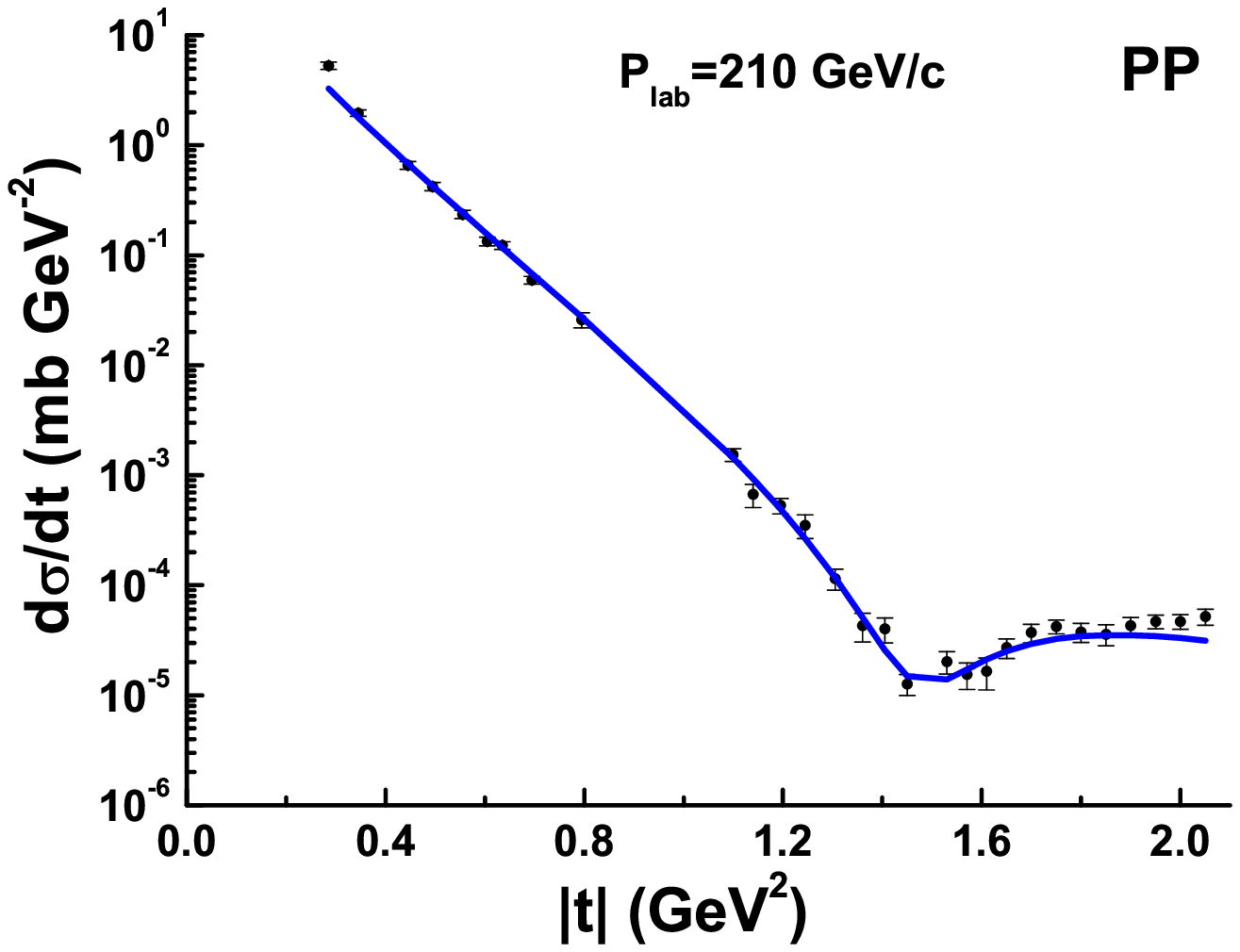}\hspace{5mm}\includegraphics[width=75mm,height=66mm,clip]{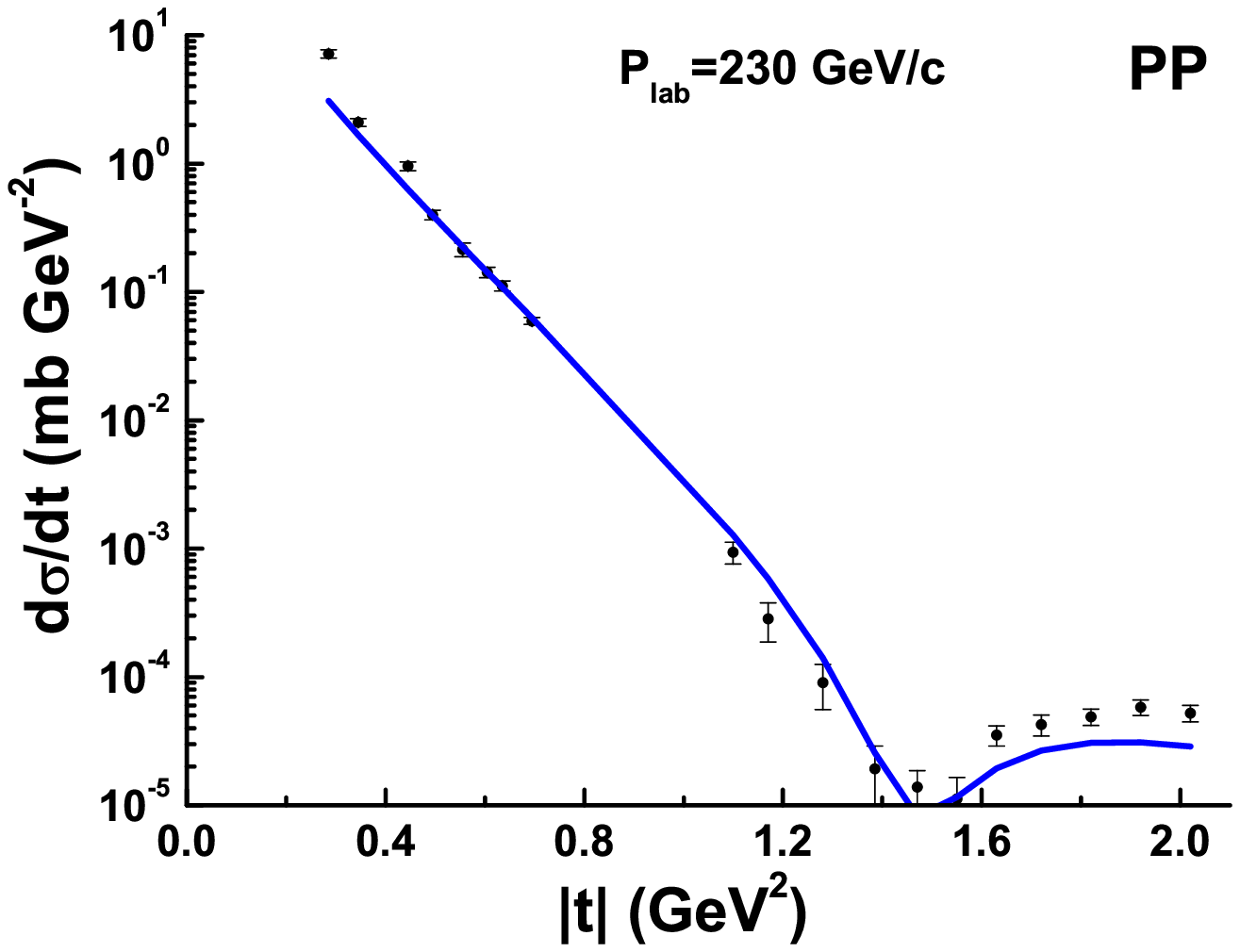}
\begin{minipage}{75mm}
{
\caption{The points are the experimental data by R. Rusack et al., Phys. Rev. Lett. {\bf 41} (1978) 1632.}
}
\end{minipage}
\hspace{5mm}
\begin{minipage}{75mm}
{
\caption{The points are the experimental data by R. Rusack et al., Phys. Rev. Lett. {\bf 41} (1978) 1632.}
}
\end{minipage}
\includegraphics[width=75mm,height=66mm,clip]{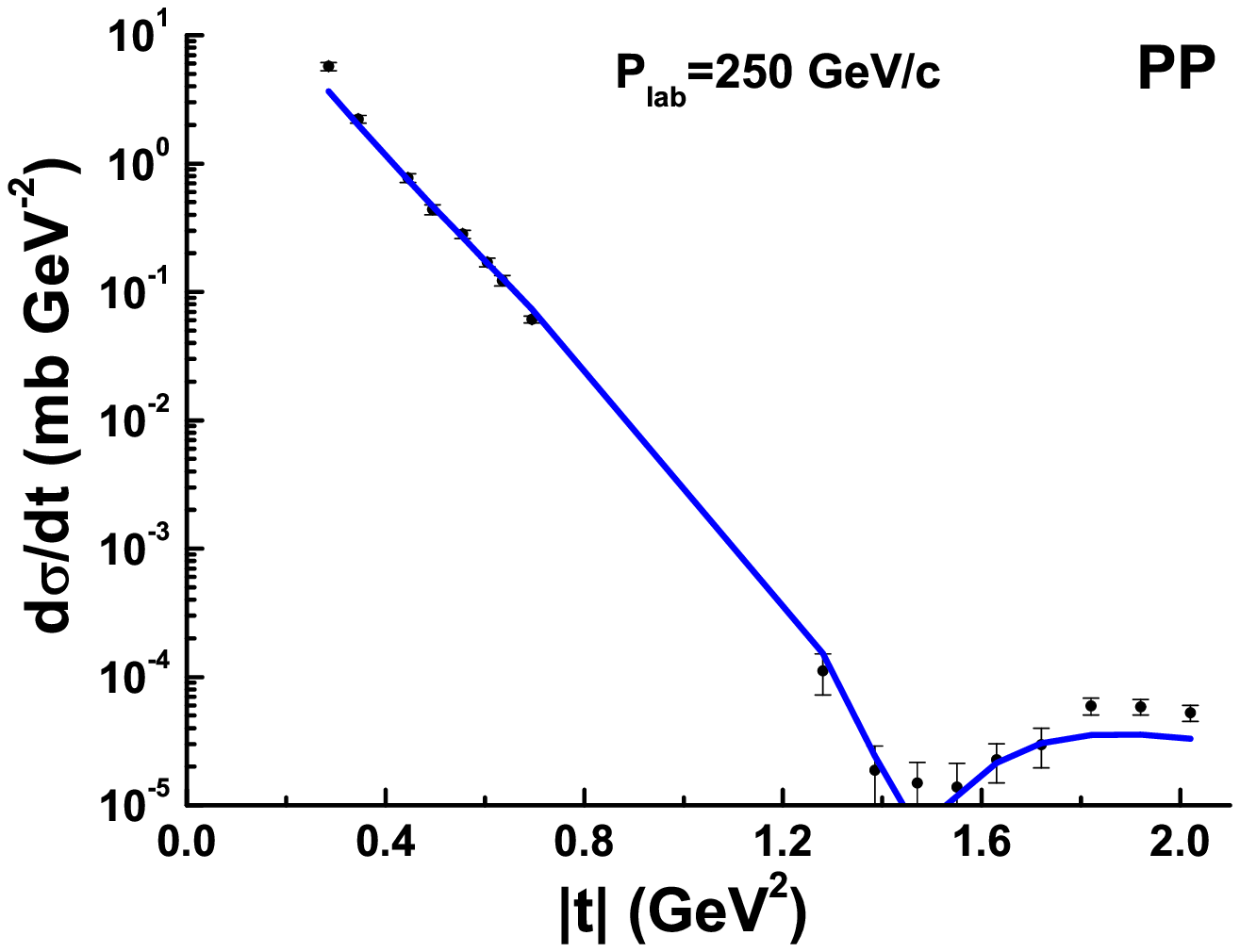}\hspace{5mm}\includegraphics[width=75mm,height=66mm,clip]{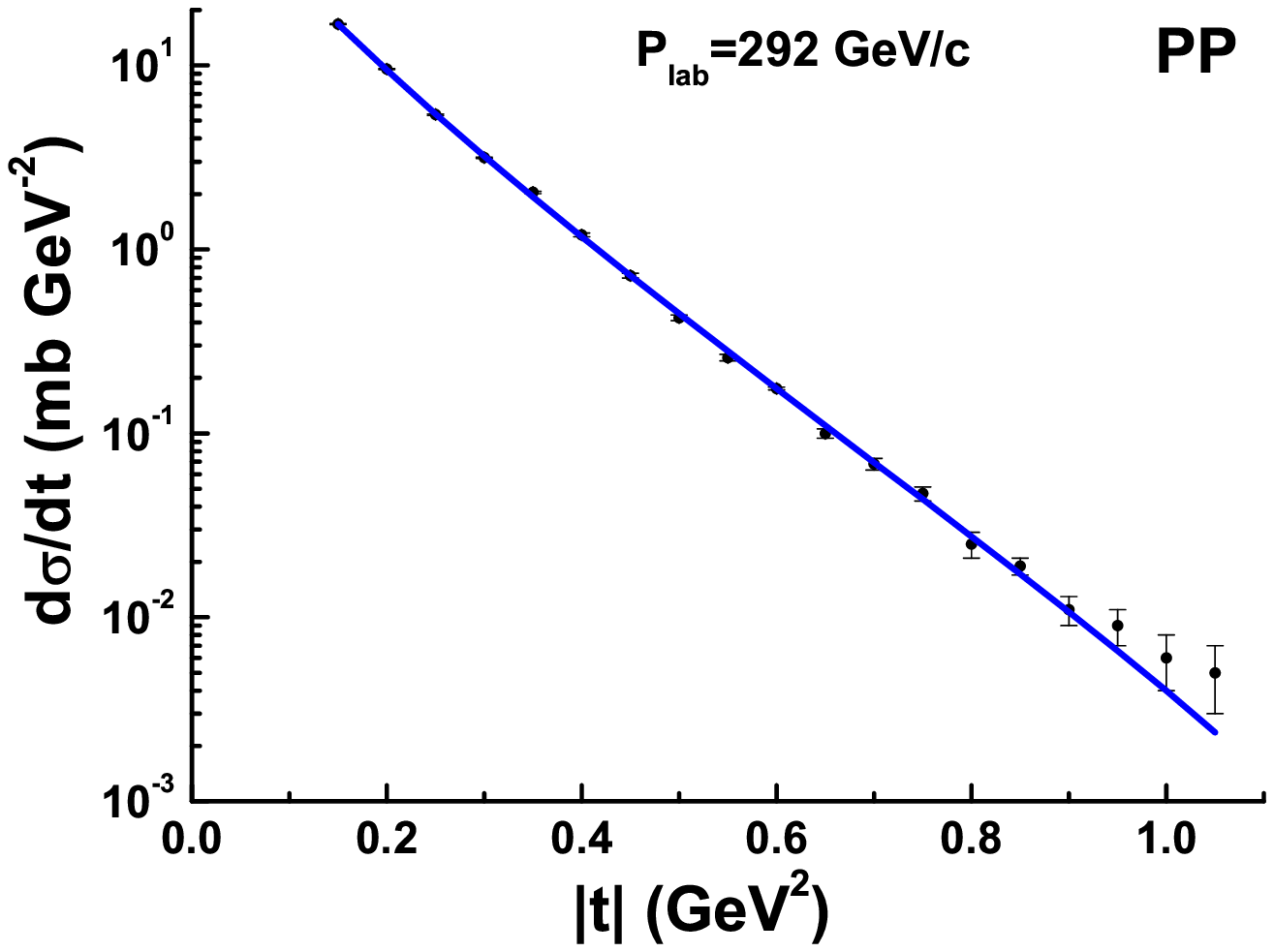}
\begin{minipage}{75mm}
{
\caption{The points are the experimental data by R. Rusack et al., Phys. Rev. Lett. {\bf 41} (1978) 1632.}
}
\end{minipage}
\hspace{5mm}
\begin{minipage}{75mm}
{
\caption{The points are the experimental data by M.G. Albrow et al., Nucl. Phys. {\bf B108} (1976) 1.}
}
\end{minipage}
\end{figure}

\begin{figure}[cbth]
\includegraphics[width=75mm,height=66mm,clip]{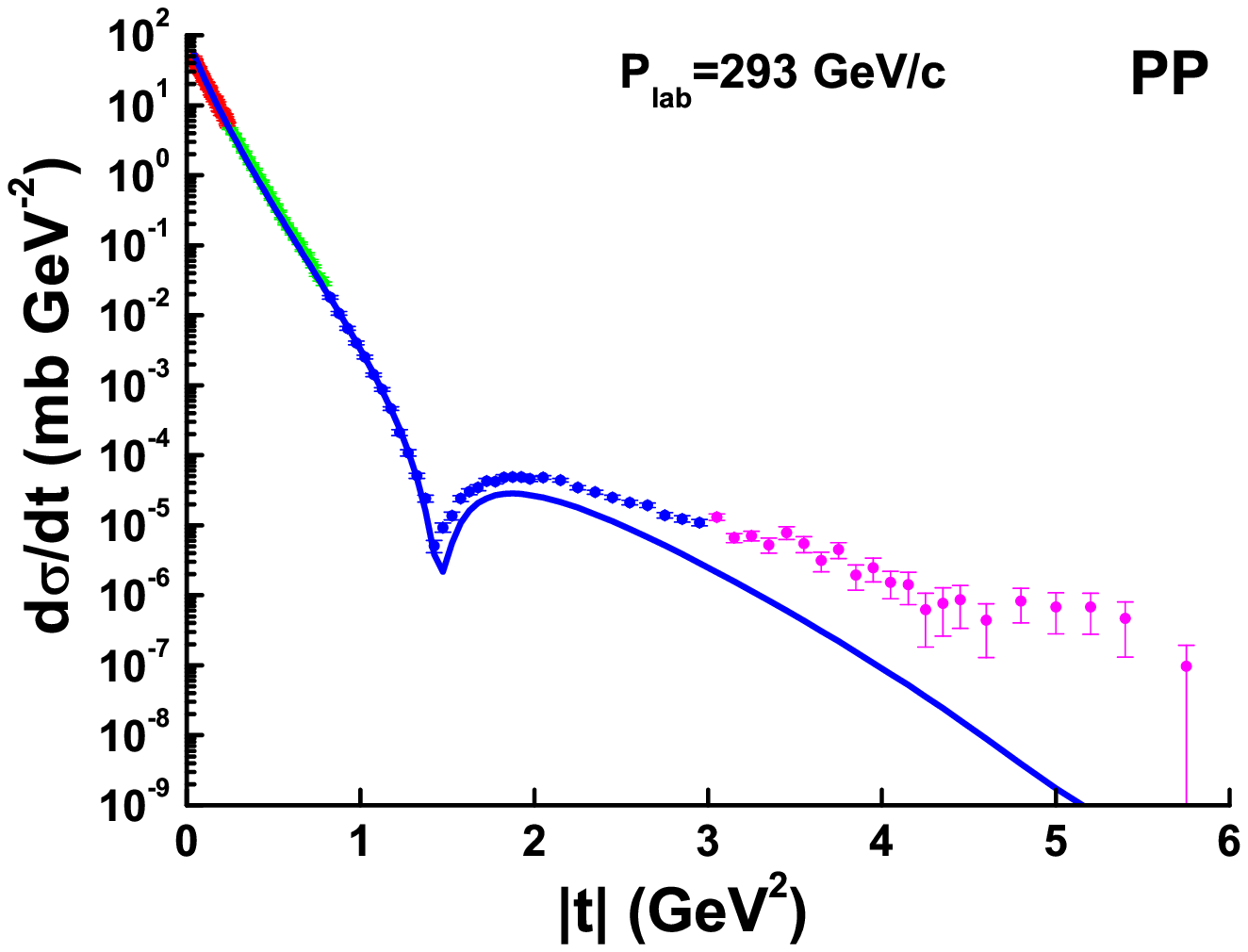}\hspace{5mm}\includegraphics[width=75mm,height=66mm,clip]{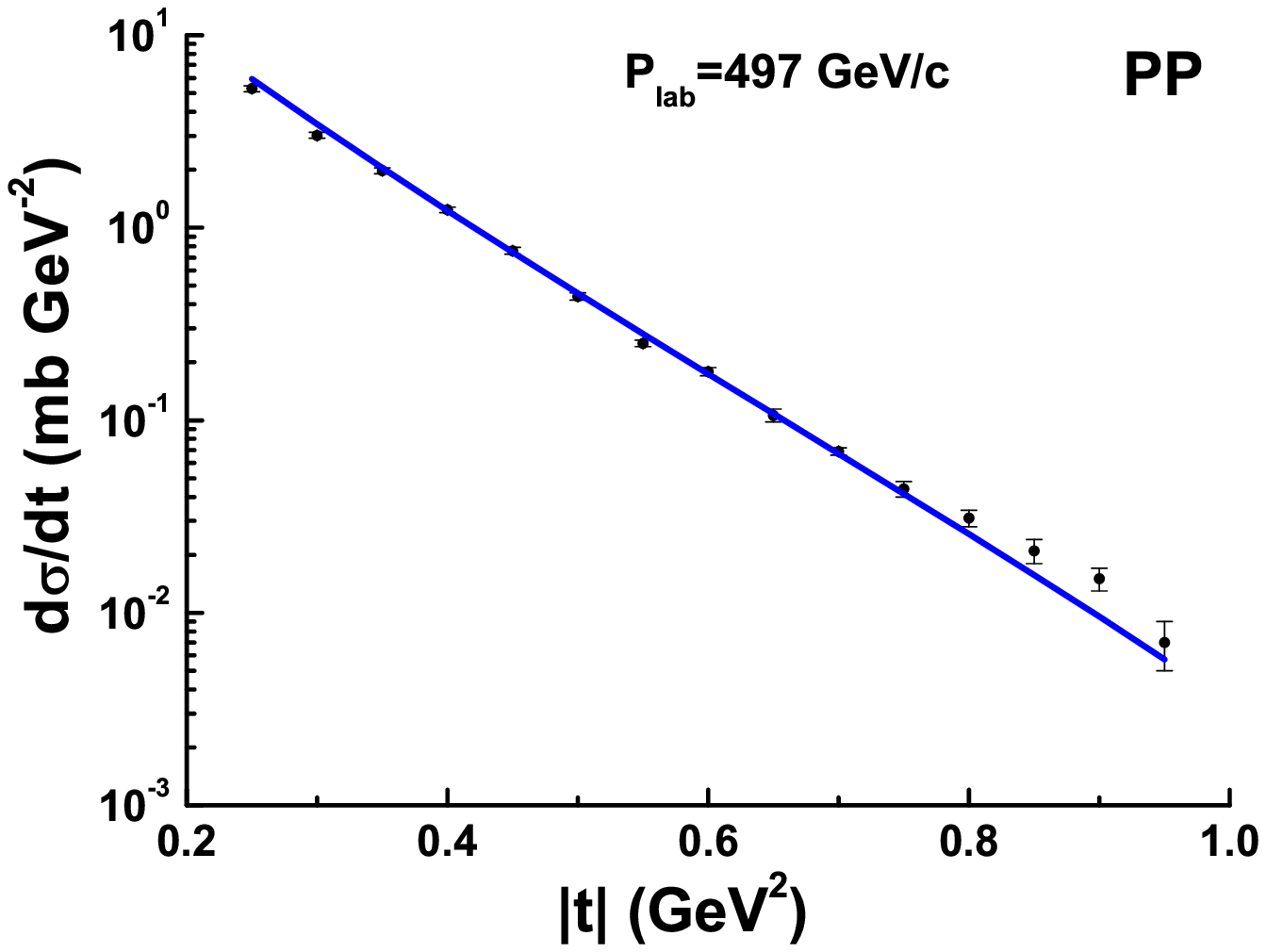}
\begin{minipage}{75mm}
{
\caption{The points are the experimental data by U. Amaldi and K.R. Schubert, Nucl. Phys. {\bf B166} (1980) 301.}
}
\end{minipage}
\hspace{5mm}
\begin{minipage}{75mm}
{
\caption{The points are the experimental data by M.G. Albrow et al., Nucl. Phys. {\bf B108} (1976) 1.}
}
\end{minipage}
\includegraphics[width=75mm,height=66mm,clip]{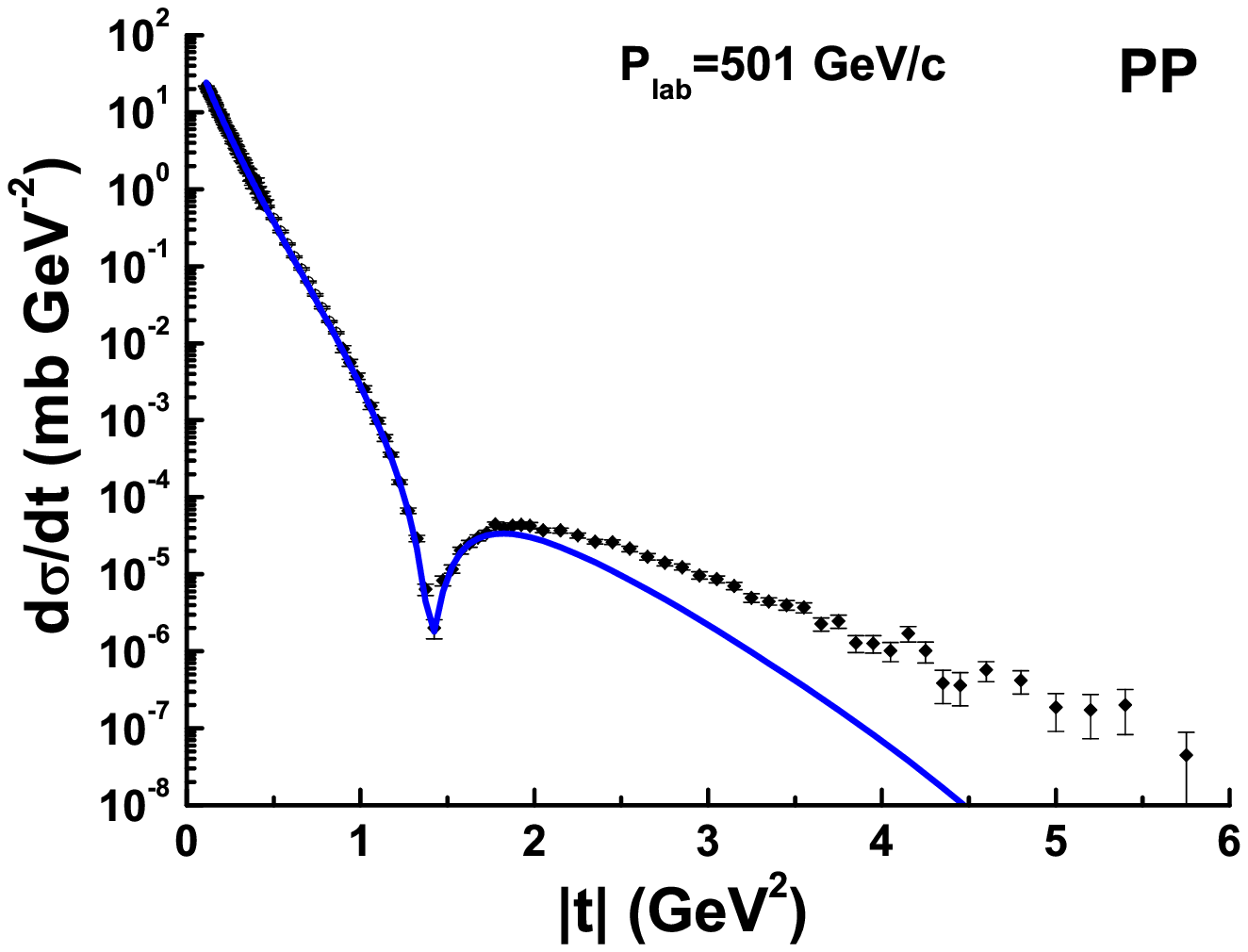}\hspace{5mm}\includegraphics[width=75mm,height=66mm,clip]{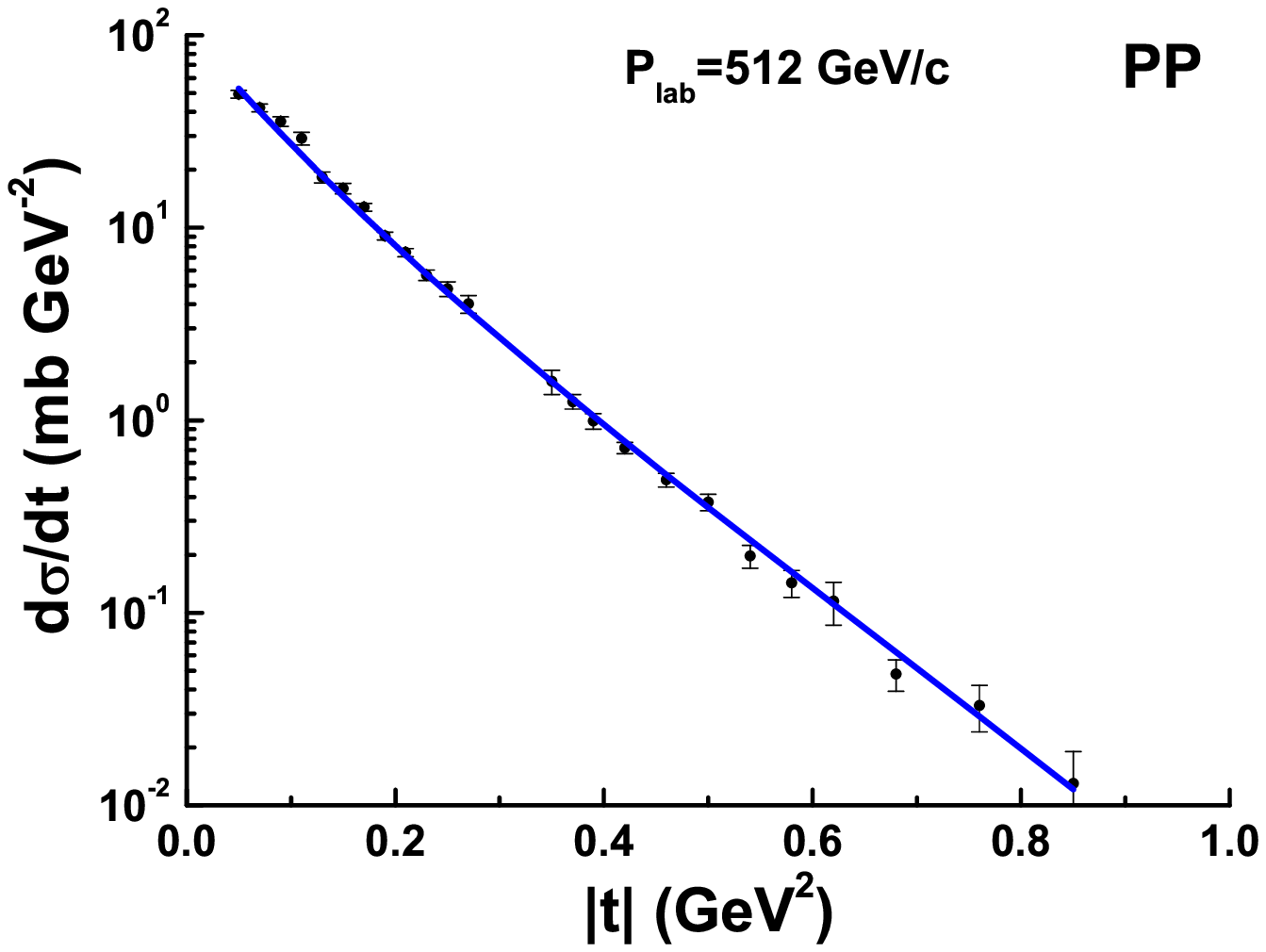}
\begin{minipage}{75mm}
{
\caption{The points are the experimental data by U. Amaldi and K.R. Schubert, Nucl. Phys. {\bf B166} (1980) 301.}
}
\end{minipage}
\hspace{5mm}
\begin{minipage}{75mm}
{
\caption{The points are the experimental data by A. Breakstone et al. Nucl. Phys. {\bf B248} (1984) 253.}
}
\end{minipage}

\includegraphics[width=75mm,height=66mm,clip]{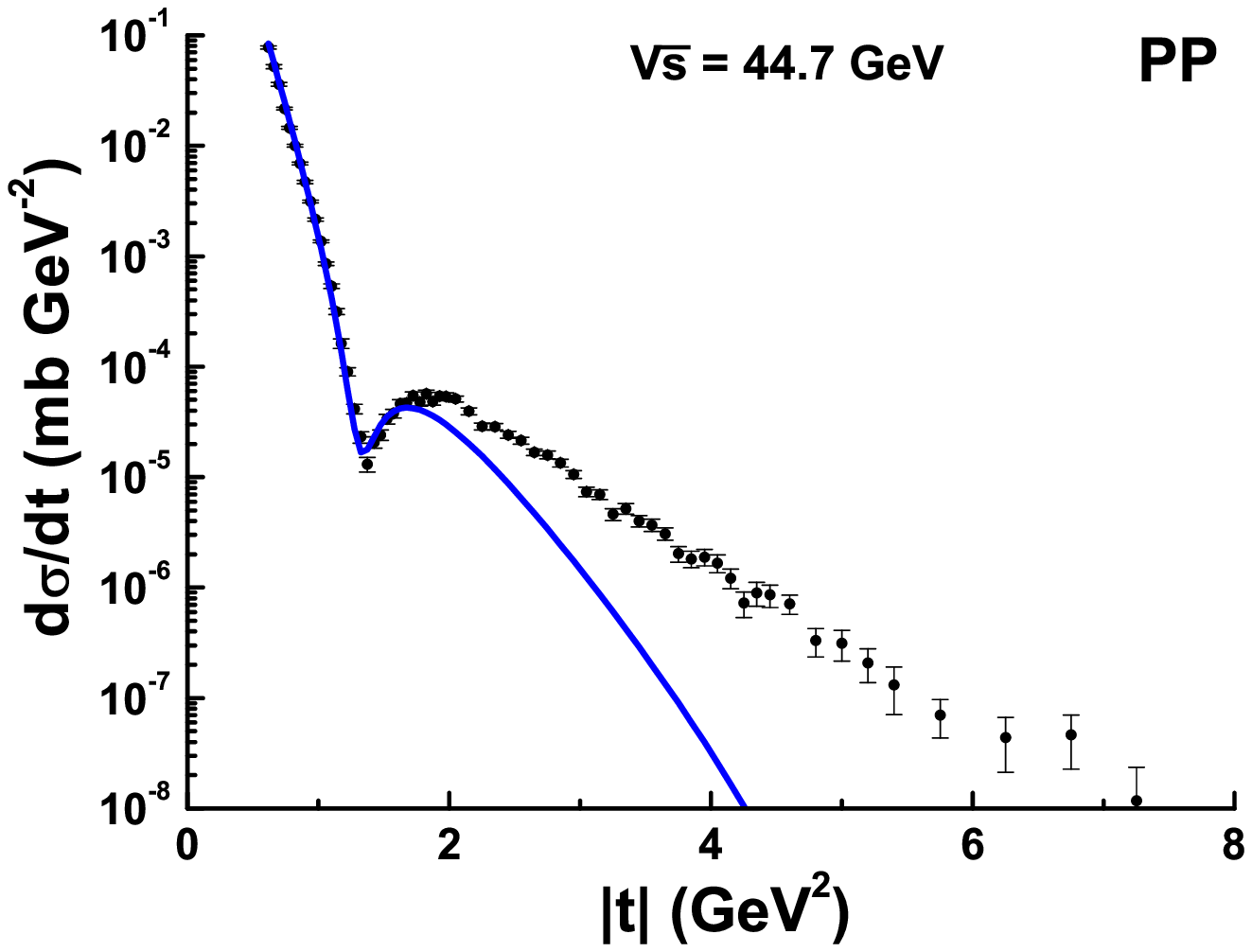}\hspace{5mm}\includegraphics[width=75mm,height=66mm,clip]{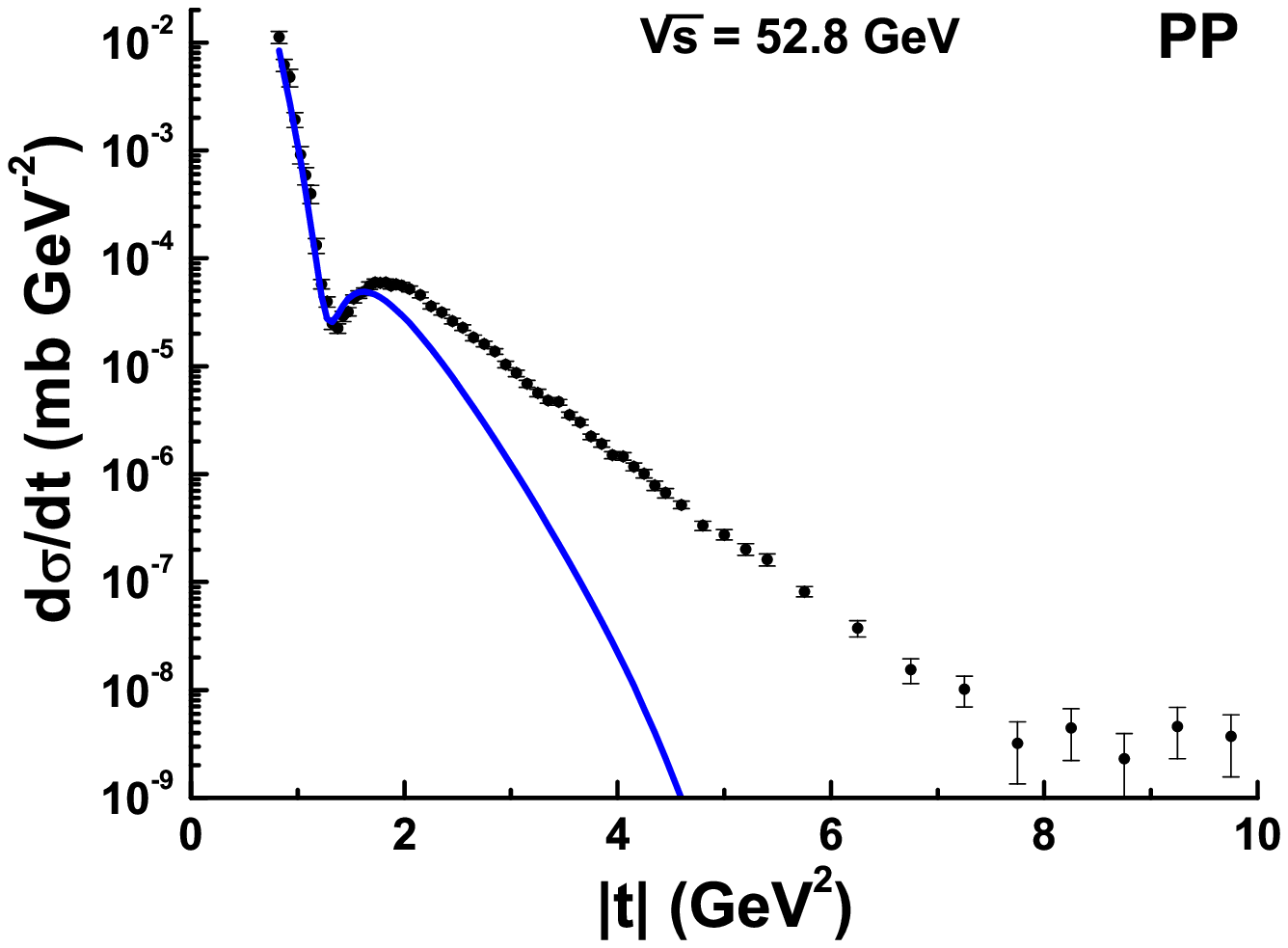}
\begin{minipage}{75mm}
{
\caption{The points are the experimental data by U. Amaldi and K.R. Schubert, Nucl. Phys. {\bf B166} (1980) 301.}
}
\end{minipage}
\hspace{5mm}
\begin{minipage}{75mm}
{
\caption{The points are the experimental data by E. Nagy et al., Nucl. Phys. {\bf B150} (1979) 221.}
}
\end{minipage}
\end{figure}

\begin{figure}[cbth]

\includegraphics[width=75mm,height=66mm,clip]{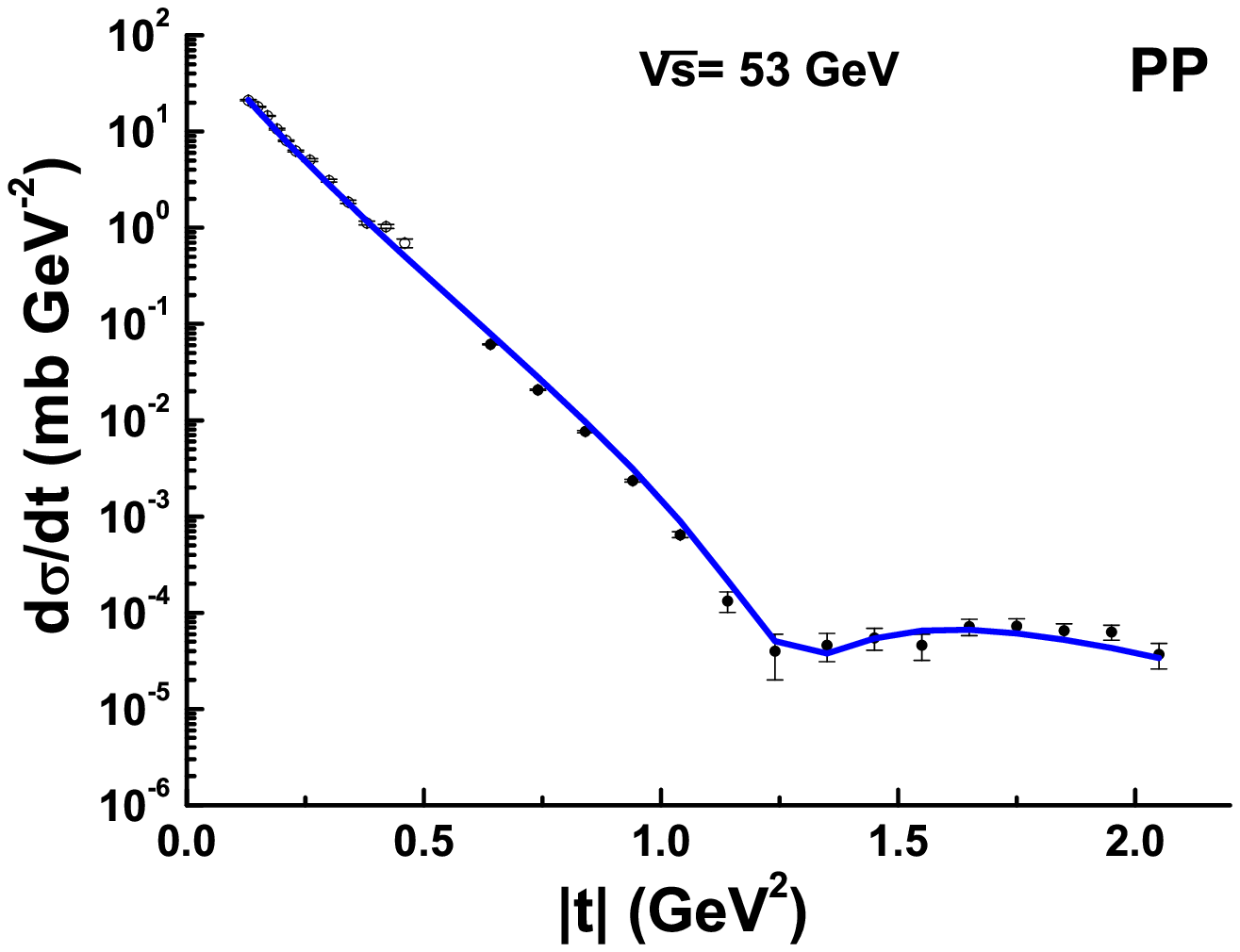}\hspace{5mm}\includegraphics[width=75mm,height=66mm,clip]{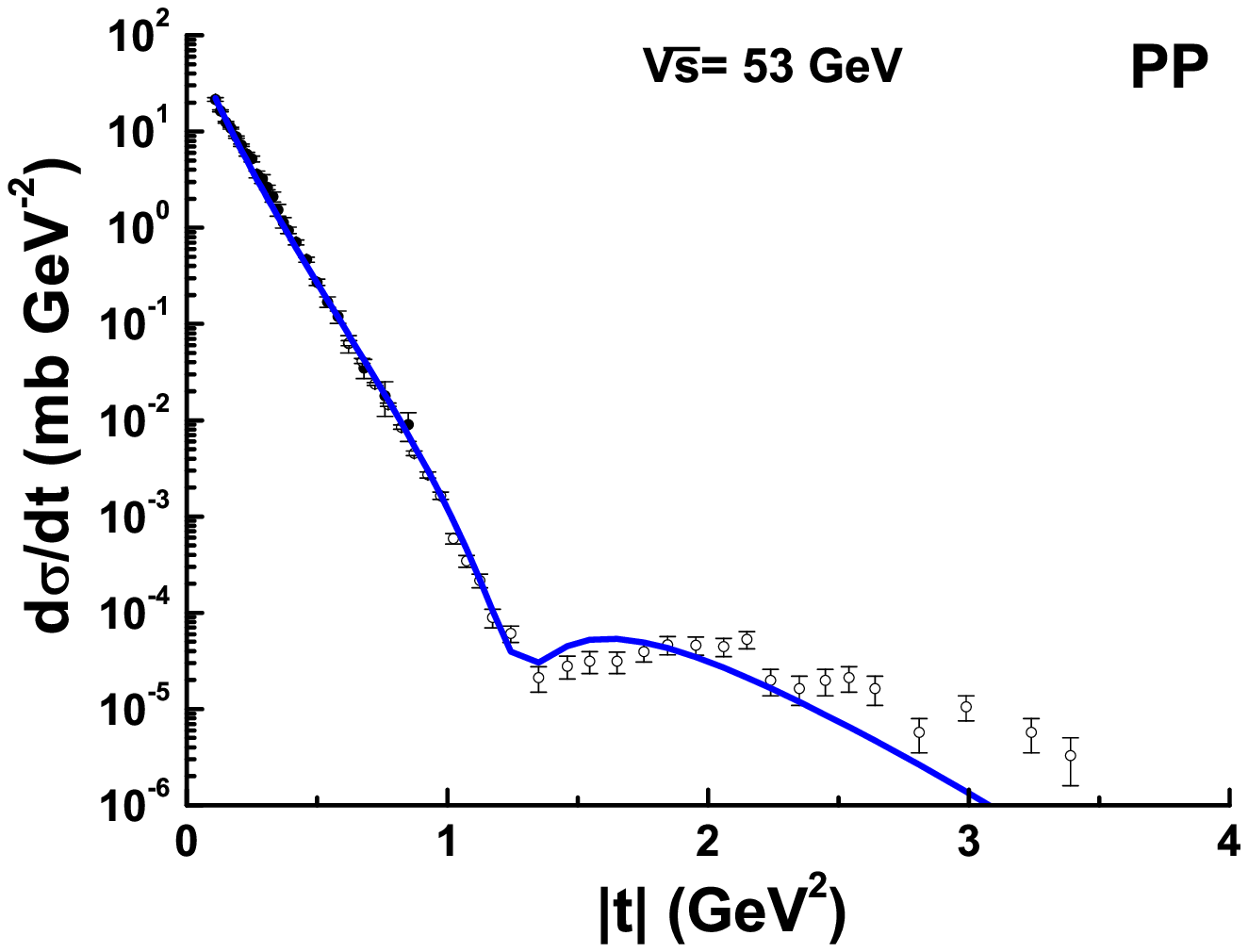}
\begin{minipage}{75mm}
{
\caption{The points are the experimental data by A. Breakstone et al. Nucl. Phys. {\bf B248} (1984) 253;
 Phys. Rev. Lett. {\bf 54} (1985) 2180.}
}
\end{minipage}
\hspace{5mm}
\begin{minipage}{75mm}
{
\caption{The points are the experimental data by S. Erhan et al., Phys. Lett. {\bf B152} (1985) 131;
J.C.M. Armitage et al., Nucl. Phys. {\bf B132} (1978) 365.}
}
\end{minipage}
\includegraphics[width=75mm,height=66mm,clip]{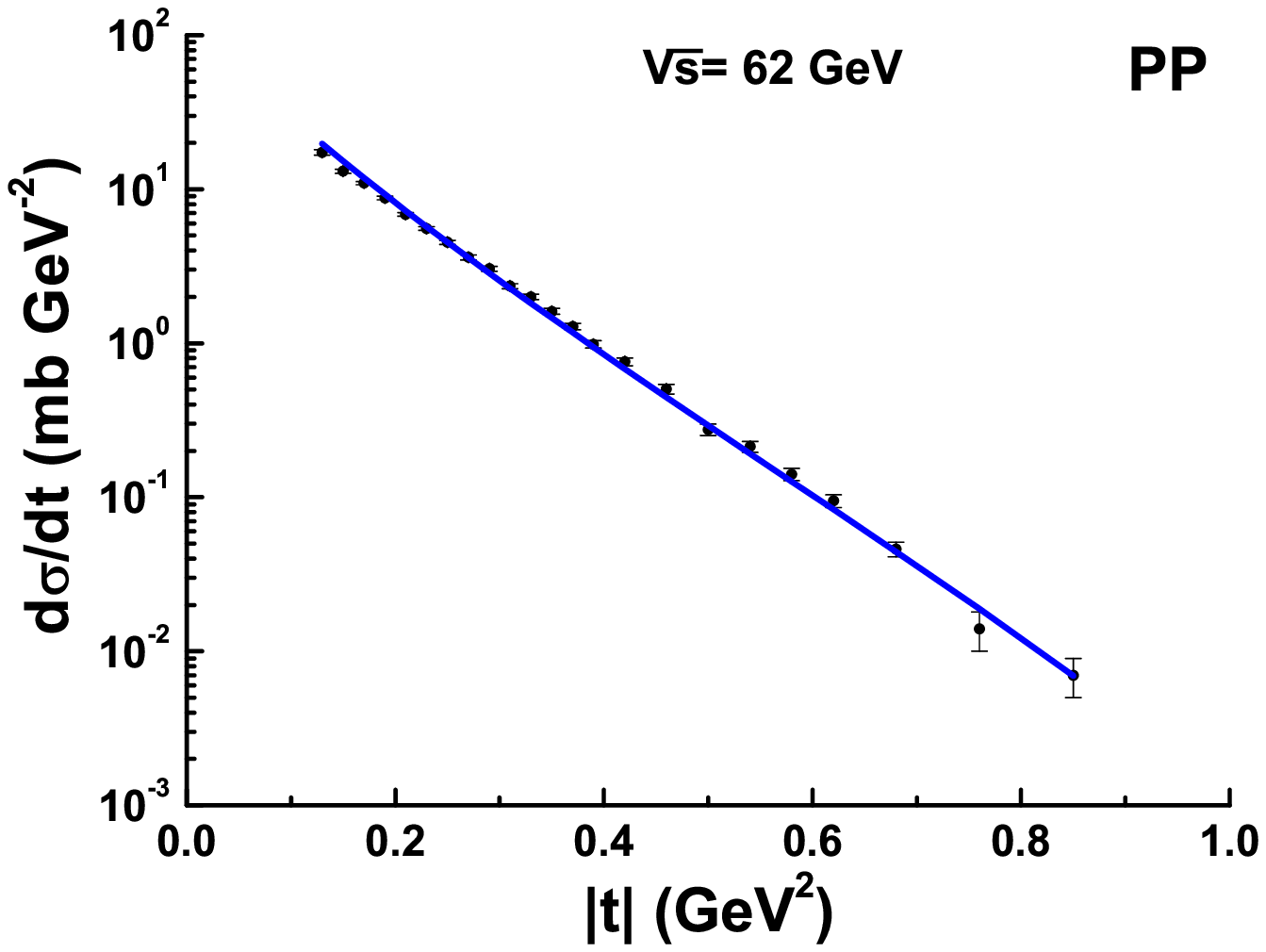}\hspace{5mm}\includegraphics[width=75mm,height=66mm,clip]{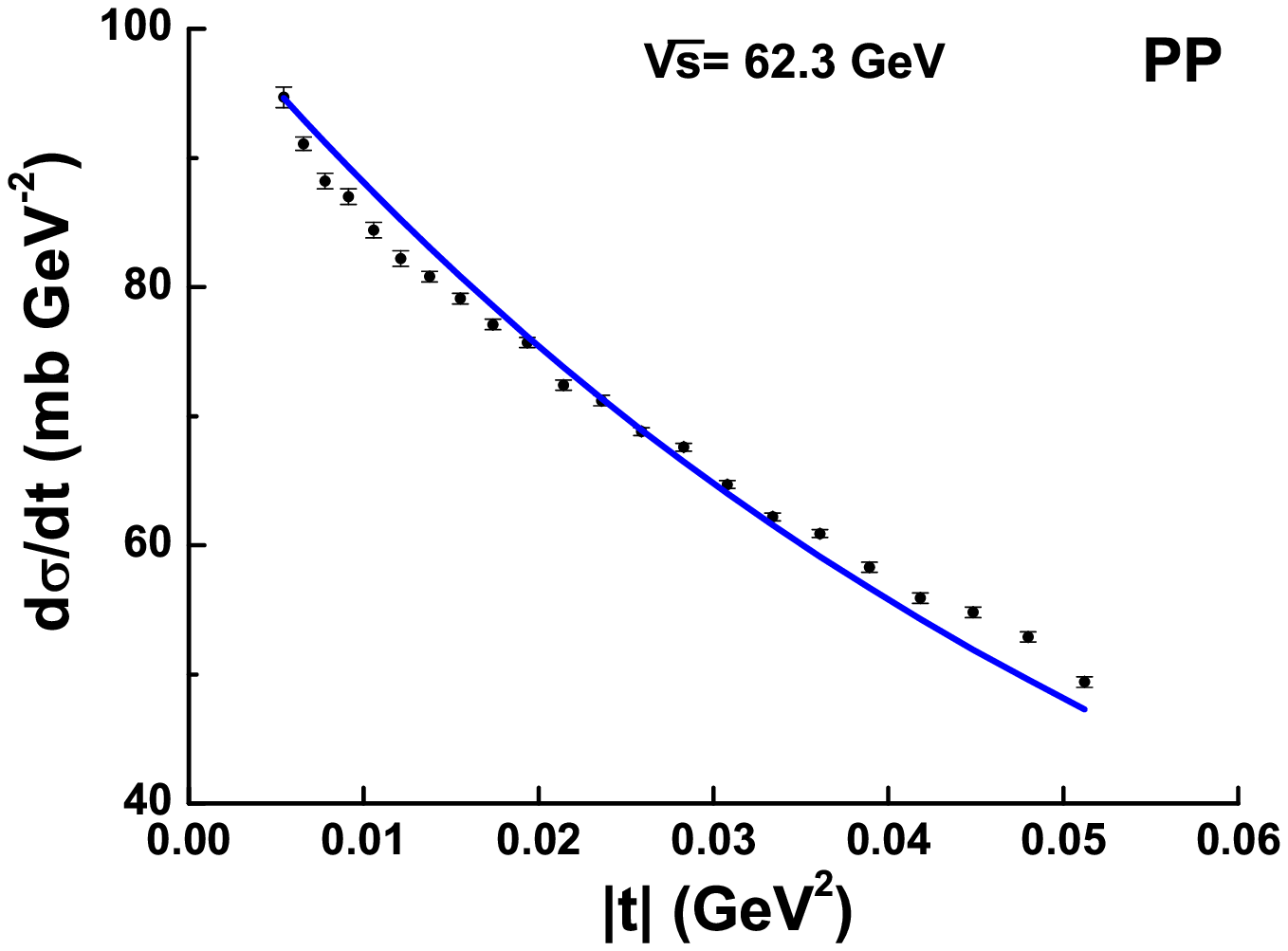}
\begin{minipage}{75mm}
{
\caption{The points are the experimental data by A. Breakstone et al. Nucl. Phys. {\bf B248} (1984) 253.}
}
\end{minipage}
\hspace{5mm}
\begin{minipage}{75mm}
{
\caption{The points are the experimental data by N. Amos et al., Nucl. Phys. {\bf B262} (1985) 689.}
}
\end{minipage}

\includegraphics[width=75mm,height=66mm,clip]{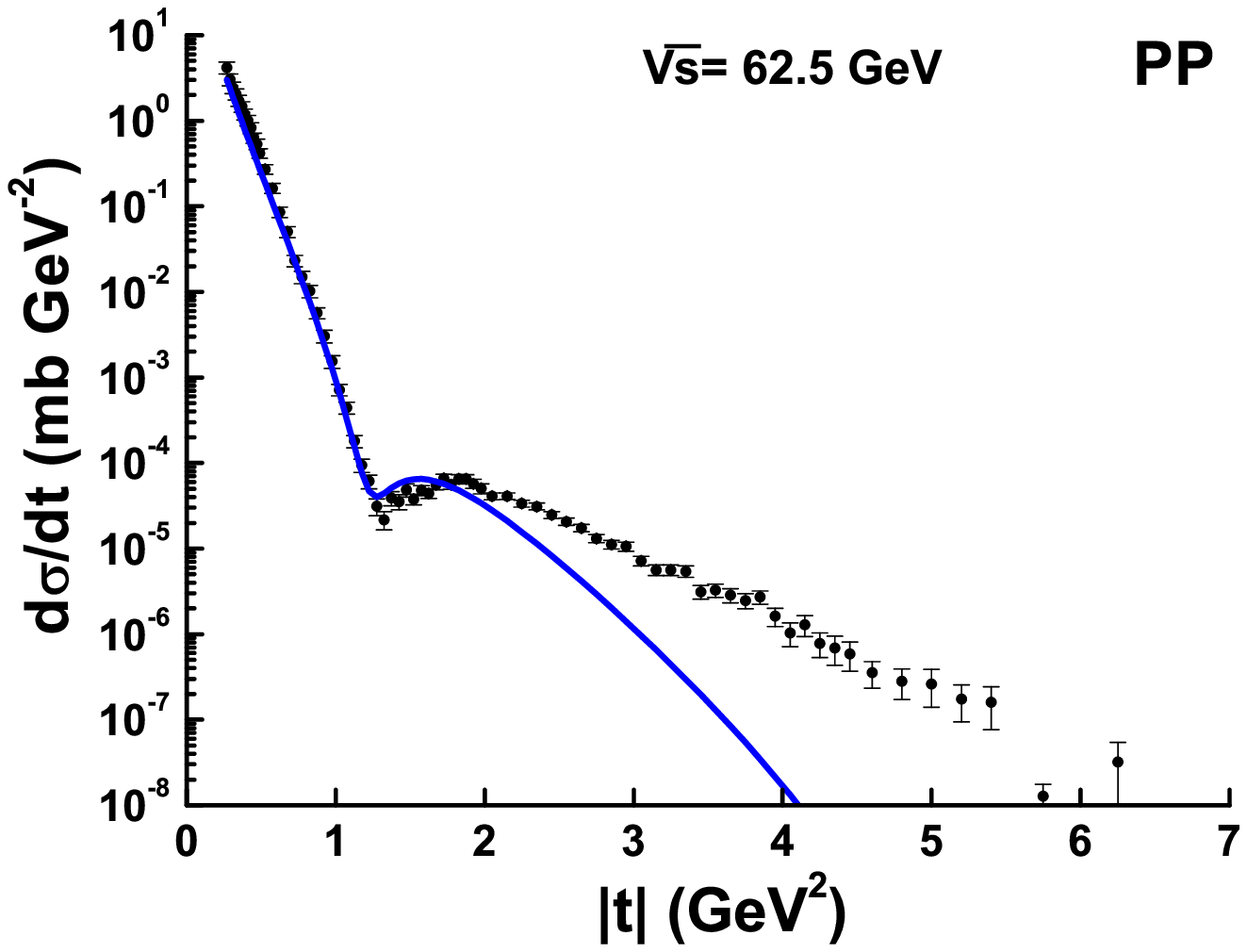}\hspace{5mm}\includegraphics[width=75mm,height=66mm,clip]{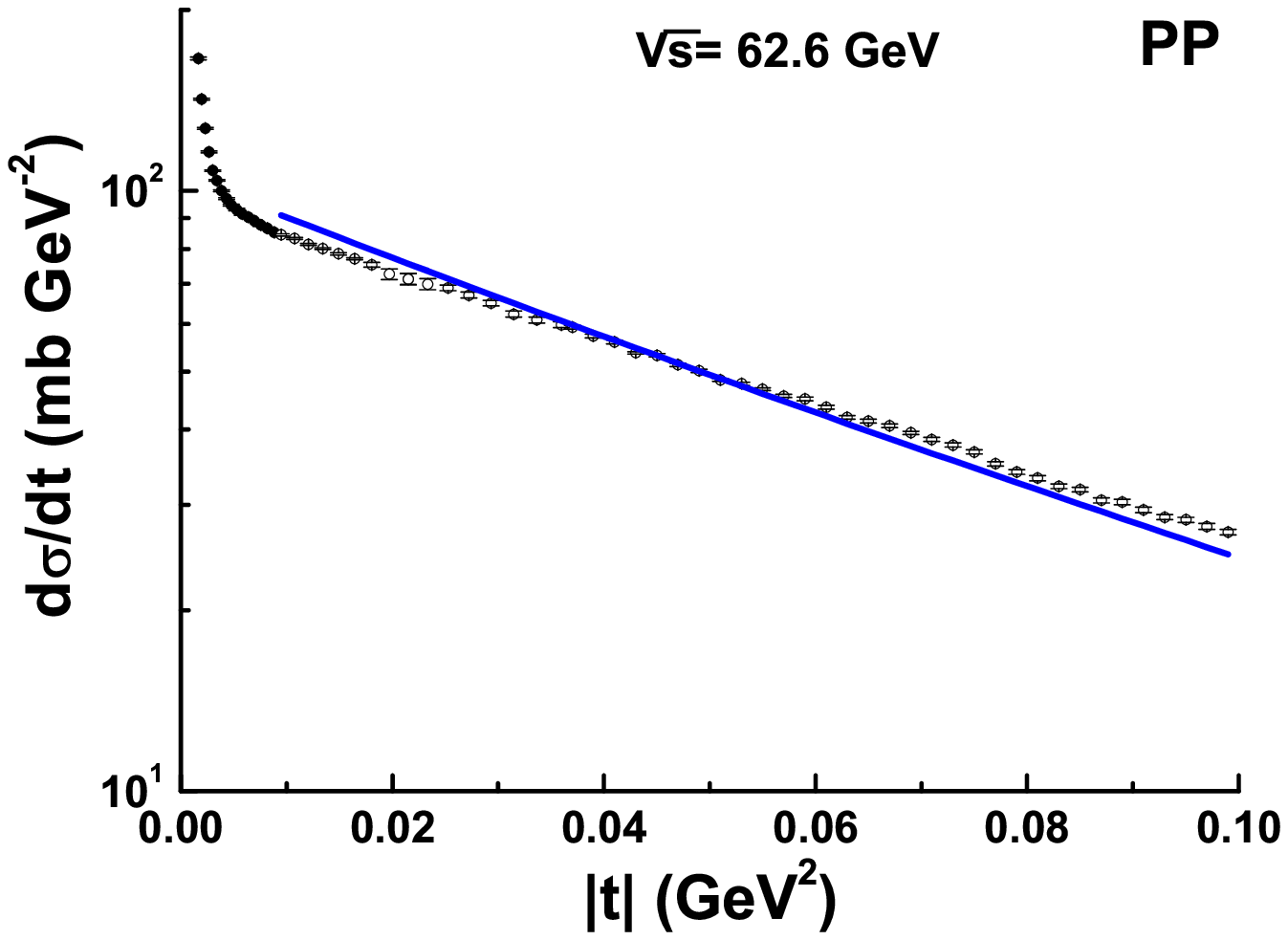}
\begin{minipage}{75mm}
{
\caption{The points are the experimental data by U. Amaldi and K.R. Schubert, Nucl. Phys. {\bf B166} (1980) 301.}
}
\end{minipage}
\hspace{5mm}
\begin{minipage}{75mm}
{
\caption{The points are the experimental data by U. Amaldi and K.R. Schubert, Nucl. Phys. {\bf B166} (1980) 301.}
}
\end{minipage}
\end{figure}

\noindent{\bf Appendix B: Comparison of experimental data on $\bar pp$-interactions with USESD parameterization}

\begin{figure}[cbth]
\includegraphics[width=75mm,height=70mm,clip]{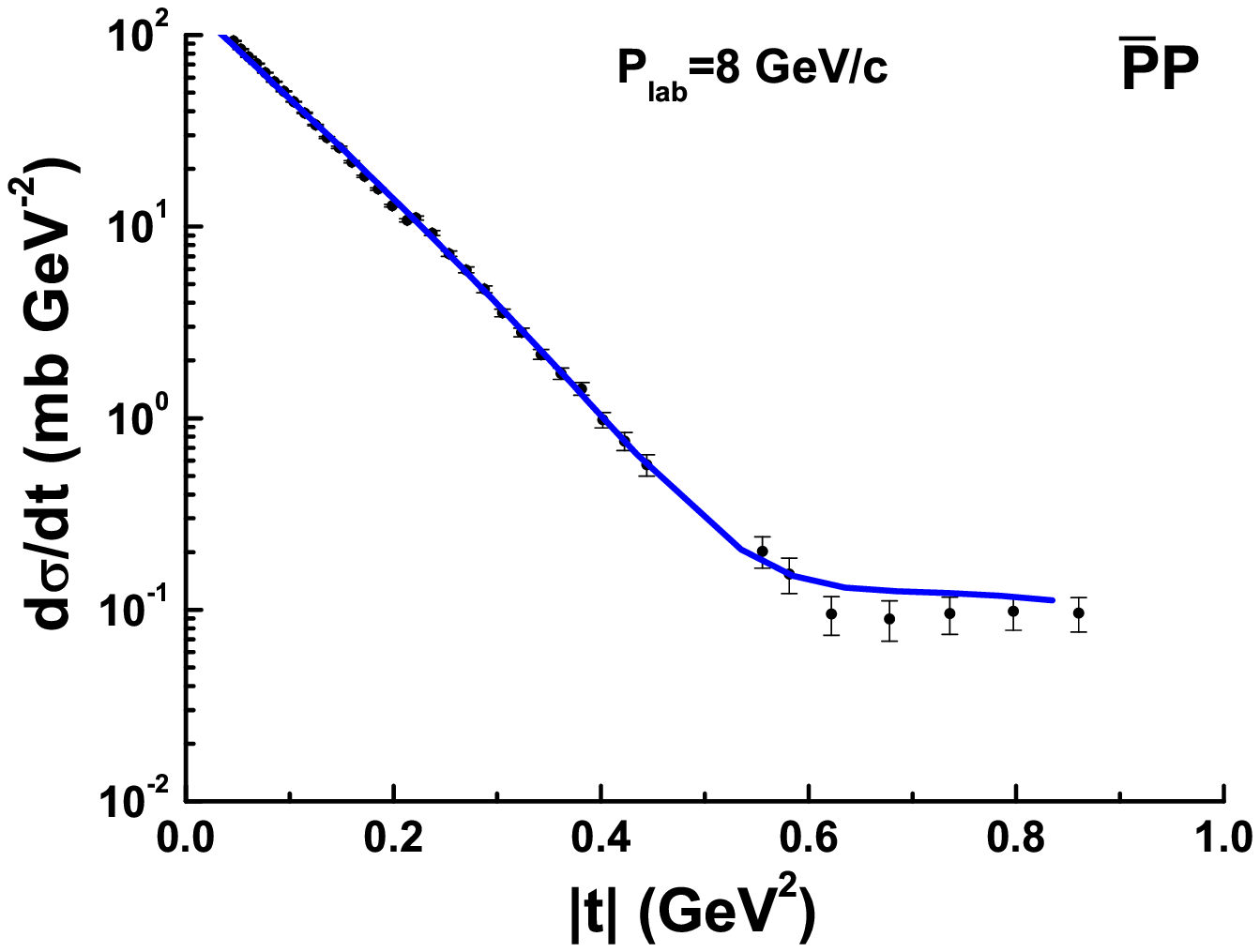}\hspace{5mm}\includegraphics[width=75mm,height=70mm,clip]{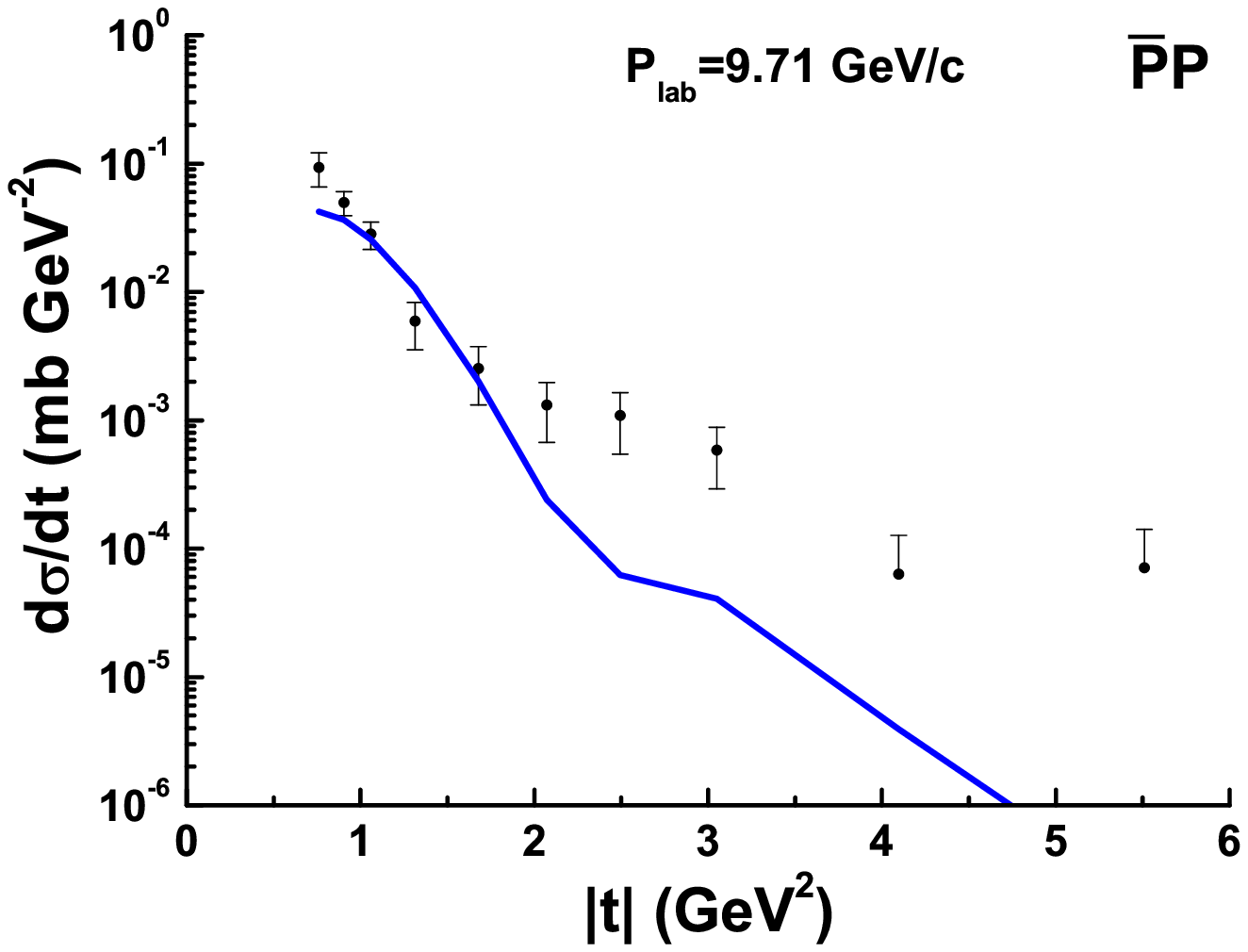}
\begin{minipage}{75mm}
{
\caption{The points are the experimental data by J.S.~Russ et al., Phys. Rev. {\bf D15} (1977) 3139.}
}
\end{minipage}
\hspace{5mm}
\begin{minipage}{75mm}
{
\caption{The points are the experimental data by D.P.~Owen et al., Phys. Rev. {181} (1969) 1794.}
}
\end{minipage}
\includegraphics[width=75mm,height=70mm,clip]{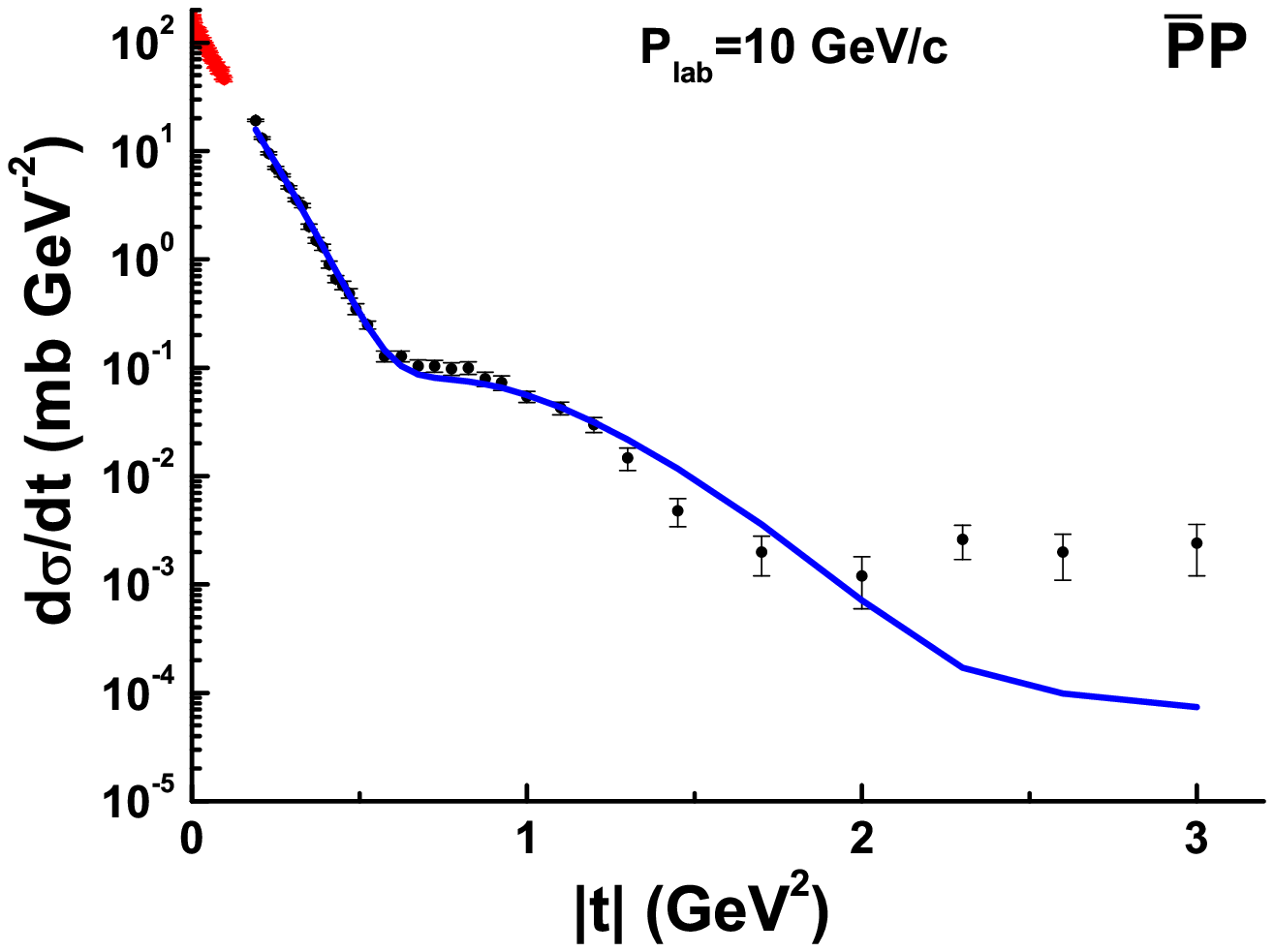}\hspace{5mm}\includegraphics[width=75mm,height=70mm,clip]{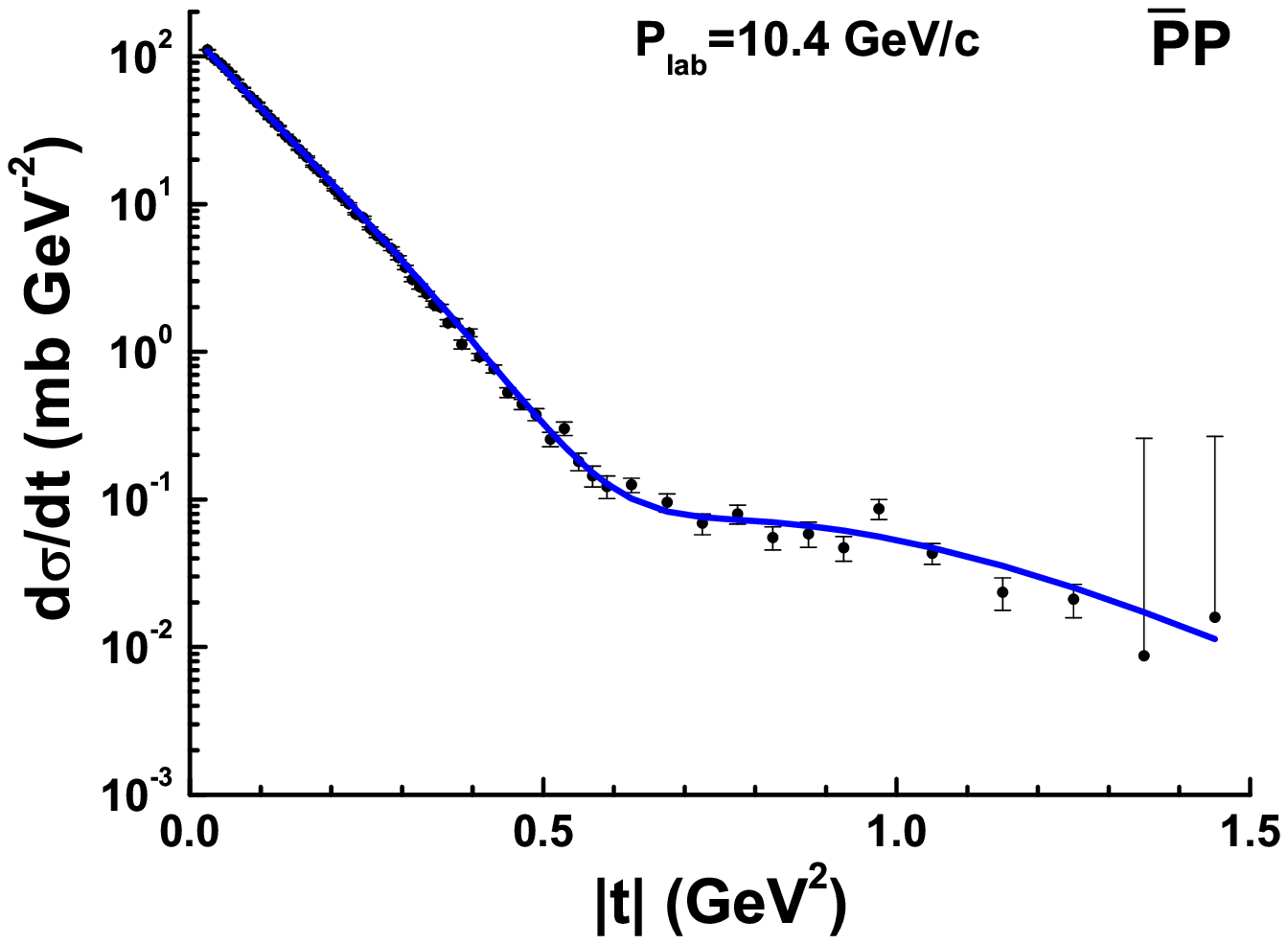}
\begin{minipage}{75mm}
{
\caption{The points are the experimental data by A.~Berglund et al., Nucl. Phys. {\bf B176} 1980) 346.}
}
\end{minipage}
\hspace{5mm}
\begin{minipage}{75mm}
{
\caption{The points are the experimental data by G.~Brandenburg et al., Phys. Lett. {\bf 58B} (1975) 367.}
}
\end{minipage}

\end{figure}

\begin{figure}[cbth]

\includegraphics[width=75mm,height=66mm,clip]{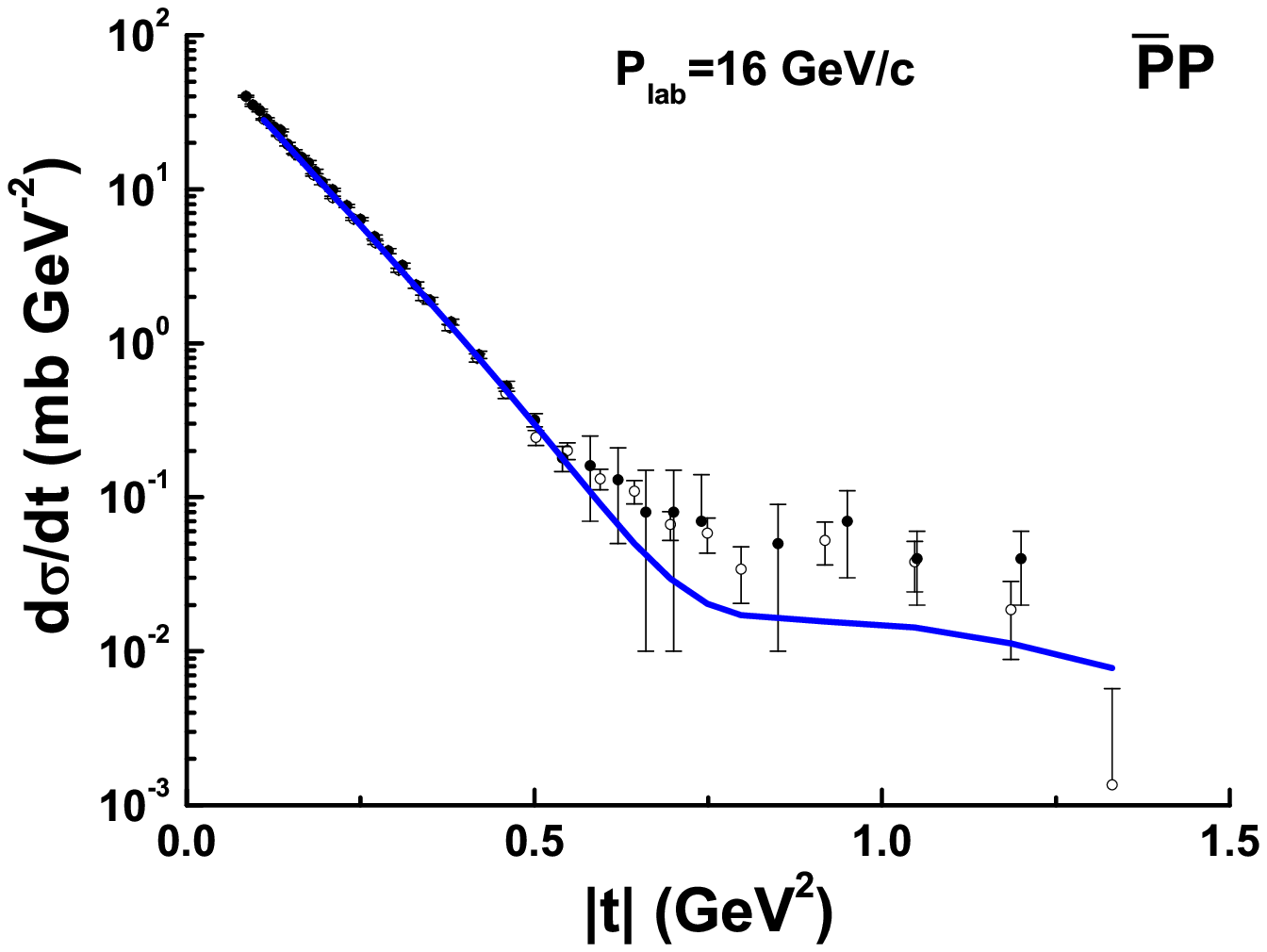}\hspace{5mm}\includegraphics[width=75mm,height=66mm,clip]{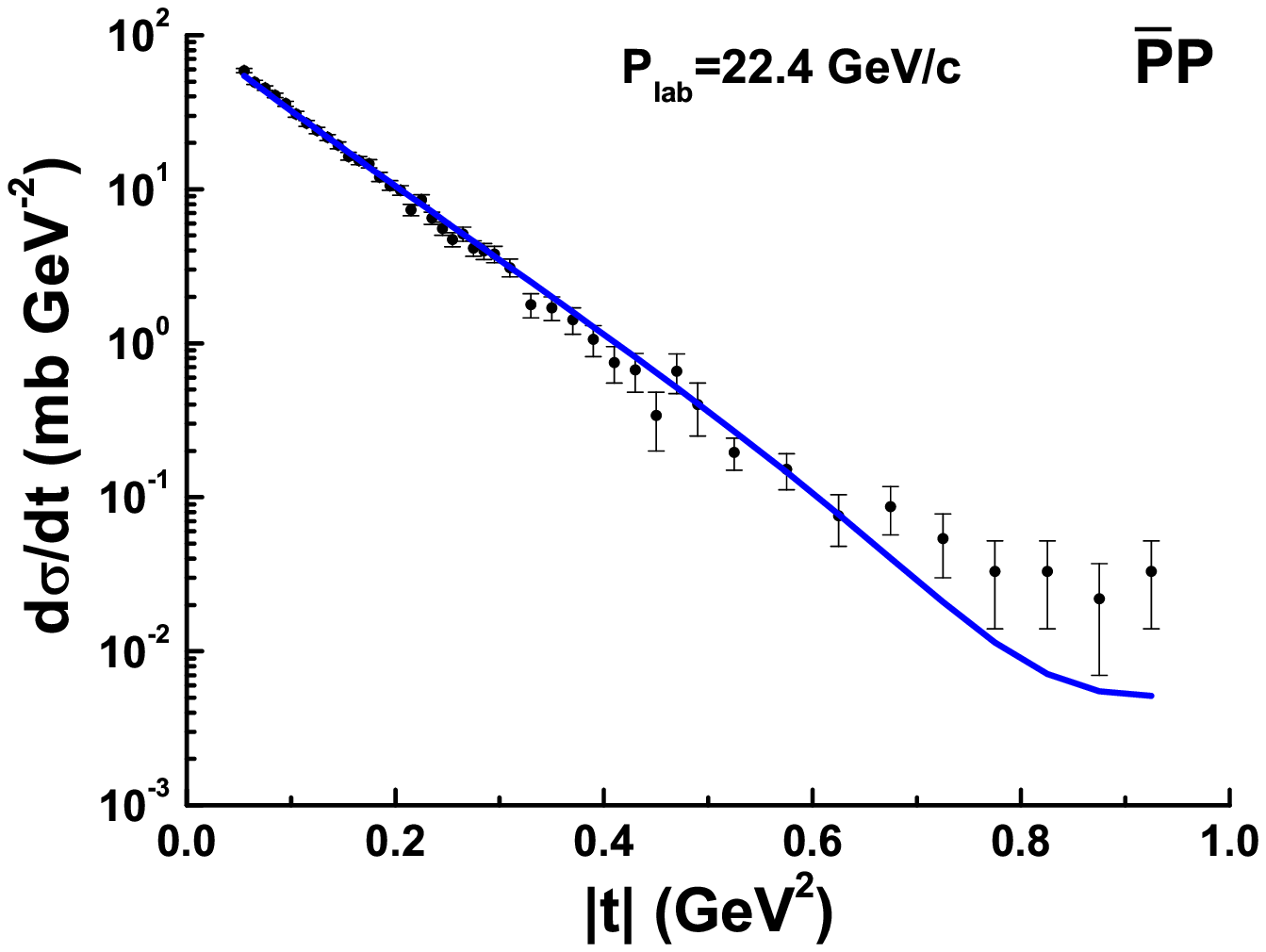}
\begin{minipage}{75mm}
{
\caption{The points are the experimental data by K.J. D. Birnbaum et al., Phys. Rev. Lett. {\bf 23} (1969) 663.}
}
\end{minipage}
\hspace{5mm}
\begin{minipage}{75mm}
{
\caption{The points are the experimental data by B.~Batyunya et al. Yad. Fiz. {\bf 44} (1986) 1489.}
}
\end{minipage}
\includegraphics[width=75mm,height=66mm,clip]{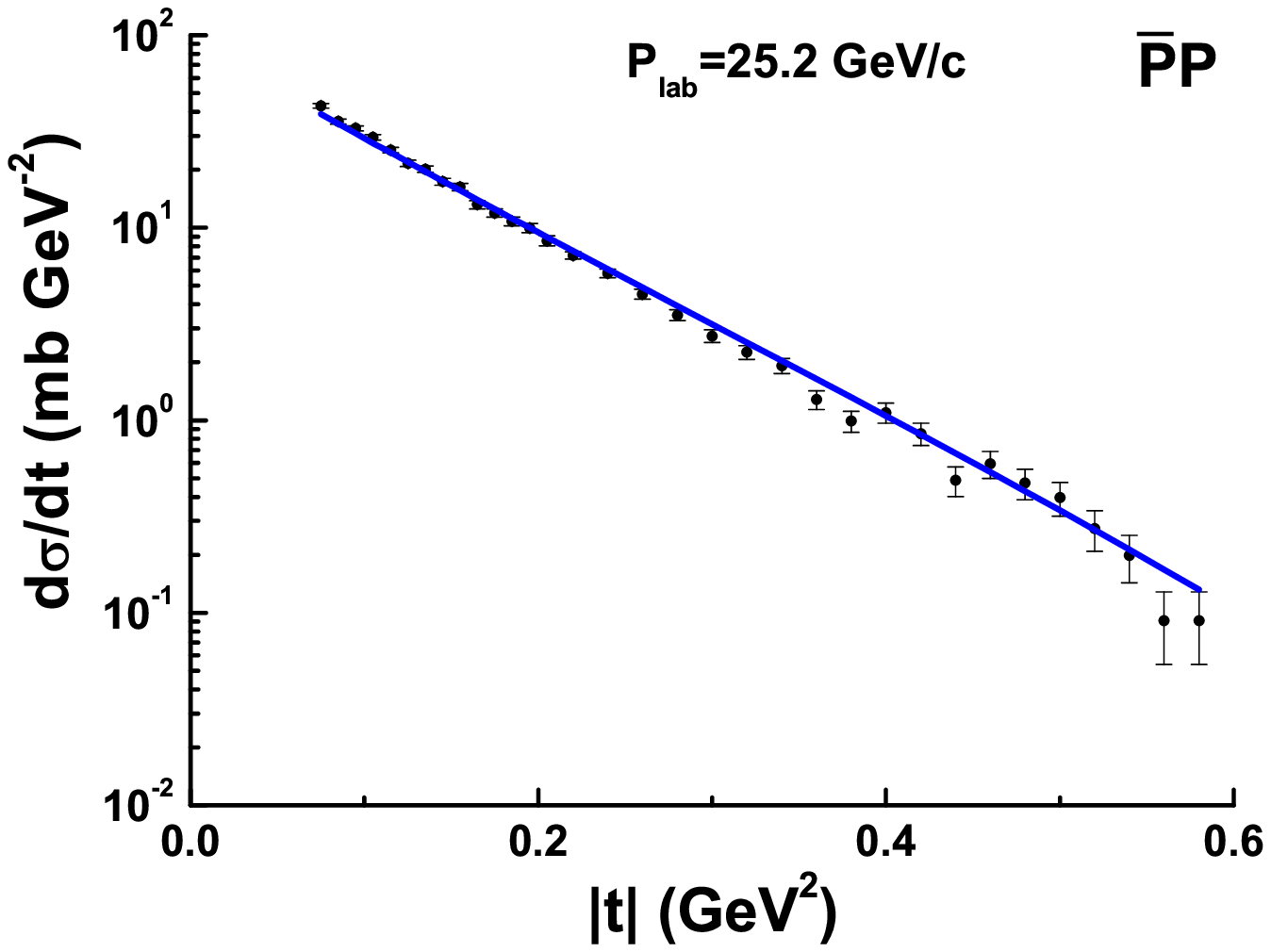}\hspace{5mm}\includegraphics[width=75mm,height=66mm,clip]{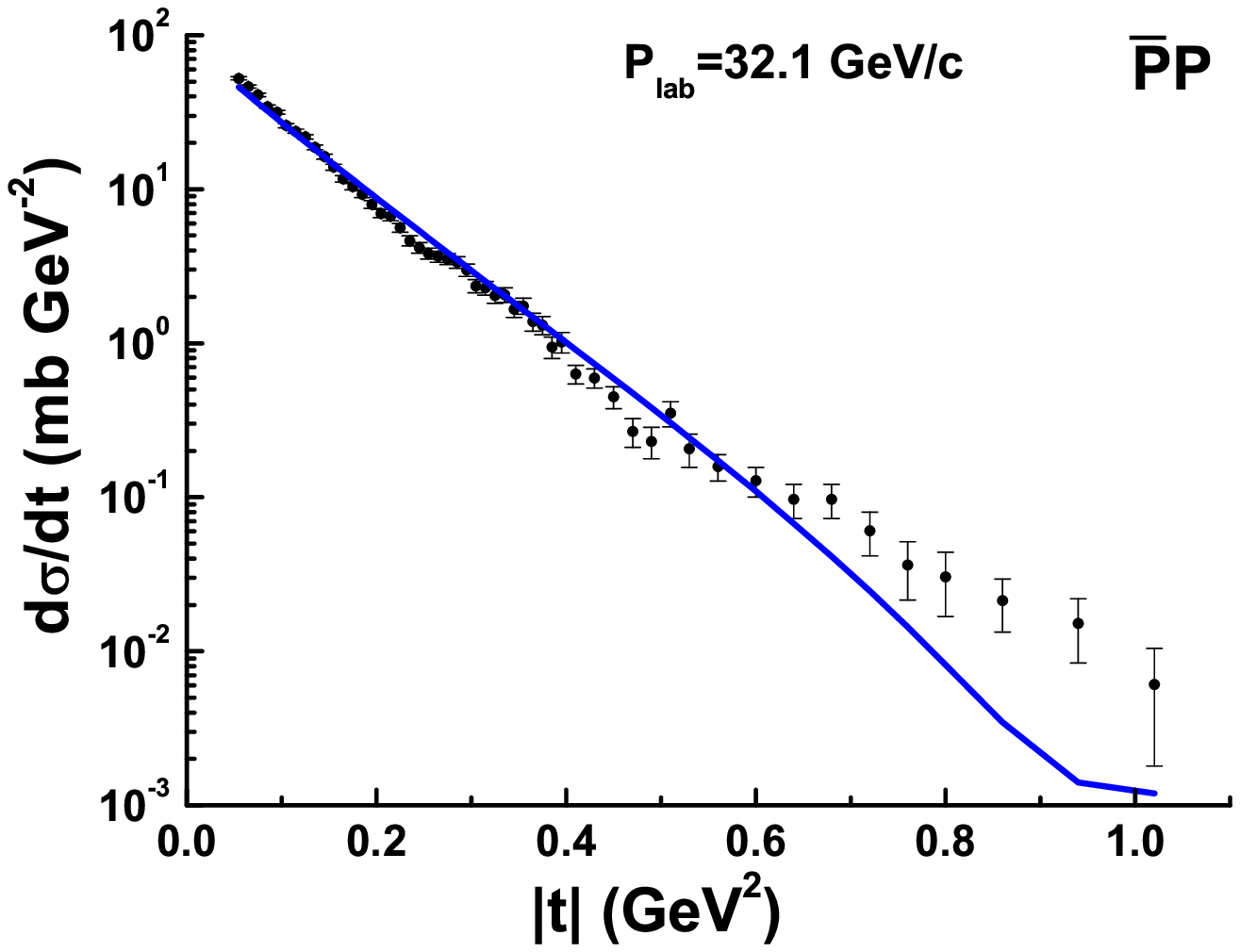}
\begin{minipage}{75mm}
{
\caption{The points are the experimental data by Yu.M.~Antipov et al., Nucl. Phys. {\bf B57} (1973) 333.}
}
\end{minipage}
\hspace{5mm}
\begin{minipage}{75mm}
{
\caption{The points are the experimental data by M.Y. Bogolyubsky et al., Yad. Fiz. {\bf 41} (1985) 1210.}
}
\end{minipage}
\includegraphics[width=75mm,height=66mm,clip]{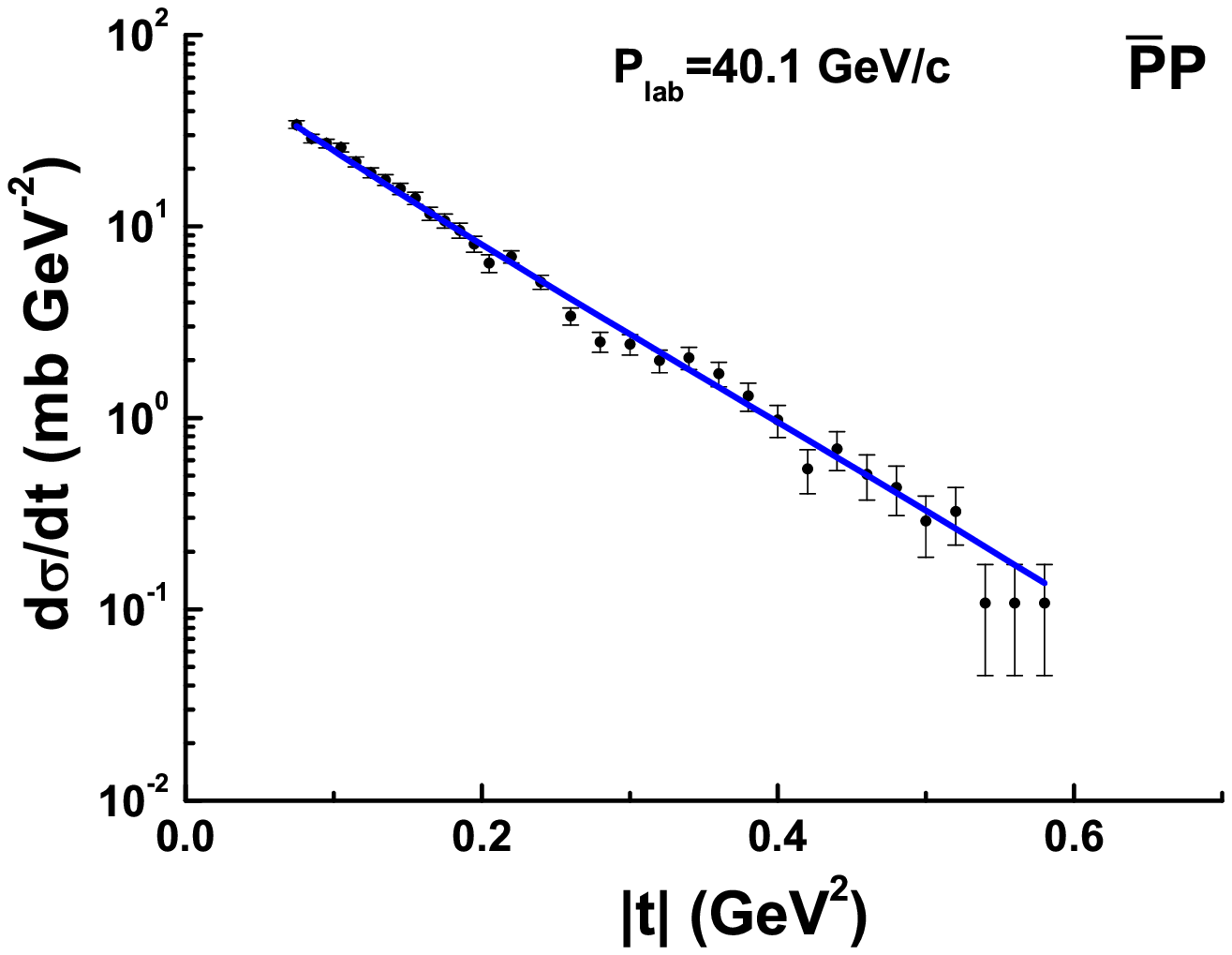}\hspace{5mm}\includegraphics[width=75mm,height=66mm,clip]{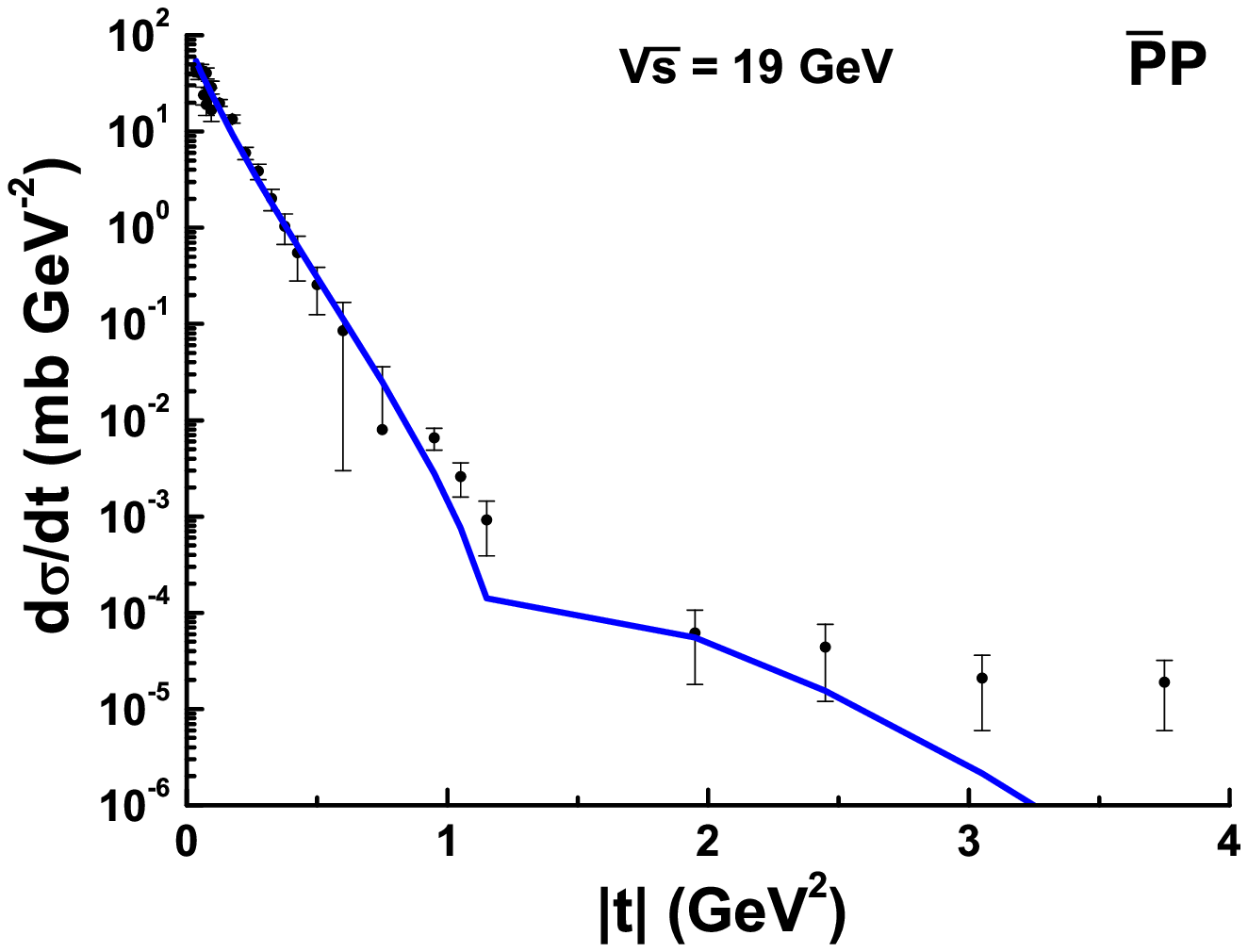}
\begin{minipage}{75mm}
{
\caption{The points are the experimental data by Yu.M.~Antipov et al., Nucl. Phys. {\bf B57} (1973) 333.}
}
\end{minipage}
\hspace{5mm}
\begin{minipage}{75mm}
{
\caption{The points are the experimental data by R.L. Cool et al., Phys. Rev. {\bf D24} (1981) 2821.}
}
\end{minipage}
\end{figure}

\begin{figure}[cbth]
\includegraphics[width=75mm,height=66mm,clip]{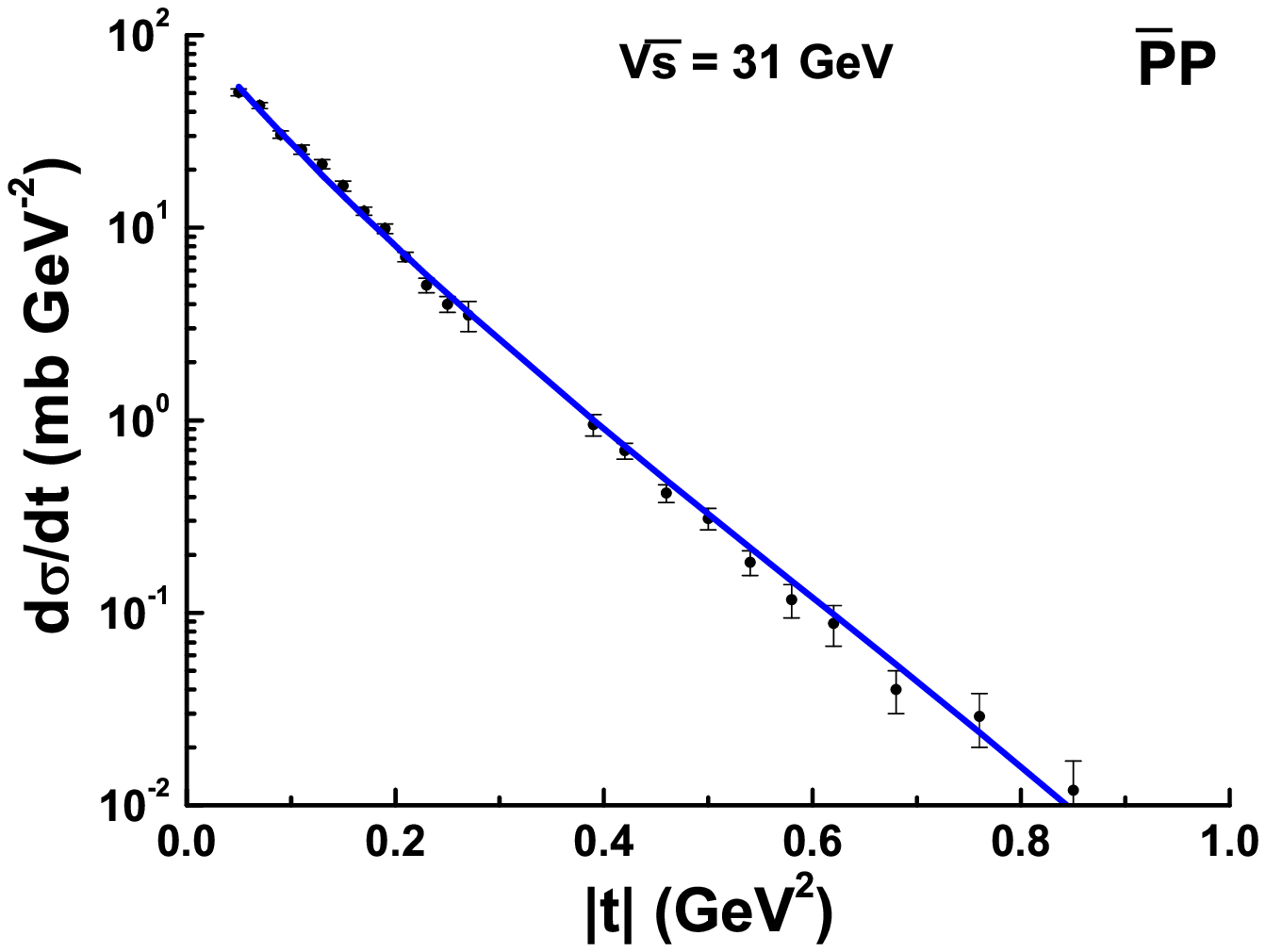}\hspace{5mm}\includegraphics[width=75mm,height=66mm,clip]{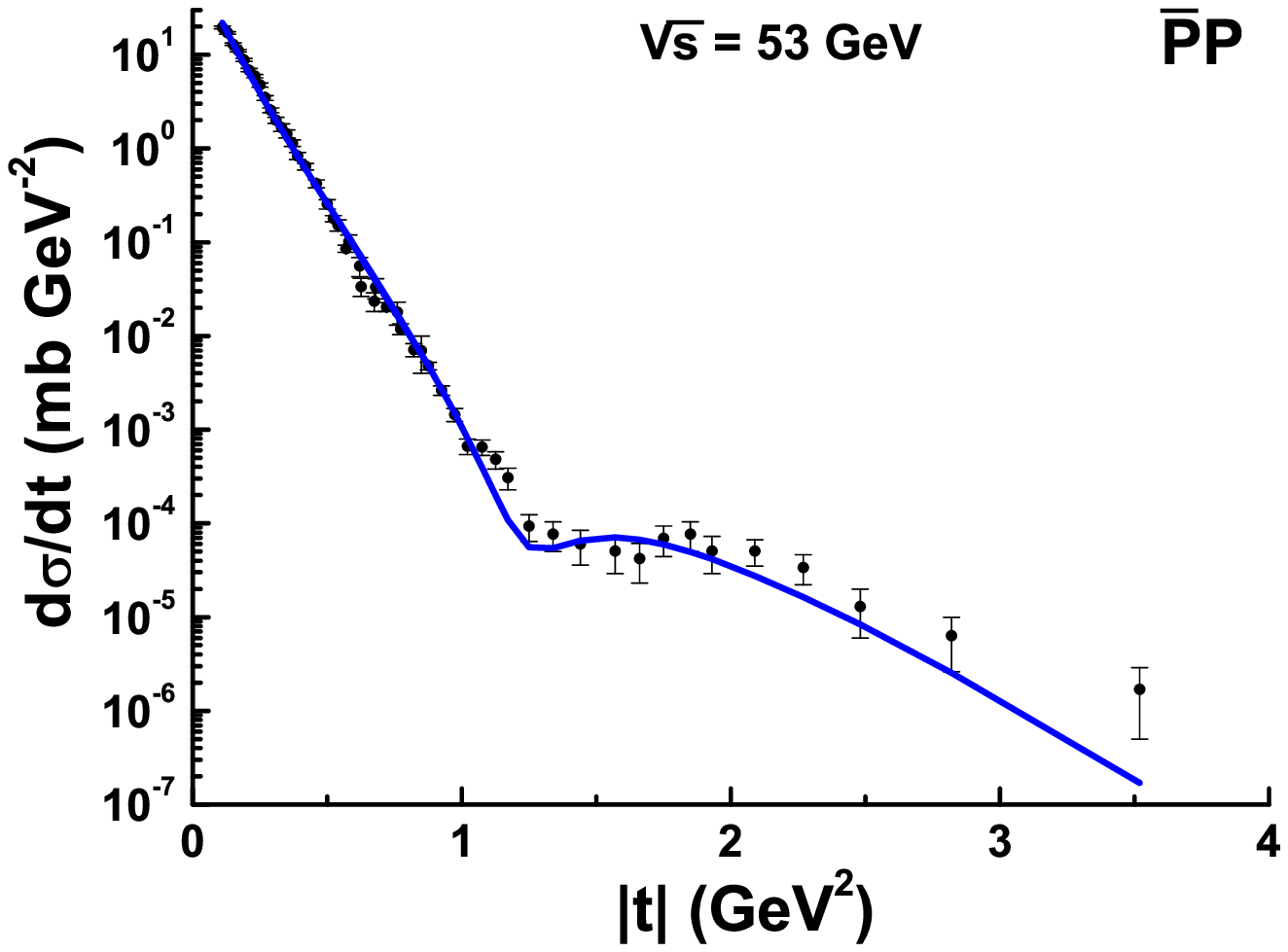}
\begin{minipage}{75mm}
{
\caption{The points are the experimental data by A. Breakstone et al., Nucl. Phys. {\bf B248} (1984) 253,
Phys. Rev. Lett. {\bf 54} (1985) 2180.}
}
\end{minipage}
\hspace{5mm}
\begin{minipage}{75mm}
{
\caption{The points are the experimental data by A. Breakstone et al., Nucl. Phys. {\bf B248} (1984) 253,
Phys. Rev. Lett. {\bf 54} (1985) 2180.}
}
\end{minipage}
\includegraphics[width=75mm,height=66mm,clip]{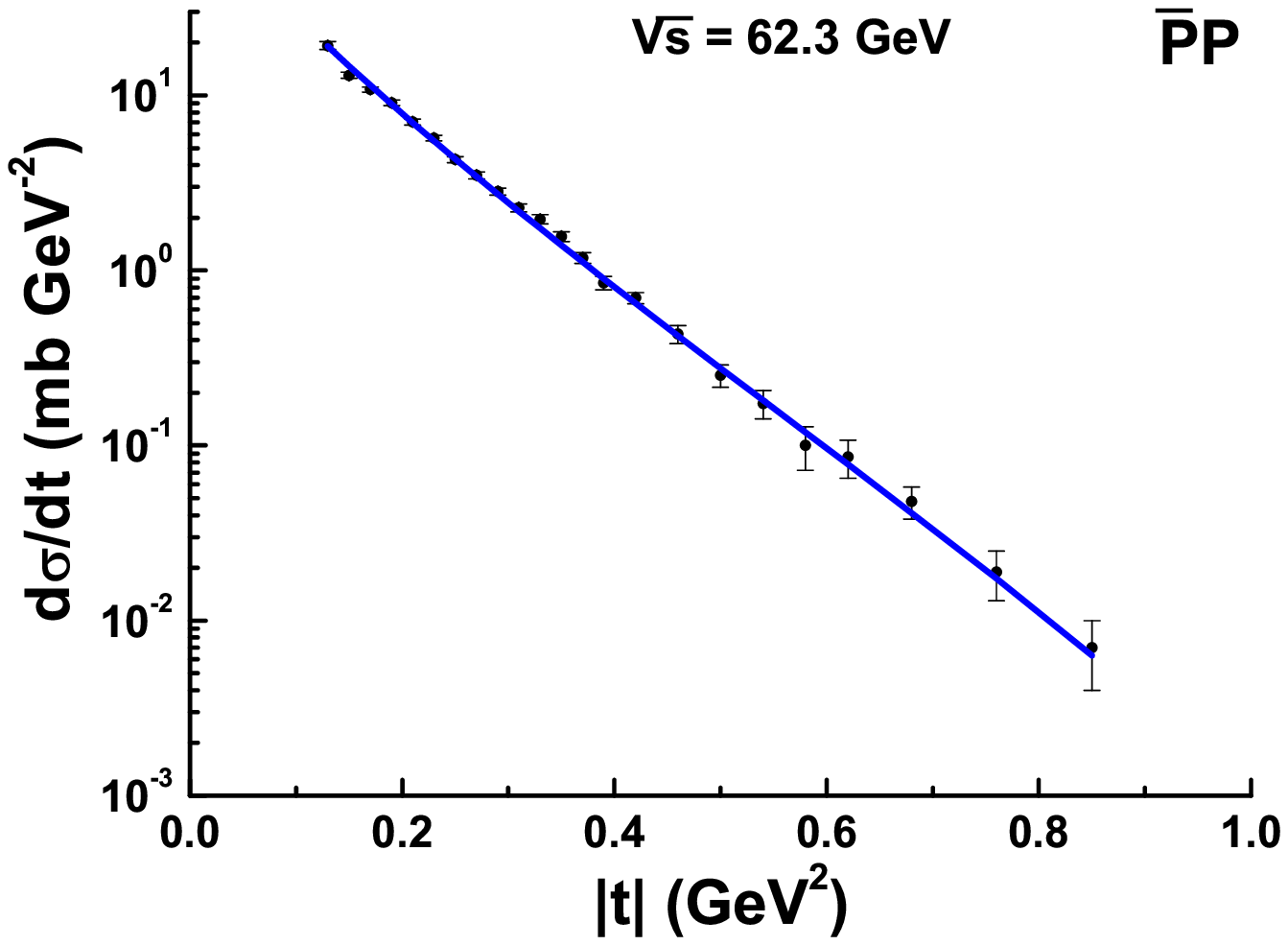}\hspace{5mm}\includegraphics[width=75mm,height=66mm,clip]{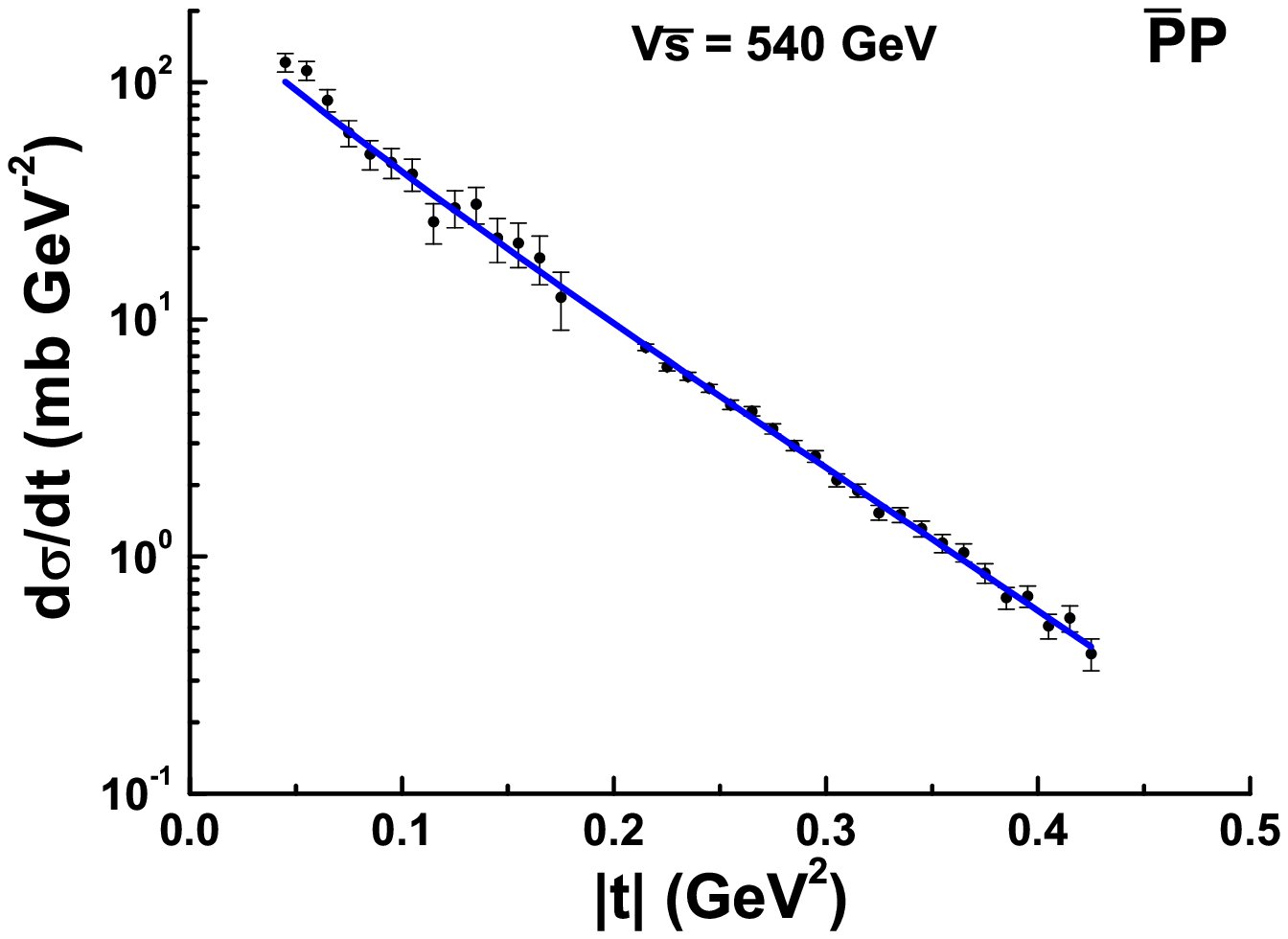}
\begin{minipage}{75mm}
{
\caption{The points are the experimental data by A. Breakstone et al., Nucl. Phys. {\bf B248} (1984) 253,
Phys. Rev. Lett. {\bf 54} (1985) 2180.}
}
\end{minipage}
\hspace{5mm}
\begin{minipage}{75mm}
{
\caption{The points are the experimental data by G. Arnison et al., Phys. Lett. {\bf B128} (1983) 336.}
}
\end{minipage}
\includegraphics[width=75mm,height=66mm,clip]{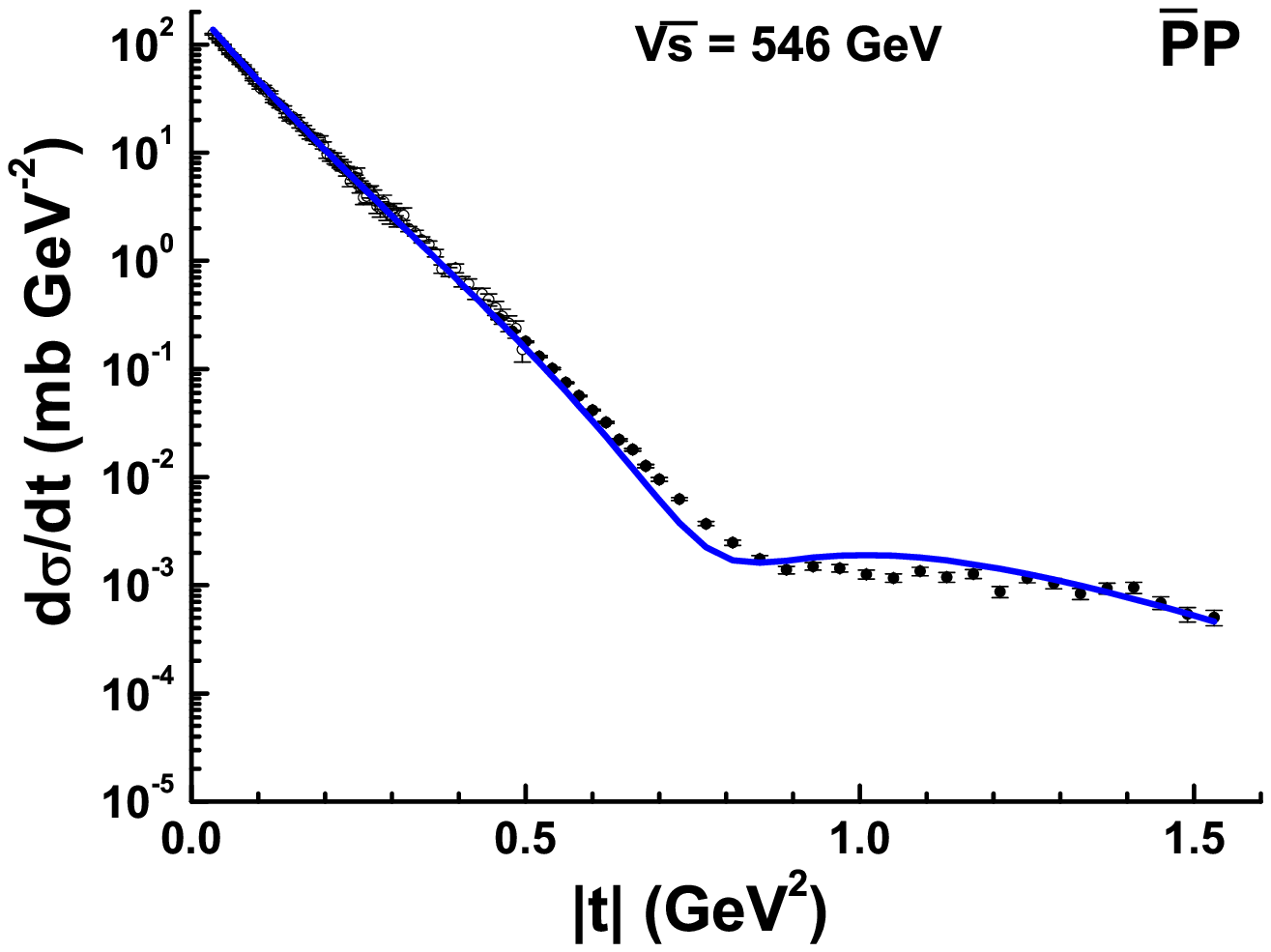}\hspace{5mm}\includegraphics[width=75mm,height=66mm,clip]{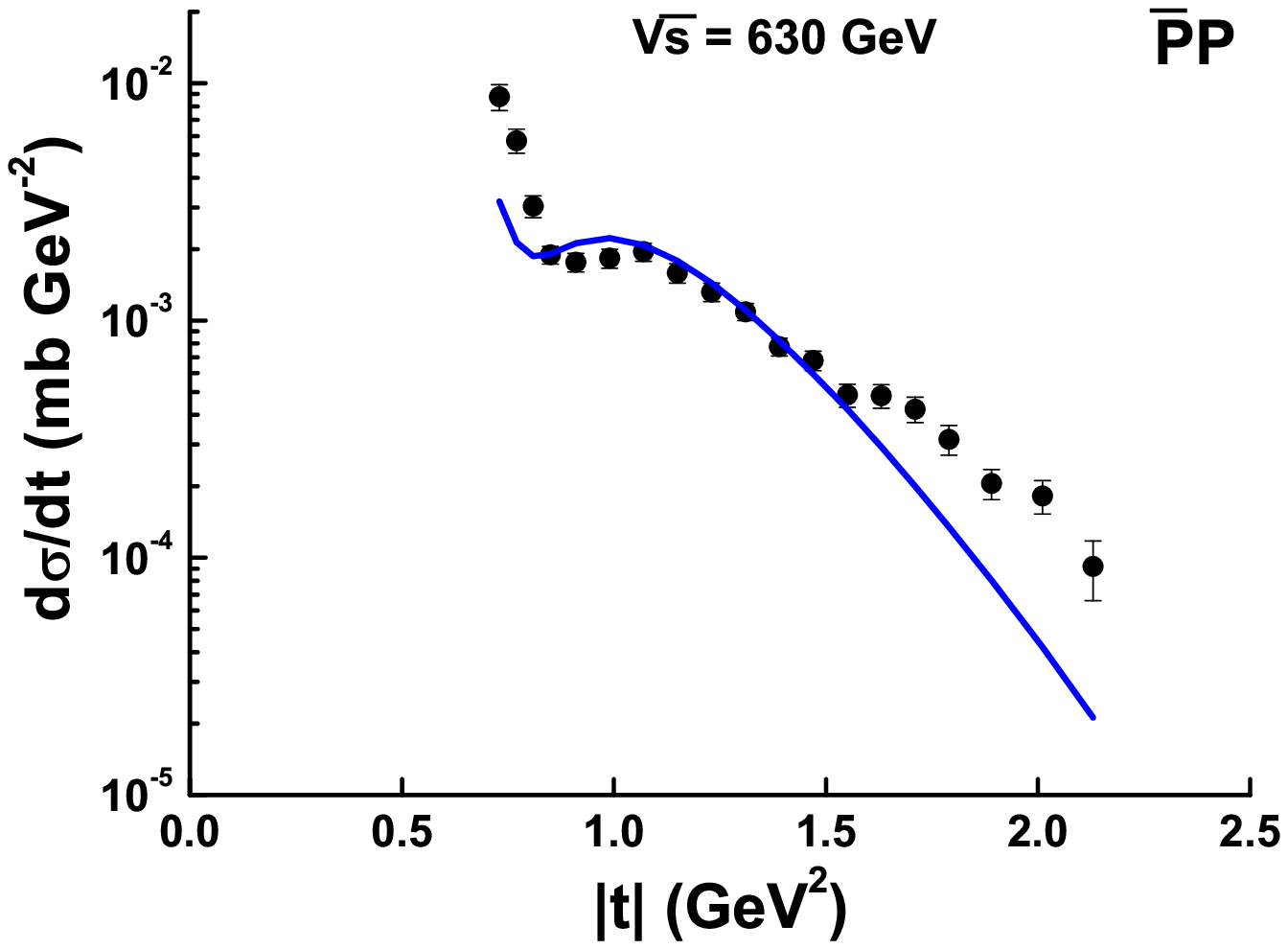}
\begin{minipage}{75mm}
{
\caption{The points are the experimental data by M. Bozzo et al., Phys. Lett. {\bf B155} (1985) 197,
Phys. Lett. {\bf B147} (1984) 385.}
}
\end{minipage}
\hspace{5mm}
\begin{minipage}{75mm}
{
\caption{The points are the experimental data by D. Bernard et al., Phys. Lett. {\bf B198} (1987) 583,
Phys. Lett. {\bf B171} (1986) 142.}
}
\end{minipage}
\end{figure}

\begin{figure}[cbth]
\includegraphics[width=75mm,height=66mm,clip]{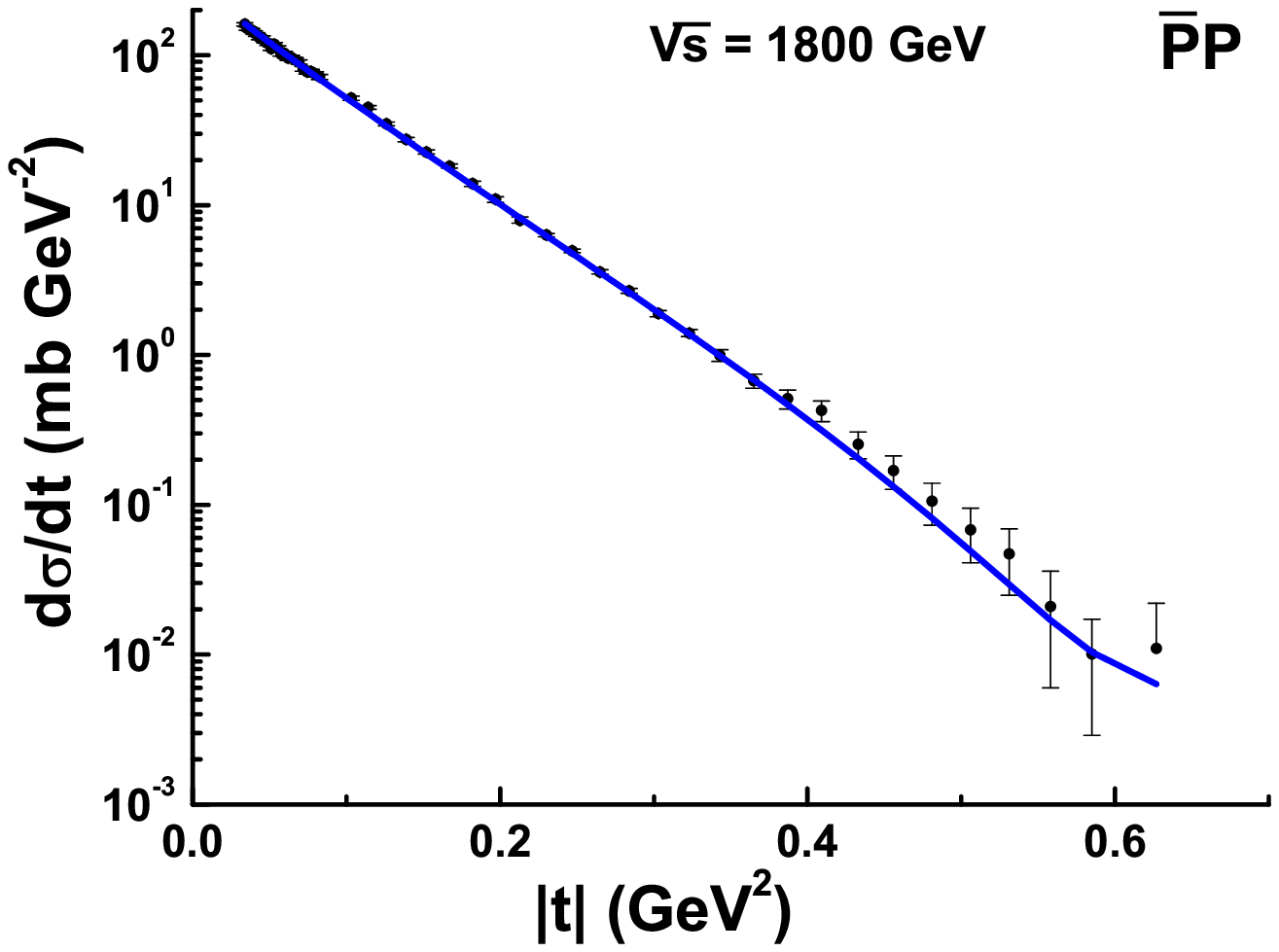}\hspace{5mm}\includegraphics[width=75mm,height=66mm,clip]{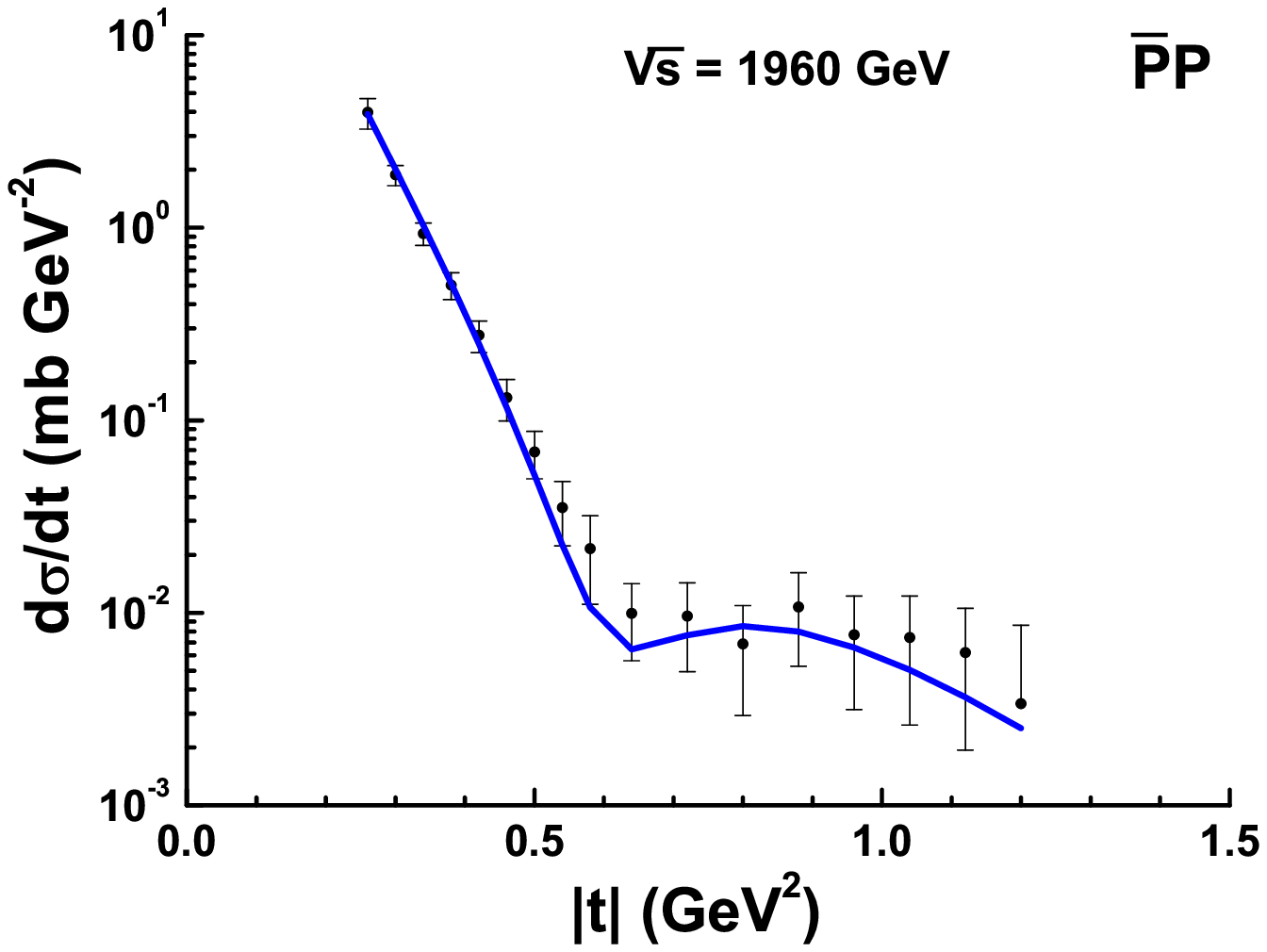}
\begin{minipage}{75mm}
{
\caption{The points are the experimental data by N. Amos et al., Nucl. Phys. {\bf B262} (1985) 689,
Phys. Lett. {\bf B247} (1990) 127.}
}
\end{minipage}
\hspace{5mm}
\begin{minipage}{75mm}
{
\caption{The points are the experimental data by the D0 Collaboration -- D0 Note 6056-CONF.}
}
\end{minipage}
\end{figure}

\begin{table}[cbth]
\noindent{\bf Appendix C: Quality of the data fitting}\\
$\chi^2$ presented below is obtained with values of the parameters
given by Eqs. \ref{Eq12} -- \ref{Eq16}, \ref{Eq18}. $N$ is number of experimental points.
$N_F$ - the number of points included in the fitting.

\begin{center}
\caption{Fitting results for $pp$-interactions}

\begin{tabular}{|r|r|r|r|r|c|c|c|c|c|}
\hline
$P_{lab}$&$\sqrt{s}$ & $N$ & $N_F$ & $\chi^2/NoF$ &  A   & $R$  & $d$ & $\rho$ & $a_1$ \\
(GeV/c)  &  (GeV)    &     &       &              &      & (fm) & (fm)&        & (fm$^2$)  \\ \hline
9.9      & 4.523     &   28&  28   &  0.560        & 1.180$\pm$ 0.018&0.547& 0.249  & -0.323& 0.0244 \\
10       & 4.543&   24&  21&  8.66&  0.746 $\pm$ 0.016& 0.547& 0.250& -0.322& 0.0242 \\
10.4     & 4.625&   97&  97&  2.56&  1.244 $\pm$ 0.002& 0.545& 0.252& -0.319& 0.0234 \\
10.8     & 4.705&   13&  13&  2.70&  1.177 $\pm$ 0.012& 0.544& 0.254& -0.315& 0.0226 \\
10.9     & 4.733&    9&   9&  0.72&  1.108 $\pm$ 0.014& 0.544& 0.254& -0.313& 0.0223 \\
12       & 4.939&   25&  18&  8.48&  0.845 $\pm$ 0.019& 0.540& 0.259& -0.303& 0.0205 \\
12.4     & 5.014&   22&  21&  0.72&  1.065 $\pm$ 0.012& 0.539& 0.260& -0.299& 0.0199 \\
12.8     & 5.088&   13&  13&  2.96&  1.180 $\pm$ 0.013& 0.538& 0.262& -0.296& 0.0193 \\
14.2     & 5.34 &   22&  12&  4.32&  0.818 $\pm$ 0.021& 0.535& 0.266& -0.283& 0.0175 \\
14.8     & 5.444&   12&  12&  1.98&  1.159 $\pm$ 0.014& 0.533& 0.268& -0.278& 0.0169 \\
15.1     & 5.496&   32&  25&  0.22&  1.200 $\pm$ 0.019& 0.533& 0.269& -0.276& 0.0166 \\
16.7     & 5.762&   12&  12&  1.22&  1.154 $\pm$ 0.014& 0.530& 0.273& -0.263& 0.0151 \\
18.4     & 6.033&   20&  18&  1.32&  1.076 $\pm$ 0.016& 0.527& 0.277& -0.251& 0.0137 \\
19.2     & 6.156&   27&  19&  14.4&  0.899 $\pm$ 0.011& 0.526& 0.279& -0.245& 0.0132 \\
19.6     & 6.217&   11&  11&  1.82&  1.132 $\pm$ 0.014& 0.526& 0.280& -0.243& 0.0129 \\
20~      & 6.277&   30&  28&  2.92&  1.043 $\pm$ 0.016& 0.525& 0.280& -0.240& 0.0127 \\
21.12    & 6.442&   17&  15& 12.04&  0.946 $\pm$ 0.011& 0.524& 0.282& -0.233& 0.0121 \\
24~      & 6.849&   35&  20&  5.62&  1.000 $\pm$ 0.015& 0.521& 0.287& -0.217& 0.0107 \\
29.7     & 7.591&   29&  25&  1.60&  1.060 $\pm$ 0.018& 0.517& 0.294& -0.190& 0.0087 \\
35~      & 8.221&   21&  19&  1.30&  1.120 $\pm$ 0.022& 0.514& 0.299& -0.171& 0.0074 \\
44.5     & 9.243&   27&  25&  7.28&  1.093 $\pm$ 0.004& 0.511& 0.306& -0.143& 0.0059 \\
50~      & 9.787&   33&  27& 11.06&  1.106 $\pm$ 0.006& 0.510& 0.309& -0.130& 0.0052 \\
55.4     & 10.29&   20&  17&  1.38&  1.047 $\pm$ 0.022& 0.510& 0.311& -0.119& 0.0047 \\
65~      & 11.13&   20&  17&  1.14&  1.134 $\pm$ 0.024& 0.509& 0.315& -0.102& 0.0040 \\
70~      & 11.55&   17&  17&  2.26&  1.136 $\pm$ 0.006& 0.509& 0.317& -0.095& 0.0037 \\
100      & 13.78&   81&  69&  1.22&  1.147 $\pm$ 0.007& 0.509& 0.324& -0.062& 0.0026 \\
100      & 13.78&   37&  34&  1.62&  1.125 $\pm$ 0.006& 0.509& 0.324& -0.062& 0.0026 \\
130      & 15.69&   11&  11&  0.34&  1.137 $\pm$ 0.030& 0.510& 0.328& -0.040& 0.0020 \\
140      & 16.28&   19&  19&  3.56&  1.124 $\pm$ 0.006& 0.510& 0.329& -0.034& 0.0019 \\
150      & 16.85&   31&  26&  0.60&  1.107 $\pm$ 0.021& 0.511& 0.330& -0.029& 0.0018 \\
170      & 17.93&   29&  23&  0.60&  1.154 $\pm$ 0.024& 0.512& 0.332& -0.020& 0.0016 \\
175      & 18.19&   15&  15&  2.82&  1.139 $\pm$ 0.007& 0.512& 0.333& -0.018& 0.0015 \\
190      & 18.95&   30&  24&  1.18&  1.098 $\pm$ 0.023& 0.513& 0.334& -0.012& 0.0014 \\
200      & 19.44&   90&  58&  1.44&  1.104 $\pm$ 0.007& 0.513& 0.334& -0.009& 0.0013 \\
200      & 19.44&  167& 156&  2.48&  1.080 $\pm$ 0.002& 0.513& 0.334& -0.009& 0.0013 \\
210      & 19.91&   29&  23&  0.72&  1.095 $\pm$ 0.025& 0.514& 0.335& -0.006& 0.0013 \\
230      & 20.84&   19&  16&  2.34&  1.064 $\pm$ 0.030& 0.515& 0.336& 0.000 & 0.0012 \\
250      & 21.72&   17&  14&  0.68&  1.167 $\pm$ 0.033& 0.516& 0.337& 0.005 & 0.0011 \\
292      & 23.47&   19&  19&  0.26&  1.186 $\pm$ 0.024& 0.518& 0.339& 0.014 & 0.0009 \\
293.5    & 23.53&  133&  97&  5.56&  1.067 $\pm$ 0.003& 0.518& 0.339& 0.015 & 0.0009 \\
497      & 30.6 &   15&  15&  0.56&  1.249 $\pm$ 0.027& 0.527& 0.344& 0.041 & 0.0005 \\
501.3    & 30.73&  124&  88&  2.32&  1.135 $\pm$ 0.003& 0.527& 0.344& 0.042 & 0.0005 \\
512      & 31.05&   24&  24&  0.48&  1.100 $\pm$ 0.015& 0.528& 0.344& 0.043 & 0.0005 \\
1064     & 44.74&   65&  26&  3.80&  1.032 $\pm$ 0.008& 0.544& 0.349& 0.070 & 0.0002 \\
1486     & 52.87&   63&  19&  3.48&  0.982 $\pm$ 0.014& 0.553& 0.351& 0.080 & 0.0002 \\
1497     & 53.07&   55&  41&  1.20&  1.027 $\pm$ 0.013& 0.553& 0.351& 0.080 & 0.0002 \\
1497     & 53.07&   27&  24&  3.82&  1.143 $\pm$ 0.009& 0.553& 0.351& 0.080 & 0.0002 \\
2048     & 62.06&   23&  23&  0.58&  1.094 $\pm$ 0.013& 0.562& 0.352& 0.087 & 0.0001 \\
2077     & 62.5 &   74&  37&  4.20&  1.000 $\pm$ 0.014& 0.563& 0.352& 0.088 & 0.0001 \\
2081     & 62.56&   49&  49&  13.6&  0.996 $\pm$ 0.001& 0.563& 0.352& 0.088 & 0.0001 \\ \hline

\end{tabular}
\end{center}
\end{table}

\begin{table}[cbth]

\begin{center}
\caption{Fitting results for $\bar pp$-interactions}

\begin{tabular}{|r|r|r|r|r|c|c|c|c|c|}
\hline
$P_{lab}$&$\sqrt{s}$ & $N$ & $N_F$ & $\chi^2/NoF$ &  A   & $R$  & $d$ & $\rho$ & $a_1$ \\
(GeV/c)  &  (GeV)    &     &       &              &      & (fm) & (fm)&        & (fm$^2$)  \\ \hline
8.00 &   4.111& 41&  41&  4.42& 0.833 $\pm$ 0.0008& 0.895& 0.238& -0.320& -0.005916 \\
9.71 &   4.483& 10&  5 &  1.90& 0.618 $\pm$ 0.0427& 0.855& 0.249& -0.320& -0.004975 \\
10.1 &   4.564& 35&  31&  5.04& 0.880 $\pm$ 0.0042& 0.847& 0.250& -0.320& -0.004801 \\
10.4 &   4.625& 61&  61&  3.68& 0.881 $\pm$ 0.0008& 0.842& 0.252& -0.320& -0.004675 \\
16.0 &   5.647& 23&  23&  2.38& 0.834 $\pm$ 0.0023& 0.766& 0.272& -0.265& -0.003136 \\
22.4 &   6.626& 44&  44&  1.28& 0.910 $\pm$ 0.0047& 0.716& 0.285& -0.206& -0.002277 \\
25.2 &   7.012& 33&  33&  2.00& 0.885 $\pm$ 0.0040& 0.701& 0.289& -0.187& -0.002034 \\
32.1 &   7.882& 52&  52&  4.98& 0.893 $\pm$ 0.0035& 0.673& 0.297& -0.152& -0.001609 \\
40.1 &   8.785& 33&  33&  1.08& 0.888 $\pm$ 0.0068& 0.650& 0.303& -0.122& -0.001296 \\
191.1&   19.1 & 27&  23&  2.28& 0.974 $\pm$ 0.0182& 0.562& 0.334&  0.016& -2.770E-4 \\
510.2&   31.0 & 22&  22&  1.44& 1.076 $\pm$ 0.0080& 0.550& 0.344&  0.062& -1.041E-4 \\
     &   53.0 & 51&  44&  3.94& 1.015 $\pm$ 0.0053& 0.560& 0.351&  0.092& -3.560E-5 \\
     &   62.3 & 23&  23&  1.12& 1.068 $\pm$ 0.0063& 0.566& 0.352&  0.099& -2.576E-5 \\
     &   540  & 36&  36&  0.84& 1.090 $\pm$ 0.0062& 0.720& 0.359&  0.131& -3.429E-7 \\
     &   546  &121& 121& 17.24& 1.143 $\pm$ 0.0018& 0.721& 0.359&  0.131& -3.354E-7 \\
     &   630  & 19&  15&  7.66& 1.124 $\pm$ 0.0147& 0.734& 0.359&  0.131& -2.520E-7 \\
     &  1800  & 51&  51&  1.04& 1.085 $\pm$ 0.0028& 0.830& 0.360&  0.134& -3.086E-8 \\
     &  1960  & 17&  17&  0.36& 1.095 $\pm$ 0.0323& 0.838& 0.360&  0.134& -2.603E-8 \\ \hline
\end{tabular}
\end{center}
\end{table}

\noindent{\bf Appendix D: Attempt to fit the COSY data on $pp$-interactions -- D. Albers et al.,Phys. Rev. Lett. {\bf 78} (1997) 1652.}\\
We represent the elastic scattering amplitude in the energy range, 1 -- 3 GeV/c as
$$
F_{pp}=f(\theta)+f(\pi -\theta),
$$
where $f(\theta)$ is given by Eq. \ref{Eq11}. Below the lines are results of a fit with 4 free parameters --
$A$, $R$, $d$ and $a_1$.

\begin{figure}[cbth]
\includegraphics[width=75mm,height=70mm,clip]{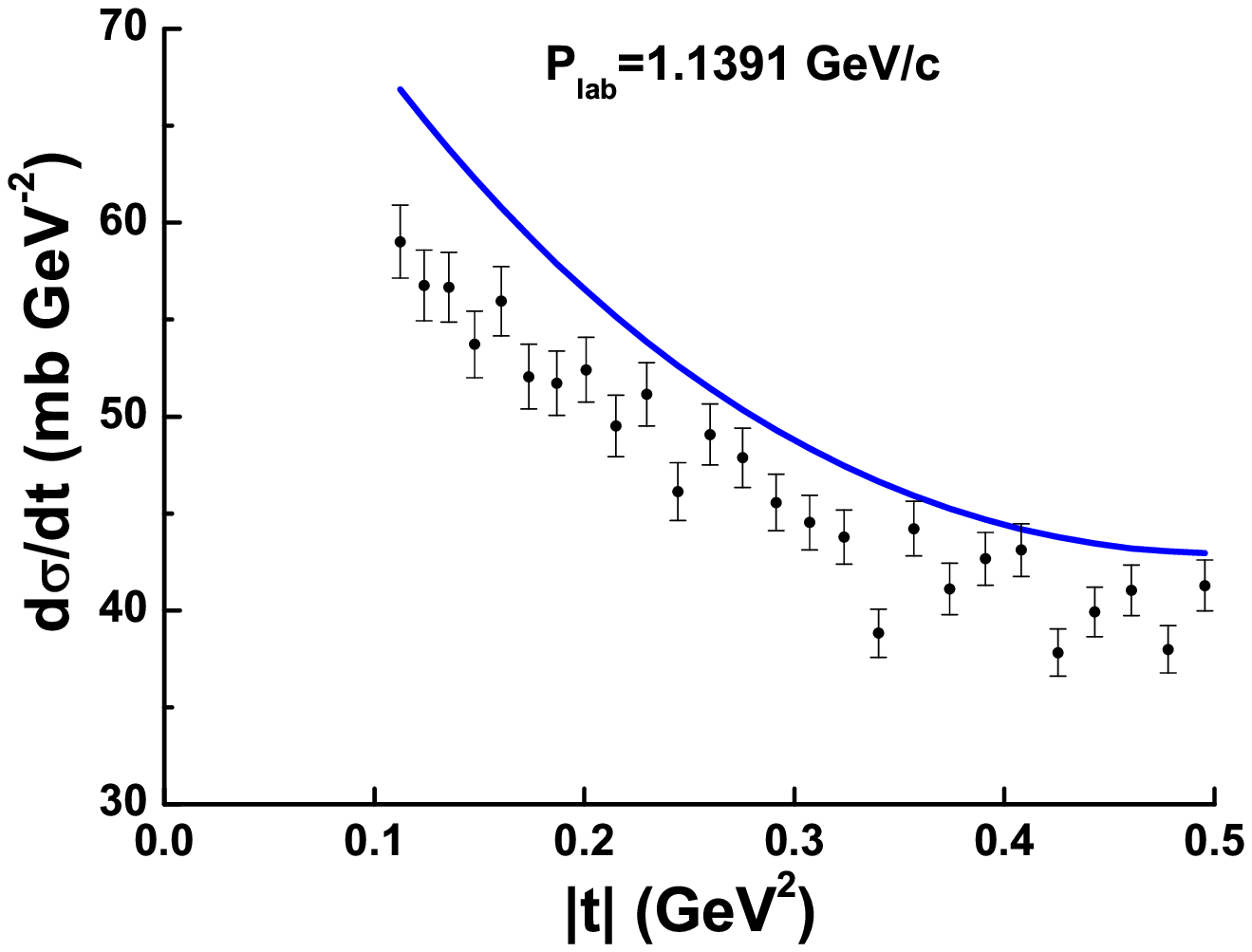}\hspace{5mm}\includegraphics[width=75mm,height=70mm,clip]{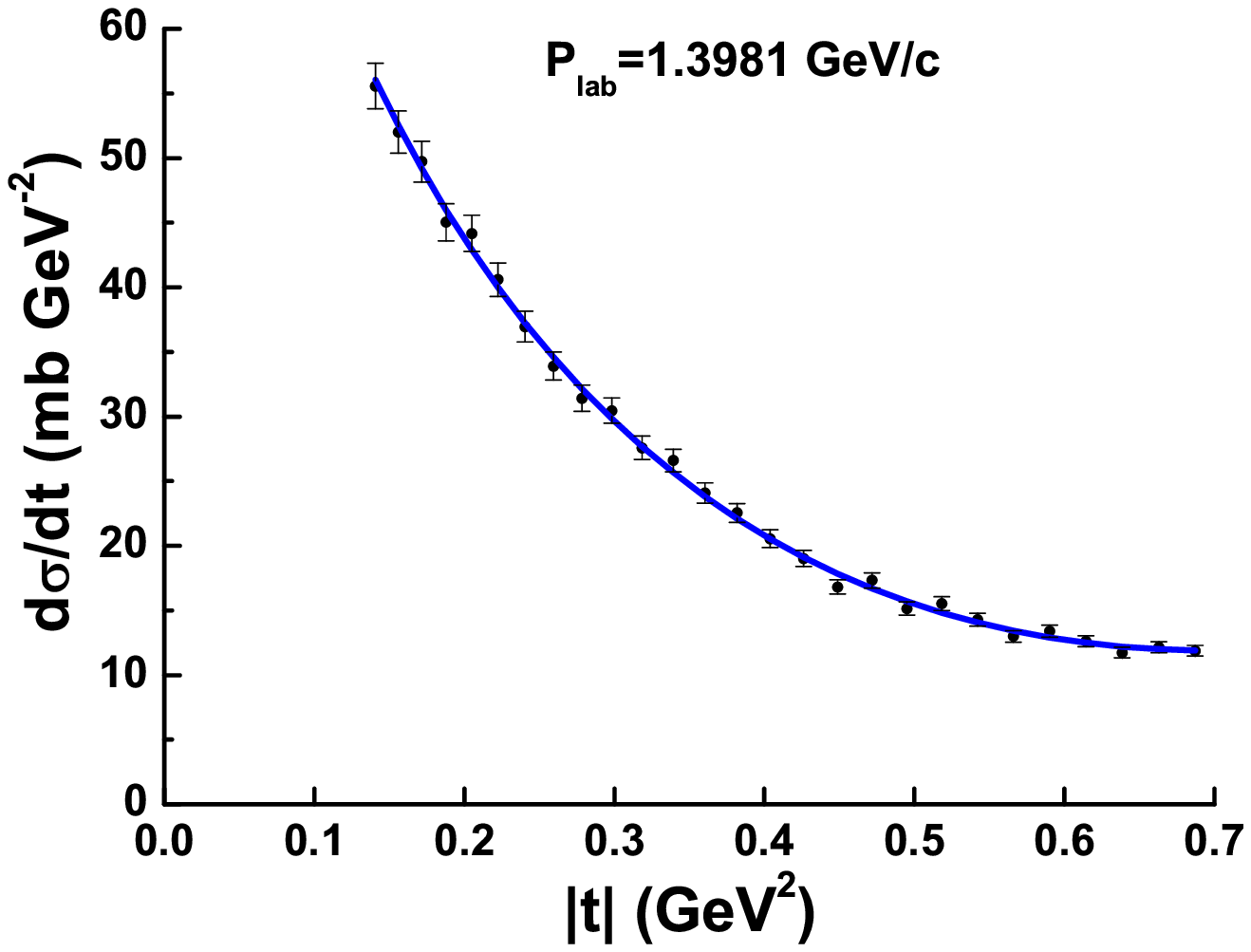}
\begin{minipage}{75mm}
{
\caption{The points are the experimental data.}
}
\end{minipage}
\hspace{5mm}
\begin{minipage}{75mm}
{
\caption{The points are the experimental data.}
}
\end{minipage}
\includegraphics[width=75mm,height=70mm,clip]{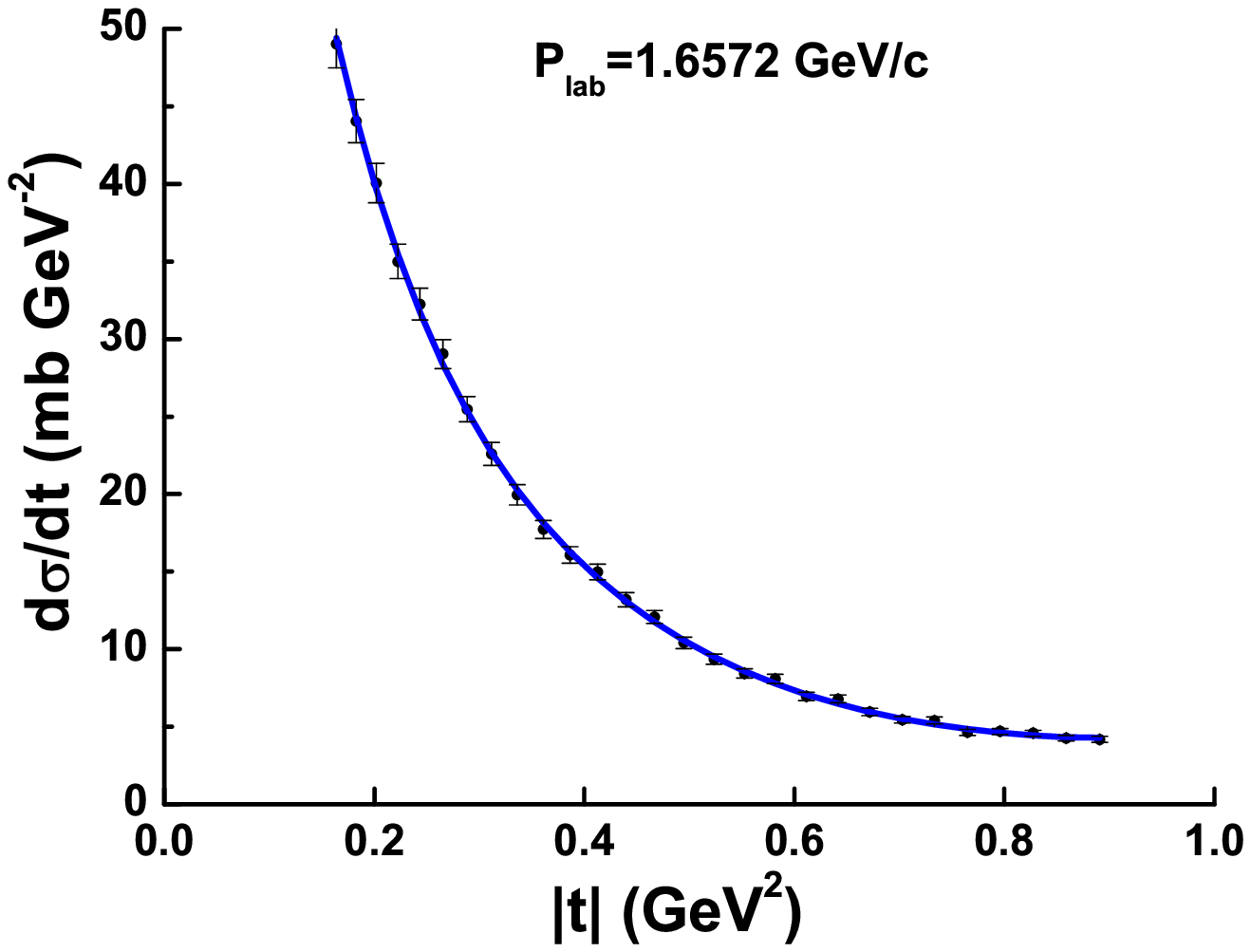}\hspace{5mm}\includegraphics[width=75mm,height=70mm,clip]{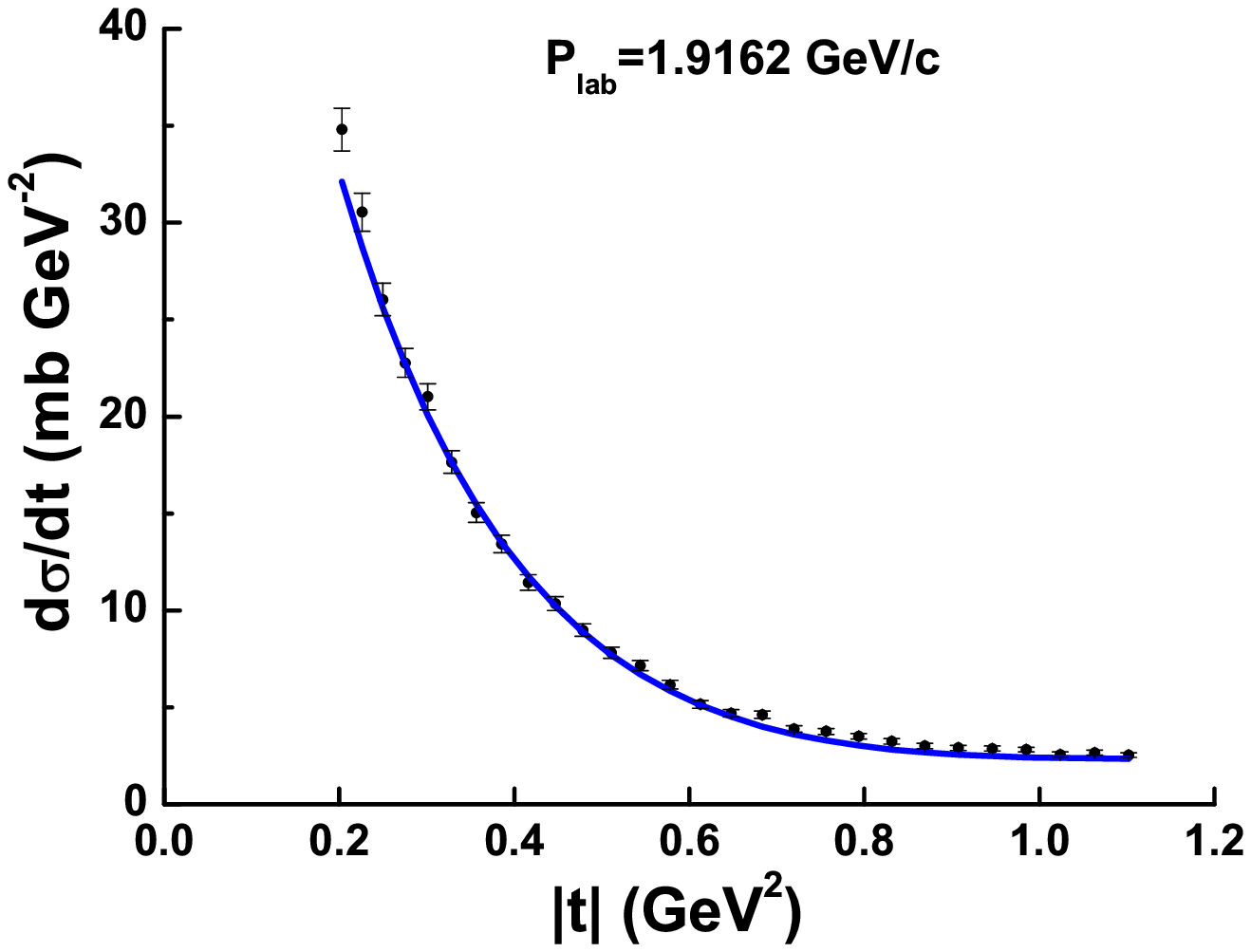}
\begin{minipage}{75mm}
{
\caption{The points are the experimental data.}
}
\end{minipage}
\hspace{5mm}
\begin{minipage}{75mm}
{
\caption{The points are the experimental data.}
}
\end{minipage}

\end{figure}

\begin{figure}[cbth]

\includegraphics[width=75mm,height=66mm,clip]{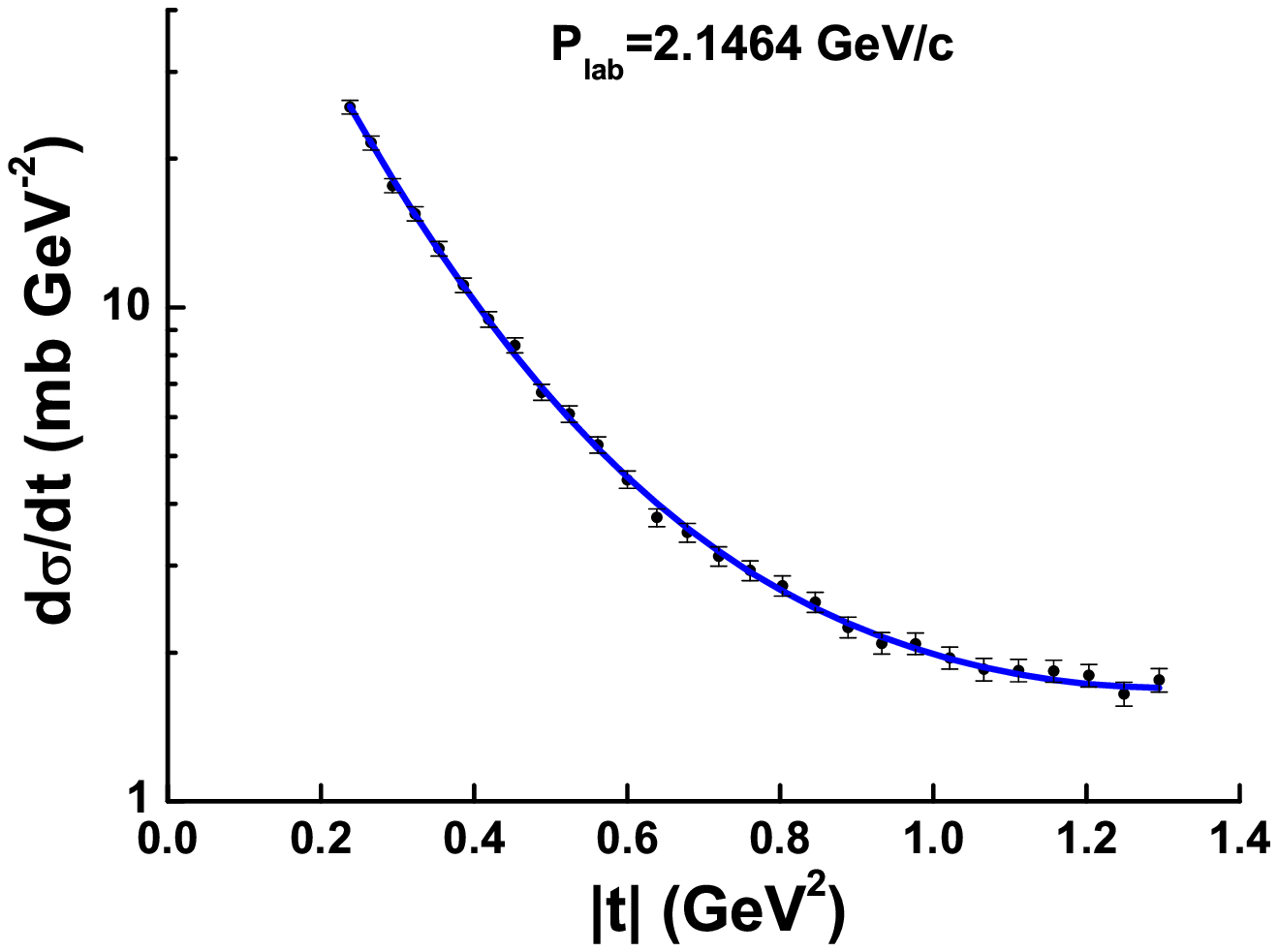}\hspace{5mm}\includegraphics[width=75mm,height=66mm,clip]{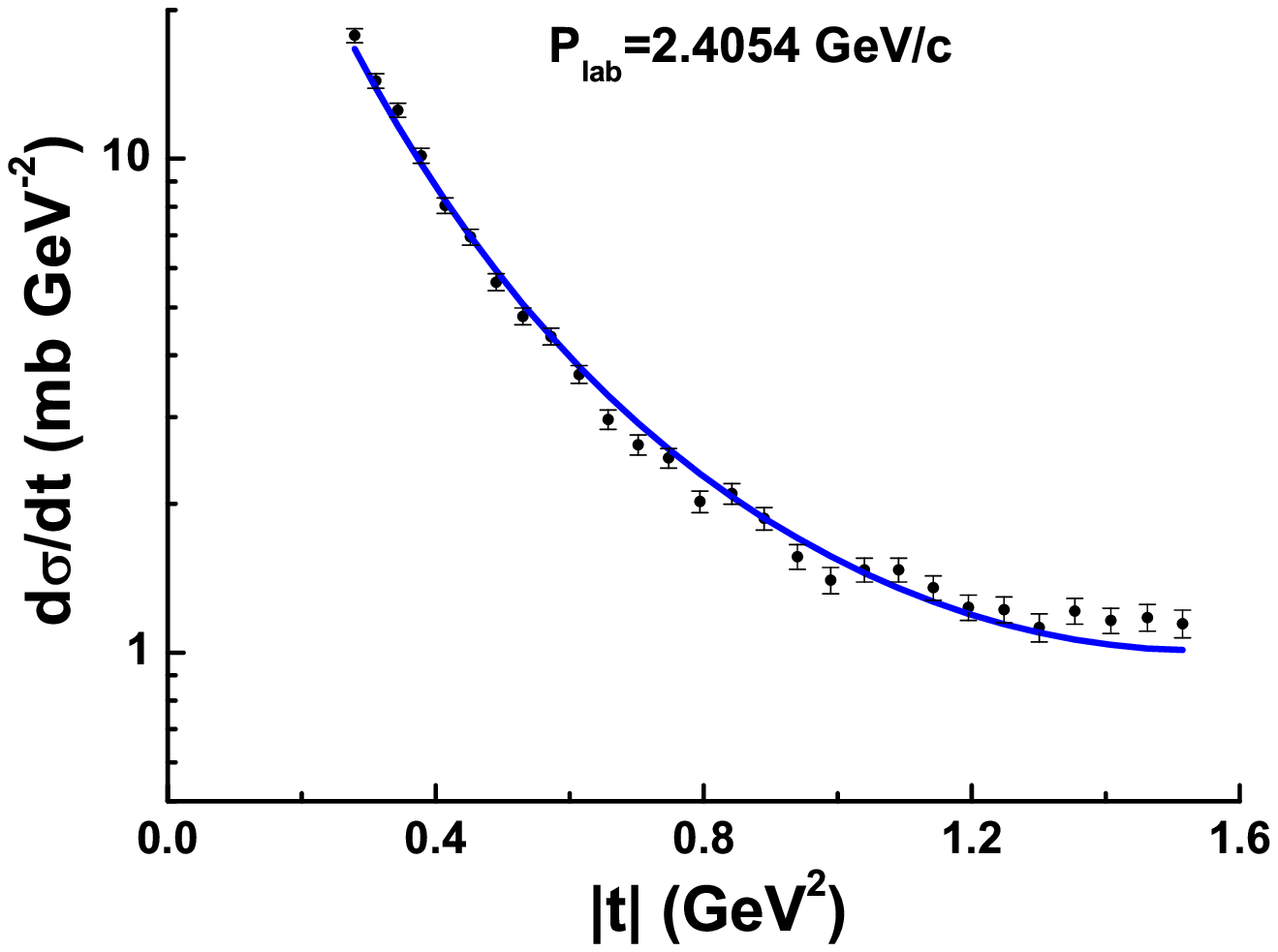}
\begin{minipage}{75mm}
{
\caption{The points are the experimental data.}
}
\end{minipage}
\hspace{5mm}
\begin{minipage}{75mm}
{
\caption{The points are the experimental data.}
}
\end{minipage}
\includegraphics[width=75mm,height=66mm,clip]{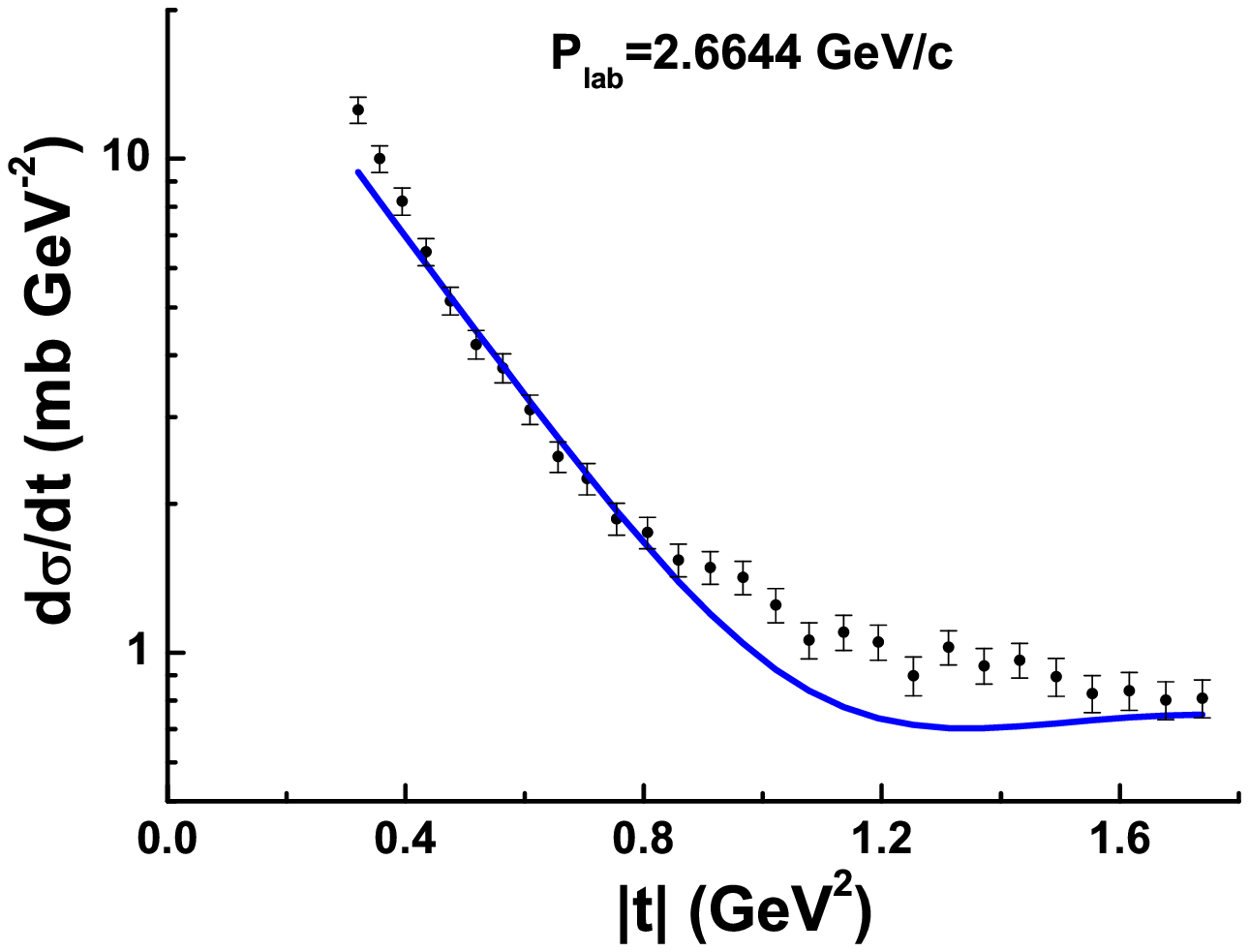}\hspace{5mm}\includegraphics[width=75mm,height=66mm,clip]{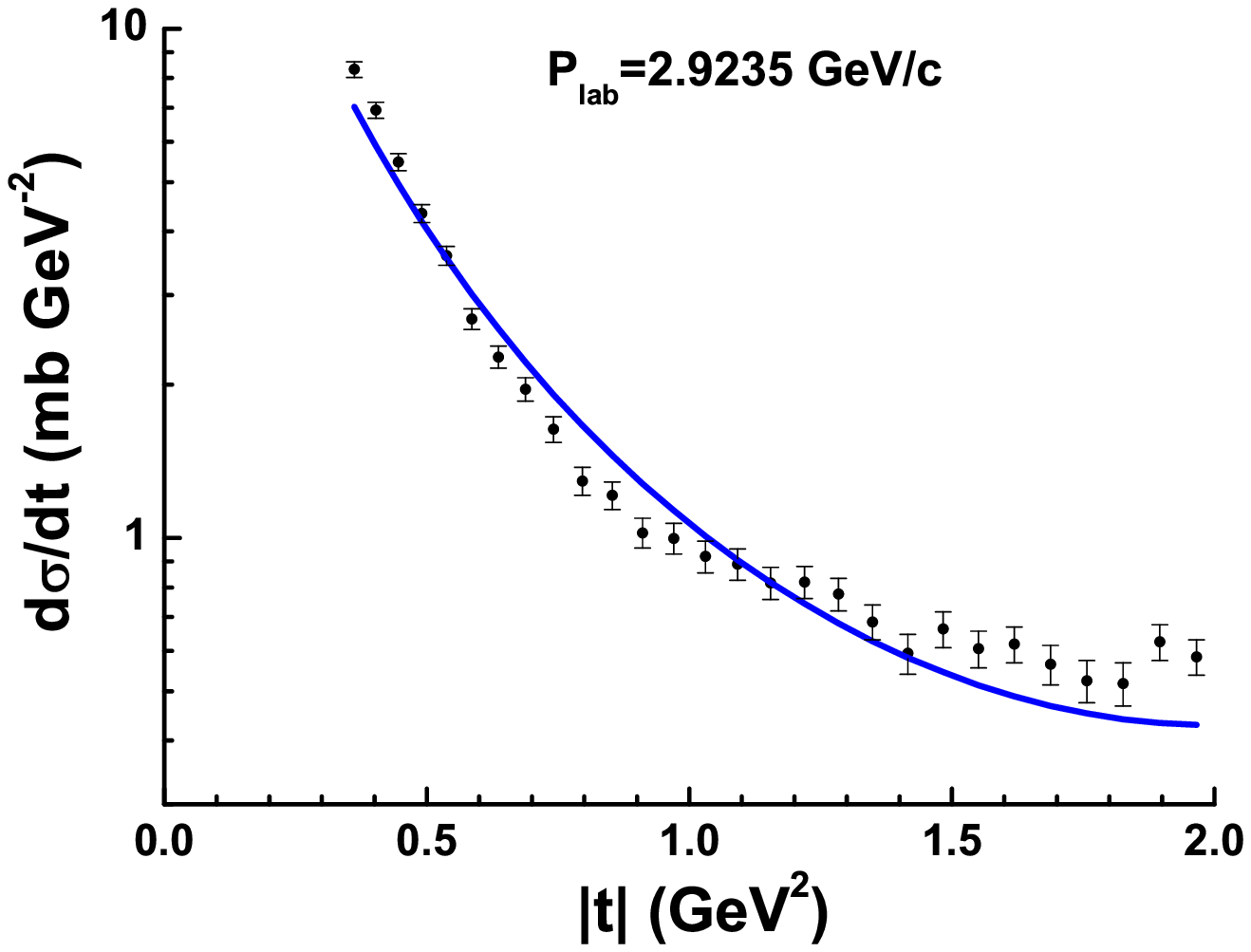}
\begin{minipage}{75mm}
{
\caption{The points are the experimental data.}
}
\end{minipage}
\hspace{5mm}
\begin{minipage}{75mm}
{
\caption{The points are the experimental data.}
}
\end{minipage}
\includegraphics[width=75mm,height=66mm,clip]{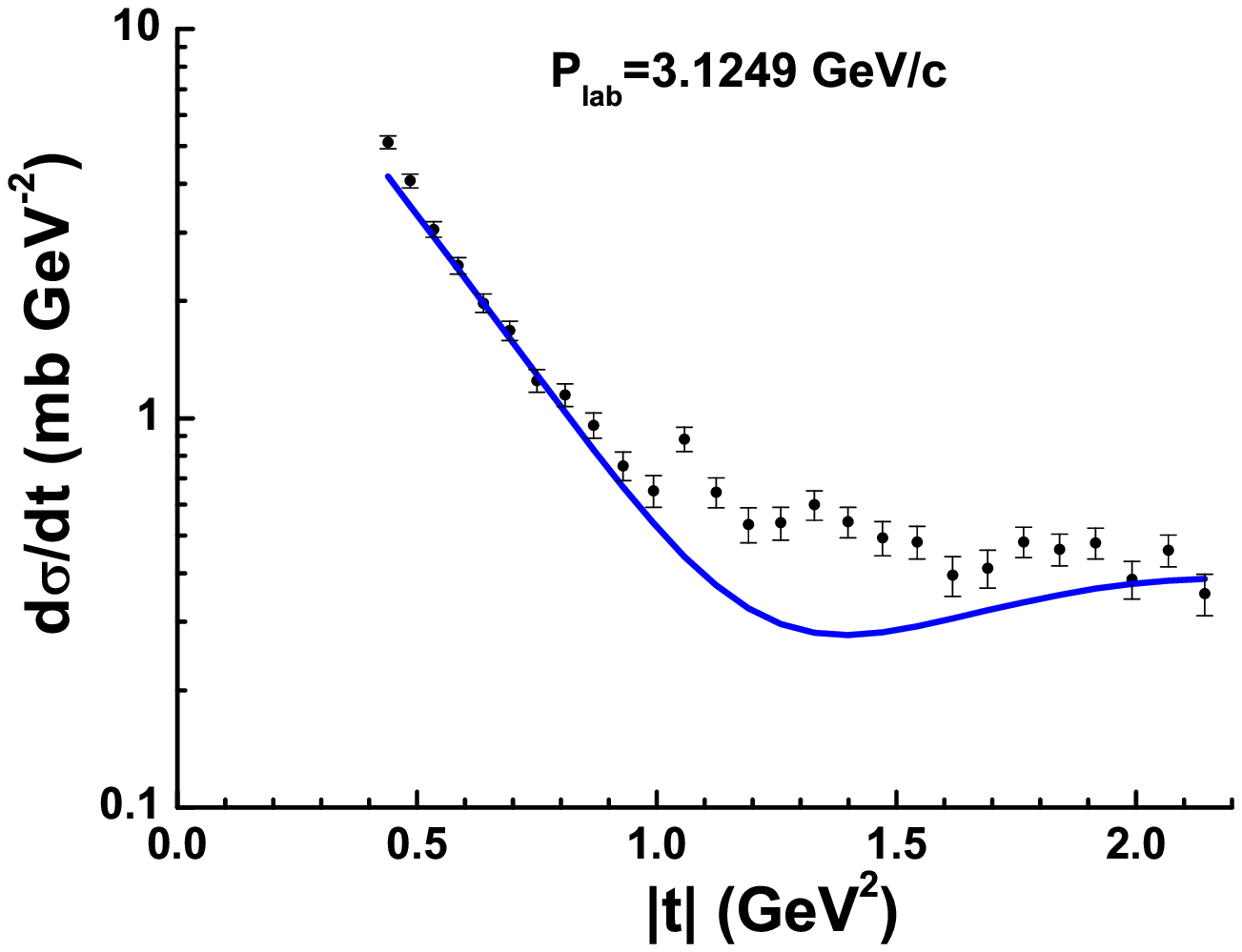}\hspace{5mm}\includegraphics[width=75mm,height=66mm,clip]{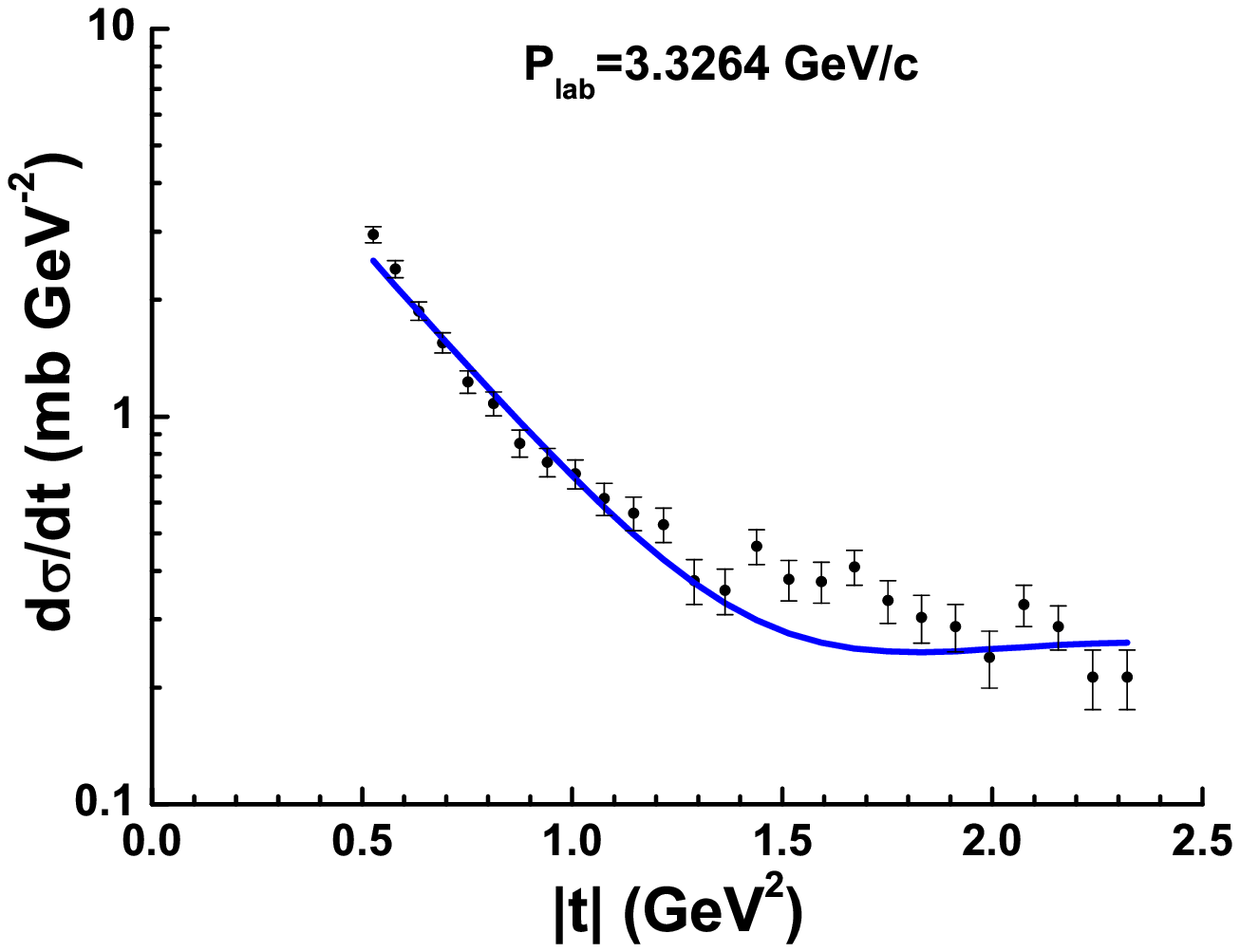}
\begin{minipage}{75mm}
{
\caption{The points are the experimental data.}
}
\end{minipage}
\hspace{5mm}
\begin{minipage}{75mm}
{
\caption{The points are the experimental data.}
}
\end{minipage}
\end{figure}

\end{document}